\documentclass[11pt,a4paper]{oxarticle}
\pdfoutput=1

\usepackage{titlesec}
\titlespacing*{\subsubsection}
{0pt}{5pt plus 1pt minus 1pt}{5pt plus 1pt}

\usepackage[usenames, dvipsnames]{xcolor}
\usepackage{graphicx, float, array, xspace, amscd, amsthm, amssymb, latexsym, longtable, bbm, fancyhdr, slashed, pifont, accents}
\usepackage[titletoc]{appendix}
\usepackage[letterpaper, body={6.5in, 9.25in}, top=1.0in, left=1.5in]{geometry}
\usepackage[bbgreekl]{mathbbol}
\usepackage[utf8]{inputenc}
\usepackage[sort&compress, comma, square, numbers]{natbib}
\usepackage{braket}
\usepackage{multirow}
\usepackage{subcaption}
\usepackage{bm}
\usepackage{mathtools}
\usepackage{tikz}
\usepackage{tikz-cd}
\usepackage[pdfpagelabels=false]{hyperref}

\usepackage{thmtools, thm-restate}

\usepackage{booktabs}
\usepackage{pgfplotstable}
\usepackage{comment}
\usepackage{adjustbox}

\usepackage{tabularx, caption, boldline}
\usepackage{cellspace}
\DeclareSymbolFontAlphabet{\mathbb}{AMSb}
\DeclareSymbolFontAlphabet{\mathbbl}{bbold}

\let\SS=\S % save \S before it is redefined

\renewcommand{\a}{\alpha}
\renewcommand{\b}{\beta}
\newcommand{\g}{\gamma}
\renewcommand{\d}{\delta}\newcommand{\D}{\Delta}
\newcommand{\e}{\epsilon}

\newcommand{\m}{\mu}

\newcommand{\p}{\pi}

\renewcommand{\S}{\Sigma}

\renewcommand{\u}{\upsilon}

\DeclareFontFamily{OT1}{pzc}{}
\DeclareFontShape{OT1}{pzc}{m}{it}{<-> s * [1.200] pzcmi7t}{}
\DeclareMathAlphabet{\mathpzc}{OT1}{pzc}{m}{it}

\newcommand{\cA}{\mathcal{A}}
\newcommand{\cB}{\mathcal{B}}
\newcommand{\cC}{\mathcal{C}}
\newcommand{\cD}{\mathcal{D}}

\newcommand{\cF}{\mathcal{F}}

\newcommand{\cL}{\mathcal{L}}
\newcommand{\cM}{\mathcal{M}}

\newcommand{\cO}{\mathcal{O}}
\newcommand{\cP}{\mathcal{P}}

\newcommand{\cS}{\mathcal{S}}

\newcommand{\cW}{\mathcal{W}}\newcommand{\ccW}{\mathpzc W}
\newcommand{\cX}{\mathcal{X}}\newcommand{\ccX}{\mathpzc X}
\newcommand{\cY}{\mathcal{Y}}
\newcommand{\cZ}{\mathcal{Z}}\newcommand{\ccZ}{\mathpzc Z}

\DeclareFontFamily{U}{bbold}{}
\DeclareFontShape{U}{bbold}{m}{n}
{  <-5.5> s*[1.05] bbold5
	<5.5-6.5> s*[1.05] bbold6
	<6.5-7.5> s*[1.05] bbold7
	<7.5-8.5> s*[1.05] bbold8
	<8.5-9.5> s*[1.05] bbold9
	<9.5-11.5> s*[1.05] bbold10
	<11.5-16> s*[1.05] bbold12
	<16-> s*[1.05] bbold17
}{}

\newcommand{\IA}{\mathbbl{A}}

\newcommand{\IC}{\mathbbl{C}}

\newcommand{\II}{\mathbbl{I}}

\newcommand{\IP}{\mathbbl{P}}

\newcommand{\IR}{\mathbbl{R}}

\newcommand{\IT}{\mathbbl{T}}

\newcommand{\IW}{\mathbbl{W}}

\newcommand{\IZ}{\mathbbl{Z}}

\newcommand{\IDelta}{\mathbbl{\Delta}}
\newcommand{\Imu}{\mathbbl{\bbmu}}
\newcommand{\Inu}{\mathbbl{\bbnu}}

\font\elevenrmfromseventeenrm = cmr17 at 11pt
\newcommand{\inbar}{\vrule height6.9pt depth-0.2pt width0.35pt}
\newcommand{\zero}{\hbox{{\elevenrmfromseventeenrm 0}\kern-3.5pt\inbar\kern1pt\inbar\kern2pt}}

\font\eightrmfromseventeenrm = cmr17 at 8pt
\newcommand{\ssinbar}{\vrule height5pt depth-0.1pt width0.3pt}
\newcommand{\sszero}{\hbox{{\eightrmfromseventeenrm 0}\kern-2.53pt\ssinbar\kern0.7pt\ssinbar\kern2pt}}

\newcommand{\fp}{\mathfrak{p}}

\newcommand{\fS}{\mathfrak{S}}

\font\eightrm=cmr8 at 8pt
\font\csc=cmcsc10

\newcommand{\beq}{\begin{equation}}
\newcommand{\eeq}{\end{equation}}
\newcommand{\beqnn}{\begin{equation*}}
\newcommand{\eeqnn}{\end{equation*}}
\newcommand{\bea}{\begin{eqnarray}}
\newcommand{\eea}{\end{eqnarray}}
\newcommand{\bean}{\begin{eqnarray*}}
	\newcommand{\eean}{\end{eqnarray*}}

\newcommand{\fref}[1]{Figure~\ref{#1}}
\newcommand{\tref}[1]{Table~\ref{#1}}
\newcommand{\sref}[1]{\SS\ref{#1}}

\newcommand{\nn}{\nonumber}

\newcommand{\defineas}{\buildrel\rm def\over =}

\newcommand{\ee}{\text{e}}
\newcommand{\ii}{\text{i}}
\newcommand{\dd}{\text{d}}

\newcommand{\place}[3]{\vbox to0pt{\kern-\parskip\kern-7pt
		\kern-#2truein\hbox{\kern#1truein #3}
		\vss}\nointerlineskip}

\DeclareFontFamily{U}{wncy}{}
\DeclareFontShape{U}{wncy}{m}{n}{<->wncyr10}{}
\DeclareSymbolFont{mcy}{U}{wncy}{m}{n}
\DeclareMathSymbol{\sha}{\mathord}{mcy}{"58}

\newcommand{\capt}[3]{\parbox{#1}{\renewcommand{\baselinestretch}{1.0}
		\caption{\label{#2}\small\it #3}}}

\newcommand{\del}{\partial}

\newcommand{\SL}{\text{SL}}
\newcommand{\Sp}{\text{Sp}}

\newcommand{\2}{\mathbf{2}}
\newcommand{\1}{\mathbf{1}}

\newcommand{\cys}{Calabi-Yau manifolds\xspace}

\newcommand{\cicy}[2]{\begin{matrix} #1\end{matrix}\!\left[\begin{matrix}#2 \end{matrix}\right]}

\renewcommand{\Re}{\text{Re~}}

\newcommand{\+}{\phantom{-}}

\renewcommand{\=}{\;=\;}

\renewcommand{\Re}{\text{Re}}
\newcommand{\wt}[1]{\widetilde{#1}}
\newcommand{\wh}[1]{\widehat{#1}}

\newcommand{\me}{\text{e}}
\newcommand{\bme}{\text{\textbf{e}}}

\newcommand{\HV}{\text{HV}}%
\newcommand{\MHV}{\text{H}\Lambda}

\newcommand\circbm[1]{\accentset{\;\circ}{\bm #1}}

\makeatletter
\g@addto@macro\bfseries{\boldmath}

\def\blindfootnote{\xdef\@thefnmark{}\@footnotetext}
\makeatother

\newenvironment{smallarray}[1]
 {\null\,\vcenter\bgroup\scriptsize
  \renewcommand{\arraystretch}{0.7}%
  \arraycolsep=.13885em
  \hbox\bgroup$\array{@{}#1@{}}}
 {\endarray$\egroup\egroup\,\null}
 
\pgfplotsset{compat=1.16}
\renewcommand{\baselinestretch}{1.1}
\numberwithin{equation}{section}
\begin{document}
	\proofmodefalse
	
	%%%%%%%%%%%%%%%%%%%%%%%%%%%%%%%%%%%%%%%%%%%
	%%%         Title Page
	%%%%%%%%%%%%%%%%%%%%%%%%%%%%%%%%%%%%%%%%%%%
	
	\thispagestyle{empty}      
	\begin{center}
		\null\vskip0.5in
		{\Huge Mirror Symmetry for Five-Parameter Hulek-Verrill Manifolds \\[12pt]}
\vskip1cm
{\csc Philip Candelas${}^1$, Xenia de la Ossa${}^2$, Pyry Kuusela${}^3$,\\
and\\
Joseph McGovern${}^4$\\[1cm]}
\blindfootnote{\null\hskip-10pt$^1\,$candelas@maths.ox.ac.uk \hfill $^2\,$delaossa@maths.ox.ac.uk \hfill 
$^3\,$pyry.r.kuusela@gmail.com\hfill
$^4\,$mcgovernjv@gmail.com\kern10pt}
{\it Mathematical Institute\\
University of Oxford\\
Andrew Wiles Building\\
Radcliffe Observatory Quarter\\
Oxford, OX2 6GG, UK\\}
\vfill
{\bf Abstract}
\end{center}
\vskip-5pt
\begin{minipage}{\textwidth}
\baselineskip=15pt
\noindent 
We study the mirrors of five-parameter Calabi-Yau threefolds first studied by Hulek and Verrill in the context of observed modular behaviour of the zeta functions for Calabi-Yau manifolds. Toric geometry allows for a simple explicit construction of these mirrors, which turn out to be familiar manifolds. These are elliptically fibred in multiple ways. By studying the singular fibres, we are able to identify the rational curves of low degree on the mirror manifolds. This verifies the mirror symmetry prediction obtained by studying the mirror map near large complex structure points. We undertake also an extensive study of the periods of the Hulek-Verrill manifolds and their monodromies. On the mirror, we compute the genus-zero and -one instanton numbers, which are labelled by 5 indices, as $h^{1,1}\!=5$. There is an obvious permutation symmetry on these indices, but in addition there is a surprising repetition of values. We trace this back to an $S_{6}$ symmetry made manifest by certain constructions of the complex structure moduli space of the Hulek-Verrill manifold. Among other consequences, we see in this way that the moduli space has six large complex structure limits. It is the freedom to expand the prepotential about any one of these points that leads to this symmetry in the instanton numbers. An intriguing fact is that the group that acts on the instanton numbers is larger than $S_6$ and is in fact an infinite hyperbolic Coxeter group, that we study. The group orbits have a `web' structure, and with certain qualifications the instanton numbers are only nonzero if they belong to what we term `positive webs'. This structure has consequences for instanton numbers at all genera.
\vspace*{10pt}
\end{minipage}
\clearpage
\newpage
{\baselineskip=16pt
\tableofcontents}
\newpage
\section{Introduction}
\vskip-10pt
\subsection{Preamble}
\vskip-10pt
In this paper, we study mirror symmetry for a family of Calabi-Yau manifolds associated to the root lattice $A_4$, a family first investigated in relation to the modularity of its zeta-function by Hulek and Verrill \cite{hulek2005}. Apart from the modular properties, these manifolds are of considerable interest due to their high degree of symmetry: the five parameter complex structure moduli space of these manifolds has an $S_6$ symmetry, which leads to an infinite group of symmetries among the instanton numbers, realised in terms of a Coxeter group. The symmetries also allow many simplifications which make computations that are usually too computationally expensive for multiparameter manifolds accessible. 

\textit{Singular Hulek-Verrill varieties} comprise a five-parameter family, parametrised by \begin{align} \nonumber
    \phi \= (\varphi^{0},\varphi^1,\varphi^2,\varphi^3,\varphi^4,\varphi^5)
\end{align} 
and denoted\footnote{We often do not display the parameter $\phi$ explicitly.} ~$\wh \HV_{\phi}$. They are embedded in the projective torus $\IT^4 = \IP^4 \setminus \{X_1 \cdots X_5 = 0 \}$ as the vanishing loci of
\begin{align} \label{eq:singular_HV}
\left(X_1 + X_2 + X_3 + X_4 + X_5 \right) \left(\frac{\varphi^1}{X_1}+\frac{\varphi^2}{X_2}+\frac{\varphi^3}{X_3}+\frac{\varphi^4}{X_4}+\frac{\varphi^5}{X_5}\right) \= \varphi^0~.
\end{align} 
These varieties admit toric compactifications, which we will review briefly in \sref{sect:Toric_Geometry}. Of particular interest are small projective resolutions $\HV$ of $\wh \HV$, which have smooth projective Calabi-Yau models~\cite{hulek2005}. We concentrate mostly on analysing these, and call them simply \textit{Hulek-Verrill manifolds}.

A reformulation of \eqref{eq:singular_HV} turns out to be very useful, whereby these manifolds are realised as a toric compactification of the locus $P_{1}=P_{2}=0$, the intersection of two polynomials
\begin{align}\label{eq:TwoPolynomials} 
P^1(\mathbf{X}) \= \sum_{\mu=0}^5 X_\mu~, \hskip30pt P^2(\mathbf{X};\phi) \= \sum_{\mu=0}^5 \frac{\varphi^\mu}{X_\mu}~,
\end{align}
on a torus $\IT^5$. This is seen by eliminating $X_{0}$, which immediately returns us to \eqref{eq:singular_HV}. We denote these manifolds by $\HV_{(\varphi^0, \dots, \varphi^5)}$, or more compactly by $\HV$.

These manifolds have received attention in the physics literature, since the periods of these manifolds (and their analogues in each dimension) are related to the banana Feynman graphs \cite{Bonisch:2020qmm}. The particular manifolds $\HV_{(1,1,1,1,1,1/\varphi)}$ and their quotients exhibit, for certain values of $\varphi$, rank-two attractor points with interesting number theoretic properties. These attractor varieties were identified in \cite{Candelas:2019llw}. 

The mirror-symmetric counterpart to the work of \cite{Candelas:2019llw} involves a IIA setup. In \cite{Candelas:2021mwz}, nonperturbative solutions were given to the attractor equations which involved instanton numbers, or Gromov-Witten invariants, giving a hint of microstate counting. This motivates us to study the geometry of the mirror Hulek-Verrill manifold focusing especially on aspects related to counting microstates of D4-D2-D0 brane systems on the manifold. 

In studying the periods of $\HV$ we are naturally led to consider integrals of products of Bessel functions, similar to those considered in \cite{BroadhurstElliptic,Acres:2021sss}. We find additional motivation for the present work in the connection between the manifolds $\HV$ and this topic.

While this paper was in preparation we received \cite{Bonisch:2020qmm}, which has overlap with the present work.

\subsection{Outline of the paper}
\vskip-10pt
The analysis of the Hulek-Verrill manifolds presented in this paper occasionally becomes somewhat involved. This being so, we give below a brief overview of the contents and main results of each section. Where possible, we strive to keep different sections relatively independent.

\subsubsection*{A comment on indices}
\vskip-5pt
We adopt specific index conventions in various sections of the paper. While these conventions are strictly followed in their respective sections, they are not consistently applied throughout the paper. These are set out in \tref{tab:indices}.
\begin{table}[H]
	\renewcommand{\arraystretch}{1.49}
	\centering
	\begin{tabularx}{16.5cm}{|>{\hsize=.1\hsize\linewidth=\hsize}X|
			>{\hsize=0.9\hsize\linewidth=\hsize}X|}
		\hline
		\textbf{Section} & \hfil \textbf{Index Convention} \\[3pt]
		\hline	\hline
		\hfil \sref{sect:Toric_Geometry} & Varies by subsection.    \\[4pt] \hline	
		\hfil \sref{sect:Periods} & Greek indices run from 0 to 5. Latin indices run from 1 to 5.  \\[4pt] \hline	
		\hfil \sref{sect:Mirror_Map} &  Greek indices run from 0 to 5. Latin indices run from 1 to 5.\\[4pt] \hline
		\hfil \sref{sect:Curve_Counting} & Latin indices run from 0 to 4. Distinct indices are understood to take distinct values.\\[4pt] \hline
	\end{tabularx}
	\vskip10pt
	\capt{6in}{tab:indices}{Index conventions in each section. }
	\vskip10pt
\end{table}
\subsubsection*{Toric geometry of mirror Hulek-Verrill manifolds}
\vskip-5pt
In \sref{sect:Toric_Geometry}, we briefly review the toric construction of the singular Hulek-Verrill manifolds $\wh \HV$ as first discussed in \cite{hulek2005}. Then we proceed to find a toric description of its small resolution. We use the method of Batyrev and Borisov \cite{Batyrev:1994pg,borisov1993towards} to find the toric description of the mirror Hulek-Verrill manifolds $\MHV$. Somewhat surprisingly, these mirror manifolds turn out to be familiar spaces \cite{Candelas:2008wb,Candelas:2015amz}, given by the complete intersection
\begin{align}
\cicy{\IP^1\\\IP^1\\\IP^1\\\IP^1\\\IP^1}{1 & 1\\1 & 1\\1 & 1\\ 1 & 1\\ 1 & 1}~.
\end{align}
Parenthetically, we note that this manifold is itself a remarkable split \cite{Candelas:1987kf} of the tetraquadric,
\begin{align}
\cicy{\IP^1\\\IP^1\\\IP^1\\\IP^1}{2 \\2 \\2 \\2 }~.
\end{align}
Subfamilies exist that admit a $\IZ_5 {\times} \IZ_2 {\times} \IZ_2$ symmetry, or a subgroup thereof. The symmetry has a simple description: denoting the coordinates in each of these projective spaces by $Y_{i,0}$ and $Y_{i,1}$, the symmetries act for all $i$ as
\begin{align}
S \, : \, Y_{i,a} \mapsto Y_{i+1,a}~, \qquad U \, : \, Y_{i,a} \mapsto (-1)^{a}Y_{i,a}~, \qquad V \, : \, Y_{i,0} \leftrightarrow Y_{i,1}~.
\end{align}
We write the most general expressions for the polynomials defining manifolds invariant under these symmetries. In particular, the manifold invariant under $\IZ_5 {\times} \IZ_2 {\times} \IZ_2$ is given as the simultaneous vanishing locus of
\begin{align}
\begin{split}
Q^1 &\= \frac{A_0}{5} \, m_{00000}+A_{1} \, m_{11000}+A_{2} \, m_{10100}+A_{3} \, m_{11110}~, \\[5pt]
Q^2 &\= \frac{A_0}{5} \, m_{11111}+A_{1} \, m_{11100}+A_{2} \, m_{11010}+A_{3} \, m_{10010}~,
\end{split}
\end{align}
where $m_{abcde}$ are $\IZ_5$ invariant multidegree $(1,1,1,1,1)$ polynomials:
\begin{align}
m_{abcde} \= \sum_{i\in\IZ_{5}} Y_{i,a} Y_{i+1,b} Y_{i+2,c} Y_{i+3,d} Y_{i+4,e}~.
\end{align} 
It will turn out to be occasionally useful to consider the singular mirror Hulek-Verrill manifolds $\wh \MHV$, which can be obtained by using the contraction procedure of \cite{Candelas:1987kf}, or equivalently by blowing down 24 degree-1 lines which are parallel to one of the $\IP^1$'s. In this way, we obtain a family of singular varieties, which are birational to mirrors of the singular Hulek-Verrill manifolds $\wh \HV$ found by using Batyrev's~method~\cite{Batyrev:1994hm}.

\subsubsection*{Periods of the five-parameter family}
\vskip-5pt
Section \ref{sect:Periods} deals with the periods of $\HV$, which describe the variation of the Hodge structure as a function of moduli space coordinates. We study the five-parameter family \eqref{eq:complete_intersection_HV}. The overall scaling of coordinates $\varphi^\mu$ does not affect the vanishing locus, and thus we can identify the moduli space\footnote{Note that two points in $\IP^5$ can correspond to biholomorphic manifolds. There exists a `fundamental domain' in $\IP^5$, where the points are in one-to-one correspondence with distinct biholomorphism classes. This issue does not affect our studies.} with $\IP^5$. The manifolds are singular on the loci where one of the coordinates vanishes,
\begin{align}
E_\mu \= \left\{(\varphi^0,\varphi^1,\varphi^2,\varphi^3,\varphi^4,\varphi^5) \in \IP^5 \; \big| \; \varphi^\mu = 0 \right\},
\end{align}
and also on the conifold locus
\begin{align}
\IDelta \;\defineas\; \prod_{\epsilon_i \in \{\pm 1\}} \big(\sqrt{\varphi^0} + \epsilon_1 \sqrt{\varphi^1} + \epsilon_2 \sqrt{\varphi^2} + \epsilon_3 \sqrt{\varphi^3} + \epsilon_4 \sqrt{\varphi^4} + \epsilon_5 \sqrt{\varphi^5}\big)\=0~.
\end{align}
Often it is necessary to work on an affine patch, for which we most often choose $\varphi^0 = 1$. Results obtained in this patch apply in any patch $\varphi^{i}=1$, with Latin indices running from 1 to 5, after making a suitable permutation of indices.

We begin the investigation by recalling a series expansion for the fundamental period \cite{Verrill1996Root,Verrill2004SumsOS},
\begin{equation}\label{eq:fund}
\varpi^{(0);0}({\bf \varphi}) \= \sum_{n=0}^{\infty} \sum_{\deg(\bm{p})=n} \binom{n}{\bm{p}}^2 \bm \varphi^{\bm{p}}~,
\end{equation}
where $\bm{p}=(p_1,\dots,p_5)$ is a five-component multi-index, $\deg(\bm{p})$ is the sum $p_1 + \dots + p_5$, and 
\begin{align}
\binom{n}{\bm{p}} \= \frac{n!}{p_1!p_2!p_3!p_4!p_5!}
\end{align}
is the multinomial coefficient. By $\bm x^{\bm{p}}$ we mean the product $x_1^{p_1}x_2^{p_2}x_3^{p_3}x_4^{p_{4}}x_5^{p_5}$. The superscript $(0)$ in $\varpi^{(0);0}$ refers to the coordinate patch $\varphi^{0}=1$. 

On seeking the differential equations obeyed by this period, we are led to the system
\begin{equation}\nn
 \cL_i \varpi^{(0);0}(\bm \varphi)\; \defineas \; \left(\frac{1}{\varphi^{0}}\left(\Theta+1\right)^{2}-\frac{1}{\varphi^{i}}\theta_{i}^{2}\right)\varpi^{(0);0}(\bm \varphi)\= 0~, \quad \text{with} \quad \theta_{i} \= \varphi^{i}\frac{\partial}{\partial \varphi^{i}}~,\quad \Theta \= \sum_{i=1}^{5}\theta_{i}~.
\end{equation}  These constitute a partial Picard-Fuchs system, giving 32 solutions among which we find the 12 periods\footnote{12 is the dimension of the third cohomology of $\HV$.}. These are the components of the vector
\begin{equation}
\varpi^{(0)}\=(\varpi^{(0);0},\varpi^{(0);i},\varpi^{(0)}_{j},\varpi^{(0)}_0)^{T}~, \qquad i \= 1,\dots,5~.
\end{equation}
By a simple separation-of-variables argument, it can be shown that integrals of Bessel functions of the following form furnish a basis of solutions:
\begin{equation}\label{eq:Bessel_gen}
\frac{\varphi^0}{\ii \pi} \int_{0}^{\infty} \dd z \, z \, B_0\left(\sqrt{\varphi^0} z\right) \prod_{i=1}^5 B_{i}\left(\sqrt{\varphi^i}z\right)~,
\end{equation}
where $B_{i}(\zeta)$ is either $K_{0}(\zeta)$ or $\ii \pi I_{0}(\zeta)$. Na\"ively there are $2^6=64$ integrals of this type. However, at a generic point in the moduli space there are exactly 32 such integrals that converge. The analytic continuation of each integral outside of its domain of convergence can be written as a linear combination of integrals of the general form \eqref{eq:Bessel_gen} that converge in the new region.

There is an additional differential operator which, together with those above, completely fixes the periods. After setting $\varphi^{0}=1$, this takes the form of a polynomial in $\Theta$ with coefficients that are polynomials in $\varphi_{\mu}$. In principle this operator is determined by the recurrence methods of \cite{Verrill2004SumsOS}, but for fully general $\varphi^{i}$ these recurrence relations cannot be solved in a practical amount of time.  It is possible, however, to choose constants $s^{i}$ and specialise the parameters to $\varphi^{i}=s^{i}\varphi$, thus restricting to lines in the moduli space, and write a differential operator in terms of $\varphi$ that governs the variation of the periods along these lines. In many cases, it is possible to find this remaining operator on these lines, and in our examples this operator obtained via the methods of \cite{Verrill2004SumsOS} turns out to factorise\footnote{This is a consequence of the fact that while the procedure in \cite{Verrill2004SumsOS} gives a recurrence of minimal order, the degrees of the polynomial coefficients are not minimised.}. We give an example of such an operator in \sref{sect:ODE}.

Despite lacking the explicit form of the general Picard-Fuchs system, we can fix the 12 periods among the 32 solutions of the partial system by imposing boundary conditions. These are found by matching the asymptotics of the solutions to the asymptotics near the large complex structure point predicted by mirror symmetry. We also give explicit series expansions for these periods near the large complex structure point. 

\subsubsection*{Mirror map and large complex structure}
\vskip-5pt
The large complex structure points are located at the loci where all but one of the coordinates $a_\mu$ vanish. Near the large complex structure point with $\varphi^{0}\neq0$, the period vector in the integral basis can be written in terms of the prepotential $\cF$ as
\begin{align} \nn
\Pi^{(0)} \= \begin{pmatrix}
\Pi^{(0)\phantom{;0}}_{0}\\[3pt]
\Pi^{(0)\phantom{;0}}_{i}\\[3pt]
\Pi^{(0);0}\\[3pt]
\Pi^{(0);i}
\end{pmatrix} \= \begin{pmatrix}
\frac{\partial}{\partial z^0} \cF\\[3pt]
\frac{\partial}{\partial z^i} \cF\\[3pt]
z^0\\[3pt]
z^i
\end{pmatrix}~, \quad \cF(z^0, \dots, z^5) = -\frac{1}{3!} \sum_{a,b,c=0}^5 Y_{abc} \frac{z^a z^b z^c}{z^0} + (z^0)^2 \sum_{\bm{p} \neq \bm{0}} n_{\bm{p}} \, \text{Li}_3(q^{\bm{p}})~.
\end{align}
The $Y_{abc}$ are topological quantities which we compute in \sref{sect:Mirror_Map} and the $n_{\bm{p}}$ are the genus-0 instanton numbers of multidegree $\bm{p}$. We find the following relation between the integral basis period vector $\Pi^{(0)}$ and the period vector $\varpi^{(0)}$ in the Frobenius basis of\sref{sect:Periods}: 
\begin{align} \label{eq:change_of_basis_Pi_varpi_intro}
\Pi^{(0)} \= \rho \nu^{-1}\, \varpi^{(0)},
\end{align} 
with matrices
\begin{align} \nn
\rho = \left(\begin{array}{cccc}
-\frac{1}{3}Y_{000} & \+\bm 1^T  & \+\bm 0^T & \+1~~\\[2pt]
\bm 1 & \zero & \!\!- \II & \bm 0\\[2pt]
\+1~~ & \+\bm 0^T & \+\bm 0^T & \+0~~\\[2pt]
\bm 0& \II & \zero & \bm 0
\end{array}\right)
\qquad\text{and}\qquad
\nu \= \text{diag}(1,(2\pi \ii) \bm 1, (2\pi \ii)^2 \bm 1, (2\pi \ii)^3 )~.
\end{align}
Here, and in what follows, $\bm 1$ denotes the vector $(1,1,1,1,1)^T$ and $\bm 0$ the vector $(0,0,0,0,0)^T$. The unit matrix is denoted by $\II$, while $\zero$ is a $5 {\times} 5$ zero matrix.

With the period vectors in the integral basis in hand, we can compute the instanton numbers by studying the Yukawa couplings $y_{ijk}$. These are given by the formula
\begin{align}
y_{ijk} \= - (\Pi^{(0)})^T \Sigma \,\partial_{ijk} \Pi^{(0)}~,
\end{align}
but also have the following expansions in terms of the instanton numbers:
\begin{align}
y_{ijk} \= Y_{ijk} + \sum_{n=1}^\infty \sum_{\deg(\bm{p}) = n} \frac{p_i \, p_j \, p_k \, n_{\bm{p}} \, \bm{q}^{\bm{p}}}{1-\bm{q}^{\bm{p}}}~,\qquad \text{where} \qquad q^{i}\=\ee^{2\pi\ii t^i}~.
\end{align}
Due to the permutation symmetry of the parameters $\varphi^i$, we can express many quantities in terms of the elementary symmetric polynomials. This results in a significantly less complicated series expressions which are far more amenable to computation. While we are still unable to reach the degrees possible in one-parameter computations, we find genus-0 the instanton numbers up to a total degree of 29, which we collect in appendix \ref{app:Instanton_Numbers}.

In addition, we are able to compute the genus-1 instanton numbers by constructing the genus-1 prepotential using the expressions in \cite{Bershadsky:1993ta}. Rather pleasantly, the prepotential turns out to be conceptually simpler than on the quotients studied in \cite{Candelas:2019llw}. This is largely due to the fact that the distinct singular points on the moduli space of the quotient are replaced by the irreducible singular locus $\IDelta = 0$ on the moduli space of $\HV$. The limiting factor is the number of genus-0 instanton numbers we are able to compute, since those are needed to extract the genus-1 numbers from the prepotential. We are thus able to compute the genus-1 instanton numbers up to total degree 29, and we give these in \sref{sect:genus1instantons}. The instanton numbers to order 16 are tabulated in the text, while longer tables giving the numbers up to degree 29 are to be found in appendix \ref{app:Instanton_Numbers}.

Having computed the instanton  numbers to a high degree, a rich array of patterns becomes evident.
%, many of which we are able to prove. 
We are able to explain some of the observed invariances of instanton numbers
using the freedom to expand the prepotential about any one of the six large complex structure limits. 
%together with
%the fact that there are many ways to do the analytic continuation of the periods around the different large complex %structure limits. 

\subsubsection*{Duality webs}
\vskip-5pt
In this section we elaborate on the symmetries discovered amongst the instanton numbers. We find that these symmetries correspond to an infinite Lorentzian 
Coxeter group $\cW \subset \Sp(12,\IZ)$. An $S_{5}$ subgroup is immediately manifest, as there is a permutation symmetry in the five K\"ahler structure moduli of~$\MHV$. The complex structure moduli space of $\HV$ can be parametrised with six homogeneous coordinates, which leads to additional identities between the instanton numbers given not by permuting indices, but the duality operation
\begin{equation}
(i,j,k,l,m)\mapsto (-i+j+k+l+m,j,k,l,m)~.
\end{equation}
By acting on a single multi-index $\bm{I}=(i,j,k,l,m)$ with a sequence of permutations and this duality, we can form orbits that we term a `web'. A multi-index is said to be positive if all of its entries are nonnegative and at least one is positive. A web is positive if every multi-index is positive. We conjecture that these webs have a surprisingly simple description: up to a permutation they are in one-to-one correspondence with \textit{source vectors} $\bm I$, which are defined by the condition $\deg \bm I \geqslant 3 \max \bm I$.

We observe and prove that the genus-0 numbers $n_{\bm{I}}$ are non-vanishing only if $\bm{I}$ belongs to a positive web, or a certain exceptional half web $\IW_{+}$, the positive elements of the web containing $(1,0,0,0,0)$. At genus one and beyond, the instanton number for a degree $\bm{I}$ is nonzero only if $\bm{I}$ belongs to a positive web.

\subsubsection*{Monodromies}
\vskip-5pt
In \sref{sect:Monodromies}, we turn to computing the monodromies around the singular loci $\varphi^0 \varphi^1 \varphi^2 \varphi^3 \varphi^4 \varphi^5 = 0$ and $\IDelta = 0$. As sugested by the fact that \eqref{eq:Bessel_gen} is a function of $\sqrt{\varphi^{\mu}}$, this is most conveniently done by first classifying the singularities in coordinates $\sqrt{\varphi^\mu}$. Then the singular locus $\IDelta = 0$ becomes a reducible union of codimension-1 hyperplanes of the form
\begin{align}
\begin{split}
    D_{I} \= \left\{ (\varphi^0,\dots,\varphi^5) \in \IP^5 \; \bigg| \; \sum_{\mu \in I} \sqrt{\varphi^\mu} = \sum_{\nu \in I^c} \sqrt{\varphi^\nu}\right\}~, \qquad I \subset \{0,\dots,5\}~.
\end{split}    
\end{align}
The monodromies around these loci can be found by numerically integrating the Picard-Fuchs equations on a path circling these loci. Alternatively, one can find the linear relations between analytically continued Bessel function integrals in different regions, and use this to compute the monodromies. While the former approach is too difficult with arbitrary paths due to the complicated nature of the complete Picard-Fuchs system, we can integrate along various lines on which the Picard-Fuchs operator can be found as discussed above. By studying various different lines and using symmetry, we can use the resulting `reduced' monodromy matrices to piece together the full monodromies. 

What makes this computation simpler than it first appears is the fact that the monodromy matrix around a conifold locus should be expressible in terms of a single vector:
\begin{align} \label{eq:monodromy_M_w_relation_intro}
\text{M}_I \= \text{I}_{12} - \bm w_I (\Sigma \bm w_I)^T~.
\end{align}
Here $\bm w$ is a 12-component vector that gives the integral basis components of the three-cycle vanishing at the conifold locus. Consequently, the vector $\bm w$ should also obey the symmetries relevant to each locus.

At first, we study the periods in the patch $\varphi^0 = 1$, although later we find it useful to consider other patches as well. To find the partial monodromy matrices, we study lines of the form 
\begin{equation}
(\varphi^0, \dots, \varphi^5) \= (1,s^1 \varphi, \dots, s^5 \varphi),
\end{equation}
where $s^1,\dots,s^5$ are constants. To make the numerical computations tractable, we take at least two $s^i$ equal. To be concrete, consider the simple case where $s^1 \neq s^2 = s^3 = s^4 = s^5$. Then, by symmetry
\begin{align}\nonumber
\begin{split} 
\Pi^{(0);2} \= \Pi^{(0);3}\= \Pi^{(0);4}\= \Pi^{(0);5}~,\\ 
\Pi^{(0)\phantom{;0}}_{2} \= \Pi^{(0)\phantom{;0}}_{3} \= \Pi^{(0)\phantom{;0}}_{4} \= \Pi^{(0)\phantom{;0}}_{5}~.
\end{split}
\end{align}
and there are 6 independent periods, which form a vector $\widehat{\Pi}^0$.
\begin{align}
\widehat{\Pi}^0 \= \begin{pmatrix}
\Pi^{(0)\phantom{;0}}_{0}\\[3pt]
\Pi^{(0)\phantom{;0}}_{1}\\[3pt]
\Pi^{(0)\phantom{;0}}_{2}\\[3pt]
\Pi^{(0);0}\\[3pt]
\Pi^{(0);1}\\[3pt]
\Pi^{(0);2}
\end{pmatrix}~.
\end{align}
In the general case the monodromy matrices $\text{M}$ can be written as
\begin{align} \label{eq:monodromy_M_definition_intro}
\text{M} \= (\bm u_0,\bm u_1,\ldots,\bm u_{10},\bm u_{11})~,
\end{align}
where $\bm u_i$ are 12-component column vectors
\begin{align}
\bm u_i \= (u_i^0,u_i^1,\ldots,u_i^{10},u_i^{11})^T~.
\end{align}
Since some of the periods are equal on the line $(\varphi^0, \dots \varphi^5) = (1,s_1 \varphi,s_2 \varphi,\dots, s_2 \varphi)$, we cannot find the full monodromy matrces $\text{M}$ directly by computing monodromies around the singular points on the line. Instead, we find reduced monodromy matrices $\wh{\text{M}}$ which give the monodromy of the vector~$\widehat{\Pi}^0$. These matrices take the form
\begin{align} \label{eq:monodromy_M_hat_definition_intro}
\widehat{\text{M}}\= (\hat{\bm u}_0,\, \hat{\bm u}_1,\, \hat{\bm u}_2+ \hat{\bm u}_3+ \hat{\bm u}_4+ \hat{\bm u}_5,\, \hat{\bm u}_6,\, \hat{\bm u}_7,\, \hat{\bm u}_8+\hat{\bm u}_9+\hat{\bm u}_{10}+\hat{\bm u}_{11})~,
\end{align}
where the $\hat{\bm u}_i$ are 6 component column vectors
\begin{align}
\hat{\bm u}_i \= (u_i^0,\, u_i^1,\, u_i^2,\, u_i^6,\, u_i^7, \, u_i^8)^T~.
\end{align}
By considering several lines and using symmetry arguments to simplify the computations, we are able to gain enough information to completely fix the full monodromy matrices.

Around a conifold locus, given the vector $\bm w$
\begin{align}
\bm w \= (w_0, \, w_1, \, w_2,\dots,\,w_2,\,w_7,\,w_8,\,w_9,\dots,\,w_9)~,
\end{align}
the reduced $6\times6$ matrix $\wh{\text{M}}$ takes the form
\begin{align}
\wh{\text{M}} \= \text{I}_{6} - \wh{\bm w} \big(\wh{\Sigma} \wh{\bm w} \big)^T~, \qquad \wh{\bm w} = (w_0,\,w_1,\,w_2,\,w_7,\,w_8,\,w_9)~.
\end{align}
The reduced intersection matrix $\wh \Sigma$ is given by
\begin{align}
\wh \Sigma \= \begin{pmatrix}
\+0& \+0 & \+0 & \+1 & \+0 & \+0 \\
\+0& \+0 & \+0 & \+0 & \+1 & \+0 \\
\+0& \+0 & \+0 & \+0 & \+0 & \+4\\
-1& \+0 & \+0 & \+0 & \+0 & \+0 \\
\+0& -1 & \+0 & \+0 & \+0 & \+0 \\
\+0& \+0 & -4 & \+0 & \+0 & \+0 \\
\end{pmatrix}~.
\end{align}
In this way we find 16 of the 32 vectors corresponding to the vanishing loci:
\begin{align}
\begin{split}
\bm w_{\{0\}} &\= (\+0,0,0,0,0,0,\+1,\+0,0,0,0,0)~,\\
\bm w_{\{0,1\}} &\= (-2,0,0,0,0,0,\+1,-1,0,0,0,0)~,\\
\bm w_{\{0,1,2\}} &\= (\+4,0,0,2,2,2,-1,\+1,1,0,0,0)~,
\end{split}
\end{align}
with the vectors of the form $\bm w_{\{0,i\}}$ obtained by effecting the permutation $(2,i+1)(8,i+7)$ on the components of $\bm w_{\{0,1\}}$. Similarly, the vectors of the form $\bm w_{\{0,i,j\}}$ are obtained from $\bm w_{\{0,1,2\}}$ by using the permutation $(2,i+1)(3,i+2)(8,i+7)(9,i+8)$. The remaining 16 vectors are most conveniently obtained by studying the other patches where $\varphi^i = 1$. For example, consider the patch $\varphi^1 = 1$. Near the large complex structure point at $\varphi^0 = \varphi^2 = \dots = \varphi^5 = 0$, we have, in the natural integral basis, the period vector $\Pi^1$, which is obtained by replacing the $\varphi^1$-dependence in $\Pi^0$ by $\varphi^0$ and vice versa. By symmetry, in this basis, the monodromy around this locus is
\begin{align}
\bm w_{\{1\}} \= (0,0,0,0,0,0,1,0,0,0,0,0)~.
\end{align}
To find the corresponding monodromy matrix in the original basis of $\Pi^0$, we just need to find the relation between these two bases. We find the transition matrix $\text{T}_{\Pi^1 \Pi^0}$ \eqref{eq:defn_T_Pi^0Pi^1} which takes us from one base to another. With this, we are able to find the monodromy matrix $\text{M}_{\{1\}}$ in the original basis:
\begin{align}
\text{M}_{\{1\}} \= \text{T}_{\Pi^1 \Pi^0}^{-1} \bigg(\text{I}_{12} - \bm w_{\{1\}} \big(\Sigma \bm w_{\{1\}}\big)^T \bigg) \text{T}_{\bm\Pi^1 \bm\Pi^0} \= \text{T}_{\Pi^1 \Pi^0}^{-1} \text{M}_{\{0\}} \text{T}_{\Pi^1 \Pi^0}~.
\end{align}
The other monodromy matrices of the form $\text{M}_{\{i\}}$, $\text{M}_{\{i,j\}}$ and $\text{M}_{\{i,j,k\}}$ are found in a similar manner.
\subsubsection*{Counting curves on the mirror Hulek-Verrill manifold}
\vskip-5pt
In \sref{sect:Curve_Counting} we use use elementary geometric methods in tandem with the Kodaira classification of singular elliptic fibres \cite{Kodaira1,Kodaira2} to directly count curves of certain multidegrees on generic manifolds in the family $\MHV$. 

Counting of these curves is based on the observation that $\MHV$ can be viewed as an elliptic fibration with base $\IP^1 \times \IP^1$. While the generic fibre is an elliptic curve, it is possible to find the discriminant locus corresponding to base points above which the fibres are singular. According to Kodaira's classification, the fibres over nodes of the discriminant locus are unions of two rational curves. By classifying these fibres, we find all rational curves of degrees $\leq 3$, and some of the higher-degree curves. 

As the discriminant of the elliptic fibration is relatively simple for tetraquadrics, it is often beneficial to consider the singular manifolds $\wh \MHV_i$ obtained by blowing down 24 lines along $i$'th copy of $\IP^1$ in the ambient space. On a generic manifold $\wh \MHV_i$, the discriminant locus has 200 nodes, of which $3 \times 24 = 72$ correspond to lines, $72$ to quadrics, and $56$ to cubics. We obtain all curves up to degree $3$ in this way. In addition, the fibres containing lines and quadrics also contain degree 5 and 4 curves, respectively, as the second component. These account for all rational curves with multidegrees $(0,0,1,2,2)$, $(0,0,1,1,2)$, and permutations thereof.

In this way we confirm the predictions from mirror symmetry, and provide details of the elliptic fibrations $\wh \MHV_{i}$ that may see future use in M/F-theory compactifications.

We collect some symbols that appear in multiple sections, together with their definitions, in \tref{tab:notation}.
\newpage

\begin{table}[H]
	\renewcommand{\arraystretch}{1.35}
	\centering
	\begin{tabularx}{\textwidth}{|>{\hsize=.095\hsize\linewidth=\hsize}X|
			>{\hsize=0.815\hsize\linewidth=\hsize}X|>{\hsize=0.09\hsize\linewidth=\hsize}X|}
		\hline
		\textbf{Symbol} & \hfil \textbf{Definition/Description} & \hfil \textbf{Ref.}\\[3pt]
		\hline	\hline
		
$\mathbf{\varphi}$
& 
The coordinates $(\varphi^{1},\varphi^{2},\varphi^{3},\varphi^{4},\varphi^{5})$ on the complex structure space of $\HV$.
&
\eqref{eq:Singular_HV_Polynomial}
\\[4pt] \hline

$\HV$
& 
The family of Hulek-Verrill manifolds.  
&
\eqref{sect:HV+M}
\\[4pt] \hline

$\MHV$
& 
The family of mirror Hulek-Verrill manifolds, which are complete intersections in $\left(\IP^{1}\right)^{5}$.
&
\eqref{sect:HV+M}
\\[4pt] \hline

$\wh\HV$
& 
Family of singular manifolds birational to $\HV$. 
&
\eqref{sect:HV+M}
\\[4pt] \hline

$\wh\MHV$
& 
Family of singular manifolds birational to $\MHV$. 
&
\eqref{sect:HV+M}
\\[4pt] \hline

$\wh{\MHV}_{i}$
& 
Families of singular manifolds birational to $\MHV$, obtained by projecting out the $i$'th $\IP^{1}$ coordinate axis. 
&
\eqref{eq:MHV_Contraction_Matrices}
\\[4pt] \hline

$P$
& 
Laurent polynomial defining $\wh\HV$ in $\mathbb{T}^{4}$. 
&
\eqref{eq:Singular_HV_Polynomial}
\\[4pt] \hline

$P^{1}$, $P^{2}$
& 
Laurent polynomials defining the small resolution of $\wh\HV$ in $\IP^{5}$.
&
\eqref{eq:complete_intersection_HV}
\\[4pt] \hline

$Q^{1}$, $Q^{2}$
& 
Multidegree $(1,1,1,1,1)$ polynomials that together define $\MHV$ in $\left(\IP^{1}\right)^{5}$. 
&
\eqref{eq:Q1Q2}
\\[4pt] \hline

$\wh Q$
& 
A multidegree $(2,2,2,2)$ polynomial defining $\wh\MHV$ in $\left(\IP^{1}\right)^{4}$. 
&
\eqref{eq:Qhat}
\\[4pt] \hline

$\wh Q^{i}$
& 
A multidegree $(2,2,2,2)$ polynomial defining $\wh\MHV_{i}$ in $\left(\IP^{1}\right)^{4}$. 
&
\eqref{eq:Q}
\\[4pt] \hline

$E$
& 
The locus in $\IP^{5}$ where any of the homogeneous coordinates vanish.
&
\eqref{eq:E}
\\[4pt] \hline

$E_{\mu}$
& 
The irreducible component of $E$ on which the $\mu$'th homogenous coordinate vanishes. 
&
\eqref{eq:singular_loci_E}
\\[4pt] \hline

$D_{I}$
& 
Irreducible components of the discriminant locus $\IDelta=0$ in variables $\sqrt{\varphi^{\mu}}$. 
&
\eqref{eq:D_I_loci}
\\[4pt] \hline

$\Pi$
& 
The period vector of the Hulek-Verrill manifold expressed in the integral symplectic basis. A superscript as in $\Pi^{(\mu)}$ denotes the expansion about the $\mu$'th large complex structure point.
&
\eqref{eq:Pi0}
\\[4pt] \hline

$\varpi$
& 
The $\HV$ period vector in the Frobenius basis. A superscript as in $\varpi^{(\mu)}$ denotes the expansion about the $\mu$'th large complex structure point.
&
\eqref{eq:period_varpi}
\\[4pt] \hline

$\pi^{(\mu)}$
& 
The $\HV$ period vector in the $\mu$'th ``Bessel integral basis". 
&
\eqref{eq:period_bib}
\\[4pt] \hline

$\text{T}_{u\,v}$
& 
The matrix effecting the basis change between period vectors $u$, $v$.  
&
Various
\\[4pt] \hline

$\text{M}_{s}$
& 
Matrix giving the monodromy transformation of $\Pi$ about the locus $s$.
&
\sref{sect:Monodromies}
\\[4pt] \hline

$\Delta$
& 
In \sref{sect:Toric_Geometry} and appendix \ref{app:Toric_Geometry_Data}, a polytope. In \sref{sect:Curve_Counting}, the discriminant of an elliptic fibration. 
&
Various
\\[4pt] \hline

$\IDelta$
& 
The discriminant. $\IDelta=0$ is the conifold locus in the moduli space of~$\HV$. 
&
\eqref{eq:discriminant_first}
\\[4pt] \hline
	\end{tabularx}
	\vskip10pt
	\capt{6in}{tab:notation}{Some symbols that are used throughout the paper with references to where they are defined. }
\end{table}
\newpage
\section{Toric Geometry and Mirror Symmetry} \label{sect:Toric_Geometry}
\vskip-10pt
We review the construction of Hulek and Verrill's manifold \cite{hulek2005} following in part \cite{Candelas:2019llw}. The starting point of their analysis is the five-parameter family $\widehat{\HV}_{(\varphi^0,\dots,\varphi^5)}$ of singular varieties embedded in the projective torus $\IT^4 = \IP^4 \setminus \{X_1 \cdots X_5 = 0 \}$ as the vanishing locus of
\begin{align} \label{eq:Singular_HV_Polynomial}
P(\bm X; \bm \varphi) \= \left(X_1 + X_2 + X_3 + X_4 + X_5 \right) \left(\frac{\varphi^1}{X_1}+\frac{\varphi^2}{X_2}+\frac{\varphi^3}{X_3}+\frac{\varphi^4}{X_4}+\frac{\varphi^5}{X_5}\right) - \varphi^0~.
\end{align}
These varieties can be compactified by using the standard methods of toric geometry (see for example \cite{MR2810322}), giving in general a variety with 30 singularities. Outside of the discriminant locus\footnote{The situation is a little more involved on the discriminant locus, for details see \cite{hulek2005}.} these have small resolutions, which constitute a smooth family that we call Hulek-Verrill manifolds~$\HV_{(\varphi^0,...,\varphi^5)}$. 

Particularly interesting examples of such manifolds are provided by a highly symmetric one-parameter subfamily, where $\varphi^0 = 1$ and ${\varphi^1= \dots= \varphi^5 = \varphi}$. These are characterised by a $\IZ_{5} {\,\times \,} \IZ_2$ symmetry, with the group action on the coordinates generated by
\begin{equation} \notag
\mathfrak{A} \,: \, X_{i} \mapsto X_{i+1}~, \qquad \mathfrak{B} \, : \, X_{i} \mapsto \frac{1}{X_{i}}~,
\end{equation}
where the indices are understood to take values in $\IZ_5$. The action on the manifold is free outside of the points $\varphi\in\{\frac{1}{25}\,,\,\frac{1}{9}\,,\,1\}$ in moduli space where fixed points are present. This allows one to take a quotient with respect to these symmetries to get a one-parameter family of Calabi-Yau manifolds, which are smooth for moduli outside these isolated points. 

As noted in \cite{hulek2005}, the varieties on $\IT^4$ defined by \eqref{eq:Singular_HV_Polynomial} are birational to complete intersection varieties in $\IP^5$ defined as the vanishing locus of two polynomials:
\begin{align} \label{eq:complete_intersection_HV}
P^1(\mathbf{X}) \= \sum_{\mu=0}^5 X_\mu~, \hskip30pt P^2(\mathbf{X};\mathbf{\varphi}) \= \sum_{\mu=0}^5 \frac{\varphi^\mu}{X_\mu}~.
\end{align}
This innocuous transformation turns out to be useful for finding the (non-singular) mirror manifolds $\MHV$ of the (non-singular) Hulek-Verrill Manifolds $\HV$. Combined with the methods of Batyrev and Borisov \cite{Batyrev:1994hm,Batyrev:1994pg,borisov1993towards}, which we briefly review in \sref{sect:Batyrev_Borisov_Review}, this allows finding the mirror Calabi-Yau manifold as a subvariety of a suitable toric variety.

By standard methods of toric geometry, we can find the mirror manifolds $\wh \MHV$ and $\MHV$ of $\wh \HV$ and~$\HV$. As expected, we find that $\wh \MHV$ is singular and birational to $\MHV$. \fref{fig:Resolutions_and_Mirror_Symmetry} outlines the pairings.
\begin{figure}[H]
	\centering
	\begin{tikzcd}
	& \text{HV} \arrow[leftrightarrow]{r}{} \arrow{d}{}
	& \text{H}\Lambda \arrow{d}{}  \\		
	& \widehat{\text{HV}} \arrow[leftrightarrow]{r}{} \arrow{u}{} 
	& \widehat{\text{H}\Lambda}	 \arrow{u}
	\end{tikzcd}
	\vskip10pt
	\capt{6in}{fig:Resolutions_and_Mirror_Symmetry}{Relations between the various families of manifolds we study: the singular Hulek-Verrill manifolds are denoted by $\wh \HV$, Hulek-Verrill manifolds by $\HV$, the singular mirror Hulek-Verrill manifolds by $\wh \MHV$, and mirror Hulek-Verrill manifolds by $\MHV$. The horizontal arrows denote mirror maps, and the vertical arrows birational equivalences (blow-ups/-downs).}	
\end{figure}
\subsection{The polytopes corresponding to singular varieties} \label{sect:Singular_Polytopes}
\vskip-10pt
\begin{table}[!ht]
	\renewcommand{\arraystretch}{1.49}
	\centering
	\begin{tabularx}{12cm}{|>{\hsize=.4\hsize\linewidth=\hsize}X|
			>{\hsize=0.3\hsize\linewidth=\hsize}X|>{\hsize=0.3\hsize\linewidth=\hsize}X|}
		\hline
		\textbf{Quantity} & \hfil $\wh N$, $N$ & \hfil $\wh M$, $M$ \\[3pt]
		\hline	\hline
		Basis & $\bm{e}_i$  & $\bm{e}^i$ \\[4pt] \hline	
		Coordinates on $\IT$ & $X_i$ & $Y_i$ \\[4pt] \hline	
		Coordinates on $N_{\IR}$/$M_{\IR}$ & $x_i$ & $y_i$ \\[4pt] \hline	
		Cox coordinates & $\xi_n$ & $\eta_n$ \\[4pt] \hline		
		Polytopes & $\wh \Delta^*$, $\Delta_1$, $\Delta_2$, $\Delta$, $\nabla^*$ & $\wh \Delta$, $\nabla_1$, $\nabla_2$, $\nabla$, $\Delta^*$ \\[4pt] \hline	Polytope vertex labels & $v_n$ & $u_n$ \\[4pt] \hline	
		Polytope face labels & $\rho_n$ & $\tau_n$ \\[4pt] \hline
	\end{tabularx}
	\vskip10pt
	\capt{6in}{tab:lattices}{Quantities associated to the lattices $\wh N$, $N$, $\wh N^* = \wh M$, and $N^* = M$.}
	\vskip10pt
\end{table}
We group the symbols denoting various polytopes, Cox coordinates, and other related information by their associated lattices in \tref{tab:lattices}. The lattices $\wh N$ and $\wh M$ associated to the singular varieties $\wh \HV$ and $\wh \MHV$ are four-dimensional, and consequently for them the index $i$ runs from $1$ to $4$. The lattices $N$ and $M$ are five-dimensional and for them the indices take values $i=1,\dots,5$.
\subsubsection*{Five-dimensional description}
\vskip-5pt
The polynomial $P(\mathbf{X};\mathbf{a})$ contains 21 monomials in coordinates $X_1,\dots,X_5$. Writing these monomials using multi-index notation defines 21 vectors $v_n=(v_n^{1},v_n^{2},v_n^{3},v_n^{4},v_n^{5})$, $n=0,\dots,20$, in $\IZ^{5}$. We write $X^{v_n}$ for the monomial
\begin{align} \notag
X^{v_n} \= X_{1}^{v_n^{1}}X_{2}^{v_n^{2}}X_{3}^{v_n^{3}}X_{4}^{v_n^{4}}X_{5}^{v_n^{5}}~.
\end{align}
The vectors $v_n$ make up the set
\begin{align} \notag
\big\{ (0,0,0,0,0) \big\} \cup \big\{\bme_i - \bme_j \; | \; i,j=1,\dots,5~,~~ i \neq j \big\}~.
\end{align}
These vectors in fact lie in a four-dimensional sublattice 
\begin{align} \notag
A_4 = \left\{ (n_1,n_2,n_3,n_4,n_5) \in \IZ^5 \; \; \Bigg| \; \;  \sum_{i=1}^5 n_i = 0 \right\} \subset N \simeq \IZ^5~,
\end{align}
with $\mathbf{e}_{i}$ denoting the standard orthonormal basis for $\IZ^{5}$. We take as basis for the sublattice $A_4$ the vectors
\begin{align} \notag
\mathbf{e}_{2,1}~, \mathbf{e}_{3,2}~,\mathbf{e}_{4,3}~,\mathbf{e}_{5,4}~, \qquad \text{where} \qquad \mathbf{e}_{i,j} = \mathbf{e}_i - \mathbf{e}_j~.
\end{align}
The dual lattice can be realised as a sublattice of $M = N^* \simeq \IZ^5$, with the basis given by
\begin{align} \notag
\mathbf{e}^{i+1,i} \= \frac{i}{5} \sum_{t=i+1}^5 \mathbf{e}^{t} - \frac{5-i}{5} \sum_{t=1}^i \mathbf{e}^t~,
\end{align}
where $\mathbf{e}_i$ and $\mathbf{e}^i$ are the canonical bases of $N \simeq \IZ^5$ and $M \simeq \IZ^5$. With these definitions we have that the canonical inner product gives a non-degenerate pairing:
\begin{align} \notag
\langle \mathbf{e}_{i+1,1}, \mathbf{e}^{j+1,j} \rangle \= \delta_{ij}~.
\end{align}
To find a convenient four-dimensional description for these lattices, we project $N \mapsto \wh N \simeq \IZ^4$ and $M \mapsto \wh M \simeq (\IZ^4)^*$ with
\begin{align} \label{eq:Lattice_Projections}
\begin{split}
\mathbf{e}_i &\mapsto \mathbf{e}_i~, \quad i = 1, \dots, 4~, \qquad \mathbf{e}_5 \mapsto 0~, \\
\mathbf{e}^i &\mapsto \mathbf{e}^i~, \quad i = 1, \dots, 4~, \qquad \mathbf{e}^5 \mapsto -\bme^1 - \bme^2 -\bme^3 - \bme^4~.
\end{split}
\end{align}
\subsubsection*{Four-dimensional description of $\wh \Delta$}
\vskip-5pt
An equivalent way of arriving at the form of the four-dimensional polytope starts with going to an affine patch, say $X_5 = 1$, where the polynomial $P(\mathbf{X};\mathbf{\varphi})$ contains 21 monomials that are now of the form
\begin{equation}\label{eq:monomials}
1~, \qquad X_i~, \qquad \frac{1}{X_i}~, \qquad \frac{X_{i}}{X_{j}}~,\qquad i\neq j~,\qquad i,j\neq5~.
\end{equation}
These monomials correspond to lattice points in $\wh N^4$ that are given by the 21 vectors in the set
\begin{align} \notag
\big\{(0,0,0,0) \big\} \cup \big\{ \pm \bme_i \; \big| \; i = 1,\dots,4 \big\} \cup \big\{\bme_i - \bme_j \; \big| \; i,j=1,\dots,4~,~~ i \neq j \big\}~.
\end{align}
For the numbering of these lattice points, see appendix \ref{app:Toric_Geometry_Data}. The convex hull of these points in the real span $\wh N_{\IR}$ of $\wh N$,
\begin{align} \label{eq:wt_Delta_vertices}
\wh \Delta \= \text{Conv}(u_0,\dots,u_{20})~,
\end{align}
is a four-dimensional reflexive polytope. The only internal lattice point is the origin $u_0$, and the vertices are exactly $u_1,\dots,u_{20}$, which are the only lattice points in $\wh \Delta$. The faces of $\wh \Delta$ consist of 10 tetrahedra and 20 triangular prisms lying on the boundary planes defined by the equations
\begin{align} \notag
\delta_1 y_1+\delta_2 y_2 + \delta_3 y_3 + \delta_4 y_4 + \epsilon_0 \= 0~, \qquad \delta_i \in \{0,1\}~, \qquad \epsilon_0 \in \{-1,1\}.
\end{align}
For the labelling of the faces, see appendix \ref{app:Toric_Geometry_Data}. The 20 triangular prisms break up into two $\IZ_{5}{\,\times\,}\IZ_{2}$ transitive orbits under the actions $\mathfrak{A}$ and $\mathfrak{B}$ given in \eqref{eq:4d_Z10}, and the tetrahedra form one such orbit. The facets meet as displayed in \fref{fig:fit}.
\begin{figure}[!ht]
\begin{center}
\includegraphics[width=6cm, height=6cm]{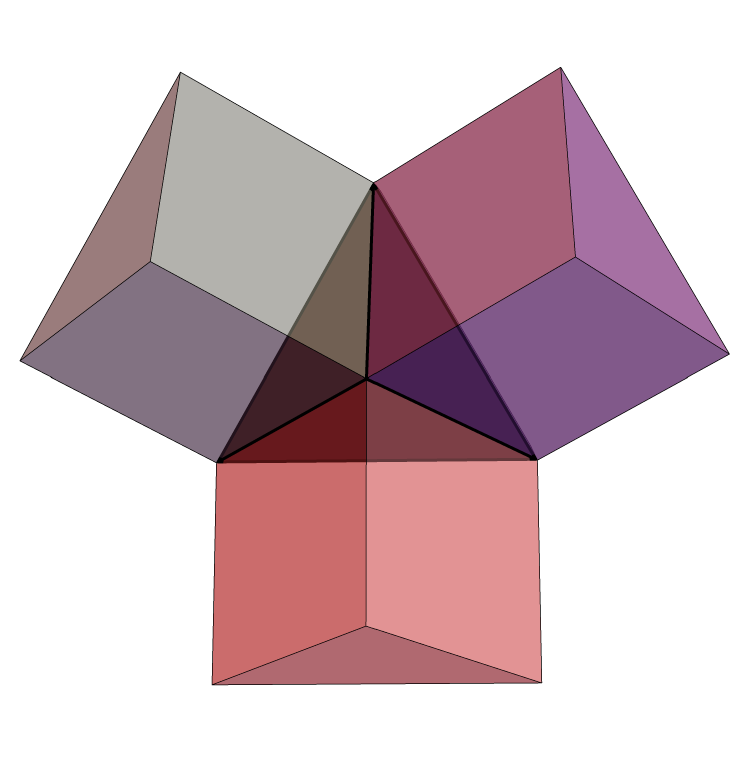}
\hskip1cm
\includegraphics[width=6cm, height=6cm]{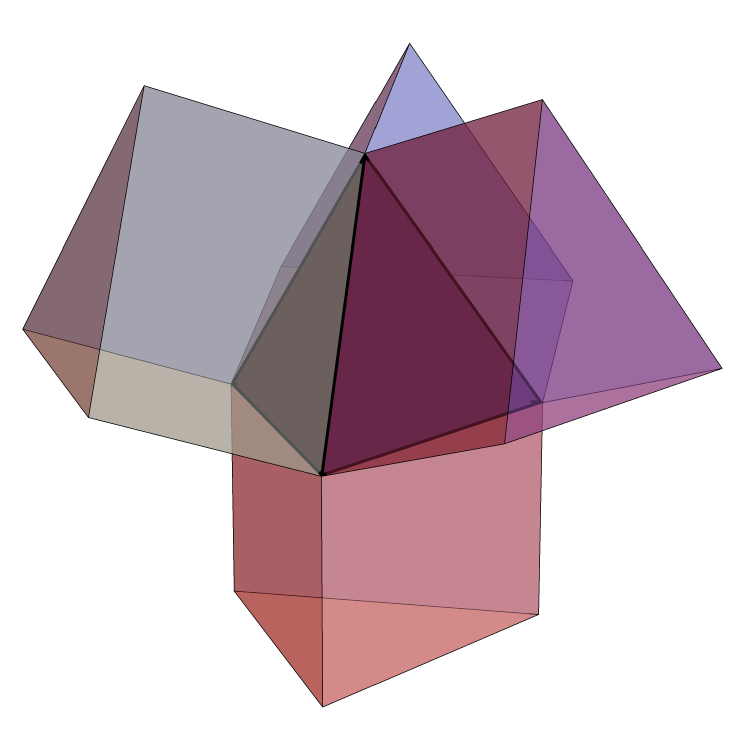}
\vskip10pt
\capt{6in}{fig:fit}{At each of the ten tetrahedra, four of the twenty prisms meet in the above configuration (depicted twice). For each pair of prisms above there is a third (not pictured) sharing a face with both. The altitudes of prisms that share a face are perpendicular. This figure corrects Fig.\ 13 of \cite{Candelas:2019llw}.}
\end{center}
\end{figure}
The polytope $\wh \Delta$ defines a fan whose cones are exactly those supported by the faces of $\wh \Delta$. This fan, however, is not simplicial, and consequently we wish to find a triangulation of $\wh \Delta$, which corresponds to a smooth fan. We find that there are two triangulations that respect the $\IZ_5 {\,\times \,} \IZ_2$ symmetry. For the purposes of this work, the choice of triangulations does not make a difference. In particular, the family of mirror manifolds one finds does not depend on the triangulation, nor are any of the properties that we consider here, such as the location of singularities, affected by this choice. Therefore we will give just the first of these triangulations.

In the four-dimensional description, the action is a composition of the $\IZ_5 {\, \times \,} \IZ_2$ action in five dimensions and the projection to four dimensions. This gives
\begin{align}\label{eq:4d_Z10}
\begin{split}
\mathfrak{A} &\, : (m_1,m_2,m_3,m_4) \mapsto \begin{cases}
(\+0,m_1,m_2,m_3)~, \qquad \text{if } \sum_{i=1}^4 m_i \= 0~,\\[3pt]
(\pm 1,m_1,m_2,m_3)~, \qquad \text{if } \sum_{i=1}^4 m_i \= \mp 1~,
\end{cases}\\
\mathfrak{B} &\, : \, \bme^i \mapsto - \bme^i~.
\end{split}
\end{align}
The cones in the first triangulation are given by
\begin{align} \label{eq:Triangulation_wh_Delta}
\begin{split}
\sigma_1 &\= \Big \langle (1,0,0,0),\;(1,0,\+0,-1),\;(1,0,-1,\+0),\;(0,1,-1,0) \Big\rangle~,\\[3pt]
\sigma_2 &\= \Big\langle (1,0,0,0),\;(0,1,\+0,\+0),\;(0,1,\+0,-1),\;(0,1,-1,0) \Big\rangle~,\\[3pt]
\sigma_3 &\= \Big\langle (1,0,0,0),\;(1,0,\+0,-1),\;(0,1,\+0,-1),\;(0,1,-1,0) \Big\rangle~,\\[7pt]
\sigma_4 &\= \Big \langle (1,0,0,0),\;(1,0,-1,\+0),\;(0,1,-1,\+0),\;(0,0,-1,1) \Big \rangle~,\\[3pt]
\sigma_5 &\= \Big \langle (1,0,0,0),\;(0,1,\+0,\+0),\;(0,1,-1,\+0),\;(0,0,\+0,1) \Big\rangle~,\\[3pt]
\sigma_6 &\= \Big \langle (1,0,0,0),\;(0,0,\+0,\+1),\;(0,1,-1,\+0),\;(0,0,-1,1) \Big\rangle~,
\end{split}
\end{align}
and their images under $\IZ_2 \times \IZ_5$, together with the 10 simplicial cones supported by the tetrahedra. The cones $\sigma_1, \sigma_2$ and $\sigma_3$ correspond to the triangulation of the triangular prism $y_1 +  y_2 = 1$ and $\sigma_4,\sigma_5$ and $\sigma_6$ give a triangulation of the prism $y_3 +  y_5 = -1$. 
\subsubsection*{The dual polytope $\wh \Delta^*$}
\vskip-5pt
The polytope $\wh \Delta$ has a dual reflexive polytope $\wh \Delta^*$ which is bounded by 20 planes 
\begin{align} \notag
\epsilon_0 + x_i \= 0~, \qquad 1 + x_i - x_j \= 0~, \qquad i,j\in\{1,2,3,4\}~,\quad i \; \neq \; j~, \qquad \epsilon_0 \in \{-1,1\}~.
\end{align}
These planes intersect $\wh \Delta^*$ in 20 cubical faces. For the explicit numbering of the faces, which manifests the explicit duality between these faces and the vertices of $\wh \Delta$, see appendix \ref{app:Toric_Geometry_Data}. It follows that $\wh \Delta^*$ is a convex hull of 31 lattice points that we label $v_0,\dots,v_{30}$.
\begin{align} \label{eq:wt_Delta^*_vertices}
\wh \Delta^* \= \text{Conv}(v_0,\dots,v_{30}) \= \text{Conv}\left(\big\{ \pm (\delta_1,\delta_2,\delta_3,\delta_4) \,\, \big| \,\, \delta_i \in \{0,1\} \big\} \right)~.
\end{align}
The corresponding fan is again non-simplicial, and thus requires triangulation to give a non-singular ambient toric variety. Hulek and Verrill \cite{hulek2005} consider a triangulation that is invariant under the $S_5$ permutations of the lattice coordinates. All of the top-dimensional cones in this triangulation are obtained by acting on the vertices of a single cone with $S_5$:
\begin{align} \notag
\big \{ \varsigma \langle (1,0,0,0),(1,1,0,0),(1,1,1,0),(1,1,1,1) \rangle \; \big| \; \varsigma \in S_5 \big \}~.
\end{align}
Note that the action of $\varsigma \in S_5$ on $\wh N$ is subtle: we have to consider the action of $S_5$ on the five-dimensional lattice and then project this to back to the four-dimensional lattice. Doing this, one is left with the following action on the basis
\begin{align} \notag
\varsigma(\bme_i) = \begin{cases}
\bme_{\varsigma(i)} \qquad \qquad \qquad \qquad \; \hskip1pt \text{if } \varsigma(i) \neq 5~,\\
-\bme_1 - \bme_2 - \bme_3 - \bme_4 \qquad \text{if } \varsigma(i) = 5~.
\end{cases} 
\end{align}
The triangulation data serves as input for Batyrev's formula \cite{Batyrev:1994hm} for the Hodge numbers of smooth members of the families of Calabi-Yau manifolds corresponding to the polytopes $\wh \Delta$ and $\wh \Delta^*$:
\begin{align} \notag
\begin{split}
h^{12} &\= \text{pts}\big(\wh \Delta^*\big) - \hskip-10pt \sum_{\text{codim}\, \wh \Theta^* = 1} \hskip-10pt \text{int}\big(\wh \Theta^*\big) \+ + \hskip-10pt \sum_{\text{codim}\, \wh \Theta = 2} \hskip-10pt \text{int}\big(\wh \Theta^*\big) \, \text{int}\big(\wh \Theta\big)-5~, \\[10pt]
h^{11} &\= \text{pts}\big(\wh \Delta\big)^{\phantom{*}} - \hskip-10pt \sum_{\text{codim}\, \wh \Theta = 1} \hskip-10pt \, \, \text{int}\big(\wh \Theta\big)^{\phantom{*}} \+ + \hskip-10pt \sum_{\text{codim}\, \wh \Theta = 2} \hskip-10pt \text{int}\big(\wh \Theta^*\big) \, \text{int}\big(\wh \Theta \big)-5~,
\end{split}
\end{align}
where $\text{pts}(\wh \Theta)$ and $\text{int}(\wh \Theta)$ denote the number of lattice points and interior lattice points of $\wh \Theta$. $\wh \Theta$ and $\wh \Theta^*$ are faces of $\wh \Delta$ and $\wh \Delta^*$, respectively.
These formulae are manifestly compatible with mirror symmetry. From the toric descriptions for the manifolds $\wh \HV$ and $\wh \MHV$, we find the Hodge numbers
\begin{align} \notag
h^{p,q}\left(\wh \HV \right) &= \begin{matrix}
& & & 1 & & & \\
& & 0 &  & 0 & & \\
& 0 &  & 26  &  & 0 & \\ 
1 &  & 16 &   & 16 &  & 1 \\ 
& 0 &  & 26  &  & 0 & \\
& & 0 &  & 0 & & \\
& & & 1 & & &  
\end{matrix}~, \qquad h^{p,q}\left( \wh \MHV \right) = \begin{matrix}
& & & 1 & & & \\
& & 0 &  & 0 & & \\
& 0 &  & 16  &  & 0 & \\ 
1 &  & 26 &   & 26 &  & 1 \\ 
& 0 &  & 16  &  & 0 & \\
& & 0 &  & 0 & & \\
& & & 1 & & &  
\end{matrix}~.
\end{align}
\subsection{The method of Batyrev and Borisov} \label{sect:Batyrev_Borisov_Review}
\vskip-10pt
To find the small resolutions $\HV$ and $\MHV$ of the singular manifolds related to the polytopes discussed above, we use the toric geometry methods pioneered by Batyrev and Borisov \cite{Batyrev:1994hm,Batyrev:1994pg,borisov1993towards}. We briefly review this approach\footnote{To keep the notation consistent throughout the paper, we adopt here notation that is slightly different from that of \cite{Batyrev:1994pg}. For example, their $\IP_\Delta$ corresponds to our $\wh \IP_{\Delta^*}$.}.

Given a variety defined as a vanishing locus of the set of $n$ Laurent polynomials $\{P^i\}_{i=1}^n$, one can study the intersection of affine hypersurfaces ${ V'(P^i) \defineas \{P^i = 0\} \subset \IT}$. If the polytopes $\{\Delta_i\}_{i=1}^n$ corresponding to the polynomials $P^i$ form a nef-partition of a reflexive polytope $\Delta$, we can define an ambient space $\wh \IP_{\Delta^{*}} \supset \IT$ corresponding to the fan associated to $\Delta^{*}$. The toric variety $\wh\IP_{\Delta^{*}}$ has a partial desingularisation $\IP_{\Delta^{*}}$, corresponding to a maximal projective triangulation of $\Delta^{*}$. The surfaces $V'(P^i)$ have closures $\wh V(P^i)\subset\wh \IP_{\Delta^{*}}$ and $V(P^i)\subset\IP_{\Delta^{*}}$, and we can define the closures of the intersections $\wh \cM = \wh V(P^1) \cap ... \cap \wh V(P^n)$ and $\cM = V(P^1) \cap ... \cap V(P^n)$. It can be shown \cite{Batyrev:1994pg} that if $\wh \cM$ is non-empty and irreducible, and also $\dim \cM \geq 3$, then $\cM$ defined in this way is a smooth manifold\footnote{The reader conversant in toric geometry will recognise this as the MPCP-desingularisation. For the present purposes it is enough to note that this desingularisation is obtained from a triangulation of the polytope, and preserves the canonical class of the manifold. }.

To find the mirror variety of the smooth manifold constructed in this way, we note that by the definition of a nef-partition 
\begin{align} \notag
\Delta \= \text{Mink}\big(\{\Delta_i\}_{i=1}^{n}\big)~,
\end{align}
where Mink denotes the Minkowski sum. In addition, we can define the convex hull of the union of the polytopes $\Delta_i$:
\begin{align} \notag
\nabla^* \; \defineas \; \text{Conv}\big( \{\Delta_i\}_{i=1}^{n} \big).
\end{align}
One can show \cite{Batyrev:1994pg} that the polytope $\nabla^*$ so defined is also a reflexive polytope. In particular, it has a well-defined dual polytope $\nabla$. This, and the dual polytope $\Delta^*$ of $\Delta$, can be shown to be expressible in terms of $n$ smaller polytopes $\{\nabla_i\}_{i=1}^n$:
\begin{align} \notag
\nabla \= \text{Mink}\big(\{\nabla_{i}\}_{i=1}^n \big)~, \qquad \Delta^* \= \text{Conv}\big(\{\nabla_{i}\}_{i=1}^n \big)~,
\end{align}
where the sum is again a Minkowski sum, and $\{\nabla_i\}_{i=1}^n$ gives a nef-partition of $\nabla$. Now we can define the mirror manifold of $\wh \cM$ as follows: first we use the polytopes $\nabla_i$ to define a set of polynomials $\{Q^i\}_{i=1}^n$ and a desingularisation $\IP_{\nabla^{*}}$ corresponding to a maximal projective triangulation of $\nabla^{*}$. Then the mirror manifold $\cW$ of $\cM$ can be expressed as the closure $V(Q^1) \cap \dots \cap V(Q^n)$ of the variety $\{Q^1 = \dots = Q^n = 0 \} \subset \IT$. Due to the way $\cW$ is constructed, it follows that it is smooth and irreducible if and only if $\cM$ is \cite{Batyrev:1994pg}.

There is an algorithm for computing the Hodge numbers of varieties defined in this way \cite{Batyrev:2007cq,BatyrevNillMolp}. In the case of complete intersection varieties, it is more complicated than Batyrev's original formulae for the Hodge numbers \cite{Batyrev:1994hm}. We will not review the details here, and simply note that some computer algebra packages, such as PALP \cite{Braun_2012}, provide an implementation of the algorithm.
\subsection{The polytopes corresponding to small resolutions}
\vskip-10pt
\subsubsection*{Small polytopes $\Delta_1$, $\Delta_2$}
\vskip-5pt
To find the toric descriptions of the non-singular manifolds $\HV$ and $\MHV$, we study the polytopes $\Delta_1, \Delta_2 \subset \IZ^5$. Their vertices correspond to monomials in the polynomials $P^1$ and $P^2$, defined in \eqref{eq:complete_intersection_HV}, that define on $\IP^5$ a variety birational to $\wh \HV$. We work directly in an affine patch where $X_0 = 1$. Then the two polytopes can be expressed as 
\begin{align} \notag
\Delta_1 \= \text{Conv}(\bm0,\,\bme_1,\,\bme_2,\,\bme_3,\,\bme_4,\,\bme_5)~, \qquad \Delta_2 \= \text{Conv}(\bm0,\,-\bme_1,\,-\bme_2,\,-\bme_3,\,-\bme_4,\,-\bme_5) \= -\Delta_1~.
\end{align}
These, and the other polytopes defined this subsection, are schematically represented in two dimensions in \fref{fig:Polytopes}. Using these two, we can construct two larger polytopes as their Minkowski sum and the convex hull of their union
\begin{align} \notag
\Delta \; \defineas \; \text{Mink}(\Delta_1,\,\Delta_2)~, \qquad \nabla^* \; \defineas \; \text{Conv}(\Delta_1 ,\, \Delta_2)~.
\end{align}
From the definition of convex hull, it follows immediately that the vertices of $\nabla^*$ are exactly $\pm \bme_{i}$ with $i = 1, \dots, 5$. Its 32 faces are the four-dimensional simplices of the form 
\begin{align} \notag
\tau_n \= \text{Conv}\left( \epsilon_1 \bme_1, \, \epsilon_2  \bme_2, \, \epsilon_3  \bme_3, \, \epsilon_4 \bme_4, \, \epsilon_5  \bme_5 \right)~, \qquad \epsilon_i \in \{-1,1\}~,
\end{align}
given by intersections of $\nabla^*$ with bounding planes
\begin{align} \notag
\epsilon_1 x_1 + \epsilon_2 x_2 + \epsilon_3 x_3 + \epsilon_4 x_4 + \epsilon_5 x_5 \= 1~, \qquad \epsilon_i \in \{-1,1\}~.
\end{align}
The polytope $\Delta$ contains in total 31 lattice points,
\begin{align} \notag
\big\{(0,0,0,0) \big\} \cup \big\{\pm \bme_i \; \big| \; i = 1,\dots,5 \big\} \cup \big\{\bme_i - \bme_j \; \big| \; i,j = 1,\dots,5~,~~i \neq j \big\}~.
\end{align}
Thus it can be written as a convex hull of 30 lattice points
\begin{align} \notag
\Delta \= \text{Conv}(v_1,\dots,v_{30})~.
\end{align}
Its only internal point is the origin, and it has 62 faces that are hypercubes, given by intersections with planes
\begin{align} \notag
\epsilon_0 + \delta_1 x_1 + \delta_2 x_2 + \delta_3 x_3 + \delta_4 x_4 + \delta_5 x_5 \= 0~, \qquad \text{with} \qquad \epsilon_0 \in \{-1,1\}~, \; \; \delta_i \in \{0,1\}~.
\end{align}
It can be shown that $\{\Delta_1,\Delta_2\}$ is a nef-partition of $\Delta$. 
\subsubsection*{Small polytopes $\nabla_1$, $\nabla_2$}
\vskip-5pt
Finally, to find the equations defining the mirror Hulek-Verrill manifold, we need the two polytopes $\nabla_1$ and $\nabla_2$. These can be obtained by first finding the duals of $\nabla^*$ and $\Delta$. The polytope $\nabla$ is a hypercube centred at the origin. Its vertices are given by the 32 points of the form
\begin{align} \notag
\nabla= \text{Conv}\left( \{(\epsilon_1,\epsilon_2,\epsilon_3,\epsilon_4,\epsilon_5)\} \; | \; \epsilon_i \in \{-1,1\} \right)~.
\end{align}
The faces are the 10 four-dimensional hypercubes given by intersections with the planes
\begin{align} \notag
y_i = \pm 1~.
\end{align}
The remaining polytope $\Delta^*$ has a slightly more complicated structure. It can be written as the convex hull of 62 vertices of the form
\begin{align} \notag
\Delta^* \= \text{Conv}\left(\left\{\pm (\d_1,\d_2,\d_3,\d_4,\d_5) \; | \; \d_i \in \{0,1\}\right\} \right)~.
\end{align}
The labelling of all vertices in given in appendix \ref{app:Toric_Geometry_Data}. It has 30 faces, given by intersections with the planes\vskip-30pt
\begin{align} \notag
1\pm y_i \= 0~, \qquad 1 + y_i - y_j \= 0~.
\end{align}
Like their duals, $\nabla$ and $\Delta^*$ can be given in terms of two smaller polytopes $\nabla_1$ and $\nabla_2$:
\begin{align} \notag
\nabla = \text{Mink}(\nabla_1, \,\nabla_2)~, \qquad \Delta^* = \text{Conv}(\nabla_1 ,\, \nabla_2)~.
\end{align}
Here $\nabla_1$ and $\nabla_2$ are hypercubes with one vertex at origin, given by
\begin{align} \notag
\nabla_1 &\= \text{Conv}\left(\{(\delta_2,\dots,\delta_5)\, | \,\delta_i\in\{0,1\}\} \right)~, \qquad
\nabla_2 \= \text{Conv}\left(\{-(\delta_1,\dots,\delta_5)\,| \,\delta_i \in\{0,1\}\} \right) \;= - \nabla_1~.
\end{align}
By the prescription of Batyrev and Borisov, the ambient variety $\IP_{\Delta^*}$ for the Hulek-Verrill manifold is given by triangulating $\Delta^*$. We leave most of the details to the reader, but the upshot is that, as in \cite{hulek2005}, we can take the triangulation to be invariant under permutations $\varsigma \in S_5$ of the coordinates $X_i$ as well as under the $\IZ_2$ inversion symmetry $X_i \mapsto \frac{1}{X_i}$.

The fan associated to $\Delta^*$ consists of 720 top-dimensional cones. There are three simplicial cones $\sigma_1,\sigma_2$ and $\sigma_3$, whose images under $S_5$ and $\IZ_2$ generate the whole fan. These are given by
\begin{align}\begin{split} \label{eq:Delta_Star_triangulation}
\sigma_1 &\= \Big\langle  (\+1,0,0,0,0),\,(\+1,\+1,0,0,0),\,(1,1,1,0,0),\,(1,1,1,1,0),\,(1,1,1,1,1) \Big\rangle~,\\[3pt] 
\sigma_2 &\= \Big\langle  (-1,0,0,0,0),\,(\+0,\+0,0,0,1),\,(0,0,0,1,1),\,(0,0,1,1,1),\,(0,1,1,1,1) \Big\rangle~,\\[3pt] 
\sigma_3 &\= \Big\langle  (-1,0,0,0,0),\,(-1,-1,0,0,0),\,(0,0,0,0,1),\,(0,0,0,1,1),\,(0,0,1,1,1) \Big \rangle ~.
\end{split}\end{align}
The first cone, together with the 119 distinct cones generated by permuting the coordinates, $\{\varsigma(\sigma_{1}) \;| \; \varsigma\in S_{5}\}$, give a triangulation of the hypercube $\nabla_1$. The $\IZ_2$ inversion symmetry acts on these cones by $\varsigma(\sigma_{1}) \mapsto \varsigma(-\sigma_{1})$. The hypercube $\nabla_2$ is triangulated by the $\IZ_2$ image of $\{\varsigma(\sigma_{1}) \;| \; \varsigma\in S_{5}\}$. The rest of the polytope $\nabla$ is triangulated by $\sigma_2$, $\sigma_3$, and their images under~$S_5 {\,\times\,} \IZ_2$. There are additional triangulations, but as in the four-dimensional case, the choice of triangulation does not affect the discussion in this paper.

\begin{figure}[H]
	\centering
    \scalebox{0.9}{
	\begin{tikzpicture}
	\usetikzlibrary{arrows.meta}
	
	\draw[step=1.5,black,thin,dashed] (-3,-3) grid (3,3);

	\shade [] (0,0) -- (1.5,0) -- (0,1.5) -- cycle;	
	\draw[fill=gray!30] (0,0) -- (1.5,0) -- (0,1.5) -- cycle;
	
	\shade [] (0,0) -- (-1.5,0) -- (0,-1.5) -- cycle;	
	\draw[fill=gray!30] (0,0) -- (-1.5,0) -- (0,-1.5) -- cycle;
	
	\draw [line width=1.2pt] (1.5,0) -- (0,1.5) -- (-1.5,0) -- (0,-1.5) -- cycle;			
	
	\foreach \Point in {(0,0), (1.5,0), (0,1.5), (-1.5,0), (0,-1.5)}{
		\node at \Point {\textbullet};
	}
	
	\node at (0.5,0.5) {$\Delta_1$};
	\node at (-0.5,-0.5) {$\Delta_2$};	
	\end{tikzpicture}}
	\hskip20pt
    \scalebox{0.9}{
	\begin{tikzpicture}
	\usetikzlibrary{arrows.meta}
	
	\draw[step=1.5,black,thin,dashed] (-3,-3) grid (3,3);

	\shade [] (0,0) -- (1.5,0) -- (0,1.5) -- cycle;	
	\draw[fill=gray!30] (0,0) -- (1.5,0) -- (0,1.5) -- cycle;
	
	\shade [] (0,0) -- (-1.5,0) -- (0,-1.5) -- cycle;	
	\draw[fill=gray!30] (0,0) -- (-1.5,0) -- (0,-1.5) -- cycle;
	
	\draw[line width=1.2pt] (1.5,0) -- (1.5,-1.5) -- (0,-1.5) -- (-1.5,0) -- (-1.5,1.5) -- (0,1.5) -- cycle;			
	
	\foreach \Point in {(0,0), (1.5,0), (0,1.5), (-1.5,0), (0,-1.5),(1.5,-1.5),(-1.5,1.5)}{
		\node at \Point {\textbullet};
	}
	
	\node at (0.5,0.5) {$\Delta_1$};
	\node at (-0.5,-0.5) {$\Delta_2$};	
	
	\end{tikzpicture}}
	\vskip20pt
    \scalebox{0.9}{
	\begin{tikzpicture}
	\usetikzlibrary{arrows.meta}
	
	\draw[step=1.5,black,thin,dashed] (-3,-3) grid (3,3);

	\shade [] (0,0) -- (1.5,0) -- (1.5,1.5) -- (0,1.5) -- cycle;	
	\draw[fill=gray!30] (0,0) -- (1.5,0) -- (1.5,1.5) -- (0,1.5) -- cycle;
	
	\shade [] (0,0) -- (-1.5,0) -- (-1.5,-1.5) -- (0,-1.5) -- cycle;	
	\draw[fill=gray!30] (0,0) -- (-1.5,0) -- (-1.5,-1.5) -- (0,-1.5) -- cycle;
	
	\draw [line width=1.2pt] (1.5,0) -- (1.5,1.5) -- (0,1.5) -- (-1.5,0) -- (-1.5,-1.5) -- (0,-1.5) -- cycle;			
	
	\foreach \Point in {(0,0), (1.5,0), (0,1.5), (-1.5,0), (0,-1.5), (1.5,1.5),(-1.5,-1.5)}{
		\node at \Point {\textbullet};
	}
	
	\node at (0.75,0.75) {$\nabla_1$};
	\node at (-0.75,-0.75) {$\nabla_2$};	
	\end{tikzpicture}}
	\hskip20pt
    \scalebox{0.9}{
	\begin{tikzpicture}
	\usetikzlibrary{arrows.meta}
	
	\draw[step=1.5,black,thin,dashed] (-3,-3) grid (3,3);
	
	\shade [] (0,0) -- (1.5,0) -- (1.5,1.5) -- (0,1.5) -- cycle;	
	\draw[fill=gray!30] (0,0) -- (1.5,0) -- (1.5,1.5) -- (0,1.5) -- cycle;
	
	\shade [] (0,0) -- (-1.5,0) -- (-1.5,-1.5) -- (0,-1.5) -- cycle;	
	\draw[fill=gray!30] (0,0) -- (-1.5,0) -- (-1.5,-1.5) -- (0,-1.5) -- cycle;
	
	\draw[line width=1.2pt] (1.5,0) -- (1.5,-1.5) -- (0,-1.5) -- (-1.5,-1.5) -- (-1.5,0) -- (-1.5,1.5) -- (0,1.5) -- (1.5,1.5) -- cycle;			
	
	\foreach \Point in {(0,0), (1.5,0), (0,1.5), (-1.5,0), (0,-1.5),(1.5,-1.5),(-1.5,1.5),(1.5,1.5),(-1.5,-1.5)}{
		\node at \Point {\textbullet};
	}
	
	\node at (0.75,0.75) {$\nabla_1$};
	\node at (-0.75,-0.75) {$\nabla_2$};	
	
	\end{tikzpicture}}
	\vskip10pt
	\capt{6in}{fig:Polytopes}{Two-dimensional analogues of the polytopes $\Delta,\nabla$, their duals, and their nef-partitions. Clockwise from top-left, we have $\text{Conv}(\Delta_{1},\,\Delta_{2}),\,\text{Mink}(\Delta_{1},\,\Delta_{2}),\,\text{Mink}(\nabla_{1},\,\nabla_{2}),$ and $\text{Conv}(\nabla_{1},\,\nabla_{2})$.}	
\end{figure}
\subsection{The Hulek-Verrill manifolds and their mirrors}\label{sect:HV+M}
\vskip-10pt
Having studied the relevant lattice geometry, we are ready to turn to the toric geometry associated to the triangulations of the fans corresponding to the triangulated polytopes that were found in the previous sections. We will give both the singular manifolds $\wh \HV$ and $\wh \MHV$ and their resolutions $\HV$ and $\MHV$ as vanishing loci of a set of polynomials inside the relevant ambient toric variety. We also find some basic properties of these manifolds, which will be relevant in the following sections. The quantities associated to each manifold are summarised in \tref{tab:CYs_and_Lattices}.
\begin{table}[H]
	\renewcommand{\arraystretch}{1.49}
	\centering
	\begin{tabular}{|l|c|c|c|c|}
		\hline
		\hfil\textbf{Quantity} &  $\wh \HV$ &  $\wh \MHV$ &  $\HV$ &  $\MHV$ \\[3pt]
		\hline	\hline
		Defining polynomials & $P$  & $Q$ & $P^1,P^2$ & $Q^1,Q^2$ \\[4pt] \hline	Polytopes defining monomials & $\wh \Delta$  & $\wh \Delta^*$ & $\Delta_1, \Delta_2$ & $\nabla_1$, $\nabla_2$ \\[4pt] \hline
		Ambient toric variety & $\IP_{\wh \Delta^*}$ & $\IP_{\wh \Delta}$ & $\IP_{\Delta^*}$ & $\IP_{\nabla^*}$  \\[4pt]	\hline	
        Coordinates & $X_1, \dots X_5$ & $Y_1, \dots, Y_4$ & $X_0, \dots, X_5$ & $Y_0,\dots, Y_4$  \\[4pt]	\hline			
	\end{tabular}\\[10pt]
	\capt{6in}{tab:CYs_and_Lattices}{Quantities associated to the manifolds $\wh \HV$, $\wh \MHV$, $\HV$, and $\MHV$.}\vskip10pt
\end{table}
\subsubsection*{The singular Hulek-Verrill Manifold $\wh \HV$}
\vskip-5pt
The ambient toric variety $\IP_{\wh \Delta ^{*}}$ in which $\wh \HV$ can be embedded corresponds to the polytope $\wh \Delta ^{*}$. To the vertices we associate Cox coordinates $\xi_1, \dots, \xi_{30}$. The ambient variety can then be given by the usual construction as
\begin{align}\label{eq:P_DeltaHat}
\IP_{\wh \Delta ^{*}} \= \frac{\IC^{30} \setminus F}{(\IC^*)^{26}}~.
\end{align}
The scalings $(\IC^*)^{26}$ correspond to linear relations between the vectors corresponding to the vertices of $\wh\Delta^{*}$. $F$ is the union of sets given by the simultaneous vanishing of Cox coordinates associated to rays not lying in the same cone. Excising this from $\IC^{30}$ prior to quotienting in \eqref{eq:P_DeltaHat} ensures a well-defined toric variety\footnote{For technical details that we omit see the textbooks \cite{MR2810322,0813.14039}, or the more physicist-oriented notes \cite{Closset:2009sv}.}. 

To study the Calabi-Yau manifold $\wh \HV \subset \IP_{\wh \Delta^*}$, we identify the coordinates $X_1,\dots,X_4$ with the coordinates $\Xi_1, \dots, \Xi_4$ on the torus, which we define in terms of Cox coordinates in appendix \ref{app:Toric_Geometry_Data}. Then the Calabi-Yau manifold can be written as as a subset
\begin{align}\label{eq:HVhatinPDeltahat}
\bigg \{\sum_{i\neq j} a_{i,j} \; \frac{X_i}{X_j} + a_0 \= 0 \bigg\} \subset \IP_{\wh \Delta ^{*}}.
\end{align} 
We are chiefly concerned with the five-parameter subfamily
\begin{align} \notag
a_{i,j} = a_j \qquad \text{for all } i \neq j~,
\end{align}
where the polynomial in \eqref{eq:HVhatinPDeltahat} takes the form $P$ given in \eqref{eq:singular_HV}. The generic manifold in this family contains 30 nodal singularities on $\wh \HV \setminus \IT^4$, which can be seen by considering the local patches corresponding to the triangulation of the polytope $\wh \Delta^*$ \cite{hulek2005}. These singular varieties have resolutions $\HV_{(a_0,...,a_5)}$, which are smooth Calabi-Yau manifolds. We will discuss the toric description of these manifolds later in this section, where we show that the resolutions we find using toric methods have the same Hodge numbers and various other properties as the small resolutions studied by Hulek and Verrill. It may be possible to identify these, although we lack a formal proof. However, for the study of mirror symmetry in this paper it is enough to consider manifolds up to birational equivalence. In particular, the periods of the Hulek-Verrill manifold do not depend on the resolution of the subvariety of $\IT^5$. Thus, in this paper, we \emph{define} the Hulek-Verrill manifolds as the toric resolutions $\HV_{(a_0,...,a_5)}$ described in the following subsections. We then see that these are birational to the small resolutions studied by Hulek and Verrill and that their mirrors are given, via the Batyrev-Borisov construction, as the complete intersection manifolds we denote $\MHV$.
\subsubsection*{The singular mirror Hulek-Verrill Manifold $\wh \MHV$}
\vskip-5pt
We can use Batyrev's construction \cite{Batyrev:1994hm} to find the mirror manifolds of the singular Hulek-Verrill manifolds. These are of interest to us since some of the manifolds that concern us turn out to be singular. However, they are birational to the mirror manifolds of the small resolutions mentioned above. The construction of the resolved manifold in this way is somewhat complicated, but in \sref{sect:HV} we give another method of finding this resolution.

We have already found the vertices of the dual polytope $\wh \Delta^*$ in \eqref{eq:wt_Delta^*_vertices}. These, together with the interior point, correspond to the monomials
\begin{align} \label{eq:singular_MHV_monomials}
1~, \quad Y_i~, \quad Y_i Y_j~, \quad Y_i Y_j Y_k~, \quad Y_i Y_j Y_k Y_l~, \quad \frac{1}{Y_i}~, \quad \frac{1}{Y_i Y_j}~, \quad \frac{1}{Y_i Y_j Y_k}~, \quad \frac{1}{Y_i Y_j Y_k Y_l}~.
\end{align}
Each of the indices $i,j,k,l$ are distinct and take values in $\{1,2,3,4\}$. The intersection of a generic mirror singular Hulek-Verrill manifold with the torus $\IT^4$ is given by the closure of the vanishing~locus
\begin{align}\label{eq:Qhat}
\wh Q \; \defineas\; \sum_{i,j,k,l=0}^2 A_{i,j,k,l} Y_1^i Y_2^j Y_3^k Y_4^l \= 0~.
\end{align}
One obtains this by taking the most general polynomial with monomials \eqref{eq:singular_MHV_monomials} and multiplying through by $Y_{1}Y_{2}Y_{3}Y_{4}$, which gives the same variety on $\IT^4$.

Given the triangulation \eqref{eq:Triangulation_wh_Delta} of $\wh \Delta$ discussed in \sref{sect:Singular_Polytopes}, we can consider the local affine patches $\IA_{\sigma_i}$ corresponding to the simplicial cones $\sigma_i$. Equivalently, we can choose suitable 4-tuples of the Cox coordinates $\eta_i$ to act as the local coordinates on patches isomorphic to $\IA^4$. It is only necessary to study the six local patches related to the fans given in \eqref{eq:Triangulation_wh_Delta} and a single patch generated by any tetrahedron. The other local patches are obtained from these by $\IZ_5 \times \IZ_2$ symmetry. 

As an example, let us consider the cone $\sigma_1$. The coordinates associated to the generators of this cone are
\begin{align} \notag
    x \; \defineas \; \eta_{20}~, \quad y \; \defineas \; \eta_{19}~, \quad z \; \defineas \; \eta_{18}~, \quad w  \;\defineas \; \eta_{14}~.
\end{align}
Since the generators corresponding to these coordinates belong to the same simplicial cone, we can set the other coordinates to unity, and thus identify the local coordinates with those on the torus~as
\begin{align} \notag
    H^1 \= xyz~, \qquad H^2 \= w~, \qquad H^3 \= \frac{1}{wz}~, \qquad H^4 \= \frac{1}{y}~.
\end{align}
We can immediately find the local coordinates on
\begin{align} \notag
    \mathfrak{A} \; \sigma_1 \= \big\langle (-1,1,0,0),~(0,1,0,0),~(0,1,0,-1),~(0,0,1,-1) \big\rangle~
\end{align}
by noting that the $\IZ_5$ action on the Cox coordinates inherited from the action on the vertices maps
\begin{align} \notag
    \eta_{20} \mapsto \eta_{4} \= x~, \qquad \eta_{19} \mapsto \eta_{16} \= y~, \qquad \eta_{18} \mapsto \eta_{15} \= z~, \qquad \eta_{14} \mapsto \eta_{12} \= w.
\end{align}
The equalities denote identifications with the coordinates on the affine patch $\IA_{\mathfrak{A}\sigma_1}$. Thus on this patch, we can make the identifications with the torus coordinates as
\begin{align} \notag
     H_1 \= \frac{1}{x}~, \qquad H_2 \= xyz~, \qquad H_3 \= w~, \qquad H_4 \= \frac{1}{wz}~.
\end{align}
Note that this corresponds to $\IZ_5$ acting on the global coordinates as
\begin{align} \notag
    H_i \mapsto H_{i+1}~, \quad i \neq 4~, \qquad \qquad H_4 \mapsto \frac{1}{H_1 H_2 H_3 H_4}~,
\end{align}
which of course corresponds to the $\IZ_5$ action $\bme^i \mapsto \bme^{i+1}$ of the five-dimensional lattice $M$, projected down to four dimensions by \eqref{eq:Lattice_Projections}.

Writing the polynomial $Q$ in global coordinates gives, for generic values of the moduli, an irreducible multidegree $(2,4,4,4)$ polynomial. A member of this family is generically smooth, but smooth members are not birational to mirrors of Hulek-Verrill manifolds $\MHV_{(a_0,\dots,a_5)}$. 

Instead, it turns out that we must only consider those whose defining polynomials can be written in the form
\begin{align} \notag
\wh Q \= \alpha\, \delta - \beta\, \gamma~,
\end{align}
where $\alpha,\beta,\gamma,$ and $\delta$ are multidegree $(1,1,1,1)$ polynomials in the coordinates $Y_1,\dots,Y_4$. A manifold with this property has exactly 24 singularities, which can be resolved in order to obtain a smooth variety.
\subsubsection*{The Hulek-Verrill manifold $\HV$} \label{sect:HV}
\vskip-5pt
As we have already remarked, Hulek and Verrill noted that the singular variety $\wh \HV_{(\varphi^0,\dots,\varphi^5)}$ defined by the equation
\begin{align}\label{eq:HVsing}
\left(\sum_{i=1}^5 X_i \right) \left(\sum_{i=1}^5 \frac{\varphi^i}{X_i} \right) \= a_0 
\end{align}
on the toric variety $\IP_{\wh \Delta^*}$ is birational to the subvariety of $\IP^5$ defined by the two polynomials
\begin{align} \notag
P^1 \; \defineas \; \sum_{\mu=0}^5 X_\mu~, \qquad P^2 \; \defineas \; \sum_{\mu=0}^5 \frac{\varphi^\mu}{X_\mu}~.
\end{align}
It is possible to develop this further by studying the two equations $P^1 = P^2 = 0$ on the torus $\IT^5$ and finding the toric closure of this variety. This can be achieved using the techniques reviewed briefly in \sref{sect:Batyrev_Borisov_Review}. In \sref{sect:Singular_Polytopes}, we have studied the polytopes $\Delta_1$ and $\Delta_2$ whose vertices correspond to the monomials in $P^1$ and $P^2$, and found the polytope $\Delta^{*}$ which gives the ambient space $\IP_{\Delta^{*}}$. The Cox coordinates and coordinate scalings defining the ambient variety are given in appendix \ref{app:Toric_Geometry_Data}.

We can analyse this variety further by specialising to various local patches. We only need to analyse the patches that are not related by symmetry. 

The Cox coordinates associated to the generators of the cone $\sigma_1$ in \eqref{eq:Delta_Star_triangulation} are
\begin{align} \notag
\xi_{1} \, \defineas \, x~, \qquad \xi_{5} \, \defineas \, y~, \qquad \xi_{13}  \, \defineas \, z~, \qquad \xi_{29} \, \defineas \, w~, \qquad \xi_{61} \, \defineas \, v~.
\end{align}
Using the leftover scalings to set the other 57 Cox coordinates $\xi$ to unity, we can identify the invariants $\Xi_1, \dots, \Xi_5$ as
\begin{align} \notag
\Xi_1 \= xyzwv~, \qquad \Xi_2 \= yzwv~, \qquad \Xi_3 \= zwv~, \qquad \Xi_4 \= wv~, \qquad \Xi_5 \= v~.
\end{align}
By further identifying these $\Xi_{i}$ with the coordinates $X_i$ on the torus, we can write the polynomials $P^1$ and $P^2$ as
\begin{align} \notag
\begin{split}
P^1 &\= 1 + v + wv + zwv + yzwv + xyzwv~,\\
P^2 &\= \varphi^{0} + \frac{\varphi^1}{xyzwv}+\frac{\varphi^2}{yzwv}+\frac{\varphi^3}{zwv}+\frac{\varphi^4}{wv}+\frac{\varphi^5}{v}~.
\end{split}
\end{align}
The analogous relations for the remaining cones, $\sigma_2$ and $\sigma_3$, can be found in a similar manner. 

By studying the equations $P^1 = P^2 = \dd P^1 \wedge \dd P^2 = 0$, it is not difficult to see that, generically, the variety $\HV$ does not have singularities. As in the original analysis of Hulek and Verrill~\cite{hulek2005}, we find that there are singularities if and only if  
\begin{align} \label{eq:discriminant_first}
    \IDelta \;\defineas\; \prod_{\epsilon_i \in \{\pm 1\}}\Big( \sqrt{\varphi^0} + \epsilon_1 \sqrt{\varphi^1} + \epsilon_2 \sqrt{\varphi^2} + \epsilon_3 \sqrt{\varphi^3} + \epsilon_4 \sqrt{\varphi^4} + \epsilon_5 \sqrt{\varphi^5} \Big)\= 0~.
\end{align}
The algorithm in \cite{Batyrev:2007cq,BatyrevNillMolp}, implemented in PALP \cite{Braun_2012}, gives the Hodge numbers of this variety as
\begin{align} \notag
h^{p,q} &\= \begin{matrix}
& & & 1 & & & \\
& & 0 &  & 0 & & \\
& 0 &  & 45  &  & 0 & \\ 
1 &  & 5 &   & 5 &  & 1 \\ 
& 0 &  & 45  &  & 0 & \\
& & 0 &  & 0 & & \\
& & & 1 & & &  
\end{matrix}~.
\end{align}

When $\varphi^0 = 1$ and $\varphi^i = \varphi$ for $i \neq 0$, the manifold admits a $\IZ_{5} \times \IZ_2 \subset S_5 \times \IZ_2$ symmetry group, which acts freely outside of the singular locus $\IDelta = 0$. The actions of $\IZ_5$ and $\IZ_2$ on the coordinates can be written as
\begin{align} \label{eq:symmetry_actions_HV}
    \mathfrak{A}~:~X_i \mapsto X_{i+1}~, \qquad \mathfrak{B}~: X_i~ \mapsto \frac{1}{X_i}~,
\end{align}
with the indices understood mod 5. The Hodge numbers of the varieties obtained by taking the quotients are given in \tref{tab:Hodge_Numbers_HV}.
\begin{table}[H]
    \vskip20pt
	\renewcommand{\arraystretch}{1.3}
	\begin{center}
		\begin{tabular}{|c||m{45pt}|m{45pt}|m{45pt}|c|c|c|}
			\hline
			Manifold          & \hfil $\HV$ & \hfil$\HV/\IZ_5$ & \hfil $\HV/\IZ_{10}$\\[3pt] \hline
			$(h^{11},h^{12})$ & \hfil  (45,5)      & \hfil (9,1)    &  \hfil (5,1)   \\[3pt] \hline
		\end{tabular}
		\vskip10pt
		\capt{5in}{tab:Hodge_Numbers_HV}{The Hodge numbers $h^{11}$ and $h^{12}$ for the free quotients of $\HV$.}
	\end{center}
\end{table}
\subsubsection*{The mirror Hulek-Verrill manifold $\MHV$} \label{sect:MHV}
\vskip-5pt
The mirror Hulek-Verrill manifold can be defined as the vanishing locus of two polynomials corresponding to the polytopes $\nabla_1$ and $\nabla_2$ inside the ambient variety $\IP_{\nabla^*}$ associated to the triangulated polytope $\nabla^*$.

The monomials associated to the vertices of $\nabla_1$ are
\begin{align} \notag
1~, \qquad Y_i~, \qquad Y_i Y_j ~, \qquad Y_i Y_j Y_k~, \qquad Y_i Y_j Y_k Y_l~, \qquad Y_i Y_j Y_k Y_l Y_m~,
\end{align}
with distinct indices understood to take distinct values. The monomials associated to $\nabla_2$ are simply the inverses of these.
\begin{align} \notag
1~, \qquad \frac{1}{Y_i}~, \qquad \frac{1}{Y_i Y_j}~, \qquad \frac{1}{Y_i Y_j Y_k}~, \qquad \frac{1}{Y_i Y_j Y_k Y_l}~, \qquad \frac{1}{Y_i Y_j Y_k Y_l Y_m}~.
\end{align}
Looking at the vertices of $\nabla^*$ listed in appendix \ref{app:Toric_Geometry_Data}, we see that the ambient variety $\IP_{\nabla^*}$ is nothing but the product $(\IP^1)^5$. The Cox coordinates are the homogeneous coordinates on each $\IP^{1}$, which we often denote by $\IP^1_i$ with $i=0,\dots,4$ if there is a need to distinguish between different factors in the product $(\IP^1)^5$ . The coordinates $Y_i$ on the torus are identified with the affine coordinates
\begin{align} \notag
Y_i = \frac{Y_{i,1}}{Y_{i,0}},
\end{align}
with $[Y_{i,0}:Y_{i,1}]$ giving the homogeneous coordinates on $\IP^1_i$. It is convenient to intoduce the following monomials of homogeneous coordinates
\begin{align} \notag
M_{abcde} \= Y_{1,a} Y_{2,b} Y_{3,c} Y_{4,d} Y_{5,e}~,
\end{align} 
where $a,b,c,d,e \in \{0,1\}$. Using these, the most general polynomials associated to $\nabla_1$ and $\nabla_2$ can be written as
\begin{align} \label{eq:Q1Q2}
Q^1 = \sum_{a,b,c,d,e} A_{abcde} M_{abcde}~, \qquad Q^2 = \sum_{a,b,c,d,e} B_{abcde} M_{abcde}~.
\end{align}
For a special choice of coefficients $A$ and $B$, the simultaneous vanishing locus of $Q^1$ and $ Q^2$ admits $\IZ_5$, $\IZ_5 \times \IZ_2$ or $\IZ_5 \times \IZ_2 \times \IZ_2$ symmetry \cite{Candelas:2008wb}. These act freely, and thus can be used to obtain smooth quotient manifolds. Denoting the generator of the $\IZ_5$ as $S$, the generator of the first $\IZ_2$ as $U$ and the second $\IZ_2$ as $V$, we can take the symmetry transformations to act on the coordinates as
\begin{align} \label{eq:Z_5Z_2_generators}
S \, : \, Y_{i,a} \mapsto Y_{i+1,a}~, \qquad U \, : \, Y_{i,a} \mapsto (-1)^{a}Y_{i,a}~, \qquad V \, : \, Y_{i,0} \leftrightarrow Y_{i,1}~,
\end{align}
where addition is again understood modulo 5. The symmetries $S$ and $V$ can be seen to descend from the $\IZ_5$ and $\IZ_2$ symmetries acting on the polytope $\nabla^*$. To write down the polynomials invariant under there symmetries, it is convenient to introduce the $\IZ_5$ invariant combinations of the monomials~$M_{abcde}$,
\begin{align} \notag
m_{abcde} \= \sum_{i=1}^5 Y_{i,a} Y_{i+1,b} Y_{i+2,c} Y_{i+3,d} Y_{i+4,e}~.
\end{align} 
The polynomials defining the $\IZ_5$ symmetric manifolds can be found by specialising the coefficients $A$ and $B$ so that the vanishing locus $Q^1 = Q^2 = 0$ is invariant under $\IZ_5$, or equivalently by finding the $\IZ_5$ orbits of $Q^1$ and $Q^2$. In this manner, we find
\begin{align} \label{eq:Z_5_Quotient_Polynomials}
\begin{split}
Q^1 \=&A_{00000} \, m_{00000}+A_{10000} \, m_{10000}+A_{11000} \, m_{11000}+A_{10100} \, m_{10100}+A_{11100} \, m_{11100}\\
&\hskip30pt+A_{11010} \, m_{11010}+A_{11110} \, m_{11110}+A_{11111} \, m_{11111}~,\\[3pt]
Q^2 \=&B_{00000} \, m_{00000}+B_{10000} \, m_{10000}+B_{11000} \, m_{11000}+B_{10100} \, m_{10100}+B_{11100} \, m_{11100}\\
&\hskip30pt+B_{11010} \, m_{11010}+B_{11110} \, m_{11110}+B_{11111} \, m_{11111}~.
\end{split}
\end{align}
To find the defining polynomials in the $\IZ_5 \times \IZ_2$ symmetric case, we can further demand that the vanishing locus of the polynomials is invariant under the $\IZ_2$ generated by $V$, which gives us two polynomials of the form
\begin{equation} \notag
\begin{aligned}
Q^1 \= &A_0 \, m_{00000}{+}A_{1} \, m_{10000}{+}A_{2} \, m_{11000}{+}A_{3} \, m_{10100}{+}A_{4} \, m_{11100}{+}A_{5} \, m_{11010}{+}A_{6} \, m_{11110}{+}A_{7} \, m_{11111}~, \\[3pt]
Q^2 \= &A_0 \, m_{11111}{+}A_{1} \, m_{11110}{+}A_{2} \, m_{11100}{+}A_{3} \, m_{11010}{+}A_{4} \, m_{11000}{+}A_{5} \, m_{10100}{+}A_{6} \, m_{10000}{+}A_{7} \, m_{00000}~.
\end{aligned}
\end{equation}
Alternatively, we can demand that the vanishing locus is invariant under the second $\IZ_2$ generated by $U$. In this case, the polynomials can be written as
\begin{align} \notag
\begin{split}
Q^1 &\= A_0 \, m_{00000}{+}A_{1} \, m_{11000}{+}A_{2} \, m_{10100}{+}A_{3} \, m_{11110}~, \\[3pt]
Q^2 &\= B_0 \, m_{11111}{+}B_{1} \, m_{11100}{+}B_{2} \, m_{11010}{+}B_{3} \, m_{10010}~.
\end{split}
\end{align}
Note that the actions of $U$ and $V$ are exchanged under a suitable redefinition of coordinates, and therefore we can choose either of these two forms for the polynomials defining the $\IZ_5 \times \IZ_2$ invariant variety. Note also that in the latter case the polynomials $Q^1$ and $Q^2$ are not each $\IZ_2$ invariant, but instead are mapped to each other under the action on $\IZ_2$, thus keeping their mutual vanishing locus invariant.

Finally, we can consider the variety invariant under the full $\IZ_5 \times \IZ_2 \times \IZ_2$. In this case we can write the defining polynomials as
\begin{align}\label{eq:Z5Z2Z2_Variety}
\begin{split}
Q^1 &\= \frac{A_0}{5} \, m_{00000}{+}A_{1} \, m_{11000}{+}A_{2} \, m_{10100}{+}A_{3} \, m_{11110}~, \\[5pt]
Q^2 &\= \frac{A_0}{5} \, m_{11111}{+}A_{1} \, m_{11100}{+}A_{2} \, m_{11010}{+}A_{3} \, m_{10010}~.
\end{split}
\end{align}
It turns out that the varieties defined in this way and their quotients under their respective symmetry groups are smooth Calabi-Yau manifolds, which we can identify as mirror manifolds of the five-parameter family $\HV_{(a_0, \dots, a_5)}$. We call these \textit{mirror Hulek-Verrill manifolds} $\MHV$. The Hodge number of the corresponding quotient varieties were already found in \cite{Candelas:2008wb}. We reproduce these in \tref{tab:Hodge_Numbers_MHV}.
\begin{table}[H]
    \vskip20pt
	\renewcommand{\arraystretch}{1.3}
	\begin{center}
		\begin{tabular}{|c|c|c|c|c|c|c|}
			\hline
			Manifold          & $\MHV$ & $\MHV/\IZ_2$ & $\MHV/\IZ_2 {\times} \IZ_2$ & $\MHV/\IZ_5$ &  $\MHV/\IZ_5 {\times} \IZ_2$ & $\MHV/\IZ_5 {\times} \IZ_2 {\times} \IZ_2$\\[3pt] \hline
			$(h^{11},h^{12})$ &   (5,45)       & (5,25)    &  (5,15)       &  (1,9)   & (1,5) & (1,3)   \\[1pt] \hline
		\end{tabular}
		\vskip10pt
		\capt{5in}{tab:Hodge_Numbers_MHV}{The Hodge numbers $h^{11}$ and $h^{12}$ for the different free quotients that $\MHV$ allows.}
	\end{center}
\end{table}\vskip-30pt
Counting the parameters in the polynomials seems naïvely to produce too many parameters compared to the Hodge numbers. However, on taking into account rescalings; remaining automorphisms of the ambient variety $(\IP^1)^5$; and $\SL(2,\IC)$ transformation of the polynomials, we find that the number of free parameters in the defining polynomials agrees with the Hodge numbers. We leave the details to appendix \ref{app:Parameters}.

Finally, we note that this variety is birational to the singular $\wh \MHV$. This is most easily seen by observing that the intersection $\wh \MHV \cap \IT^4$ can be obtained from $\MHV$ by blowing up a suitable set of degree-1 rational lines, as we will discuss in detail in \sref{sect:Curve_Counting}.

\newpage
\section{The Periods of Hulek-Verrill Manifolds} \label{sect:Periods}
\vskip-10pt
The periods of the $\HV$ manifold are essential for understanding both the geometry and physics of the Hulek-Verrill manifolds as well as their mirrors. The series expansions of periods about large complex structure points allow for a mirror-symmetry computation of the instanton numbers for the manifold $\MHV$. In this section, We derive series expressions that we utilise to perform this computation in \sref{sect:Mirror_Map}. Additionally, the periods as functions of the complex structure moduli of $\HV$ are instrumental in describing string theory compactifications on $\HV$. We hope to return to this point in future work, to study flux vacua in type IIB string theory compactified on $\HV$.

Our approach begins with investigating some differential equations satisfied by the fundamental period $\varpi^{0}$, which is long known to admit concise descriptions \cite{hulek2005,Verrill2004SumsOS}. We find a set of PDEs which, together with asymptotic data coming from mirror-symmetry considerations, allow us to find all periods within the large complex structure regions of moduli space. We go further by using the methods of \cite{Verrill2004SumsOS} to study an ODE satisfied by the fundamental, and indeed all, periods. This latter equation is used to analytically continue the periods, and with the data we obtain from this, we can give expressions for the periods in all regions of moduli space. 

We derive also formulae that express all periods using integrals of products of Bessel functions. To our knowledge, this is the first appearance of such expressions and we anticipate that these also have applications in the study of banana graph amplitudes. For instance, the expansion (4.16) of \cite{Bonisch:2020qmm} expresses the full non-equal mass 4-loop banana integral in the large momentum region of parameter space, where the simplest available expression (their equation~(2.10)) does not converge. The authors gave the first few terms of the series expansions of the functions that are used as a basis. The integral expressions that we use to describe the periods also fit this purpose after a change of basis. Appropriate generalisations of our expressions relevant to higher-dimensional Hulek-Verrill manifolds will perform the same task for higher-loop banana diagrams.
\vskip-10pt
\subsection{Moduli space}
\vskip-10pt
The parameters $\varphi^0,\cdots,\varphi^5$ in the equation \eqref{eq:complete_intersection_HV} defining the manifold $\HV$ constitute a set of projective coordinates for $\IP^5$. The parameters $\varphi^0,\cdots,\varphi^5$ appear symmetrically, which we can use to great effect to describe different regions in the moduli space. A convenient atlas for $\IP^{5}$ is given by the six sets where one of the projective coordinates is nonvanishing. In the following sections, we mostly work in the patch where $\varphi^{0}\neq0$, but the arguments go through in the other five patches mutatis mutandis. Accordingly, the Latin subscripts $i,j,k,\dots$ are always understood to run from $1$ to $5$, whereas the Greek subscripts $\mu,\nu,\lambda,\dots$ are taken to run from $0$ to $5$.

It can be seen that the manifold $\HV$ is singular on the locus
\begin{align} \label{eq:E}
E \= \varphi^0\, \varphi^1\, \varphi^2\, \varphi^3\, \varphi^4\,\varphi^5 \= 0~.
\end{align}
We denote the irreducible components in this locus by
\begin{align} \label{eq:singular_loci_E}
E_\mu \= \big\{(\varphi^0,\varphi^1,\varphi^2,\varphi^3,\varphi^4,\varphi^5) \in \IP^5 \; \big| \; \varphi^\mu = 0 \big\}
\end{align}
The intersections of $5$ of these hypersurfaces turn out to be large complex structure points, or points of maximal unipotent monodromy, as we will verify in \sref{sect:Mirror_Map} by computing the monodromies around these hypersurfaces explicitly.

As we have reviewed earlier in \sref{sect:MHV}, the Hulek-Verrill manifold has conifold singularities on the locus
\begin{align} \label{eq:disc}
\IDelta \;\defineas\; \prod_{\epsilon_i \in \{ \pm 1 \}} \Big(\sqrt{\varphi^0} + \epsilon_1 \sqrt{\varphi^1} + \epsilon_2 \sqrt{\varphi^2} + \epsilon_3 \sqrt{\varphi^3} + \epsilon_4 \sqrt{\varphi^4} + \epsilon_5 \sqrt{\varphi^5}\Big)\=0~.
\end{align}
It is often useful to consider the square roots $\sqrt{\varphi^i}$ as coordinates on the moduli space. This of course gives rise to a multiple cover. We can, to start, choose branches for the square roots with $\Re[\sqrt{\varphi^i}]>0$. The functions that we study are related to those in other branches via monodromy transformations $\varphi^i \mapsto \me^{2 \pi \ii} \varphi^i$ around the large complex structure point. 

In the coordinates $\sqrt{\varphi^{i}}$ it is convenient to study the vanishing loci of the individual factors in $\IDelta$. Let $I$ be a subset of indices in $\{0,\dots,5\}$ and $I^c$ be its complement in $\{0,\dots,5\}$. Then we define the following closed components $D_I$ corresponding to each set $I$, sketched in \fref{fig:Moduli_Space_2D}:
\begin{align} \label{eq:D_I_loci}
\begin{split}
    D_{I} \= \left\{ (\varphi^0,\cdots,\varphi^5) \in \IP^5 \; \bigg| \; \sum_{i \in I} \Re \left[\sqrt{\varphi^i} \; \right] = \sum_{j \in I^c} \Re \left[\sqrt{\varphi^j} \; \right]\right\}.
\end{split}    
\end{align}
\begin{figure}[H]
\begin{center}
\includegraphics[width=8cm, height=8cm]{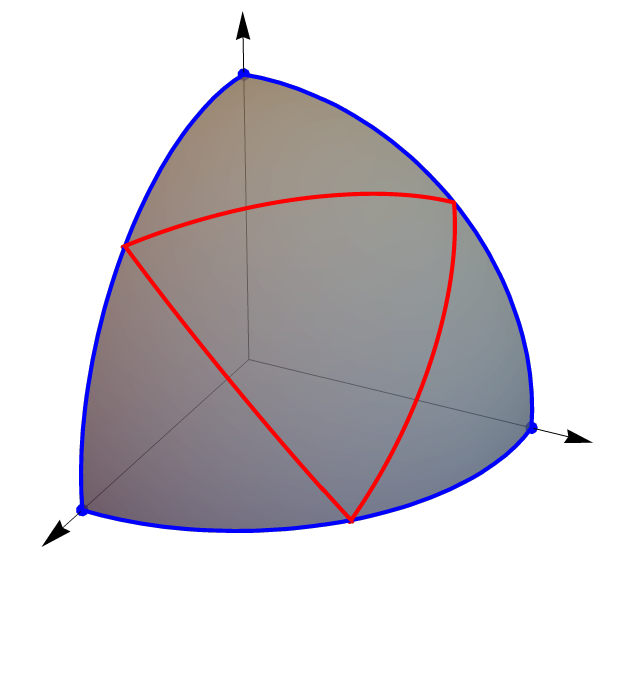}
\vskip-10pt
\capt{6in}{fig:Moduli_Space_2D}{A heuristic sketch of the moduli space in coordinates $\sqrt{\varphi^\mu}$, with the branch choice of $\sqrt{\varphi^i}>0$. The grey shell is the moduli space $\IP^2$, the red lines represent the irreducible components $D_{\{\mu\}}$ of the discriminant locus $\IDelta = 0$, the blue lines are the loci $E_{\mu}$, and the blue points are the large complex structure points. The four triangular regions between these lines correspond to the sets $U_{\{\mu\}}$ and $U_{\{0,1\}} \cap U_{\{0,2\}} \cap U_{\{1,2\}}$, which we define in \eqref{eq:sets_U_I_definition}.}		
\end{center}
\end{figure}
\subsection{The fundamental period}
\vskip-10pt
The holomorphic period for $\HV_{(\varphi^0,...,\varphi^5)}$ can be found by integrating the holomorphic three form over the torus. We briefly review this procedure. As we consider the torus, we can use the equation \eqref{eq:HVsing} defining $\wh \MHV$ in order to obtain this period by the Dwork-Katz-Griffiths method \cite{CoxKatz}.
\begin{align} \notag
\varpi^{(0);0}({\bm \varphi}) 
\;= - \frac{\varphi^0}{(2\pi\ii)^{5}} \int \prod_{i=1}^{5} \frac{\dd X_{i}}{X_{i}} 
\frac{1}{P({\bf X}; {\bm \varphi}) }
\;= - \frac{\varphi^0}{(2\pi\ii)^{5}} \int \prod_{i=1}^{5}\frac{\dd X_{i}}{X_{i}} \Bigg[\sum_{i=1}^5 \frac{\varphi^i}{X_i} \sum_{j=1}^5 X_j - \varphi^0 \Bigg]^{-1}~.
\end{align}
The first superscript $(0)$ is to remind us that we are working in the coordinate patch of $\IP^{5}$ where $\varphi^{0}=1$. The second superscript $0$ indicates that this is the holomorphic period near the large complex structure point at $\varphi^1 = \varphi^2 = \cdots = \varphi^5 = 0$, where one finds the series expansion
\begin{align} \label{eq:Fundamental_Period_Series}
\begin{split}
\varpi^{(0);0}({\bm \varphi}) &\;\= \sum_{n=0}^{\infty} \, \frac{(\varphi^{0})^{-n}}{(2\pi\ii)^{5}}\, \int \prod_{i=1}^{5}\frac{\dd X_{i}}{X_{i}}~\Bigg(\sum_{i=1}^5 \frac{\varphi^i}{X_i} \sum_{j=1}^5 X_j\Bigg)^{n}
\\[5pt]
&\= \sum_{n=0}^{\infty} \, \frac{(\varphi^{0})^{-n}}{(2\pi\ii)^{5}}
\sum_{\deg(\bm{p}) = n}
\sum_{\deg(\bm{q}) = n} \binom{n}{\bm{p}}\binom{n}{\bm{q}} 
\prod_{i=1}^{5}  \int \frac{\dd X_{i}}{X_{i}}\,  \bm{X}^{\bm{p}- \bm{q}}~ \bm \varphi^{\bm{q}}
\\[5pt]
&\=\sum_{n=0}^{\infty} \, (\varphi^{0})^{-n}\sum_{\deg(\bm{p}) = n} \binom{n}{\bm{p}}^{2}\bm \varphi^{\bm{p}} \\[5pt]
&\; \defineas \; \sum_{n=0}^\infty c_n (\bm \varphi) \, (\varphi^0)^{-n}~.
\end{split}
\end{align}
We will next identify a set of differential operators that annihilate this fundamental period, the expectation being that the other periods should satisfy the same equations. Although this set of equations is demonstrably \textit{not} the full Picard-Fuchs system, we can proceed using the high degree of symmetry and the asymptotics for the periods found from mirror symmetry considerations. In this way, we are able to find expressions for the periods using the 32 solutions to this partial Picard-Fuchs system. As a very non-trivial check, we are able to compute several genus-0 instanton numbers in \sref{sect:Mirror_Map}, the first few of which match the numbers that we find from geometric arguments~in~\sref{sect:Curve_Counting}. 

In principle one could also obtain the periods and the full Picard-Fuchs system from the toric data using the methods described in \cite{Batyrev:1993wa,Hosono:1994ax} to derive the Gel'fand-Kapranov-Zelevinski system satisfied by the fundamental period \eqref{eq:Fundamental_Period_Series}. This system could then be factored to obtain the Picard-Fuchs system. However, we have shown that it is possible to find the periods with a partial Picard-Fuchs system together with suitable boundary conditions, and therefore we do not need to use this more cumbersome procedure here.

\subsection{The ordinary differential equation obeyed by the fundamental period}\label{sect:ODE}
\vskip-10pt
Consider the sequence of $c_n$ which gives the coefficients in the series \eqref{eq:Fundamental_Period_Series}. In principle, one could use a recurrence relation that $c_{n}$ satisfies in order to write an ODE --- containing derivatives only with respect to $\varphi^{0}$, but coefficients functions of all $\varphi^{\mu}$ --- which is satisfied by the fundamental period. Such recurrence relations which themselves depend on the $\varphi^{i}$ were studied by Verrill in \cite{Verrill2004SumsOS}, wherein a method for determining such a recurrence was given. It was shown that $c_{n}$ is a holonomic sequence, solving a linear recurrence with polynomial coefficients. 

Unfortunately, the fully general case with all $\varphi^{i}$ set to indeterminates is not amenable to a computer analysis as the rational functions of the $\varphi^{i}$ that appear are prohibitively large. Nonetheless, finding this recurrence for fixed values of $\varphi^{i}$ is possible with the methods of \cite{Verrill2004SumsOS}. Although this recurrence can be used to obtain a differential equation annihilating $\varpi^{0}$, this is not generally of minimal order\footnote{The recurrence provided by this method is of lowest possible order, but without any constraint on the degree of the polynomial coefficients therein. This leads to extraneous factors in the differential equation provided.}. We get around this by using the recurrence relation to generate a large number of terms in the series efficiently, and then use these to fix a lower-degree differential equation. One example we will use later is given by the line $(\varphi^0,\dots,\varphi^5) = (1,\varphi,\varphi/{20},\dots,\varphi/{20})$, where the differential operator takes the form
\begin{align} \label{eq:L_6_operator_example}
\cL^{(6)} \= S_6 \, \theta^6 + S_5 \, \theta^5 + S_4 \, \theta^4 + S_3 \, \theta^3 + S_2 \, \theta^2 + S_1 \, \theta + S_0~,
\end{align}
with coefficients polynomials $S_n$ of degree 11.
\begin{align}
\begin{split} \notag
{\scriptstyle S_6} & {\scriptstyle \= -4393216 \, \varphi ^{11}+367906816 \, \varphi ^{10}+2766668800 \, \varphi ^9-39077007900 \, \varphi ^8+206484873000 \, \varphi ^7-612252422500 \, \varphi ^6}\\ 
&\hskip12pt{\scriptstyle +898848500000 \, \varphi^5 - 698473812500\, \varphi^4+301613125000 \, \varphi ^3-63023437500 \, \varphi ^2+1968750000 \, \varphi +781250000}~,\\
{\scriptstyle S_5} &{\scriptstyle \= -52718592 \, \varphi ^{11}+2701502528 \, \varphi ^{10}+35940053200 \, \varphi
   ^9-311032483500 \, \varphi ^8+1552596065500 \, \varphi ^7-3847452445000 \, \varphi
   ^6}\\
&\hskip12pt{\scriptstyle +3932465125000 \, \varphi ^5 -1862764937500 \, \varphi ^4+296554687500 \, \varphi
   ^3+111468750000 \, \varphi ^2-33250000000 \, \varphi -1562500000}~,\\
{\scriptstyle S_4} &{\scriptstyle \= -254806528 \, \varphi ^{11}+7499038076 \, \varphi^{10}+150742085265 \, \varphi^9-1014941685775 \, \varphi ^8+5431523295000 \, \varphi^7-11316503848750 \, \varphi^6}\\
&\hskip12pt{\scriptstyle +9307004090625 \, \varphi ^5 -4235035421875 \, \varphi ^4+739773593750 \, \varphi^3+87307812500 \, \varphi ^2+10281250000 \, \varphi +781250000}~,\\
{\scriptstyle S_3} &{\scriptstyle \= -632623104 \, \varphi ^{11}+9348961064 \, \varphi ^{10}+303965630550 \, \varphi^9-1813508252350 \, \varphi ^8+10679775875000 \, \varphi ^7-18511281897500\, \varphi ^6}\\
&\hskip12pt{\scriptstyle +13297668268750 \, \varphi ^5 -5361295718750 \, \varphi ^4+157421562500\, \varphi ^3+235818750000 \, \varphi ^2-5250000000 \, \varphi}~,\\
{\scriptstyle S_2} &{\scriptstyle \= -847890688 \, \varphi ^{11}+4174802636 \, \varphi^{10}+326073152765 \, \varphi^9-1845417676975\, \varphi ^8+11974702116500\, \varphi ^7-17568183998750\, \varphi ^6}\\
&\hskip12pt{\scriptstyle +11730618440625 \, \varphi ^5-4223423609375 \, \varphi ^4-204717031250 \, \varphi ^3+170156250000 \, \varphi ^2-937500000 \, \varphi}~,\\
{\scriptstyle S_1} &{\scriptstyle \= -579904512 \,\varphi ^{11}-1001197360 \,\varphi ^{10}+180117501740 \,\varphi^9-1000845945900 \,\varphi ^8+7134958504500 \,\varphi ^7-9085888402500 \,\varphi^6}\\
&\hskip12pt{\scriptstyle +5823431700000 \,\varphi ^5-1860707500000 \,\varphi ^4-250435000000 \, \varphi^3+61875000000 \, \varphi ^2}~,\\
{\scriptstyle S_0} &{\scriptstyle \= -158155776 \,\varphi ^{11}-992481296 \,\varphi ^{10}+40441278660 \,\varphi^9-224468019900 \,\varphi ^8+1746333878500 \,\varphi ^7-1982812512500 \,\varphi^6}\\
&\hskip12pt{\scriptstyle +1243608875000 \,\varphi ^5-349708500000 \,\varphi ^4-79745000000 \,\varphi^3+9250000000 \,\varphi ^2}~.
\end{split}
\end{align}
\subsection{Partial differential equations obeyed by the fundamental period}
\vskip-10pt
We adopt the following notation:
\begin{equation}\label{eq:thetas}
\partial_{i} \= \frac{\partial}{\partial \varphi^{i}},\qquad \theta_{i} \= \varphi^{i}\partial_{i},\qquad \Theta \= \sum_{i=1}^{5}\theta_{i}~.
\end{equation}
Note that on a single term $\bm \varphi^{\bm{p}} \, (\varphi^0)^{-n}$, where $\deg(\bm{p})=n$, the action of the operator $\Theta$ is the same as that of $-\varphi^{0} \, \partial_{0}$. Using this fact, we find that the fundamental period $\varpi^{0}$ obeys the following five differential equations:
\begin{equation} \label{eq:system2}
\cL_i \varpi^0(\bm \varphi)\; \defineas \; \left(\frac{1}{\varphi^{0}}\left(\Theta+1\right)^{2}-\frac{1}{\varphi^{i}}\theta_{i}^{2}\right)\varpi^{0}(\bm \varphi) \= 0~.
\end{equation}
These equations are, after a change of variables, equivalent to  the differential equations (4.8) of \cite{Bonisch:2020qmm}. In addition, we have equations obtained by taking differences of the above equations, or by directly inspecting \eqref{eq:fund}:
\begin{equation} \label{eq:system1}
\cL_{i,j} \varpi^0(\bm \varphi)\; \defineas \; \left(\frac{1}{\varphi^{i}}\theta_{i}^{2}-\frac{1}{\varphi^{j}}\theta_{j}^{2}\right)\varpi^{0}(\bm \varphi) \= 0~.
\end{equation}
These equations \eqref{eq:system1} are manifestly separable, which is suggestive of a route to the other periods.
\subsection{The Frobenius method}\label{sect:frobenius}
\vskip-10pt
We prove that the fundamental period $\varpi^{0}(\bm \varphi)$ is (up to scale) the sole holomorphic power series solution to \eqref{eq:system2}. Make the projective choice $\varphi^{0}=1$, and then suppose that one has a series 
\begin{equation} \notag
f(\bm \varphi) \= \sum_{\bm{p}} f_{\bm{p}} \, \bm \varphi^{\bm{p}}~    
\end{equation} 
that solves \eqref{eq:system2}. After comparing powers of $\varphi^{i}$ in $\cL_{1}\,f(\bm \varphi)=0$, one obtains
\begin{equation} \label{eq:rel}
p_{1}^{2}\,f_{(p_{1},p_{2},p_{3},p_{4},p_{5})} \= (p_{1}+p_{2}+p_{3}+p_{4}+p_{5})^{2}\, f_{(p_{1}-1,p_{2},p_{3},p_{4},p_{5})}~,\qquad n\geq1.
\end{equation}
There is a similar relation obtained from the other four equations $\cL_{i}\,f(\bm \varphi)=0$. Together these five relations \eqref{eq:rel} require
\begin{equation} \label{eq:solLi}
f_{\bm{p}} \= \left(\frac{\left(p_{1}+p_{2}+p_{3}+p_{4}+p_{5}\right)!}{p_{1}!p_{2}!p_{3}!p_{4}!p_{5}!}\right)^2 \; f_{\bm{0}}~.
\end{equation}
While the system of equations $\cL_{i}F=0$ has a unique holomorphic solution, it is shown below that the system has a solution-space of dimension greater than ${\text{dim }H^{3}\left(\HV\right)=12}$. Therefore it cannot be the entire Picard-Fuchs system, since it is not sufficiently constrained. The solution space can be suitably restricted by the differential equation discussed in \sref{sect:ODE}, but this is too difficult to write down in full generality. Perhaps more simply, one can appeal to the method of Frobenius and the homological interpretation discussed in \sref{sect:frob_homology}. Equivalently, one can fix the system of 12 periods by imposing boundary conditions consistent with mirror symmetry. 

The Frobenius method reveals that there are 32 functions, sums of power series multiplied by logarithms of the $\varphi^{i}$, which solve \eqref{eq:system2}. To see this, one sets up an indicial equation. With $f_{\bm{p}}$ from \eqref{eq:solLi} and $f_{\bm{0}}=1$, take a solution ansatz 
\begin{equation} \label{eq:frob_ansatz}
\varpi^{\bm{\e}}(\bm \varphi) \= \sum_{\bm{p}} \frac{f_{\bm{p}+\bm{\epsilon}}}{f_{\bm{\epsilon}}} \, \bm \varphi^{\bm{p}+\bm{\e}} ,
\end{equation}
where the $\bm{\e} = (\epsilon_1,\dots,\epsilon_5)$ is a five-component multi-index consisting of as-yet undetermined algebra elements, which we refer to here as Frobenius elements, and $f_{\bm{p}+\bm{\epsilon}}$ is defined by replacing $x!$ by $\Gamma(1+x)$ in \eqref{eq:solLi}. One can compute
\begin{equation} \label{eq:eps2}
\cL_{i}\,\varpi^{\bm{\epsilon}}(\bm \varphi)\=-\frac{\epsilon_{i}^{2}}{\varphi^{i}}\bigg(1+\cO(\mathbf{a})\bigg)+\cO(\epsilon_{i}^{3})~.
\end{equation}
We recover the original series solution $\varpi^{0}(\bm \varphi)$ by taking $\bm{\epsilon} \to \bm{0}$. Additionally, there are new solutions obtained by first differentiating $\varpi^{\bm{\epsilon}}(\bm \varphi)$ once with respect to any number of the $\epsilon_{i}$ and then taking $\bm{\epsilon} \to \bm{0}$. Each such derivative introduces a logarithmic dependence on the $\varphi^{i}$. There are five $\epsilon_{i}$ with respect to which we can either differentiate zero or one times. In total all such choices give us $2^{5}=32$ independent solutions. 

The Taylor expansion of $\varpi^{\bm{\epsilon}}$ in $\bm{\epsilon}$ is infinite, but only terms not containing the square of an $\epsilon_{i}$ have coefficients that are solutions of \eqref{eq:system2}. Introduce the algebraic relation $\epsilon_{i}^{2}=0$ for the $\epsilon_{i}$. The Taylor expansion of $\varpi^{\bm{\epsilon}}$ then truncates, and every coefficient left is a solution to \eqref{eq:system2}.
\begin{equation} \label{eq:pi_epsilon_series}
\varpi^{\bm{\epsilon}}=\varpi^{0}+\varpi^{i}\epsilon_{i}+\frac{1}{2!}\varpi^{ij}\epsilon_{i}\epsilon_{j}+\frac{1}{3!}\varpi^{ijk}\epsilon_{i}\epsilon_{j}\epsilon_{k} + \frac{1}{4!} \varpi^{ijkl} \epsilon_i \epsilon_j \epsilon_k \epsilon_l + \frac{1}{5!}\varpi^{ijklm}\epsilon_i \epsilon_j \epsilon_k \epsilon_l \epsilon_m~.
\end{equation}
In the above implicit sums, the terms with any two indices equal are absent. To obtain the 12 periods among these solutions, we will later in \sref{sect:frob_homology} impose that $\epsilon_i$ satisfy cohomology relations. Equivalently, we can impose boundary conditions consistent with mirror symmetry. Explicit expressions for the coefficients in the above series are obtained by differentiating \eqref{eq:frob_ansatz} and then taking $\epsilon_{i} \to 0$.

\subsection{Separation of variables}
\vskip-10pt
Upon expanding the operators $\theta_i$, the differential equations $\cL_{i,j} F = 0$ become
\begin{equation} \label{eq:sep}
\left[\partial_{i}-\partial_{j}+\varphi^{i}\partial_{i}^{2}-\varphi^{j}\partial_{j}^{2}\right]F \= 0~.
\end{equation}
Making a separation-of-variables ansatz $F(\mathbf{a})=\prod_{j=1}^{5}G_{j}(\varphi^{j})$ and simplifying $\cL_{i,j} F = 0$, one obtains
\begin{equation} \notag
\frac{\partial_i G_{i}(\varphi^{i})+\varphi^{i}\, \partial_i^2 G_{i}(\varphi^{i})}{G_{i}(\varphi^{i})} \= \frac{\partial_j G_{j}(\varphi^{j})+\varphi^{j}\, \partial_j^2 G_{j}(\varphi^{j})}{G_{j}(\varphi^{j})}~.
\end{equation}
Employ the traditional separation of variables logic: both sides of this equation respectively depend only on $\varphi^{i}$ and $\varphi^{j}$, and so both must equal a constant. With a certain prescience, we denote this constant by $z^{2}/{4}$. Attention should then be turned to the ordinary differential equation that the $G_i$ satisfy:
\begin{equation} \notag
x\,\frac{\dd^{2}}{\dd x^{2}}G(x)+\frac{\dd}{\dd x}G(x)\=\frac{z^{2}}{4}G(x).
\end{equation}
This has the following general solution: 
\begin{equation} \notag
G(x)\=C_{1}(z)I_{0}(z\sqrt{x})+C_{2}(z)K_{0}(z\sqrt{x})~.
\end{equation}
$C_{1}(z) $ and $C_{2}(z)$ are arbitrary functions of the parameter $z$. Therefore, for any choice of the functions $C_{1}(z),\,C_{2}(z),$ the equations $\cL_{i,j}F=0$ for $i,j=1,\dots,5$ have solutions of the form 
\begin{equation} \label{eq:sepsol}
F(\bm \varphi)\=\int \text{d}z\,C(z)\prod_{j=1}^{5}B_{j}\Big(z\sqrt{\varphi^{j}}\Big)~,
\end{equation}
where the five functions $B_{j}$ are each taken to be modified Bessel functions $I_{0}$ or $K_{0}$. This brings us closer to the periods, but at this stage of our reasoning, only looking at the system $\cL_{i,j}F=0$, there is still a considerable degree of ignorance as to what the function $C$ should be and which combinations of these solutions we should take to give the periods.

In the regime
\begin{align} \notag
\text{Re}\left[\sum_{i=1}^{5}\sqrt{\varphi^{i}}\right]<\text{Re}\left[\sqrt{\varphi^{0}}\right]
\end{align}
the following expression for the fundamental period is valid:
\begin{equation} \label{eq:Bessel_pi0}
\varpi^{(0);0}(\bm \varphi) \= \varphi^{0}\int_{0}^{\infty}\!\!\dd z \, z \, K_0\left(\sqrt{\varphi^0} z \right) \prod_{i=1}^5 I_0 \left(\sqrt{\varphi^i}z \right)~. 
\end{equation}
We give a proof of this claim in appendix \ref{app:Bessels}. The identity \eqref{eq:Bessel_pi0} suggests that $C(z)$ should be taken to be $K_0(\sqrt{\varphi^0} z)$. Indeed, by replacing the $I_{0}$ functions in the above integral with $K_{0}$ functions, we can form 32 functions $f$ that obey the equations $\cL_{i,j}\,f=0$. These 32 functions can be seen to satisfy the system $\cL_{i}f=0$, and therefore must furnish a basis of series solutions of the system $\cL_i f = \cL_{i,j} f = 0$. To be sure, the 32 functions obtained in this way have powers series that form a basis for the linear span of the 32 Frobenius solutions given by the construction in \sref{sect:frobenius}.

On symmetry grounds, there will be a role for functions obtained by replacing the $K_{0}$ with an $I_{0}$ in patches $\varphi^i = 1$ in the moduli space. The reason for this is that, from the global perspective, $\varphi^{0}$ is not distinguished from the $\varphi^{i}$.
\subsection{Determining closed form expressions for all periods}
\vskip-10pt
We have seen that the partial Picard-Fuchs system given by \eqref{eq:system1} and \eqref{eq:system2} should have exactly 32 solutions. Furthermore, we have seen that the integrals of Bessel functions of the form 
\begin{equation} \label{eq:Bessel_gen2}
\frac{\varphi^0}{\ii \pi} \int_{0}^{\infty}\!\! \dd z \, z \, B_0(\sqrt{\varphi^0} z) \prod_{i=1}^5 B_{i}(\sqrt{\varphi^i}z)
\end{equation}
furnish a set of solutions to our partial differential equations. The $B_{\mu}(\sqrt{\varphi^\nu}\; z)$ above are replaced by a conveniently normalised modified Bessel function: either $K_{0}(\sqrt{\varphi^\mu}\; z)$ or $\ii \pi\; I_{0}(\sqrt{\varphi^\mu}\; z)$. Naïvely it seems that this would give us 64 solutions. However, not all of these converge simultaneously. Indeed, an integral of this form converges in the region of the moduli space where
\begin{align} \label{eq:Bessel_Integeral_Convergence_Criterion}
    \sum_{\mu = 0}^5 \epsilon_\mu \Re \left[\sqrt{\varphi^\mu} \, \right] < 0~,
\end{align}
where $\epsilon_\mu = \pm1$, and the negative sign is chosen when $B_0(z \sqrt{\varphi^\mu}) = K_0(z \sqrt{\varphi^\u})$, and the positive sign when $B_0(z \sqrt{\varphi^\mu}) = \ii\pi I_0(z \sqrt{\varphi^\mu})$. This follows from demanding that the product of Bessel functions decays exponentially in the limit $z \to \infty$ and recalling the asymptotics of the Bessel functions for large $z$:
\begin{align} \notag
K_0(z) \; \sim \; \sqrt{\frac{\pi}{2z}} \; \me^{-z}~, \; \qquad \; \ii\pi I_0(z) \; \sim \; \ii\sqrt{\frac{\pi}{2  z}}\; \me^{z}~.
\end{align}
The boundary between the different regions of convergence is exactly the restriction of the conifold locus $\IDelta = 0$ to the real plane. 

On a generic\footnote{In addition to the restriction of the discriminant locus to the real plane, the Bessel function integrals also diverge on points whose real parts satisfy the equation \eqref{eq:Bessel_Integeral_Convergence_Boundary}.} point in the moduli space corresponding to a non-singular manifold, there are exactly 32 convergent integrals of Bessel functions of the form \eqref{eq:Bessel_gen2}. This is seen as follows: every curve of the form 
\begin{align} \label{eq:Bessel_Integeral_Convergence_Boundary}
    \Re \left[\sqrt{\varphi^0} \, \right] + \sum_{i = 1}^5 \epsilon_i \Re \left[\sqrt{\varphi^i} \, \right]\= 0~
\end{align}
divides the space into two regions, those `above' and `below'. The curve itself belongs to the discriminant locus. There is exactly one Bessel function integral of the form \eqref{eq:Bessel_gen2} that converges almost everywhere above the curve \eqref{eq:Bessel_Integeral_Convergence_Boundary} and exactly one converging almost everywhere below the curve. As there are 32 such curves we find exactly 32 convergent integrals at any given point. We can find an almost\footnote{The open sets cover the moduli space apart from points which satisfy \eqref{eq:Bessel_Integeral_Convergence_Boundary}.} open covering, where every open subset of the covering corresponds to a different set of Bessel functions.

We can express these covering sets as intersections of suitably-defined sets $U_I$. Let $I$ be a set of indices in $\{0,\dots,5\}$ and $I^c$ be its complement in $\{0,\dots,5\}$. Then we define open sets in the moduli space corresponding to each set $I$:
\begin{align} \label{eq:sets_U_I_definition}
\begin{split}
    U_{I} \= \left\{ (a_0,\cdots,a_5) \in \IP^5 \; \bigg| \; \sum_{i \in I} \Re \left[\sqrt{\varphi^i} \, \right] > \sum_{j \in I^c} \Re \left[\sqrt{\varphi^j} \, \right]\right\}.
\end{split}    
\end{align}
These have the following convenient properties
\begin{align} \notag
\begin{split}
    U_I \subset U_J \quad \text{if} \quad J \subset I \subset \{0,\dots,5\}~, \qquad \qquad U_I^c \= U_{I^c}\setminus \Re \IDelta~,
\end{split}
\end{align}
where $\Re \IDelta$ denotes the space of all points that satisfy any of the equations \eqref{eq:Bessel_Integeral_Convergence_Boundary}.

In the subset of each patch where they converge, these Bessel function integrals satisfy the partial differential equations \eqref{eq:system1} and \eqref{eq:system2}. There are exactly 32 solutions to these equations, so it follows that the periods, which should solve the differential equations, can be expressed in terms of the convergent Bessel function integrals in any patch. 
In the next subsection we will present an argument, based on known asymptotics, to fix the periods as sums of these Bessel integrals in the regions $U_{\{i\}}$ and $U_{\{0\}}$. To find the correct linear combinations of these integrals to give the periods in other regions we study the ODE of \sref{sect:ODE}. Choosing values $\varphi^{i} = s_i \varphi$ in this ODE gives a differential equation that the restrictions of the periods to these lines must satisfy. Given enough lines, we can always find enough equations to completely fix the periods in terms of the Bessel integrals.

To find the relation between the bases of periods in different patches, we analytically continue the Bessel integrals from one region to another. In practice, the easiest way to do this is to numerically integrate the Picard-Fuchs equation along a line crossing multiple regions, and then find the relations between each pair of bases. By the normalisation of the Bessel function integrals, these matrices relating different bases are integral. In what follows, we will not need most of these relations, hence we do not record them here. However, an important special case that we will be using relates the basis of periods near the large complex structure point in the patch $U_{\{0\}}$ to the basis in the patch $U_{\{i\}}$, where there is another large complex structure point.

For instance, we can study the line $(1,\varphi,\frac{\varphi}{20},\dots,\frac{\varphi}{20})$ where the periods satisfy the Picard-Fuchs equation $\cL^{(6)} f = 0$, with the operator $\cL^{(6)}$ given by \eqref{eq:L_6_operator_example}. The Bessel function integrals near $\varphi^i=0$ that satisfy this equation are given by
\begin{align} \notag
    \wh{\pi}^{(0)} \= \frac{1}{\ii \pi} \int_{0}^\infty \dd z \,\, \frac{z}{\varphi} \,\,\, \left(
\begin{array}{c}
\cA_0 \cB_1 \cB^4 \\
\cB_0 \cB_1 \cA^4\\
4\, \cB_0 \cA_1 \cB \cA^3\\
6\, \cB_0 \cA_1 \cB^2 \cA^2\\
12\, \cB_0 \cB_1 \cB \cA^3 + 12 \, \cB_0 \cA_1 \cB^2 \cA^2 \\
4 \cB_0 \cA_1 \cB^3 \cA + 6 \cB_0 \cB_1 \cB^2 \cA^2
\end{array}
\right),
\end{align}
where we have used the following shorthand for the Bessel functions appearing here
\begin{equation} \notag
\begin{aligned}
\cA_0 \= \ii \pi I_0\left( \varphi^{-1/2} z \right)~, \quad \cB_0 &\= K_0\left(  \varphi^{-1/2} z \right)~, \quad &\cA_1 &\= \ii \pi I_0(z)~, \quad \cB_1 \= K_0(z)~,\\[5pt]
\cA &\= \ii \pi I_0\left(20^{-1/2}z\right)~, \quad &\cB &\= K_0\left(20^{-1/2}z\right)~.
\end{aligned}
\end{equation}
On the line, the discriminant locus $\IDelta = 0$ has singularities at five points:
\begin{align} \notag
\varphi \;\simeq\; 0.2786~, \quad \varphi \;\simeq\; 0.4775~, \quad \varphi \= 1~, \quad \varphi \;\simeq\; 3.2725~, \quad \text{and} \quad \varphi \;\simeq\; 89.7214~.
\end{align}
The region $|\varphi| > 89.7214$ lies in the region $U_{\{1\}}$, which contains the large complex structure point at $\varphi^0 = \varphi^2 = \varphi^3 = \varphi^4 = \varphi^5 = 0$. By symmetry, we can deduce that the Bessel function integrals giving a basis of solutions to the Picard-Fuchs equation $\cL^{(6)} f = 0$ are
\begin{align} \notag
     \wh{\pi}^{(1)} \= \frac{1}{\ii \pi} \int_{0}^\infty \dd z \,\, \frac{z}{\varphi} \,\,\, \left(
\begin{array}{c}
\cB_1 \cA_0 \cA^4 \\
\cB_1 \cB_0 \cA^4\\
4\, \cB_1 \cA_0 \cB \cA^3\\
6\, \cB_1 \cA_0 \cB^2 \cA^2\\
12\, \cB_1 \cB_0 \cB \cA^3 + 12\, \cB_1 \cA_0 \cB^2 \cA^2\\
4 \cB_1 \cA_0 \cB^3 \cA + 6 \cB_1 \cB_0 \cB^2 \cA^2
\end{array}
\right).
\end{align}
Given the operator $\cL^{(6)}$ it is indeed easy to check that these integrals satisfy the equation.

By integrating the Picard-Fuchs operator $\cL^{(6)}$ numerically, we can find the continuation of the period vector $\wh{\pi}^{(0)}$ to the region $|\varphi| > 89.7214$, giving the following relation between the vectors $\wh{\pi}^{(0)}$ and $\wh{\pi}^{(1)}$:
\begin{align} \label{eq:pi0_pi1_relation}
 \wh{\pi}^{(0)} \=  \wh{\text{T}} \, \wh{\pi}^{(1)}~, \qquad \text{with} \qquad \wh{ \text{T}} \= \left(
\begin{array}{rrrrrr}
 1 & 0 & 0 & 0 & 0 & 0 \\
 0 & 1 & 0 & 0 & 0 & 0 \\
 -6 & -3 & -1 & -3 & 0 & -6 \\
 -4 & 0 & 0 & -1 & 0 & -4 \\
 -4 & -3 & -2 & -3 & -1 & -4 \\
 0 & 0 & 0 & 0 & 0 & -1 \\
\end{array}
\right).
\end{align}
We have written the Bessel function integrals in $\wh{\bm{\pi}}^0$ and $\wh{\bm{\pi}}^1$ in this particular way because these are natural restrictions of the 12 periods to the line $(a_0,\dots,a_5) \= (1,\varphi,\frac{\varphi}{20},\dots,\frac{\varphi}{20})$. The generic 12-component period vectors are given by
\begin{equation} \label{eq:period_bib}
\pi^{{(0)}} \= \frac{\varphi^0}{\ii \pi} \int_0^\infty \dd z \, z \,\, \left(\pi^{(0);0},\pi^{(0);1},\dots,\pi^{(0);5},\pi^{(0)}_{1},\dots,\pi^{(0)}_{5},\pi^{(0)}_0\right)^T~,
\end{equation}
in which
\begin{equation} \notag
    \begin{aligned}
    \pi^{(0);0} &\= \cB_0 \cA_1 \cA_2 \cA_3 \cA_4 \cA_5~, \quad& \pi^{(0);i} &\= \cB_0 \cB_i \prod_{j \neq i} \cA_j~, \\[5pt]  \pi^{(0)}_{i} &\= \sum_{\substack{m<n\\m,n \neq i}} \cB_0 \cB_m \cB_n \prod_{j \neq m,n} \cA_j~, \quad& \pi^{(0)}_0 &\= \sum_{\substack{l<m<n}} \cB_0 \cB_l \cB_m \cB_n \prod_{j \neq l,m,n} \cA_j~,\\[5pt]
    \mathcal{A}_{\mu}&\=\ii \pi \; I_{0}\left(\sqrt{a_{\mu}} \; z\right)~,\quad& \mathcal{B}_{\mu}&\=K_{0}\left(\sqrt{a_{\mu}} \; z\right)~.
    \end{aligned}
\end{equation}
The vector $\pi^{(1)}$ is given by interchanging the indices $0$ and $1$. In terms of these quantities, restricted to the line, we have a natural way of writing the relations \eqref{eq:pi0_pi1_relation} in a symmetric form. For example, the relation corresponding to the third row of the matrix can be written as
\begin{align} \notag
\pi^{(0);2} + \pi^{(0);3} + \pi^{(0);4} + \pi^{(0);5}\= -4 \pi^{(1);0} - 4 \pi^{(1);1} - \pi^{(1);2} - \pi^{(1);3} - \pi^{(1);4} - \pi^{(1);5}~.
\end{align}
The coordinates $\varphi^2$, $\varphi^3$, $\varphi^4$, and $\varphi^5$ must appear symmetrically in all of these relations. Thus we are able to guess that the relations in the case where all of the coordinates are unequal are
\begin{align} \notag
\pi^{(0);j} \= - \pi^{(1);0} - \pi^{(1);1} - \pi^{(1);j}~.
\end{align}
We can verify this expectation by studying the line $(\varphi^0,\dots,\varphi^5) = (1,\varphi,\frac{\varphi}{50},\frac{\varphi}{100},\dots,\frac{\varphi}{100})$, which singles out the period $\pi^{(0);2}$, and allows us to verify the above relation in the case $j=2$. The other relations then follow by symmetry. Working in this way, we find that in general the period vectors $\pi^{(0)}$ and $\pi^{(1)}$ are related by
\begin{align} \label{eq:change_of_basis_01}
\pi^{(0)} \= \text{T}_{\pi^{(0)} \pi^{(1)}} \pi^{(1)}~, \qquad \text{with} \qquad \text{T}_{\pi^{(0)} \pi^{(1)}} \= \left(
\begin{smallarray}{rrrrrrrrrrrr}
 -1 & 0 & 0 & 0 & 0 & 0 & 0 & 0 & 0 & 0 & 0 & 0 \\
 0 & 1 & 0 & 0 & 0 & 0 & 0 & 0 & 0 & 0 & 0 & 0 \\
 1 & -1 & -1 & 0 & 0 & 0 & 0 & 0 & 0 & 0 & 0 & 0 \\
 1 & -1 & 0 & -1 & 0 & 0 & 0 & 0 & 0 & 0 & 0 & 0 \\
 1 & -1 & 0 & 0 & -1 & 0 & 0 & 0 & 0 & 0 & 0 & 0 \\
 1 & -1 & 0 & 0 & 0 & -1 & 0 & 0 & 0 & 0 & 0 & 0 \\
 -6 & 6 & 3 & 3 & 3 & 3 & \+ 1 & -1 & -1 & -1 & -1 & 0 \\
 -3 & 3 & 0 & 2 & 2 & 2 & 0 & -1 & 0 & 0 & 0 & 0 \\
 -3 & 3 & 2 & 0 & 2 & 2 & 0 & 0 & -1 & 0 & 0 & 0 \\
 -3 & 3 & 2 & 2 & 0 & 2 & 0 & 0 & 0 & -1 & 0 & 0 \\
 -3 & 3 & 2 & 2 & 2 & 0 & 0 & 0 & 0 & 0 & -1 & 0 \\
 4 & -4 & -3 & -3 & -3 & -3 & 0 & 1 & 1 & 1 & 1 & -1 \\
\end{smallarray}
\right)~.
\end{align}
\subsection{The Frobenius elements as cohomology}\label{sect:frob_homology}
\vskip-10pt
It has long been known that the Frobenius elements $\epsilon_i$ have an interpretation in the (co-)homology of the mirror manifold $\MHV$ \cite{Hosono:1994ax,Braun:2007vy}. These relations are instructive in the present case, and we work through this here. The Frobenius elements $\epsilon_i$ satisfy the relations 
\begin{align} \label{eq:Frobenius_element_relations}
\epsilon_i \epsilon_j \= \epsilon_j \epsilon_i~, \qquad \epsilon_i^2 \= 0~, \qquad \epsilon_i \epsilon_j \epsilon_k \= Y_{ijk} \eta~, \qquad \eta \,\epsilon_i \= 0~.
\end{align}
The elements $\epsilon_i$ can be understood as cohomology two-forms, or as their duals, which are four-surfaces. If we think of $\epsilon_i$ as four-surfaces, then the products corresponds to intersections, the $Y_{ijk}$ are the intersection numbers, and $\eta$ is a point. If we think of $\epsilon_i$ as two-forms, the products correspond to the wedge product, and $\eta$ is the volume form. As surfaces, the intersection $\epsilon_i \epsilon_j$ is a curve. These are not all independent, since we know that $b^2(\MHV)=5$. We can choose a basis $\mu^j$ to be dual to the $\epsilon_i$ so that we have the relations 
\begin{align} \label{eq:indicial_algebra}
\epsilon_i \mu^j \= \delta_i^j \eta~, \qquad \epsilon_i \epsilon_j \= Y_{ijk} \mu^k~.
\end{align}
The form of the second relation is dictated by the intersection relation in \eqref{eq:Frobenius_element_relations}. Now the Frobenius period is
\begin{align} \label{eq:frobenius_period}
\varpi \= \varpi^0 + \varpi^{i}\, \epsilon_i + \varpi_{i}\, \mu^i + \varpi_0\, \eta~.
\end{align}
Comparing with \eqref{eq:pi_epsilon_series}, we obtain expressions for $\varpi^{i}$, $\varpi_{i}$, and $\varpi_{0}$. The $\epsilon_i$ have also explicit representations as $12 \times 12$ matrices, which follow from the monodromy matrices given in \eqref{eq:Monodromy_M_E_1}
\begin{align} \notag
\text{M}_{E_j} \= \text{I}_{12} + 2\pi\ii \, \epsilon_j~,
\end{align}
with I$_{12}$ the identity matrix. It is a pleasure to check the indicial algebra above for these matrices.
\subsection{The periods near large complex structure points}
\vskip-10pt
The set $U_{\{0\}}$ is a neighbourhood of the large complex structure point at $E_1 \cap \cdots \cap E_5$, and the $U_{\{i\}}$ are neighbourhoods of other large complex structure points. In the region $U_{\{0\}}$, according to the discussion above, the convergent integrals are of the form. 
\begin{equation} \label{eq:Bessel_gen3}
\frac{\varphi^0}{\ii \pi}\int_{0}^{\infty} \dd z \, z \, K_0 \Big(\sqrt{\varphi^0} z \Big) \prod_{i=1}^5 B_{i} \Big(\sqrt{\varphi^i}z \Big)~.
\end{equation}
A basis for the periods can be given as 12 linear combinations of these functions\footnote{Recall that for a Calabi-Yau threefold $X$, $\text{dim } H^{3}(X)=2h^{2,1}+2.$}. To fix the precise combinations, we compare the equation \eqref{eq:frobenius_period} with the result of applying the relations \eqref{eq:indicial_algebra} to the expansion \eqref{eq:pi_epsilon_series}. It is only a matter of direct comparison to express the coefficients of the $\epsilon$-series \eqref{eq:frobenius_period} in terms of Bessel integrals. The relation between the periods in the Bessel integral basis $\pi^{(\mu)}$ and the periods in the Frobenius basis $\varpi^{(\mu)}$ is
\begin{align} \label{eq:period_varpi}
\varpi^{(\mu)} \= \text{T}_{\varpi \pi} \pi^{(\mu)}, \qquad \text{with} \qquad 
\text{T}_{\varpi \pi} \= \left(
\begin{matrix*}[r]
 \frac{1}{\pi^4}~ & \bm 0^T~~ & \bm 0^T & 0~~ \\[3pt]
\bm 0~~ & -\frac{2 \ii}{\pi^3} \II~~ & \zero~~ & \bm 0~~ \\[3pt]
-\frac{4}{\pi^2} \bm 1 & \zero~~~ & -\frac{8}{\pi^2} \II & \bm 0~~ \\[3pt]
 80 \frac{\zeta(3)}{\pi^4} &-\frac{4 \ii}{\pi} \bm 1^T & \bm 0^T & \frac{16 \ii}{\pi}~ \\[3pt]
\end{matrix*}
\right).
\end{align}
Explicitly, this means that the single-logarithm periods near the large complex structure point at $\varphi^1 = \dots = \varphi^5 = 0$ are given by
\begin{equation} \label{eq:Bessel_pi1}
\varpi^{(0);j}(\bm \varphi)\= -2 \varphi^{0}\int \dd z \, z \, K_0 \Big(\sqrt{\varphi^0} z \Big) K_{0} \Big(\sqrt{\varphi^{j}}z \Big)\prod_{i\neq j} I_0 \Big(\sqrt{\varphi^i}z \Big)~.
\end{equation}
For the double-logarithm periods, we have
\begin{equation} \notag
\varpi^{(0)}_{j}(\bm \varphi) \= 8 \varphi^{0}\int \dd z \, z \sum_{\substack{m<n\\m,n\neq j}} \, K_0 \Big(\sqrt{\varphi^0} z \Big) K_{0} \Big(\sqrt{\varphi^{m}}z \Big)K_{0} \Big(\sqrt{\varphi^{n}}z \Big)\prod_{i\neq m,n} I_0 \Big(\sqrt{\varphi^i}z \Big) -4\pi^{2}\varpi^{(0)}(\bm \varphi)~.
\end{equation}
The period that is cubic in logarithms is
\begin{equation}\begin{split} \label{eq:Bessel_pi3}
\varpi^{(0)}_{0}(\bm \varphi)&\={-}16\sum_{l<m<n} \varphi^{0}\int \dd z \, z \, K_0 \Big(\sqrt{\varphi^0} z \Big) K_{0} \Big(\sqrt{\varphi^{l}}z \Big) K_{0}\Big(\sqrt{\varphi^{m}}z \Big) K_{0}\Big(\sqrt{\varphi^{n}}z \Big)\prod_{i\neq l,m,n} I_0\Big(\sqrt{\varphi^i}z \Big)\\[7pt]
&\hskip40pt-4\pi^{2}\sum_{k=1}^{5}\varpi^{(0);k}(\bm \varphi)+80\zeta(3)\;\varpi^{(0);0}(\bm \varphi)~.
\raisetag{1.2cm}\end{split}
\end{equation}
\subsubsection*{Series expansions}
\vskip-5pt
We collect some expressions below that are used to express the periods as series. Denote by $H_{n}^{(r)}$ the $n^{th}$ harmonic number of order $r$:
\begin{equation} \notag
H_{n}^{(r)}\=\sum_{k=1}^{n}\frac{1}{k^{r}}~,\qquad H_{n}\=H_{n}^{(1)}~.
\end{equation}
We express the periods using the following intermediate series:
\begin{equation} \notag
\begin{aligned}
h^{(0)}_{i}(\bm \varphi)&=\sum_{n=0}^\infty\;\sum_{\deg(\bm{p})=n} 2\left(H_{n}{-}H_{p_{i}}\right) \binom{n}{\bm{p}}^2 \bm \varphi^{\bm{p}} \; (\varphi^0)^{-n}~,\\[5pt]
h^{(0)}_{ij}(\bm \varphi)&=\sum_{n=0}^\infty\;\sum_{\deg(\bm{p})=n} \left[4\left(H_{n}{-}H_{p_{i}}\right)\left(H_{n}{-}H_{p_{j}}\right)-2H^{(2)}_{n}\right]\binom{n}{\bm{p}}^2 \bm \varphi^{\bm{p}} \; (\varphi^0)^{-n}~,\\[5pt]
h^{(0)}_{ijk}(\bm \varphi)&=\sum_{n=0}^\infty\;\sum_{ \deg(\bm{p})=n}\left[8\left(H_{n}{-}H_{p_{i}}\right)\left(H_{n}{-}H_{p_{j}}\right)\left(H_{n}{-}H_{p_{k}}\right)-4\left(3H_{n}{-}H_{p_{i}}{-}H_{p_{j}}{-}H_{p_{k}}\right)H^{(2)}_{n}\right.\\[5pt]&\left.\hskip100pt+4H^{(3)}_{n}\right]\binom{n}{\bm{p}}^2 \bm \varphi^{\bm{p}} \; (\varphi^0)^{-n}~.
\end{aligned}
\end{equation}
The periods \eqref{eq:Bessel_pi1}-\eqref{eq:Bessel_pi3} can be expressed near the point $\varphi^i = 0$ in terms of the above series. These results can be derived by considering the Frobenius expansion \eqref{eq:pi_epsilon_series}, but we also sketch a derivation from the Bessel function integrals in appendix \ref{app:Bessels}.
\begin{equation}\begin{split} \label{eq:period_series}
\varpi^{(0);j}(\bm \varphi) &\=\! \varpi^{(0);0}({\bm \varphi})\log\frac{\varphi^{j}}{\varphi^{0}}+h^{(0)}_{j}(\bm \varphi)~, \\[10pt]
\varpi^{(0)}_{j}(\bm \varphi)&\=\! 2\sum_{\substack{m<n\\m,n\neq j}}\left[\varpi^{(0);0}(\bm \varphi)\log\frac{\varphi^{m}}{\varphi^{0}}\;\log\frac{\varphi^{n}}{\varphi^{0}}+h^{(0)}_{n}(\bm \varphi)\log\frac{\varphi^{m}}{\varphi^{0}}+h^{(0)}_{m}(\bm \varphi)\log\frac{\varphi^{n}}{\varphi^{0}}+h^{(0)}_{mn}(\bm \varphi)\right]~,\\[10pt]
\varpi^{(0)}_{0}(\bm \varphi)&\=\!2\sum_{l<m<n}\left[\varpi^{(0);0}(\bm \varphi)\log\frac{\varphi^{l}}{\varphi^{0}}\;\log\frac{\varphi^{m}}{\varphi^{0}}\;\log\frac{\varphi^{n}}{\varphi^{0}}\right.\\[5pt]&\hskip20pt\left.+h^{(0)}_{n}(\bm \varphi)\log\frac{\varphi^{l}}{\varphi^{0}}\;\log\frac{\varphi^{m}}{\varphi^{0}}+h^{(0)}_{l}(\bm \varphi)\log\frac{\varphi^{m}}{\varphi^{0}}\;\log\frac{\varphi^{n}}{\varphi^{0}}+h^{(0)}_{m}(\bm \varphi)\log\frac{\varphi^{n}}{\varphi^{0}}\;\log\frac{\varphi^{l}}{\varphi^{0}}\right.\\[5pt]&\hskip20pt\left.+h^{(0)}_{mn}(\bm \varphi)\log\frac{\varphi^{l}}{\varphi^{0}}+h^{(0)}_{lm}(\bm \varphi)\log\frac{\varphi^{n}}{\varphi^{0}}+h^{(0)}_{nl}(\bm \varphi)\log\frac{\varphi^{m}}{\varphi^{0}}+h^{(0)}_{lmn}(\bm \varphi)\right]~.
\raisetag{3.3cm}
\end{split}\end{equation}
\newpage
\section{Mirror Map and Large Complex Structure}\label{sect:Mirror_Map}
\vskip-10pt
To determine the mirror map, we recall that near the large complex structure limit the period vector takes the form \cite{Candelas:1990pi}
\begin{align} \label{eq:Pi0}
\bm\Pi \= \begin{pmatrix}
\cF_0\\[3pt]
\cF_i\\[3pt]
z^0\\[3pt]
z^i
\end{pmatrix}~, \qquad i\=1,\ldots,5~,\qquad \cF_{\mu}=\frac{\partial\cF}{\partial z^{\mu}}~.
\end{align}
Here $z^i$ are the projective coordinates on the K\"ahler moduli space of $\MHV_{(\varphi^1, \dots \varphi^5)}$. We often use the corresponding affine coordinates $t^i \defineas z^i/z^0$, so that for example the complexified Kähler class of $\MHV_{(\varphi^1, \dots \varphi^5)}$ is given by
\begin{align} \notag
B + \ii J \= \sum_{i=1}^{5}t^i \,\me_i~,
\end{align}
where $\me_i$ generate the second integral cohomology $H^2(\MHV,\IZ)$. The quantities $\cF_0$ and $\cF_i$ are derivatives of the prepotential $\cF$, which near the large complex structure point is given in terms of the genus-0 instanton numbers $n_{\bm{p}}$ by
\begin{align} \notag
\cF(z^0, \dots, z^5) = -\frac{1}{3!} \sum_{a,b,c=0}^5 Y_{abc} \frac{z^a z^b z^c}{z^0} + (z^0)^2 \sum_{\bm{p} \neq \bm{0}} n_{\fp} \, \text{Li}_3(\bm{q}^{\bm{p}})~,\qquad q_{i}\; \defineas \; \exp(2\pi\ii \, t^{i}).
\end{align}
The $Y_{abc}$ are given by topological quantities related to $\MHV$:
\begin{equation}
\begin{aligned} \notag
Y_{ijk} &\=\int_{\MHV} \me_i \wedge \me_j \wedge \me_k~, & \qquad Y_{ij0} &\; \in \; \left\{0,\frac{1}{2} \right\}~, \\[10pt] Y_{i00} &\;= -\frac{1}{12} \int_{\MHV} c_2(\MHV) \wedge \me_i~, &\qquad Y_{000} &\;= -3 \chi(\MHV) \frac{\zeta(3)}{(2\pi \ii)^3}~.
\end{aligned}
\end{equation}
To find the triple intersection numbers $Y_{ijk}$, we first note that $\me_i \wedge \me_i = 0$ for every $i$. Therefore the only non-vanishing triple intersection numbers are those with all indices different. To find these numbers, we recall that $\me_i$ is dual to a hypersurface $\{Y_i - y_i = 0\} \subset \MHV$, where $y_i$ is a constant. The intersection of two of these hyperplanes gives an elliptic curve, which in turn intersects a third hyperplane generically in two points. Therefore the $Y_{ijk}$ are given by
\begin{align} \label{eq:yijk}
Y_{ijk} \= \int_{\MHV} \me_i \wedge \me_j \wedge \me_k \= \begin{cases}
2 \qquad &\text{for}~i,\,j,\,k\;\;\text{distinct,}\\
0~\,       &\text{otherwise.}
\end{cases}
\end{align}
For the quantities $Y_{i00}$, we need to compute the second Chern class of $\MHV$. Applying the adjunction formula gives the total Chern class as
\begin{align} \notag
c(\MHV) \= \frac{\prod_{r=1}^5 (1+\me_r)^2}{(1+\sum_{r=1}^5 \me_r)^2}~.
\end{align}
From this we can verify the Calabi-Yau condition $c_1(\MHV) = 0$, and find that the second Chern class $c_2(\MHV)$ can be written as
\begin{align} \notag
c_2(\MHV) \= 2 \sum_{r < s} \me_s \wedge \me_r~.
\end{align}
Integrating this against $e_i$ and using the integral computed in \eqref{eq:yijk} gives
\begin{align} \notag
Y_{i00} \;\=-\frac{1}{12} \int_{\MHV} c_2(\MHV) \wedge \me_i \;= -2~.
\end{align}
Naïvely, the numbers $Y_{ij0}$ would equal $\int_{\MHV} c_1(\MHV) \wedge \me_i \wedge \me_j$ and so vanish. This argument is not correct, and in fact $Y_{ij0}$ can in some cases take the value $1/2$. Based on the gamma class~\cite{Halverson_2014}, it is expected that in the one-parameter case one can take $Y_{110} = 0$ exactly when $Y_{111}$ is even. On the quotient $\MHV/\IZ_5$ the triple intersection number $Y_{111}$ is 24, so $Y_{110} = 0$. The five-parameter prepotential is related to the prepotential for one-parameter manifolds essentially by setting ${t^1 = \ldots = t^5 = t}$ and dividing by $5$. It follows that the quantities $Y_{ij0}$ can in fact be taken to vanish.  
\begin{align} \notag
Y_{ij0} \= 0~.
\end{align}
As we know the Hodge numbers $h^{11} = 5$ and $h^{12}=45$ of $\MHV$, the Euler characteristic is immediately given by $\chi(\MHV) = 2(h^{11}-h^{12}) = -80$. So the last quantity $Y_{000}$ is given by
\begin{align} \notag
Y_{000} \= 240 \frac{\zeta(3)}{(2\pi \ii)^3}~.
\end{align}
Note that as a consequence of the highly symmetric nature of the manifold $\MHV$, none of the couplings depend on the particular values of the indices $i,j,k$, apart from the fact that they need to be distinct for the triple intersection numbers to be nonvanishing. It is then convenient to write the non-vanishing quantities $Y_{abc}$ as 
\begin{align} \notag
Y_{ijk} \; \defineas \; Y~, \qquad \qquad Y_{i00} \; \defineas \; Y_{00}~.
\end{align}
The large complex structure points are located on loci where all but one of the parameters $\varphi^i$ vanish. For concreteness, we are going to concentrate on the large complex structure point at $\varphi^1 = \ldots = \varphi^5 = 0$ in the affine patch $\varphi^0 = 1$. We denote the integral period vector in this patch by~$\Pi^{(0)}$. The other cases are related to this one by the permutation symmetry.

The affine coordinates $t^i$ of the Kähler moduli space are related to the periods $\varpi$ by
\begin{align} \notag
t^i \= \frac{1}{2\pi \ii} \frac{\varpi^{i}}{\varpi^{0}} \= \frac{1}{2\pi \ii} \log \varphi^i + \cO(\bm \varphi)~.
\end{align}
The second relation gives the asymptotic form in the limit $\varphi^1,\dots,\varphi^5 \to 0$, and $\cO(\bm \varphi)$ denotes terms that are of order 1 or higher in any $\varphi^i$. Inverting this map order-by-order one finds the coordinates $\varphi^i$ in terms of $t^i$. It is useful to write the resulting map in terms of the elementary symmetric polynomials\footnote{Due to the identity $q_{1}^{5}-q_{1}^{4}\sigma_{1}+q_{1}^{3}\sigma_{2}-q_{1}^{2}\sigma_{3}+q_{1}\sigma_{4}-\sigma_{5}=0$, this expression is not unique. Unique expressions are obtained, for example, by using this identity to eliminate occurrences of $\sigma_{5}$, or explicit appearances of powers of $q_{1}$ higher than four.} $\sigma_i(\bm{q})$:
\begin{align} \notag
\varphi^i &\= q_i \Big[1 - \big(2 \sigma_1+2 q_i\big)+\Big(\sigma_1^2{+}2 \sigma_2 {-} 2\sigma_1 q_i {+} q_i^2\Big) - \Big(2 \sigma_1 \sigma_2 {+} 14\sigma_3{-}\left(16 \sigma_2 {+} 2 \sigma_1^2\right) q_i{+}10
\sigma_1 q_i^2{-}12 q_i^3\Big)\\ \nonumber
&+\Big(\sigma_2^2{+}26 \sigma_1 \sigma_3 {-} 174 \sigma_4+\left(2 \sigma_1^3 {-} 22 \sigma_2
\sigma_1 {+} 130 \sigma_3\right) q_i + \left(18 \sigma_1^2 {-} 136 \sigma_2\right) q_i^2 + 116 \sigma_1 q_i^3-136 q_i^4\Big)\\ \nonumber
&+\Big({-} 12 \sigma _3 \sigma _1^2 {+} 192 \sigma_4
\sigma _1 {-} 28 \sigma _2 \sigma _3+\left(4 \sigma _2 \sigma_1^2 {-} 132 \sigma_3 \sigma_1 {+} 28 \sigma_2^2 {-} 1376 \sigma_4\right) q_i \\ \nonumber
&+ \left({-}10 \sigma_1^3{+}122 \sigma_2 \sigma_1{+}1346 \sigma_3\right)q_i^2 + \left({-}128 \sigma _1^2{-}1328 \sigma_2\right) q_i^3+1488 \sigma_1
q_i^4-1350 q_i^5 \Big) \Big] + \cO(\bm{q})^7
\end{align}
Near this large complex structure point the periods in the Frobenius basis have the asymptotic~form
\begin{align}\label{eq:varpi_asymptotics}
\begin{pmatrix}
\varpi^{(0);0}\\[5pt]
\varpi^{(0);i}\\[5pt]
\varpi^{(0)\phantom{;0}}_i\\[5pt]
\varpi^{(0)\phantom{;0}}_{0}
\end{pmatrix}
= 
\begin{pmatrix}
1 \\[5pt]
\log \varphi^i\\[5pt]
\frac{1}{2!}Y_{imn} \log \varphi^m \log \varphi^n\\[5pt]
\frac{1}{3!}Y_{lmn} \log \varphi^l \log \varphi^m \log \varphi^n
\end{pmatrix}
+ \cO(\bm \varphi \log^3 \bm \varphi)
= 
\begin{pmatrix}
1\\[5pt]
2\pi \ii\, t^i\\[5pt]
\frac{(2\pi \ii)^2}{2}\, Y_{imn} t^m t^n\\[5pt]
\frac{(2\pi \ii)^3}{6}\,Y_{lmn} t^l t^m t^n
\end{pmatrix} + \cO(\bm t^3 \bm{q})\, ,
\end{align}
where repeated indices are summed over. On the other hand, the asymptotics of $\bm\Pi^0$ can be read directly from the prepotential and are given~by
\begin{align}\notag
\Pi^{(0)} 
\= \begin{pmatrix}
\cF_0\\[5pt]
\cF_i\\[5pt]
z^0\\[5pt]
z^i
\end{pmatrix} 
\= z^0 \begin{pmatrix}
Y \displaystyle\sum_{l<m<n} t^l t^m t^n - \frac{1}{2} Y_{00} \sum_n t^n - \frac{1}{3} Y_{000}\\[17pt]
-Y\displaystyle \sum_{\substack{m<n\\m,n \neq i}} t^m t^n - \frac{1}{2} Y_{00}\\[20pt]
1\\[5pt]
t^i
\end{pmatrix} + \cO(\bm{q}).
\end{align}
By requiring that the asymptotic forms match\footnote{Note that we have identified $z^{0}=\varpi^{(0);0}$, which has asymptotics $1+O(\bm{\varphi})$.}, we find that the period vectors must be related~by
\begin{align} \label{eq:change_of_basis_Pi_varpi}
\Pi^{(0)} \=  \text{T}_{\Pi^{(0)} \varpi^{(0)}}\, \varpi^{(0)} 
\= \rho\, \nu^{-1} \varpi^{(0)},
\end{align} 
with matrices 
\begin{align} \nn
\rho \= \left(\begin{array}{clcc}
-\frac{1}{3}Y_{000} & \bm 1^T  & \+\bm 0^T & \+1~~\\[2pt]
\bm 1 & \zero & \!\!-\II & \bm 0\\[2pt]
\+1~~ & \bm 0^T & \+\bm 0^T & \+0~~\\[2pt]
\bm 0& \II & \zero & \bm 0
\end{array}\right)
\qquad\text{and}\qquad
\nu \= \text{diag}(1,\,(2\pi \ii) \bm 1,\, (2\pi \ii)^2 \bm 1,\, (2\pi \ii)^3 )~.
\end{align}
\subsection{Yukawa couplings and instanton numbers}
\vskip-10pt
To find the instanton numbers, we compute the Yukawa couplings
\begin{align} \notag
y_{IJK} \;= -\int_{\MHV} \Omega \wedge   \frac{\partial^{3}\, \Omega}{\partial \varphi^{I} \partial \varphi^J \partial \varphi^{K}}~,
\end{align}
where the indices $I,J,K$ run from 1 to 5. The couplings can be computed using the relation between forms on the manifold $\MHV$ and the ring of defining polynomials modulo the Jacobian ideal~\cite{Candelas:1987se}. Alternatively, one can find $y_{ijk}$ as a series in $\bm{q}$ by a direct computation. As we are mostly interested in finding the instanton numbers, the latter method is sufficient. We express the Yukawa couplings in terms of the period vectors as 
\begin{align} \notag
y_{IJK} \;= - \Pi^{(0)\,T}\, \Sigma\, \partial_{IJK} \Pi^{(0)} 
\;= - \varpi^{(0)\,T} \, \nu^{-1} \rho^T \Sigma \rho \, \nu \, \partial_{IJK}\, \varpi^{(0)}~,
\end{align}
where $\Sigma$ is the matrix giving the standard symplectic inner product
\begin{align} \notag
\Sigma \= \begin{pmatrix}
\+0 & \II\\
-\II & 0\\
\end{pmatrix}.
\end{align} 
Using the period formulae, we start by computing $y_{IJK}$ to order 20 in the variables $\varphi^{i}$. It is expected that the Yukawa couplings are rational functions of the moduli $a$. A natural first guess for the denominator is the discriminant $\IDelta$ as given in \eqref{eq:disc}, which is a polynomial of order 16. In fact, we find that
\begin{equation} \notag
y_{IJK}=\frac{P_{IJK}}{\varphi^{I}\varphi^{J}\varphi^{K}\IDelta}~,
\end{equation}
where the $P_{IJK}$ are degree-14 polynomials in the moduli. Fully expanded, $P_{123}$ is a sum of 11628 monomials while $P_{111}$ and $P_{112}$ each have 8568. Given this size, and the fact that we have no essential need of them in what follows, we omit their display. Simpler expressions can be arrived at by exploiting symmetries. For example, $P_{111}$ is symmetric in $\varphi^{2},\,\varphi^{3},\,\varphi^{4},\,\varphi^{5}$. This allows one to write $P_{111}$ as a degree-4 polynomial in $\varphi^{1}$ with coefficients that are polynomials in the five elementary symmetric polynomials in the five $\varphi^i$. The resulting expression contains 730 monomials when fully expanded. Further simplification is obtained by reintroducing the sixth variable $\varphi^{0}$. This reduces $P_{111}$ to an expression comprising of 212 monomials. Similar simplifications are possible for $P_{112}$ and $P_{123}$.

We then express the Yukawa coupling in terms of the quantities $q_i$. The $y_{ijk}$ above is computed in the gauge $z^0 = \varpi^{(0);0}$. To be able to compare this to the expression \eqref{eq:Yukawa_coupling_instantons} we need to transform to the gauge $z^0 =1$ in addition to the tensor transformation:
\begin{align} \label{eq:Yukawa_coupling_transformation}
y_{ijk} \= \frac{(2\pi\ii)^{3} }{(\varpi^{(0);0})^2}\,q_{i}q_{j}q_{k}\,\sum_{I,J,K} \frac{\partial \varphi^I}{\partial q_i}\frac{\partial \varphi^J}{\partial q_j}\frac{\partial \varphi^K}{\partial q_k}\; y_{IJK}~.
\end{align}
Owing to the symmetries, there are only three independent Yukawa couplings up to permutation of coordinates. These can be taken to be $y_{111},\,y_{112}$ and $y_{123}$. For the purposes of finding the instanton numbers, we need only one of these, say $y_{111}$, this being somewhat simpler to compute. Expressing it as series in $\bm{q}$, we find 
\begin{align} \nonumber
\begin{split}
y_{111} &\= 24 q_1 \bigg[1 + \sigma_1 + \frac{1}{3} \left(-14 q_1 \sigma _1+17 q_1^2+14 \sigma _2\right) + \left(-36 q_1^3 + 37 q_1^2 \sigma _1-38 q_1 \sigma _2+\sigma _1 \sigma _2+43 \sigma _3 \right)\\[3pt]
&\hskip15pt + \left(-36 q_1^3 \sigma _1+37 q_1^2 \sigma _1^2-2 q_1 \left(19 \sigma _1 \sigma _2+3 \sigma _3\right)+\sigma _2^2+44 \sigma _1 \sigma _3+306 \sigma _4  +\frac{312 \sigma _5}{q_1} \right) + \cO(\bm{q}^5)\bigg]~.
\end{split}
\end{align}
Similar expressions hold for $y_{112}$ and $y_{123}$. The series expansions for the Yukawa couplings can be written in terms of the instanton numbers as
\begin{align} \label{eq:Yukawa_coupling_instantons}
y_{ijk} \= Y_{ijk} + \sum_{\bm{p}}n_{\bm{p}}\, p_i p_j p_k \,  
\frac{ \bm{q}^{\bm{p}}}{1-\bm{q}^{\bm{p}}}~.
\end{align}
By comparing this to the series expansion \eqref{eq:Yukawa_coupling_transformation}, we can identify the instanton numbers up to degree 29 as listed in appendix~\ref{app:Instanton_Numbers}. 
\subsection{Genus-1 instanton numbers}
\vskip-10pt
It is possible \cite{Bershadsky:1993ta} to define a genus-1 prepotential, which effectively counts the genus-1 curves. In the topological limit it can be expressed as
\begin{align}\notag
F_1 \= \log \left[ \left(\varpi^{(0);0}\right)^{-\left(3+h^{11}(\MHV) - \chi(\MHV)/12\right)}\, \det \left( \frac{\partial \bm \varphi}{\partial \bm{t}} \right) f \right] + \text{const.}~,
\end{align}
where $f$ is a holomorphic function which can be fixed by imposing appropriate boundary conditions. In particular, the prepotential $F_1$ must be regular at the points inside the Kähler moduli space corresponding to nonsingular manifolds. In the large complex structure limit, $F_1$ has an expansion~\cite{Gopakumar:1998jq}
\begin{align} \label{eq:F_1}
F_1 \= 2\pi \ii \sum_{i=1}^5 Y_{i00}\, t^i - 2 \sum_{\bm{p}} \left( d_{\bm{p}}  + \frac{1}{12} n_{\bm{p}}  \right)\text{Li}_{1}\left(\bm{q}^{\bm{p}}\right)+ \text{const.}
\end{align}
where $d_{\bm{p}}$ are the genus-1 instanton numbers\footnote{The reader should be aware that we are computing $F_1$ as in \cite{Gopakumar:1998jq}, which generates counts of BPS wrappings of M2-branes. This prescription differs from that of \cite{Bershadsky:1993ta}, which was used in \cite{Candelas:2019llw}, which instead counts primitive elliptic curves.}.

To get the correct growth in the large complex structure limit, $f$ must contain a factor of $\left(\prod_i \varphi^i\right)^{-3}$. Outside the locus $\varphi^\mu = 0$,  $F_1$ can be singular only on the discriminant locus \eqref{eq:disc}.
These considerations fix the form of the holomorphic ambiguity $f$, up to a multiplicative constant, as
\begin{align} \notag
f \= \frac{\IDelta^c}{\left(\prod_{i=1}^5 \varphi^i\right)^3}~.
\end{align}
In the one-parameter cases, where the singularities appear as points $\varphi_*$ in the moduli space, conifold singularities produce a factor of $(\varphi-\varphi_*)^{-1/6}$. We assume that a straightforward generalisation of this holds in the multiparameter case, and thus we take $c=-\frac{1}{6}$. With this choice we find the genus-1 instanton numbers up to degree 29. These are recorded in appendix~\ref{sect:genus1instantons}.

\subsection{Relations between the instanton numbers\texorpdfstring{: from $S_5$ to $S_6$ and beyond}{}}\label{sect:fromS5toS6}
\vskip-10pt
Inspecting the tables of appendix~\ref{app:Instanton_Numbers}, it is striking that there are many recurrences of the instanton numbers. For example, 24, 80, and many other numbers repeat. In fact,~if 
\begin{align} \label{eq:duality_relation}
\bm{I}\=(i,\,j,\,k,\,l,\,m) \quad \text{and} \quad \wt{\bm{I}}\=(\deg(\bm{I})-2i,\,j,\,k,\,l,\,m)~,
\end{align}
and all the components of $I$ and $\wt{I}$ are nonnegative, where
\begin{align} \label{eq:degree}
\deg(\bm{I})\=i+j+k+l+m
\end{align}
is the {\it degree} of $\bm{I}$, then
\begin{align}\label{eq:nDuality}
n_{\bm{I}} \= n_{\wt{\bm{I}}}~.
\end{align}
This identity has many ramifications, and a proper understanding of these devolves upon a discussion of what we term the \textit{web of indices} related to $\bm{I}$, $W[\bm{I}]$, obtained by composing the operations $\bm{I}\mapsto\wt{\bm{I}}$ with permutations, in all possible ways.
We note in passing that the only $\bm{I}$ in our tables such that $\wt{\bm{I}}$ has a negative entry is $(1,0,0,0,0)$, and its permutations.
We discuss these matters is detail in \sref{sect:webs}. Our immediate aim however is to discuss the origins of this group of symmetries.

The symmetry \eqref{eq:duality_relation}, which we will here refer to as a duality, has its origin in the presentation \eqref{eq:TwoPolynomials} of the Hulek-Verrill manifold. In this presentation the six coordinates $\varphi^\m$, $\m=0,1,\ldots,5$ are projective coordinates that enter on an equal footing. We subsequently passed to five affine coordinates $\varphi^i$ by setting $\varphi^0\,=\,1$. Clearly we could have chosen coordinates by setting any of the $\varphi^\m$ to unity and using the other $\varphi^\m$ as coordinates. Now we have taken the large complex structure point to be the point $\varphi^0 \= 1$, $\bm \varphi\,=\,(1,0,0,0,0,0)$, and a consequence of the symmetry of the presentation \eqref{eq:TwoPolynomials} is that there are in fact six large complex structure points, corresponding to taking one of the $\varphi^\m$ to be unity and the other $\varphi^\m$ to vanish. We could equally expand the Yukawa coupling about any of these points. We will study the effect of this below, but it is this choice, coupled with the fact that the form of the instanton expansion must remain invariant, that enforces~\eqref{eq:nDuality}.  

After making such a change in choice of large complex structure point, the new period vector is obtained using the transition matrix $\text{T}_{\Pi^{(0)} \Pi^{(1)}}$ given in \eqref{eq:defn_T_Pi^0Pi^1}. This matrix can be seen to act on the complex structure moduli space coordinates as 
\begin{align} \notag
z^0 \mapsto -z^{0}~, \qquad z^1 \mapsto z^1~, \qquad z^i \mapsto -z^i -z^1 - z^0 \quad i \neq 0,1~.  
\end{align}
This can be stated equivalently in terms of the quantities $t^i$ or $q_i$:
\begin{align} \label{eq:scale_inv}
t^1 \mapsto -t^1~, \quad t^i \mapsto t^i + t^1 + 1~; \qquad \text{or equivalently} \qquad q_1 \mapsto q_1^{-1}~, \quad q_i \mapsto q_i q_1~; \quad i \neq 1~.
\end{align}
We can combine these transformations with monodromies around the loci $E_i$ to find a simpler form for them. Consider the following matrix
\begin{align} \notag
\text{M}_{E_5}^{-1} \text{M}_{E_4}^{-1} \text{M}_{E_3}^{-1} \text{M}_{E_2}^{-1} \text{T}_{\Pi^{(0)} \Pi^{(1)}} \= -\left(\begin{array}{llllll} 
1 & \+0 &  \+\bm 0_4^T & \+0 & \+0 & \+\bm 0_4^T  \\
0 & -1 & \+\bm 1_4^T& \+0 & \hskip5pt 12 & \+\bm 0_4^T \\
\bm 0_4& \+\bm 0_4 & \+\hskip2pt\II_{4} &  \+\bm 0_4 & \+\bm 0_4 & \+0 \\
0 & \+0 & \+\bm 0_4^T & \+1 & \+0 & \+\bm 0_4^T \\
0 & \+0 & \+\bm 0_4^T & \+0 & -1 & \+\bm 0_4^T\\
\bm 0 & \+\bm 0 & \+\zero_4 & \+\bm 0_4 & \+\bm 1_4 & \+\hskip2pt\II_4
\end{array} \right),
\end{align}
where $\bm 0_4 = (0,0,0,0)^T$, $\bm 1 = (1,1,1,1)^T$, $\zero_4$ is a $4 {\times} 4$ zero matrix, and $\II_4$ a $4 {\times} 4$ identity matrix.
and the other matrices obtained by swapping the second and $(i+2)$'th column and row and the seventh and $(i+7)$'th column and row with each other.

Rather than discussing the Yukawa coupling to see the consequences of these symmetries on instanton numbers, it  is simpler to consider the prepotential, bearing in mind the symplectic symmetries. The genus-0 prepotential is
\begin{equation} \label{eq:prepot}
\cF^{(0)}(\bm t)\;=-\big(z^0\big)^2\left(\frac{1}{6}Y_{ijk}\,t^{i}t^{j}t^{k}+\frac{1}{2}Y_{i00}\,t^{i}-\frac{\zeta(3)}{2(2\pi\ii)^{3}}\,\chi(\MHV) +\frac{1}{(2\pi\ii)^{3}}\sum_{\bm{p}}n_{\bm{p}} \, \text{Li}_{3}\!\left(q^{\bm{p}}\right)\right)~.   
\end{equation}
The sum is over \textit{positive} multi-indices $\bm{p}$, where by \textit{positive} we mean all entries are nonnegative and at least one is positive. We now make a change of coordinates,
\begin{equation} \label{eq:changes}
t^{i}\mapsto\wt{t}^{i}~, \qquad \wt{t}^{1}\=-t^{1}~,\quad \wt{t}^{i}\=t^{i}+t^{1}~,\quad i>1~,
\end{equation}
which differs from the transformation \eqref{eq:scale_inv} by a constant term $\wt t^i \mapsto t^i - 1$, amounting to an action of monodromy transformation, which simplifies the following analysis.

Writing $\wt{q}_{i}=\ee^{2\pi\ii\wt{t}^{i}}$, $\bm{p}=(i,j,k,l,m)$, and $q^{\bm{p}}=q_{1}^{i}q_{2}^{j}q_{2}^{k}q_{4}^{l}q_{5}^{m}$, we have $q^{\bm{p}}=\wt{q}^{\wt{\bm{p}}}$, with $\wt{I}$ obtained from $I$ by \eqref{eq:duality_relation}. Writing the prepotential $\cF^{(0)}(\bm t)$ in terms of the coordinates $\wt t^i$, we find
\begin{equation}\label{eq:prepot2}
\cF^{(0)}(\bm t) \= \cF^{(0)}\big( \,\bm{\wt t} \, \big) - 4 \left(\wt t^1\right)^3 - 2 \wt t^1~.
\end{equation}
The cubic and linear terms account for the difference in the perturbative parts of the prepotentials.

In making the change of coordinates \eqref{eq:changes}, we have given the expansion of the prepotential around a different large complex structure point to that of \eqref{eq:prepot}. Due to the $S_6$ symmetry, the functional form of the prepotential should be, up to an effect of a symplectic transformation, invariant under the change of coordinates from $t^i$ to $\wt t^i$. In other words, the functional dependence of the left-hand side of \eqref{eq:prepot2} on $\bm t$ should be the same as the functional dependence of the right-hand side of \eqref{eq:prepot2} on $\bm{\wt t}$. Requiring this, we obtain
\begin{equation}\label{eq:uptosymplectic1}
4t_{1}^{3}+2t_{1}-\frac{1}{(2\pi\ii)^{3}}\sum_{\bm p}n_{\bm p}\text{Li}_{3}\!\left(q^{\wt{\bm p}}\right) \simeq -\frac{1}{(2\pi\ii)^{3}}\sum_{\bm p}n_{\bm p}\text{Li}_{3}\!\left(q^{\bm p}\right)~,
\end{equation}
with the equality holding up to a symplectic transformation. One now needs to rewrite the sum on the left-hand side. The map \eqref{eq:duality_relation} is an involution, i.e.  $\wt{\wt{\bm p}}=\bm p$, but the sums above are over positive~$\bm p$. The only $\bm p$ such that $\wt{\bm p}$ has a negative entry is $(1,0,0,0,0)$. Therefore, we have an equality
\begin{equation} \label{eq:sum}
\sum_{\bm p>0}n_{\bm p}\text{Li}_{3}\!\left(\bm q^{\wt{\bm p}}\right) = \left[\sum_{\substack{\bm p \text{ for which}\\ \bm p,\,\wt{\bm p}>0}}n_{\wt{\bm p}}\,\text{Li}_{3}\!\left(\bm q^{\bm p}\right) +n_{(1,0,0,0,0)}\,\text{Li}_{3}\!\left(q_{1}\right)\right]+n_{(1,0,0,0,0)}\left(\text{Li}_{3}\!\left(q_{1}^{-1}\right)-\text{Li}_{3}\!\left(q_{1}\right)\right).
\end{equation}
Substituting \eqref{eq:sum} into \eqref{eq:uptosymplectic1} one obtains
\begin{equation} \label{eq:uptosymplectic2}
4t_{1}^{3}+2t_{1} \simeq \frac{1}{(2\pi\ii)^{3}}n_{(1,0,0,0,0)}\left(\text{Li}_{3}\!\left(q_{1}^{-1}\right)-\text{Li}_{3}\!\left(q_{1}\right)\right)-\frac{1}{(2\pi\ii)^{3}}\sum_{\substack{\bm p \text{ for which}\\ \bm p,\,\wt{\bm p}>0}}(n_{\bm p}-n_{\wt{\bm p}})\text{Li}_{3}\!\left(\bm q^{\bm p}\right)~,
\end{equation}
This imposes that $n_{\bm p}=n_{\wt{\bm p}}$ for positive $\bm p$ for which $\wt{\bm p}$ is also positive. The left hand side can then be uniquely cancelled by taking $n_{(1,0,0,0,0)}=24$, in light of the identity
\begin{equation} \notag
\frac{24}{(2\pi \ii)^{3}}\bigg(\text{Li}_{3}\!\left(\ee^{-2\pi \ii t_{1}}\right)-\text{Li}_{3}\!\left(\ee^{2\pi\ii t_{1}}\right)\bigg)\=2t_{1}-6t_{1}^{2}+4t_{1}^{3}~,\qquad \text{Im}[t_{1}]>0~,
\end{equation}
and the fact that the quadratic term $-6t_{1}^{2}$ can be removed by a change of symplectic integral basis of $H^3(\HV,\IZ)$. It is interesting that we have `derived' the instanton number $n_{(1,0,0,0,0)}=24$ purely on the grounds of symmetry, without performing a mirror computation or curve-counting. Equation \eqref{eq:uptosymplectic2} then imposes \eqref{eq:nDuality}, which applies for all positive $\bm p$. This includes all $\bm p$ in our tables besides $(1,0,0,0,0)$ and its permutations. 

Analogous relations can be seen to hold for the genus-1 instanton numbers that we compute. Indeed, the above argument hold \textit{mutatis mutandis} for higher-genus prepotentials implying that this duality continues to hold for all higher-genus numbers.
\subsection{Recovering the results on quotient manifolds}
\vskip-10pt
Using these results, instanton numbers $\wh n_s$ and $\wh d_s$ on the quotient manifolds $\MHV_{\IZ_5}$ and $\MHV_{\IZ_{10}}$ can be recovered. The first few instanton numbers for the one-parameter manifolds are reproduced from \cite{Candelas:2019llw} in \tref{tab:instanton_numbers}. 

We fix attention here on the $\IZ_{5}$ quotient. The action of the $\IZ_5$ on the cohomology $H^2(\MHV,\IZ)$ is given by
\begin{align} \notag
\me_{i} \mapsto \me_{i+1}~,
\end{align}
where the index is understood to take values in $\IZ_5$, and we have taken this action to be consistent with the choice \eqref{eq:symmetry_actions_HV} for the action of $\IZ_5$ on $\HV$. This also induces an action on $H^4(\MHV,\IZ)$ via Hodge duality. The $\IZ_{5}$ action on the periods of $\HV$ is
\begin{align} \notag
\varpi^{0} \mapsto \varpi^{0}~, \qquad
\varpi^{i} \mapsto \varpi^{i+1}~, \qquad \varpi_{i} \mapsto \varpi_{i+1}~, \qquad
\varpi_{0} \mapsto \varpi_{0}~.
\end{align}
The locus of $\IZ_5$ symmetric Hulek-Verrill manifolds is $\varphi^1 = \dots =\varphi^5 \defineas \varphi$, and the corresponding mirror manifolds are found on the locus $t^1 = \dots = t^5 \defineas t$. Thus one identifies the generator of the second cohomology of the one-parameter manifold with
\begin{align} \notag
\me \= \me_1 + \me_2 + \me_{3} + \me_{4} + \me_{5}~.
\end{align}
The prepotential $\wh \cF$ on the one-parameter family is identified with that of the five-parameter family~by
\begin{align} \notag
\wh\cF(t) \= \frac{1}{5} \, \cF(t,t,t,t,t)~.
\end{align}
Indeed, this agrees with the following simple computation of the classical Yukawa coupling for the $\IZ_5$ quotient:
\begin{align} \notag
\wh Y_{111} \= \int_{\MHV / \IZ_5} \me \wedge \me \wedge \me \=  \frac{1}{5} \sum_{i,j,k = 1}^5 \int_{\MHV} \me_i \wedge \me_j \wedge \me_k \= \frac{1}{5} \sum_{i,j,k = 1}^5 Y_{ijk} \= 24~.
\end{align}
We can identify the other topological numbers $Y_{abc}$ and the instanton numbers in a similar fashion. Since the group $\IZ_{5}$ has no proper subgroups, curves on the manifold must either belong to an orbit of 5 curves or be mapped to themselves. If a curve with Euler character $\chi$ is mapped to itself by $\IZ_{5}$ then the quotient map will take said curve to a curve with Euler character $\chi/5$. In particular, the Euler character of a genus-0 curve is 2, and so there cannot be any genus-0 curves fixed by the $\IZ_{5}$ action. Similar considerations applied to $\IZ_{10}$ action show that every curve lies in an orbit of length 10.
\begin{equation} \label{eq:recover_genus_0}
\wh n_{s} \= \frac{\kappa}{10}\sum_{\deg (\bm p)=s} n_{\bm p}~,
\end{equation}
where, again, $\kappa=1$ for $\IZ_{10}$ and $\kappa=2$ for $\IZ_{5}$.

For the genus-1 numbers $d_{\bm{p}}$ the relation is more complex since a genus-1 curve has $\chi=0$, so there can exist genus-1 curves, invariant under the symmetry group, whose quotient is again a genus-1 curve. The formula analogous to \eqref{eq:recover_genus_0} is now
\begin{equation}\label{eq:recover_genus_1}
\wh{d}_{s} \= \frac{1}{5}\sum_{\deg{\bm p}=s}d^{\,\text{non-inv}}_{\bm p}+d^{\,\text{inv}}_{s,s,s,s,s}~,
\end{equation}
and serves to compute the numbers $d^{\,\text{inv}}_{s,s,s,s,s}$ of $\IZ_{5}$ invariant genus-1 curves of degree $k$. A small~check is that the numbers $d_{s,s,s,s,s}-d^{\,\text{inv}}_{s,s,s,s,s}$ should be divisible by 5, which they are, to the extent of the tables.

The fact that all the instanton numbers we have computed agree with those computed on the one-parameter families through increasingly intricate relations provides a non-trivial consistency check of the results of sections \sref{sect:Periods} and \sref{sect:Mirror_Map}.

\vskip6pt
\begin{table}[H]
	\begin{center}
		\begin{tabular}{|l|l|l|}
			\hline
\vrule height14pt depth7pt width0pt \hfil $k$ & \hfil $\wh{n}_k$ & \hfil $\wh{d}_k$\\ 
\hline\hline
\vrule height12pt depth0pt width0pt 
\small 1 & $\scriptstyle 12 \,\kappa$ & $\scriptstyle 20 - 10 \,\kappa$ \\ 
\small 2 & $\scriptstyle 24 \,\kappa$ & $\scriptstyle 122 - 40 \,\kappa$ \\ 
\small 3 & $\scriptstyle 112 \,\kappa$ & $\scriptstyle 1200 - 448 \,\kappa$ \\ 
\small 4 & $\scriptstyle 624 \,\kappa$ & $\scriptstyle 12218 - 4468 \,\kappa$ \\ 
\small 5 & $\scriptstyle 4200 \,\kappa$ & $\scriptstyle 133800 - 48948 \,\kappa$ \\ 
\small 6 & $\scriptstyle 31408 \,\kappa$ & $\scriptstyle 1513032 - 550744 \,\kappa$ \\ 
\small 7 & $\scriptstyle 258168 \,\kappa$ & $\scriptstyle 17647096 - 6407540 \,\kappa$ \\ 
\small 8 & $\scriptstyle 2269848 \,\kappa$ & $\scriptstyle 210213862 - 76165868 \,\kappa$ \\ 
\small 9 & $\scriptstyle 21011260 \,\kappa$ & $\scriptstyle 2545256772 - 920643890 \,\kappa$ \\ 
\small 10 & $\scriptstyle 202527600 \,\kappa$ & $\scriptstyle 31212555028 - 11273167424 \,\kappa$ \\ 
\small 11 & $\scriptstyle 2017537884 \,\kappa$ & $\scriptstyle 386727907556 - 139494386722 \,\kappa$ \\ 
\small 12 & $\scriptstyle 20654747200 \,\kappa$ & $\scriptstyle 4832557014112 - 1741106595848 \,\kappa$ \\ 
\small 13 & $\scriptstyle 216372489804 \,\kappa$ & $\scriptstyle 60820504439316 - 21890039477898 \,\kappa$ \\ 
\small 14 & $\scriptstyle 2311525544064 \,\kappa$ & $\scriptstyle 770126009447308 - 276916199510504 \,\kappa$ \\ 
\small 15 & $\scriptstyle 25115533695300 \,\kappa$ & $\scriptstyle 9802710122684812 - 3521744606430982 \,\kappa$ \\ 
\small 16 & $\scriptstyle 276942939016224 \,\kappa$ & $\scriptstyle 125345359041305658 - 44996106493639596 \,\kappa$ \\ 
\small 17 & $\scriptstyle 3093639869100240 \,\kappa$ & $\scriptstyle 1609189343845395984 - 577237489764357432 \,\kappa$ \\ 
\small 18 & $\scriptstyle 34957447938066952 \,\kappa$ & $\scriptstyle 20732103880969324866 - 7431797272240376304 \,\kappa$ \\ 
\small 19 & $\scriptstyle 399082284262216044 \,\kappa$ & $\scriptstyle 267947664660167267380 - 95989385991015664466 \,\kappa$ \\ 
\small 20 & $\scriptstyle 4598143339631725920 \,\kappa$ & $\scriptstyle 3472847998706120977380 - 1243366526906482828392 \,\kappa$ \\ 
\small 21 & $\scriptstyle 53420849666489458232 \,\kappa$ & $\scriptstyle 45126364143189924137384 - 16147280867335074115108 \,\kappa$ \\ 
\small 22 & $\scriptstyle 625334338772563692216 \,\kappa$ & $\scriptstyle 587733058797585235078306 - 210193232419243602788840 \,\kappa$ \\ 
\small 23 & $\scriptstyle 7370491340262022774308 \,\kappa$ & $\scriptstyle 7670883739613425230865660 - 274199136862302475877554 \,\kappa$ \\ 
\small 24 & $\scriptstyle 87419782094909562148112 \,\kappa$ & $\scriptstyle 100310865002094048953427112 - 35839510423715766917658440 \,\kappa$ \\ 
\small 25 & $\scriptstyle 1042868408542514775921540 \,\kappa$ & $\scriptstyle 1314072243953354318776636044 - 469285414203290732598797814 \,\kappa$ \\ 
\small 26 & $\scriptstyle 12507178017340321543927896 \,\kappa$ & $\scriptstyle 17242438907892929716931810362 - 6155022515235842521035603944 \,\kappa$ \\ 
\small 27 & $\scriptstyle 150738741242255934466584688 \,\kappa$ & $\scriptstyle 226585807117189893207597411984 - 80851004064783102896144565240 \,\kappa$ \\ 
\small 28 & $\scriptstyle 1825033540198187573067367200 \,\kappa$ & $\scriptstyle 2981765590671191125416200860324 - 1063545103192576581182654608448 \,\kappa$ \\ 
\vrule height0pt depth6pt width0pt \small 29 & $\scriptstyle 22190047278214311145705359228 \,\kappa$ & $\scriptstyle 39289668166799514883939622674020 - 14008698625940299577598837703530 \,\kappa$ \\
\hline
		\end{tabular} 
		\vskip10pt
		\capt{6in}{tab:instanton_numbers}{The constants $\wh{n}_k$ and $\wh{d}_k$ are respectively the genus-0 and {genus-1} degree-$k$ instanton numbers for the quotient manifolds. The quantity $\kappa$ is taken to equal 1 or 2 depending on whether one is working on the $\mathbb{Z}_{10}$ or $\mathbb{Z}_{5}$ quotient. Note that this differs from the table in \cite{Candelas:2019llw}, as here we are using the conventions of \cite{Gopakumar:1998jq}, which differ from the conventions of \cite{Bershadsky:1993ta} used in \cite{Candelas:2019llw}.
		}
	\end{center}
\end{table}
\newpage

\section{Duality Webs}\label{sect:webs}
\vskip-10pt
We refer to the operation in \eqref{eq:duality_relation}, together with permutations of indices, as duality operations. We denote the set of all multi-indices related to $\bm{I}$ by dualities as $W[\bm{I}]$, the \emph{duality web} containing~$\bm{I}$. The duality operations form a group that we will denote by $\cW$. The web $W[\bm{I}]$ is the group orbit $\cW \! \cdot \! \bm{I} = \{w \bm I \,|\, w \in \cW\}$ of $\bm{I}$.

We begin with some elementary properties of the webs, and will observe later that $\cW$ is a Coxeter group.
\begin{itemize}
    \item These duality relations are equivalence relations, so if $\bm{J} \in W[\bm{I}]$, then $W[\bm{J}]=W[\bm{I}]$. 
    \item  If an integer $r$ divides $\bm{I}$, that is $r$ divides each component of $\bm{I}$, then $r$ divides any dual of $\bm{I}$. It follows that the greatest common divisor of the elements of $\bm{I}$, $\gcd \bm{I}$, is preserved by the duality transformation and that $\gcd{\bm{J}}=\gcd{\bm{I}}$, for all $\bm{J} \in W[\bm{I}]$. Moreover, for an integer $r$, the web of $r\bm{I}$ is obtained by multiplying each multi-index in $W[\bm{I}]$ by $r$. We can say that $W[r\bm{I}]=r\,W[\bm{I}]$. This observation shows, for example, that $(1,1,1,0,0)$ and $(3,3,3,0,0)$ belong to different duality webs.
\end{itemize} 

The webs have interesting properties mod 3. As in \eqref{eq:degree}, we denote the sum of elements of a multi-index $\bm{I}$ by $\deg(\bm{I})$. If the operation $g_1$ takes $\bm{I}$ to~$\wt{\bm{I}}$, as in \eqref{eq:duality_relation}, then 
\begin{align}\notag
\deg\big(\,\wt{\bm{I}}\,\big)\= 2\,\deg(\bm{I})-3 I_1 \;= -\deg(\bm{I}) \mod 3~.
\end{align}
\begin{itemize}
    \item It follows that $\deg(\bm{I})^2 \mod 3$ is an invariant of the web, so either $\deg(\bm{J})=0 \mod3$, for all $\bm{J}\in W[\bm{I}]$, or $\deg(\bm{J})^2=1 \mod 3$ for all $\bm{J}\in W[\bm{I}]$. Moreover, if $\deg(\bm{I})=0 \mod3$, we see from \eqref{eq:duality_relation} that $\bm{J}=\bm{I}\mod 3$, for all $\bm{J}\in W[\bm{I}]$.
\end{itemize}
Recall that $\bm{I}$ is \emph{positive} if all the components of $\bm{I}$ are nonnegative and at least one is strictly positive. We say that $\bm{I}$ is \emph{negative} if $-\bm{I}$ is positive. A nonzero multi-index that is neither positive nor negative has both strictly positive and strictly negative components and we refer to such a multi-index as \emph{mixed}. 

We say that a \emph{web $W$ is positive} if all its multi-indices are positive. A \emph{web $W$ is negative} if $-W$ is positive. A web that is neither positive nor negative is said to be \emph{mixed}.
\begin{itemize}
    \item The only multi-index in appendix \ref{app:Instanton_Numbers} that has a duality transform that is not positive is $(1,0,0,0,0)$ which has among its transforms the multi-index $(-1,0,0,0,0)$. We have
    \vskip-20pt\begin{align} \notag 
    W[(1,0,0,0,0)]\= \IW_{+}\cup \IW_{-}~,
    \end{align}
    where 
    \begin{align}\notag
    \IW_{+}\=\Big\{ (1,0,0,0,0),~(1,1,0,0,0),~(2,1,1,0,0)~,\ldots~\Big\}
    \end{align} 
    consists entirely of positive multi-indices, and $\IW_{-}=-\IW_{+}$.
    \item It follows from the discussion of \sref{sect:fromS5toS6}, that the instanton numbers $n_{\bm{I}}$ can be nonzero precisely for multi-indices that lie in the positive webs and in $\IW_{+}$. For $\bm{J}\in \IW_{+}$ we have $n_{\bm{J}}=24$, and for $\bm{J}$ in a positive web $W[\bm{I}]$ we have~$n_{\bm{J{}}}=n_{\bm{I}}$. To the extent of the tables, all the genus-zero instanton numbers that are permitted in this way are in fact nonzero.
\end{itemize}
There are some intriguing identities that are explained by the duality operations. For example: 
\begin{align} \notag
n_{(i,\,j,\,k,\,0,\,0)}\= n_{(2-i,\,2-j,\,2-k,\,0,\,0)}~,~~ \text{for}~~ 0\leqslant i,j,k\leqslant2~,~~\text{and}~~ (i,j,k)\neq (0,0,0)~\text{or}~
(2,2,2)~.
\end{align}
For $(i,j,k)=(1,1,1)$, the identity is trivial, and for the other values of $i,j,k$ this is explained by the fact that all these multi-indices are in the web $\IW_{+}$. This equality can also be explained as a consequence of the elliptic fibration structure of the $\MHV$ manifold as we will see in \sref{sect:Curve_Counting}. 

In fact, as is evident from the fact that both permutations and duality operations keep instanton numbers invariant: if $\bm{J{}} \in W[\bm{I}]$, then the genus-0 and genus-1 instanton numbers associated to $\bm{I}$ and $\bm{J{}}$ agree, $n_{\bm{J{}}} = n_{\bm{I}}$ and $d_{\bm{J{}}} = d_{\bm{I}}$. Thus it is possible to associate to each web $W$ unique instanton numbers $n_W$ and $d_W$ by $n_W = n_{\bm{I}}$ and $d_W = d_{\bm{I}}$ for any $\bm{I} \in W$. This explains most of the repetitions in the tables in appendix \ref{app:Instanton_Numbers}.

\tref{tab:InstantonNumberRecurrences}, however, sets out an intriguing relation among the instanton numbers that are not explained in this way. There are distinct positive webs $W$ and $W'$ such that $n_W = n_{W'}$. Moreover, precisely when this happens (to the extent of the tables) we have also $d_W = d_{W'}$, where $d_W$ denotes the genus-one instanton numbers. There exists also a quadratic invariant $h(W)$, which we will define in \eqref{eq:h_invariant}, associated to each web. It is an interesting observation that the webs with equal instanton numbers, corresponding to a single row of \tref{tab:InstantonNumberRecurrences} have equal invariants $h(W)$. It is also worth noting that the vectors listed in the table all contain a zero component. The instanton numbers that are involved, in these relations, are highly nontrivial.

While it is possible that some of the webs that are stated to be distinct could in fact coincide, we should state that, in all cases, we have checked the webs up to total degree 250, and these partial webs are distinct. Moreover, in many cases, the webs are demonstrably distinct. The webs in the first two rows of the table, for example, are all distinct since each index vector $\bm{I}$ has a distinct greatest common divisor. Row 3 of the table refers to two webs that have degrees that are both zero mod 3. The index vectors of the first web are equal to $(2,2,2,0,0)$ mod 3, up to permutation, while the second web consists of index vectors that, up to permutation, are $(1,1,1,0,0)$ mod 3. Many other rows of the table correspond to webs that can be shown to be distinct, in a similar~way.

\subsection{The Coxeter group of dualities}
\vskip-10pt
Denote the operation analogous to \eqref{eq:duality_relation} acting on the $r$'th coordinate by $g_r$. These operations, combined with permutations of the indices, generate $\cW$. The $S_5$ subgroup, corresponding to the permutations, can be generated by the transpositions $(r,s)$. In fact, it can be generated by just the four transpositions $s_r=(r,r+1)$, $r=1,2,3,4$. Once we admit the permutations, then the dualities $g_r$ can all be obtained from $g_1$, say; for \hbox{example $g_2=s_1g_1s_1$}.

We have \vskip-35pt
\begin{align} \notag
g_1^2 \= s_1^2 \= s_2^2 \= s_3^2 \= s_4^2 \= 1~. 
\end{align}
and it is clear that the elements of the generating set $\fS=\{g_1,\,s_1,\,s_2,\,s_3,\,s_4\}$ commute, apart from $g_1$ with $s_1$, and $s_r$ with $s_{r+1}$, $r=1,2,3$. The operation $s_r s_{r+1}$ amounts to a cyclic permutation of the three indices $(I_r I_{r+1}I_{r+2})\to (I_{r+2} I_{r}I_{r+1})$, so $(s_r s_{r+1})^3=1$ and one checks also that $(g_1s_1)^6=1$. We have not been able to find any other relations between the elements of $\fS$ so we assume that in fact there are no further relations. If there were additional relations, the webs would break into smaller webs, giving fewer relations between the instanton numbers. Then the instanton numbers would be expected to take different values --- one corresponding to each of the smaller webs. However, we do not observe this to the extent of the tables. 

\begin{table}[p]
\renewcommand{\arraystretch}{0.70}
\begin{center}
\begin{tabular}{|>{\hskip-3pt}l<{\hskip-4pt}| >{\hskip-3pt}l<{\hskip-4pt}| >{\hskip-3pt}l<{\hskip-4pt} |
>{\hskip-2pt}c<{\hskip-3pt} | >{\hskip-4pt}c<{\hskip-4pt} >{\hskip-4pt}c<{\hskip-4pt} >{\hskip-4pt}c<{\hskip-4pt} >{\hskip-4pt}c<{\hskip-4pt} >{\hskip-4pt}c<{\hskip-12pt}|}
\hline
\vrule height13pt depth5pt width0pt $\#$ & \hfil $n$ & \hfil $d$ & $h$ & \multicolumn{5}{|c|}{Webs} \\
\hline\hline
 &&&&&&&&\\[-8pt]
 1 & 112 & \+0 & \+0 & \multicolumn{5}{|c|}{$(2k+1)\,W[(1,1,1,0,0)]~;~~k=1,2,\ldots$} \\
 2 & 80   & \+4 & \+0 & \multicolumn{5}{|c|}{$\phantom{(2+1)}k\,W[(2,2,2,0,0)]~;~~k=1,2,\ldots$}\\[2pt]\hline
 &&&&&&&&\\[-8pt]
 3 & 234048 & -5600 & -9 & (3,2,2,2,0) & (4,4,3,1,0) &  &
    &  \\
 4 & 795936 & -29136 & -11 & (3,3,2,2,0) & (4,4,4,1,0) &  &
    &  \\
 5 & 4326048 & -251520 & -14 & (3,3,3,2,0) & (5,5,5,1,0) &  &
    &  \\
 6 & 33777312 & -3031872 & -18 & (3,3,3,3,0) & (5,5,3,2,0) &
   (7,7,6,1,0) &  &  \\
 7 & 7371792 & -484896 & -15 & (4,3,3,2,0) & (6,6,5,1,0) &  &
    &  \\
 8 & 88179456 & -9395616 & -20 & (4,3,3,3,0) & (7,7,7,1,0) &  &
    &  \\
 9 & 20578560 & -1679040 & -17 & (4,4,3,2,0) & (6,6,6,1,0) &  &
    &  \\
 10 & 347078520 & -46049040 & -23 & (4,4,3,3,0) & (5,5,4,2,0) &
   (8,8,8,1,0) &  &  \\
 11 & 1935300720 & -327015680 & -27 & (4,4,4,3,0) & (6,5,5,2,0) &
   (6,6,3,3,0) &  &  \\
 12 & 14386855920 & -3110590260 & -32 & (4,4,4,4,0) & (6,6,6,2,0) &
    &  &  \\
 13 & 539120544 & -76342880 & -24 & (5,4,3,3,0) & (9,9,8,1,0) & 
   &  &  \\
 14 & 140436672 & -16170272 & -21 & (5,4,4,2,0) & (8,8,7,1,0) & 
   &  &  \\
 15 & 4392333792 & -824199120 & -29 & (5,4,4,3,0) & (6,6,5,2,0) &
    &  &  \\
 16 & 45007048752 & -11043084816 & -35 & (5,4,4,4,0) & (6,6,4,3,0) &
   (7,7,6,2,0) &  &  \\
 17 & 1272585120 & -203310240 & -26 & (5,5,3,3,0) & (5,5,5,2,0) &
   (9,9,9,1,0) &  &  \\
 18 & 193411225936 & -55127514240 & -39 & (5,5,4,4,0) & (8,7,7,2,0) &
    &  &  \\
 19 & 65215603200 & -16642969280 & -36 & (5,5,5,3,0) & (7,7,4,3,0) &
    &  &  \\
 20 & 1096632180480 & -368134832160 & -44 & (5,5,5,4,0) & (7,7,5,3,0) &
    &  &  \\
 21 & 7888589144400 & -3138370134624 & -50 & (5,5,5,5,0) & (7,7,6,3,0)
   & (9,9,9,2,0) &  &  \\
 22 & 65215569408 & -16642956928 & -36 & (6,4,4,4,0) & (8,8,6,2,0) &
    &  &  \\
 23 & 21143067840 & -4775506080 & -33 & (6,5,4,3,0) & (7,6,6,2,0) &
    &  &  \\
 24 & 391409808576 & -119442727776 & -41 & (6,5,4,4,0) & (6,6,5,3,0) &
   (8,8,7,2,0) &  &  \\
 25 & 135171775392 & -37176746592 & -38 & (6,5,5,3,0) & (7,7,7,2,0) &
    &  &  \\
 26 & 2981800050480 & -1093125957120 & -47 & (6,5,5,4,0) & (7,6,6,3,0)
   & (7,7,4,4,0) & (9,9,8,2,0) &  \\
 27 & 27765085214112 & -12215408263200 & -54 & (6,5,5,5,0) &
   (7,7,7,3,0) & (9,9,6,3,0) &  &  \\
 28 & 1096632086784 & -368134868160 & -44 & (6,6,4,4,0) & (8,8,8,2,0) &
    &  &  \\
 29 & 10848408360480 & -4429601736480 & -51 & (6,6,5,4,0) & (8,7,6,3,0)
   &  &  &  \\
 30 & 126532108859856 & -62415555336480 & -59 & (6,6,5,5,0) &
   (7,6,6,4,0) & (8,8,5,4,0) & (8,8,7,3,0) &  \\
 31 & 1535514818112 & -531223501536 & -45 & (6,6,6,3,0) & (7,6,4,4,0) &
   (8,8,5,3,0) & (9,8,8,2,0) &  \\
 32 & 725912434085952 & -405156007308576 & -65 & (6,6,6,5,0) &
   (8,7,6,4,0) & (9,8,8,3,0) &  &  \\
 33 & 51294957112992 & -23657221999872 & -56 & (7,5,5,5,0) &
   (7,7,5,4,0) & (8,7,7,3,0) &  &  \\
 34 & 553728279360 & -174588053440 & -42 & (7,6,5,3,0) & (9,9,7,2,0) &
    &  &  \\
 35 & 20350993239840 & -8738280013680 & -53 & (7,6,5,4,0) & (8,8,6,3,0)
   &  &  &  \\
 36 & 305922925426848 & -160791639748800 & -62 & (7,6,5,5,0) &
   (9,9,7,3,0) &  &  &  \\
 37 & 2235977596096128 & -1345692401785920 & -69 & (7,6,6,5,0) &
   (9,8,6,4,0) &  &  &  \\
 38 & 19503820669876800 & -13461969999093600 & -77 & (7,6,6,6,0) &
   (8,7,6,5,0) & (9,8,7,4,0) &  &  \\
 39 & 408865565088240 & -219322647849280 & -63 & (7,7,6,4,0) &
   (8,6,5,5,0) & (8,8,8,3,0) &  &  \\
 40 & 8765016259161504 & -5758034709276000 & -74 & (7,7,6,5,0) &
   (9,9,5,5,0) &  &  &  \\
 41 & 92700939550359360 & -70199768003592720 & -83 & (7,7,6,6,0) &
   (9,8,6,5,0) & (9,8,8,4,0) &  &  \\
 42 & 1692511362069504 & -1000067627051904 & -68 & (7,7,7,4,0) &
   (8,7,5,5,0) & (9,9,8,3,0) &  &  \\
 43 & 42801528146793216 & -30974226462689184 & -80 & (7,7,7,5,0) &
   (9,9,7,4,0) &  &  &  \\
 44 & 3704581973944705776 & -3435204329200397376 & -98 & (7,7,7,7,0) &
   (9,7,7,6,0) &  &  &  \\
 45 & 3881643757375656 & -2421429008571216 & -71 & (8,6,6,5,0) &
   (8,7,7,4,0) & (8,8,5,5,0) & (9,9,6,4,0) &  \\
 46 & 42801528135993600 & -30974226462442944 & -80 & (8,6,6,6,0) &
   (8,8,8,4,0) &  &  &  \\
 47 & 1810611871504105272 & -1616941505273075616 & -95 & (8,7,7,6,0) &
   (8,8,8,5,0) & (9,8,6,6,0) &  &  \\
 48 & 55456767284050560 & -40752668020556480 & -81 & (8,8,6,5,0) &
   (9,6,6,6,0) &  &  &  \\
 49 & 419093788958668992 & -345708325307878560 & -89 & (8,8,7,5,0) &
   (9,7,6,6,0) &  &  &  \\
 50 & 5101035246140064 & -3238317072282240 & -72 & (9,7,7,4,0) &
   (9,8,5,5,0) &  &  &  \\
 51 & 198280729061595552 & -156843054827785632 & -86 & (9,7,7,5,0) &
   (9,9,6,5,0) &  &  &  \\
 52 & 170193515484672 & -85781801925696 & -60 & (9,8,7,3,0) &
   (9,9,5,4,0) &  &  &  \\
\hline
\end{tabular}
\vskip10pt
\capt{5.5in}{tab:InstantonNumberRecurrences}{Pairs $(n_W, d_W)$ that arise for the distinct webs that are shown on the right. The webs indicated on the right have the same quadratic invariant $h$ that is defined in \eqref{eq:h_invariant}.}
\end{center}
\end{table}
In this way, we observe that $\cW$ is a Coxeter group corresponding to the graph in \fref{fig:coxetergraph}.
\vskip-0.0cm
\begin{figure}[H]
\begin{center}
\framebox[1.0\width]{
\begin{minipage}{4.5in}
\begin{center}
\vspace{-0.6cm}
\includegraphics[width=3in]{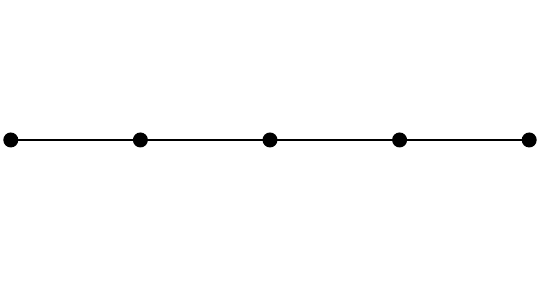}
\vspace{-0.6cm}
\end{center}
\end{minipage}}
\end{center}
\vskip-0.08in %do not remove
\place{2.13}{0.85}{6}
\place{1.74}{0.45}{$g_1$}
\place{2.47}{0.45}{$s_1$}
\place{3.2}{0.45}{$s_2$}
\place{3.93}{0.45}{$s_3$}
\place{4.64}{0.45}{$s_4$}
\vspace*{-0.2cm}
\begin{center}
\capt{4in}{fig:coxetergraph}{The Coxeter graph corresponding to $\cW$.}
\end{center}
\end{figure}
\vskip-25pt
For the following discussion it is convenient to think of multi-indices as integral elements of the vector space $\IR^5$. We shall refer to the elements of these vector spaces as \textit{index vectors}. For two index vectors $\bm{J},\bm{J}'$ let $H(\bm{J},\bm{J}')$ denote the bilinear form
\begin{align} \notag
H(\bm{J},\bm{J}') \= \frac{3}{2}\, \bm{J}{\cdot} \bm{J}' - \frac{1}{2}\, \deg(\bm{J})\deg(\bm{J}')
\end{align}
where $\bm{J}{\cdot} \bm{J}'$ denotes the standard inner product. This bilinear form is Lorentzian with signature $(-,+,+,+,+)$. The squared length of an index vector $\bm{J}$ with respect to $H$ is denoted by $h(\bm{J})$
\begin{align} \label{eq:h_invariant}
h(\bm{J}) \= H(\bm{J},\bm{J})~.
\end{align}
In addition, let us define root vectors $\a^r$, $r=0,\ldots,4$, 
\begin{align} \notag
\begin{split}
\a^0 &\= \phantom{ \frac{1}{\sqrt{3}}}\,(1,\+0,0,0,0)~,\\[2pt]
\a^1 &\= \frac{1}{\sqrt{3}}\, (-1,1,0,0,0)~,\\[2pt]
\a^2 &\= \frac{1}{\sqrt{3}}\,(0,-1,1,0,0)~,\\[2pt]
\a^3 &\= \frac{1}{\sqrt{3}}\,(0,0,-1,1,0)~,\\[2pt]
\a^4 &\= \frac{1}{\sqrt{3}}\,(0,0,0,-1,1)~.\\
\end{split}
\end{align}
These are all spacelike and normalised to unity, with respect to $H$. The action of the generators on the index vectors are realised as reflections relative to the metric~$H$
\begin{align} \notag
\begin{split}
g_1 \bm{J} &\= \bm{J} - 2H(\a^0,\bm{J}) \,\a^0\\
s_r \bm{J}  &\= \bm{J} - 2H(\a^r,\bm{J}) \,\a^r~,~r\=1,2,3,4~.\\
\end{split}
\end{align}
Clearly $H$ is invariant under permutations of the indices and so under the actions of the $s_r$. It is also an immediate check, from the first of the relations above, that 
\begin{align} \notag
H\big( g_1 \bm{J},\, g_1 \bm{J}'\big) \= H(\bm{J},\bm{J}')~.
\end{align}
Thus $h(\bm{J})$ is an invariant of a web and $h(\bm{J})=h(\bm{I})$ for all $\bm{J}\in W[\bm{I}]$.

\subsection{Positive webs}
\vskip-10pt
We wish now to describe the structure of positive webs in some detail. In particular, we show that they admit unique characterisation in terms of simple index vectors we term \textit{source vectors}. To define these, consider now a positive index vector $\bm{I}=(I_1,I_2,I_3,I_4,I_5)$, with its indices in {\it standard order}, that is with $I_1\geqslant I_2 \geqslant I_3 \geqslant I_4 \geqslant I_5$. We shall refer to such a vector as an {\it ordered source}, if
\begin{align}
I_1 \leqslant \frac{\deg(\bm{I})}{3}~.
\notag\end{align}
More generally, a {\it source} is an ordered source up to permutation of indices. For a source, the condition on the indices becomes $\max(\bm{I})\leqslant \deg(\bm{I})/3$. Examples of sources are the vectors $(1,1,1,0,0)$, which is null, and $(1,1,1,1,0)$, which has $h=-2$. Note that the special vector $(1,0,0,0,0)$ is \emph{not} a source, and has $h=1~.$

A first observation is that \emph{positive webs exist}. We will see in the following section that the dualities are generated by flops on the genus zero curves whose index vectors are permutations of $(1,0,0,0,0)$. Given these symmetries, the instanton numbers $n_{\bm I}$ can only be nonzero if $\bm I$ belongs to a positive web. We have found index vectors for which the $n_{\bm I}$ are nonzero by explicit computation. This shows that positive webs exist. 

At root, however, this observation concerns the representations and orbits of a Coxeter group and it should be possible to prove the existence of positive webs, and indeed of the conjecture that follows below, purely from the properties of the group and its representations. We gather some observations along these lines in the appendix \ref{app:properties_of_W}.

Based on extensive numerical computation we make:
\begin{restatable}{conjecture}{conj:positive_web_properties} \label{conj:positive_web_properties}
The positive webs are in one-one correspondence with ordered source vectors, and each positive web contains an ordered source vector as a vector of minimal degree. 
\end{restatable}

This conjecture has been checked by computer calculation for index vectors up to and including degree 100. Among these vectors there are 144,223 positive webs and there is a one-one correspondence between these and ordered sources. Assuming the conjecture, this characterisation of positive webs as being in one-one correspondence with ordered sources is surprisingly simple, since the ordered source vectors are easily listed.

The fact that every positive web contains an ordered source as a vector of minimal degree is easily shown. A positive web has an index vector $\bm I$ of lowest degree, which we can take to be ordered. This being so, we have
\beq
\deg(g_1 {\bm I}) \geqslant \deg({\bm I}) 
\notag\eeq 
However 
\beq
\deg(g_1 {\bm I}) - \deg({\bm I}) \= \deg({\bm I}) - 3I_1~.
\notag\eeq 
Since the right hand side of this last relation is positive, it follows that $\bm I$ is an ordered source.

Several properties follow from the conjecture, some of which admit simple proofs independent from the conjecture:

\begin{enumerate}  
\item \emph{The degree of a source $\bm{I}$, which we may take to be ordered, is not lowered by the action of any $w \in \cW$.}

This follows immediately from the conjecture. Since, if there were vectors of degree lower than $\deg({\bm I})$, then there would have to be an ordered vector of minimal degree, whose degree is lower than $\deg({\bm I})$. Such a vector would be another ordered source, contrary to the conjecture.

\item \emph{Positive webs are infinite. In particular, it is always possible to increase the degree of an index vector $\bm I$ by acting on it with an element of the Coxeter group.}

To see this, take $\bm I$ to be ordered. Then clearly $I_5\leqslant \deg({\bm I})/5$ and
\beq
\deg(g_5{\bm I}) \= 2\deg({\bm I}) - 3 I_5 \;\geqslant\: \frac{7}{5} \deg({\bm I}) \; > \; \deg({\bm I}) ~.
\notag\eeq

\item \emph{A positive web $W$ has $h{\;\leqslant\;} 0$, so every vector in $W$ is future pointing and timelike, or future pointing and null.}

To see this, note that $h$ is constant across a web, and for the unique ordered source $\bm I$ in $W$ we have
\beq\notag
h(\bm I) \= H(\bm I, \bm I) \;= -\frac{1}{2} \left(\deg( \bm I)^2 - 3 \bm I \cdot \bm I \right)
\;\leqslant -\frac{1}{2} \left(\deg( \bm I)^2 - 3 I_1 \deg( \bm I)  \right)~,
\eeq
where, in passing to the last expression we have used the definition of a source, and the right-hand side is non-positive in virtue of the properties of an ordered source.
\end{enumerate}
\subsection{Cone of positive webs}
\vskip-10pt
The index vectors belonging to positive webs form a convex cone which is specified by identifying its generators. This identification proceeds in two stages where the first is to note that the source vectors themselves form a convex cone.

Note first that the condition for an ordered source, $i\leqslant \deg(\bm{I})/3$, is linear. Thus a linear combination with positive coefficients of ordered sources is again an ordered source. The converse is also true: every ordered source can be written as a linear combination with positive coefficients of basis elements. To see this, we can proceed by adding sources to the basis as necessary as we increase the degree, say, although it is not clear a priori that such a process will lead to a finite basis. However, on the basis of explicit calculation up to the extent of the tables have the following conjecture:

\begin{restatable}{conjecture}{sources} \label{conj:sources}Each source may be written as a positive linear combination with integer coefficients of index vectors taken from the 16 sources
	\begin{align} \label{eq:sourcebasis}
	\begin{split}
	&(1,1,1,0,0) + \text{\em permutations,}\\
	&(1,1,1,1,0) + \text{\em permutations,}\\
	&(1,1,1,1,1)~.
	\end{split}
	\end{align}
\end{restatable}
These elements are certainly necessary. At degree three it is clear that none of the permutations of (1,1,1,0,0) can be written as a positive linear combination of the other permutations. Similarly at degree four the permutations of (1,1,1,1,0) cannot be written as positive linear combinations of the other permutations and the degree three sources. At degree five it is clear that a degree five vector cannot be written as a positive linear combination of vectors of degree three and four.

At degree six, we meet the sources $(2,2,2,0,0),\,(2,2,1,1,0),\,(2,1,1,1,1)$, together with their permutations, and now it \emph{is} possible to write these in terms of earlier sources: 
\begin{align} \notag
\begin{split}
(2,2,2,0,0) &\=  2\,(1,1,1,0,0)~,\\
(2,2,1,1,0) &\=  \phantom{2\,}(1,1,1,0,0) + (1,1,0,1,0)~,\\
(2,1,1,1,1) &\=  \phantom{2\,}(1,1,1,0,0) + (1,0,0,1,1)\\
\end{split}
\end{align}
We do not have a proof that all sources of degree greater than five can be written as positive linear combinations of the index vectors \eqref{eq:sourcebasis}, but we have checked that up to degree 80 all the sources can be so written.

If we accept the conjecture \ref{conj:sources}, then the cone subtended by the vectors in positive webs can be described in simple terms.
\begin{restatable}{corollary}{cone}
The index vectors that belong to the positive webs form a cone whose generators are the index vectors that lie in the union of the webs determined by the generators of the web of sources, given above, that is by 
\begin{align} \notag
W[(1,1,1,0,0)]\cup W[(1,1,1,1,0)]\cup W[(1,1,1,1,1)]~.
\end{align}
\end{restatable}
To see this, note first that all the index vectors in $W[(1,1,1,0,0)]$ are necessary. We give two arguments; the first is immediately clear, the second is an argument that can be applied to other webs also. For the first argument note that all the vectors of $W[(1,1,1,0,0)]$ are future pointing and null, and no two are linearly dependent. This being so, any nontrivial positive linear combination of vectors in this web yields a vector that is timelike so cannot be a member of this web.  For the alternative argument suppose that some vector $\bm{J}$ can be written as a positive integral combination of other vectors $\bm{K}_a$ so that 
\begin{align} \notag
\bm{J}\= \sum_a \a^a \bm{K}_a~,\qquad \bm{K}_a\in W[(1,1,1,0,0)]~,
\end{align}
with positive integral coefficients $\a^a$. By acting with some element $g\in\cW$ we can transform $\bm{J}$ to the source $(1,1,1,0,0)$. So now the relation reads
\begin{align} \notag
(1,1,1,0,0)\= \sum_a \a^a \wt{\bm{K}}_a~,
\end{align}
where $\wt{\bm{K}}_a = g \bm{K}_a$. Now the source has degree three and each $\deg(\wt{\bm{K}}_a)\geqslant 3$ for each $a$, and this relation is impossible for nontrivial combinations.

To show that all the vectors of $W[(1,1,1,1,0)]$ are needed as generators of the cone, suppose that some vector $\bm{J}\in W[(1,1,1,1,0)]$ can be expressed as a nontrivial integral linear combination of the vectors in $W[(1,1,1,0,0)]\cup W[(1,1,1,1,0)]$. Then it is possible to write
\begin{align} \notag
\bm{J}\= \sum_a \a^a \bm{K}_a + \sum_b \b^b \bm{L}_b~,\qquad \bm{K}_a\in W[(1,1,1,0,0)]~,~~\bm{L}_b\in W[(1,1,1,1,0)]~.
\end{align} 
Again, we transform $\bm{J}$ to the source, which is now $(1,1,1,1,0)$ and we see that now
\begin{align} \notag
(1,1,1,1,0)\= \sum_a \a^a \wt{\bm{K}}_a + \sum_b \b^b \wt{\bm{L}}_b~.
\end{align}
The degree of the source is four while $\deg(\wt{\bm{L}}_b)\geqslant 4$ and the $\deg(\wt{\bm{K}}_a)$ are multiples of three, so this relation is also impossible, for nontrivial combinations.

We see in a similar way that all the vectors in $W[(1,1,1,1,1)]$ are also generators.

For the remaining positive webs, we reverse the argument. Let $\circbm{I}$ be the source of such a web, so with $\deg \circbm{I} \geqslant 6$; we know that it may be written in terms of the 16 generators of the cone of sources
\begin{align} \notag
\circbm{I} \= \sum_a \a^a \circbm{K}_a + \sum_b \b^b \circbm{L}_b + \g\,(1,1,1,1,1)~,
\end{align}
where, in this relation, the $\circbm{K}_a$ are the permutations of $(1,1,1,0,0)$ and the $\circbm{L}_b$ are the permutations of $(1,1,1,1,0)$. For $\bm{J}$ any multi-index in $W\big[\,\circbm{I}\,\big]$ we choose $g_{\bm{J}}\in\cW$ that transforms $\circbm{I}$ to $\bm{J}$. Then by applying $g_{\bm{J}}$ to the relation above, we express $\bm{J}$ as a sum of vectors in the union of the webs $W[(1,1,1,0,0)]\cup W[(1,1,1,1,0)]\cup W[(1,1,1,1,1)]$.
\subsection{Birational varieties and the linear sigma model}\label{sect:birational}
\vskip-10pt
We are not the first to observe the action of an infinite group on the instanton numbers. Many aspects of the situation with respect to the action of the Coxeter group were explained in terms of birational models by Hosono and Takagi~\cite{Hosono:2017hxe} in the context of certain two-parameter families of \cys. We note in addition the recent paper by Brodie, Constantin, Lukas, and Ruehle \cite{Brodie:2021toe}, which also observes infinite groups generated by flopping curves. 

We pause to briefly recall aspects of the analysis of Hosono and Takagi, adapted to the mirror HV manifold.
Note that we may write the two polynomials $Q^f$, $f=0,1$, in terms of the coordinates $Y_{i,b}$, $i=0,\ldots,4$ of the five $\IP^1$'s, in the form
\beq\notag
Q^f \= A_{abcdef}\,Y_{0,a} Y_{1,b} Y_{2,c} Y_{3,d} Y_{4,e} ~,
\eeq
where, in the general case, the coefficients $A_{abcdef}$ have no particular symmetry. 

Let us introduce Lagrange multipliers $Y_{5,f}$ and a superpotential
\beq\notag
\ccW \= Y_{5,f}\, Q^{f} \=  A_{abcdef}\,Y_{0,a} Y_{1,b} Y_{2,c} Y_{3,d} Y_{4,e}Y_{5,f}~,
\eeq
and consider the linear sigma model for the B-model of the mirror HV manifold~\cite{Witten:1993Phases}. This corresponds to a path integral
\beq\notag
\int D\cY\, \ee^{\ii\ccW(\cY)}~,
\eeq
where $D\cY$ denotes integration over superfields $\cY_{j,a}$ corresponding to the coordinates $Y_{j,a}$ and $\ccW(\cY)$ denotes $\ccW$ with $Y_{j,a}$ replaced by $\cY_{j,a}$. In this form, the coordinates $Y_{j,a}$ enter symmetrically. Any of the six $Y_{j,a}$, $j=0,\ldots,5$, for given $a$, can be regarded as Lagrange multipliers.
In this way we see that the linear sigma model makes reference to the six manifolds $\ccX_j$, with our original manifold $\ccX$ identified with $\ccX_5$.
\vskip-20pt
\beq\begin{split}\notag
\ccX_0:\quad A_{abcdef}\,Y_{1,b}Y_{2,c}Y_{3,d}Y_{4,e}Y_{5,f}&\= 0\\
\ccX_1:\quad A_{abcdef}\,Y_{0,a}Y_{2,c}Y_{3,d}Y_{4,e}Y_{5,f}&\= 0\\ 
&\hskip-2.2cm\vdots\hskip2.46cm\vdots\hskip-5.8cm\vdots\\
\ccX_5:\quad A_{abcdef}\,Y_{0,a}Y_{1,b}Y_{2,c}Y_{3,d}Y_{4,e} &\= 0~.\\ 
\end{split}\eeq 
The polynomials corresponding to $\ccX_j$ are $\del\ccW/\del Y_{j,a}$. For generic coefficients, these are different.

The equations for $\ccX_j$ are multilinear in the remaining coordinates, so linear in $Y_{i,a}$, for given $i\neq j$. We have two equations that are linear in the two coordinates $Y_{i,a}$, $a=0,1$. So the determinant of the coefficient matrix
\beq\notag
\D_{ij} \defineas \det \left(\frac{\del^2\ccW}{\del Y_{i,a} \del Y_{j,b}}\right)_{a,b = 0,1}
\eeq
must vanish. Denote the variety that corresponds to the vanishing of $\D_{ij}$ by $\ccZ_{\,ij}$. The determinant is symmetric in the labels $i$ and $j$, so $\ccZ_{\,ij}=\ccZ_{\,ji}$. If instead we were to start with the equations for the manifold $\ccX_i$ and eliminate the coordinates $Y_{i,a}$, $a=0,1$, we would arrive at the same determinant $\D_{ij}$. The variety $\ccZ_{\,ij}$ is singular with a certain number of conifold singularities. The situation is as in \fref{fig:birmap}. The projection $\p_{ij}$ blows down copies of $\IP^1_j$ while $\p_{ji}$ blows down copies of $\IP^1_i$. The inverses of these projections are defined almost everywhere, so  proceeding around the diagram from, say, $\ccX_i$ to $\ccX_j$ via $\ccZ_{\,ij}$ furnishes a birational map. This amounts to the flopping of curves, blowing down curves parallel to $\IP^1_j$ and blowing up the singular points by copies of curves parallel to $\IP^1_i$.
\vskip-10pt
\begin{figure}[H]
\centering
\adjustbox{scale=1.2,center}{
\begin{tikzcd}
\ccX_i \arrow[rr, leftrightsquigarrow] \arrow[dr,"\p_{ij}" ']  &   & \ccX_j\\
         				  & \ccZ_{\,ij} \arrow[ur,"\p_{ji}" ', leftarrow]\\
\end{tikzcd}}
\vskip-20pt\capt{5.5in}{fig:birmap}{The birational maps between $\ccX_i$ and $\ccX_j$ derive from the projections $\p_{ij}$ and $\p_{ji}$.}
\end{figure}
\vskip-20pt
The birational map $\cX_i\rightsquigarrow\cX_j$ amounts to the familiar flopping of curves. A number of copies of $\IP^1_i$ are contracted to points and these are then blown up to copies of $\IP^1_j$. 

It is instructive to see explicitly the action of the birational map on the instanton numbers. Consider generic coefficients $A_{abcdef}$, and the map $\cX_5\rightsquigarrow\cX_0$, and let $\IA$ denote the $2{\times}2$ matrix with the following components
\vskip-10pt
\beq\label{eq:Amatrix}
\IA_{af} \= \frac{\del^2 \ccW}{\del Y_{0,a}\del Y_{5,f}} \= A_{abcdef}\, Y_{1,b} Y_{2,c} Y_{3,d} Y_{4,e}~.
\eeq
Now we have the familiar conifold story: the equations
\beq\label{eq:lineqs}
\IA_{af}\,Y_{5,f}\=0 \qquad \text{and} \qquad Y_{0,a}\,\IA_{af}\=0
\eeq
describe $\cX_5\;(=\cX)$ and $\cX_0$, respectively. In each case we have two equations in two unknowns and the $Y_{j,a}$, $a=0,1$, being the coordinates of a $\IP^1$ cannot both vanish. Thus we recover the condition
\beq
\det(\IA)\=0~.
\notag\eeq
This equation describes $\cZ_{5,0}$, which is a singular member of the family of tetraquadrics
\beq\notag
\cicy{\IP^1\\ \IP^1\\ \IP^1\\ \IP^1\\}{2\\ 2\\ 2\\ 2\\}~.
\eeq
Since the rank of the matrix $\IA$ is not 2, the generic case is that it is 1. When this is so, for each point in $\cZ_{5,0}$ the equations \eqref{eq:lineqs} determine unique points in $\cX_0$ and $\cX_5$, respectively. 

The rank of $\IA$ vanishes when the four elements of the matrix vanish simultaneously. When this happens the $Y_{j,a}$ are completely undetermined and so correspond to copies of $\IP^1_5$ and $\IP^1_0$, respectively. In this way, we see that, in passing from $\cX_5$ to $\cX_0$, copies of $\IP^1_5$ are replaced by copies of $\IP^1_0$. The points where $\IA{\=}0$ are the 24 nodes of $\cZ_{5,0}$, so 24 copies of $\IP^1_5$ are replaced by 24 copies of $\IP^1_0$ as we pass from $\cX_5$ to $\cX_0$. We will review this in somewhat more detail in \sref{sect:Curve_Counting}.

Consider now a curve $\g\subset\cX_5$, of genus zero. Such a curve may project under under $\p_{50}$ to either a node of $\cZ_{50}$, or to a curve of $\cZ_{50}$. The curves that project to a node are the 24 curves with index vector $(1,0,0,0,0)$. To start, suppose that $\g$ projects to a curve. The projection $\p_{50}(\g)$ lifts to a curve $\tilde\g$ in $\cX_0$. We want to understand the relation between the index vector $I$ of $\g$ and the index vector $\tilde I$ of $\tilde\g\subset\cX_0$. The reason that $\tilde I\neq I$ is that, owing to the fact that some curves are flopped, the homology group $H_2$ changes, and so the dual group $H_4$ must also change. The index vector is a vector of intersection numbers $\g\cdot D_i$ with a basis of divisors $D_i\in H_4$, so this also changes.

Let us take an affine coordinate $z$ for $\g$. We may take the coordinates $Y_{j,a}$ to be polynomials in $z$ which are of minimal degree in the sense that $Y_{j,1}$ and $Y_{j,0}$ have no common factors. Let us also simplify notation by 
     writing $Y_{5,a}{\=}Y_a$ and $Y_{0,a}{\=}\widetilde{Y}_a$. In virtue of \eqref{eq:lineqs} we have that
\beq\label{eq:ratios}\begin{split}
\frac{Y_1}{Y_0} 
&\;= -\frac{\IA_{00}}{\IA_{01}} \;= -\frac{\IA_{10}}{\IA_{11}}~, \\[5pt]
\frac{\widetilde{Y}_1}{\widetilde{Y}_0} 
&\;= -\frac{\IA_{00}}{\IA_{10}} \;= -\frac{\IA_{01}}{\IA_{11}} ~.
\end{split}\eeq
We see from \eqref{eq:Amatrix} that the $\IA_{af}$ each have degree $I_2 + I_3 + I_4 + I_5$. 
We want to calculate the degrees $I_1$ and $I_0$ of the $Y_a$ and the $\widetilde{Y_f}$, respectively. The various ratios of the $\IA_{af}$ in \eqref{eq:ratios} may have common factors. However, since we have assumed the $Y_a$ and $\widetilde{Y}_f$ to have minimal degree, we may write
\beq\begin{split}
\IA_{00}&\;= -Y_1 u~, \qquad\qquad \IA_{10}\;= - Y_1 v~, \\
\IA_{00}&\;= \+ Y_0 u~, \qquad\qquad \IA_{11}\;=\+ Y_0 v~, 
\end{split}\notag\eeq
where $u$ and $v$ are polynomials in $z$. From the second row in \eqref{eq:ratios} we see that we may take $u{\;=-}\widetilde{Y}_1$ and $v{\=}\widetilde{Y}_0$. In this way, we see that 
\beq
\IA\=
\begin{pmatrix}
\+ Y_1\widetilde{Y}_1 & -Y_1\widetilde{Y}_0 \\[2pt]
-Y_0\widetilde{Y}_1 & \+ Y_0\widetilde{Y}_0\\
\end{pmatrix}~.\notag\eeq
We have already observed that the degree of $\IA$ is $I_2 + I_3 + I_4 + I_5$, and this degree is seen to be equal to that of the right hand side of the above relation, which is $I_0 + I_1$.  It follows that
\beq\label{eq:I0rule}
 I_0 \;= -I_1 + I_2 + I_3 + I_4 + I_5.
\eeq
Taken together with the fact that $\tilde{I}_j{\=} I_j$, for $j{\=}2,3,4,5$, we see that we have recovered the duality transformation $\tilde{I}{\=}g_1 I$. 
Since this transformation law is a consequence of the transformation acting on $H_4$, it follows that the curves that project to the index vectors of the curves that project to the nodes are subject to the same transformation, so for example
\beq
(1,0,0,0,0) \longrightarrow (-1,0,0,0,0)~.
\notag\eeq
In order to find the relation \eqref{eq:I0rule}, we have identified curves in $\cX_0$ and $\cX_5$. Equally, we could have identified the homology groups and compared two curves $\gamma_0$ and $\gamma_5$ with the same index vectors, which would have given different curves.
\newpage
\section{Monodromies}\label{sect:Monodromies}
\vskip-10pt
We wish to find the monodromies around the loci $E_\mu$ and $D_{I}$ defined in \eqref{eq:singular_loci_E} and \eqref{eq:D_I_loci}. In the next subsection, we will compute the monodromy around the varieties $E_i$ using the series expansions for the periods around the large complex structure point. For the loci $D_{I}$, we use numerical integration of the Picard-Fuchs equation to find the monodromies. As we do not have the general five-parameter Picard-Fuchs equation and such an equation would in any case be impractical for this purpose, we use the Picard-Fuchs equations for one-parameter subfamilies as discussed in \sref{sect:ODE}. Finally, using the relation between the natural basis of periods in the patch $\varphi^0 \neq 0$ and $\varphi^i \neq 0$, we are able to compute the monodromies around $E_{0}$ in \sref{sect:monodromy_infinity}.
\subsection{Monodromies around the large complex structure points \texorpdfstring{$E_i$}{}}
\vskip-10pt
The monodromy matrices around the loci $E_{i}$ can be read directly from the asymptotics of the period vector $\Pi^{(0)}$ in the integral basis. These correspond to coordinate transformations $\varphi^i \to \me^{2\pi \ii} \varphi^i$, or alternatively $t^i \to t^i + 1$. These transformations give the following monodromies.
\begin{align} \label{eq:Monodromy_M_E_1}
\text{M}_{E_1} \= \left(\begin{array}{llllll} 1 & -1 &  \bm 0_4^T & 2 & 0 & \bm 0_4^T  \\
0 & \+1 & \bm 0_4^T& 0 & 0 & \bm 0_4^T \\
\bm 0_4 & \+\bm 0_4 & \II_{4} & \bm 0_4 & \bm 0_4 & 2 \II_{4} - \mathbbm{2}_{4} \\
0 & \+0 & \bm 0_4^T & 1 & 0 & \bm 0_4^T \\
0 & \+0 & \bm 0_4^T & 1 & 1 & \bm 0_4^T\\
\bm 0_4 & \+\bm 0_4 & \zero_{4} & \bm 0_4 & \bm 0_4 & \hskip2pt\II_{4}
\end{array} \right).
\end{align}
The monodromies around other loci $E_i$ are obtained by swapping the second and $(i+2)$'th column and row and the seventh and $(i+7)$'th column and row with each other.
\subsection{Monodromies around the \texorpdfstring{loci $D_I$}{conifold loci}}
\vskip-10pt
We now set $\varphi^i = s^i \varphi$, $\varphi^0=1$ with $s^i$ complex constants. $\IDelta$ becomes a polynomial of degree 16 in $\varphi$. This has 16 roots, which are the intersections of the singular locus $\IDelta=0$ with the plane $\varphi^i = s^i \varphi$. We will find particularly simple Picard-Fuchs operators when some of the $s^i$ are equal. In these cases some of the periods become equal, hence there exists an operator of degree $<12$, whose independent solutions are exactly the distinct periods. These differential equations can be integrated numerically, yielding the monodromy matrices for the independent periods. 

Of course the matrices found this way do not give the complete monodromy, as not all of the 12 periods are independent on the lines that we study. However, there is a natural relation between these 'reduced' matrices and the full monodromy matrices, which can be used, together with the $S_{5}$ symmetry, to find the full monodromy. To exemplify this process, let us consider the case where $s^1 \neq s^2 = s^3 = s^4 = s^5$. This leaves a set of 6 independent periods, as
\begin{align} \notag
{
\setlength\arraycolsep{0pt}
\begin{matrix*}[l] 
\varpi^{(0);2}(\varphi) &\= \varpi^{(0);3}(\varphi) &\= \varpi^{(0);4}(\varphi) &\= \varpi^{(0);5}(\varphi)~,\\[5pt]
\varpi^{(0)}_{2}(\varphi) &\= \varpi^{(0)}_{3}(\varphi)^{\phantom{;3}} &\= \varpi^{(0)}_{4}(\varphi)^{\phantom{;4}} &\= \varpi^{(0)}_{5}(\varphi)~.
\end{matrix*}
}
\end{align}
The general monodromy matrix, giving the monodromy transformation of the periods around a singularity $\varphi_*$, can be written as
\begin{align} \label{eq:monodromy_M_definition}
\text{M}_{\varphi_*} \= (\bm u_0,\bm u_1,\ldots,\bm{ u}_{10},\bm u_{11})~,
\end{align}
where $\bm u_i$ are 12-component column vectors
\begin{align} \notag
\bm u_i \= (u_i^0,u_i^1,\ldots,u_i^{10},u_i^{11})^T~.
\end{align}
Since some of the periods are equal, we cannot find their individual contributions to this matrix from the reduced monodromy matrix. Instead, the reduced matrix takes the form
\begin{align} \label{eq:monodromy_M_hat_definition}
\widehat{\text{M}}_{\varphi^*} \= (\hat{ \bm u}_0,\hat{ \bm u}_1,\hat{ \bm u}_2+ \hat{\bm u}_3+ \hat{\bm u}_4+ \hat{\bm u}_5, \hat{\bm u}_6,\hat{\bm u}_7,\hat{\bm u}_8+\hat{\bm u}_9+\hat{\bm u}_{10}+\hat{\bm u}_{11})~,
\end{align}
where now $\hat u_i$ are 6 component column vectors
\begin{align} \notag
\hat{\bm u}_i \= (u_i^0,u_i^1,u_i^2,u_i^6,u_i^7,u_i^8)^T~.
\end{align}
Relations like this constrain the full $12 \times 12$ monodromy matrices. We can construct the full matrices from this data by numerically integrating the Picard-Fuchs equation along several paths in the complex line.  

Finally, to make the computation slightly simpler, we use the fact that the singularities at $\IDelta=0$ correspond to conifolds. It is expected that the monodromies around the conifold loci take the form
\begin{align} \label{eq:monodromy_M_w_relation}
\text{M} \= \text{I}_{12} - \bm w (\Sigma \bm w)^T,
\end{align}
where $\bm w$ is a 12-component vector that gives the cycle vanishing at the conifold point. Thus we can reduce the problem to finding $16$ vectors corresponding to the different components $D_I$ of the singular locus.

To get an idea of how the computation proceeds, we briefly explain the computation of some monodromies in a relatively simple example. To be precise, we study the case
\begin{align} \notag
s^1 = 1~, \qquad s^2 \= s^3 \= s^4 \= s^5 = \frac{95}{100}~.
\end{align} 
We have 6 independent periods and so can find, using the procedure outlined in \sref{sect:ODE}, a Picard-Fuchs operator of degree~6. This operator has solutions $\varpi^{(0);0}(\varphi)$, $\varpi^{(0);1}(\varphi)$, $\varpi^{(0);2}(\varphi)$, $\varpi^{(0)}_{1}(\varphi)$, $\varpi^{(0)}_{2}(\varphi)$, and $\varpi^{(0)}_3(\varphi)$. In the ensuing discussion, we shall find use for the shorthands
\begin{equation} \label{eq:shorthands}
\Imu\=5\,\frac{81{-}4\sqrt{95}}{5041}~,\qquad \overline{\Imu}\=5\,\frac{81{+}4\sqrt{95}}{5041}~,\qquad \Inu\=5\,\frac{12{-}\sqrt{95}}{98}~,\qquad\overline{\Inu}\=5\,\frac{12{+}\sqrt{95}}{98}~.
\end{equation}
The discriminant expressed in terms of $\varphi$ is in this case, up to a multiplicative constant,
\begin{align} \nonumber
\IDelta \= \big(\varphi -1\big)^6 \big(\varphi - \Imu\big) \big(\varphi -\overline{\Imu}\big) \big(\varphi - \Inu\big)^4 \big(\varphi - \overline{\Inu}\big)^4~.
\end{align}
Each of these factors corresponds to an intersection of a component $D_I$ with the line. In this way, we can associate each factor with such a component:
\begin{alignat*}{2}
&D_{\{0\}} &&\= \left\{\varphi \= \Imu \right\}~,\\
&D_{\{0,2\}} &&\= D_{\{0,3\}} \= D_{\{0,4\}} \= D_{\{0,5\}} \= \left\{\varphi \= \Inu\right\}~,\\
&D_{\{0,1\}} &&\= \left\{\varphi \= \overline{\Imu}\right\}~,\\
&D_{\{0,2,3\}} &&\= D_{\{0,2,4\}} \= D_{\{0,2,5\}} \= D_{\{0,3,4\}} \= D_{\{0,3,5\}} \= D_{\{0,4,5\}} \=\{\varphi = 1\}~,\\
&D_{\{0,1,2\}} &&\= D_{\{0,1,3\}} \= D_{\{0,1,4\}} \= D_{\{0,1,5\}} \= \left\{\varphi \= \overline{\Inu}\right\}~.
\end{alignat*}
The monodromy matrices around these points are given by
\begin{alignat*}{2} \nonumber
\widehat{\text{M}}_{\Imu} &\= \left(
\begin{smallarray}{rrrrrr}
1 & 0 & 0 & 0 & 0 & 0 \\
0 & \+1 & 0 & 0 & 0 & 0 \\
0 & 0 & \+1 & 0 & 0 & 0 \\
-1 & 0 & 0 & \+1 & 0 & 0 \\
0 & 0 & 0 & 0 & \+1 & 0 \\
0 & 0 & 0 & 0 & 0 & \+1 \\
\end{smallarray}
\right), \qquad \widehat{\text{M}}_{\Inu} &&\hskip-45pt\= \left(
\begin{smallarray}{rrrrrr}
\phantom{1}9 & 0 & -8 & \+16 & \phantom{-1}0 & \phantom{-1}0 \\
0 & \+1 & 0 & 0 & 0 & 0 \\
0 & 0 & \phantom{-1}1 & 0 & 0 & 0 \\
-4 & 0 & 4 & -7 & 0 & 0 \\
0 & 0 & 0 & 0 & \+1 & 0 \\
1 & 0 & -1 & 2 & 0 & \+1 \\
\end{smallarray}
\right),\\[5pt]
\widehat{\text{M}}_{\overline{\Imu}} &\= \left(
\begin{smallarray}{rrrrrr}
3 & -2 & 0 & 4 & 0 & 0 \\
0 & 1 & 0 & 0 & 0 & 0 \\
0 & 0 & \+1 & 0 & 0 & 0 \\
-1 & 1 & 0 & -1 & 0 & 0 \\
1 & -1 & 0 & 2 & \+1 & 0 \\
0 & 0 & 0 & 0 & 0 & \+1 \\
\end{smallarray}
\right), \qquad \widehat{\text{M}}_{1} &&\hskip-45pt\= \left(
\begin{smallarray}{rrrrrr}
25 & 0 & -48 & 96 & 48 & 96 \\
12 & \+1 & -24 & 48 & 24 & 48 \\
6 & 0 & -11 & 24 & 12 & 24 \\
-6 & 0 & 12 & -23 & -12 & -24 \\
0 & 0 & 0 & 0 & 1 & 0 \\
3 & 0 & -6 & 12 & 6 & 13 \\
\end{smallarray}
\right), \\[5pt]
&\hskip45pt\widehat{\text{M}}_{\overline{\Inu}} \= \left(
\begin{smallarray}{rrrrrr}
 17 & -16 & -16 & 64 & 0 & 96 \\
 0 & 1 & 0 & 0 & 0 & 0 \\
 6 & -6 & -5 & 24 & 0 & 36 \\
 -4 & 4 & 4 & -15 & 0 & -24 \\
 4 & -4 & -4 & 16 & \+1 & 24 \\
 1 & -1 & -1 & 4 & 0 & 7\\
\end{smallarray}
\right)~.
\end{alignat*}
To find the full monodromy matrix corresponding to the monodromy around $D_{\{0\}}$, we use
\begin{align} \label{eq:Monodromy_Identification}
\widehat{\text{M}}_{\Imu} \= \widehat{\text{M}}_{\{0\}},
\end{align}
where $\text{M}_{\{0\}}$ is of the form \eqref{eq:monodromy_M_w_relation}, and $\widehat{\text{M}}_{\{0\}}$ of the form \eqref{eq:monodromy_M_hat_definition}. This equation allows us to partially fix the vector $\bm w$, which we denote by $\bm w_{\{0\}}$, giving conditions which can be solved by
\begin{align} \notag
\bm w_{\{0\}} \= (0,\,0,\,0,\,w^4,\,w^5,\,-w^4,\,-w^5,\,1,\,0,\,0,\,w^{10},\,w^{11},\,-w^{10}-w^{11})~.
\end{align}
To proceed, we can compute the monodromies on other similar lines, such as ${s^1 {\,=\,} s^3 {\,=\,} s^4 {\,=\,} s^5 {\,=\,} \frac{95}{100}}$, $s^2 {\,=\,} 1$. Alternatively, we could impose the $S_{5}$ symmetry, whereby all the periods related by a permutations of the indices 2,3,4 and 5 must contribute equally. In this way we see that the cycle vanishing at $D_{\{0\}}$ has sixth component 1, and all other components zero:
\begin{align} \notag
\bm w_{\{0\}} \= (0,\,0,\,0,\,0,\,0,\,0,\,1,\,0,\,0,\,0,\,0,\,0)~.
\end{align}
Next, we concentrate on the singularities at $\varphi = \overline{\Imu}$ and $\varphi = \Inu$. The latter lies on four singular loci, $D_{\{0,2\}}$, $D_{\{0,3\}}$, $D_{\{0,4\}}$, and $D_{\{0,5\}}$, while the former lies solely in $D_{\{0,1\}}$. Therefore we can use an expression of the form \eqref{eq:monodromy_M_w_relation} for the monodromy matrix around the singularity at $\Inu$, while around $\overline{\Imu}$ the monodromy is a product of four similar matrices. By comparing to $\text{M}_{\Inu}$, we find
\begin{align} \notag
\bm w_{\{0,1\}} \= (-2,\,0,\,0,\,w^4,\,w^5,\,-w^4-w^5,\,1,\,-1,\,0,\,w^{10},\,w^{11},\,-w^{10}-w^{11})~.
\end{align}
By either computing monodromies with different values of $s^i$, or by a symmetry argument, we find that the vector is given by
\begin{align} \notag
\bm w_{\{0,1\}} \= (-2,0,0,0,0,0,1,-1,0,0,0,0)~,
\end{align}
which allows us to compute the monodromy matrix $\text{M}_{\{0,1\}}$. Again, by symmetry or considering different values of weights, it can be shown that the vectors giving the monodromy matrices $\text{M}_{\{0,2\}}$, $\text{M}_{\{0,3\}}$, $\text{M}_{\{0,4\}}$ and $\text{M}_{\{0,5\}}$ are given by permuting the components of the vector $\bm w_{\{0,1\}}$:
\begin{align} \notag
\begin{split}
\bm w_{\{0,2\}} \= (-2,\,0,\,0,\,0,\,0,\,0,\,1,\,0,\,-1,\,0,\,0,\,0)~,\\
\bm w_{\{0,3\}} \= (-2,\,0,\,0,\,0,\,0,\,0,\,1,\,0,\,0,\,-1,\,0,\,0)~,\\
\bm w_{\{0,4\}} \= (-2,\,0,\,0,\,0,\,0,\,0,\,1,\,0,\,0,\,0,\,-1,\,0)~,\\
\bm w_{\{0,5\}} \= (-2,\,0,\,0,\,0,\,0,\,0,\,1,\,0,\,0,\,0,\,0,\,-1)~.
\end{split}
\end{align}
As a consistency check, it can be seen that the matrix around $\Inu$ is given by a product of reduced monodromy matrices:
\begin{align} \notag
\text{M}_{\Inu} \= \widehat{\text{M}}_{\{0,2\}}\, \widehat{\text{M}}_{\{0,3\}}\, \widehat{\text{M}}_{\{0,4\}}\, \widehat{\text{M}}_{\{0,5\}}~.
\end{align}
The matrices corresponding to the remaining loci can be found using similar techniques. This is made slightly more complicated by the fact that paths on the lines $s^1 \neq s^2 = s^3 = s^4 = s^5$ only circle intersections of multiple components. Perhaps the easiest way to circumvent this is to consider a new case where $s^1 \neq s^2 \neq s^3 = s^4 = s^5 \neq s^1$, and permutations thereof. In the case $s_{1}\neq s_{2}\neq s_{3}$, $D_{\{0,1,2\}}$ intersects the plane $\varphi^{i}=s^i \varphi$ in a point that is distinct from the other components. This computation, together with symmetry considerations, leads us to a form for the monodromy matrix where the vanishing cycle is given by
\begin{align} \notag
\bm w_{\{0,1,2\}} \= (4,\,0,\,0,\,2,\,2,\,2,\,-1,\,1,\,1,\,0,\,0,\,0)~.
\end{align}
The vectors in other cases are given by permuting the components of this vector. Again, one can check that the matrices $\widehat{\text{M}}_1$ and $\widehat{\text{M}}_{\overline{\Inu}}$  can be written in terms of the reduced matrices associated to these vectors:
\begin{align} \notag
\begin{split}
\widehat{\text{M}}_1 &\= \widehat{\text{M}}_{\{0,2,3\}}\, \widehat{\text{M}}_{\{0,2,4\}} \, \widehat{\text{M}}_{\{0,2,5\}}\, \widehat{\text{M}}_{\{0,3,4\}}\, \widehat{\text{M}}_{\{0,3,5\}}\, \widehat{\text{M}}_{\{0,4,5\}}~,\\
\widehat{\text{M}}_{\overline{\Inu}} &\= \widehat{\text{M}}_{\{0,1,2\}}\, \widehat{\text{M}}_{\{0,1,3\}}\, \widehat{\text{M}}_{\{0,1,4\}}\, \widehat{\text{M}}_{\{0,1,5\}}~.
\end{split}
\end{align}
We have found 16 matrices $M_{\{0\}}$, $M_{\{0,i\}}$, and $M_{\{0,i,j\}}$, and there remain 16 still unaccounted for. These, however, can be constructed from the known 16 by a change of indices $0 \leftrightarrow i$. By symmetry, the matrices that are related to each other by such a permutation must be equal. We must, however, take into account that the monodromy transformations obtained in this way are expressed in different bases. Changing all to a common basis, which we take to be the symplectic basis where $\Pi^{(0)}$ is given by \eqref{eq:Pi0}, gives matrices with different entries. Thus, for example
\begin{align} \notag
\text{M}_{\{1\}} \= \text{T}_{\Pi^{(1)} \Pi^{(0)}}^{-1}\, \text{M}_{\{0\}} \text{T}_{\Pi^{(1)}\, \Pi^{(0)}}~,
\end{align}
where $\text{T}_{\Pi^{(1)} \Pi^{(0)}}$, given explicitly in \eqref{eq:defn_T_Pi^0Pi^1}, is a change of basis matrix from the canonical integral basis in the patch $\varphi^0 = 1$ to the canonical integral basis in the patch $\varphi^1 = 1$. We will see another explicit example of this in the next subsection where we use this observation to compute the monodromy around the locus $E_0$ `at~infinity'. 
\subsection{Monodromy around infinity\texorpdfstring{, $E_0$}{}}\label{sect:monodromy_infinity}
\vskip-10pt
The remaining singular locus is $\varphi^0=0$, which, seen from the patch $\varphi^0 = 1$ corresponds to the monodromy around infinity. Due to the $S_{5}$ symmetry, we know that the locus $\varphi^0 = 0$ is on a par with the other loci $\varphi^i = 0$. The only essential difference to the earlier computation is the use of a different basis for the periods. 

To find the appropriate change of basis, we use the matrix $\text{T}_{\pi^{(0)} \pi^{(1)}}$ from \eqref{eq:change_of_basis_01}, which gives the relation between the period vectors $\pi^{(1)}$ and $\pi^{(0)}$, whose components give the periods as combinations of Bessel function integrals. Using the matrices $\text{T}_{\varpi^{(i)} \pi^{(i)}}$ and $\text{T}_{\Pi^{(i)} \varpi^{(i)}}$, we can change from this basis to the integral basis. Note that due to the symmetry, the relation of the vectors $\pi^{(1)}$ to the integral period vector $\Pi^{(1)}$ is same as that of $\pi^{(0)}$ to $\Pi^{(0)}$, so that $\text{T}_{\Pi^{(1)} \pi^{(1)}} = \text{T}_{\Pi^{(0)} \pi^{(0)}}$. The transformation from $\Pi^{(1)}$ to $\Pi^{(0)}$ is thus given by
\begin{equation} \label{eq:defn_T_Pi^0Pi^1}
\begin{aligned}
\text{T}_{\Pi^{(0)} \Pi^{(1)}} &\= \text{T}_{\Pi^{(0)} \varpi^{(0)}} \text{T}_{\varpi^{(0)} \pi^{(0)}} \text{T}_{\pi^{(0)} \pi^{(1)}}\big(\text{T}_{\Pi^{(1)} \varpi^{(1)}} \text{T}_{\varpi^{(1)} \pi^{(1)}}\big)^{-1}\\[8pt]
&\= \left(
\begin{smallarray}{rrrrrrrrrrrr}
 -1 & 0 & 1 & 1 & 1 & 1 & -16 & -12 & -6 & -6 & -6 & -6 \\
 0 & \+1 & -1 & -1 & -1 & -1 & 12 & 12 & 6 & 6 & 6 & 6 \\
 0 & 0 & -1 & 0 & 0 & 0 & 6 & 6 & 0 & 4 & 4 & 4 \\
 0 & 0 & 0 & -1 & 0 & 0 & 6 & 6 & 4 & 0 & 4 & 4 \\
 0 & 0 & 0 & 0 & -1 & 0 & 6 & 6 & 4 & 4 & 0 & 4 \\
 0 & 0 & 0 & 0 & 0 & -1 & 6 & 6 & 4 & 4 & 4 & 0 \\
 0 & 0 & 0 & 0 & 0 & 0 & -1 & 0 & 0 & 0 & 0 & 0 \\
 0 & 0 & 0 & 0 & 0 & 0 & 0 & 1 & 0 & 0 & 0 & 0 \\
 0 & 0 & 0 & 0 & 0 & 0 & -1 & -1 & -1 & 0 & 0 & 0 \\
 0 & 0 & 0 & 0 & 0 & 0 & -1 & -1 & 0 & -1 & 0 & 0 \\
 0 & 0 & 0 & 0 & 0 & 0 & -1 & -1 & 0 & 0 & -1 & 0 \\
 0 & 0 & 0 & 0 & 0 & 0 & -1 & -1 & 0 & 0 & 0 & -1 \\
\end{smallarray}
\right)\!.
\end{aligned}
\end{equation}
The monodromy of $\Pi^{(0)}$ around $\varphi^0 = 0$ is, by symmetry, equal to the mondromy of $\Pi^{(1)}$ around $\varphi^1 = 0$, which directly allows us to find the monodromy of $\Pi^{(1)}$ around the locus $\varphi^0 = 0$:
\begin{align} \notag
\text{M}_{E_{1}} \= \left(\text{T}_{\Pi^{(1)} \Pi^{(0)}}\right)^{-1} \left(\text{M}_{E_{0}}\right)^{-1} \text{T}_{\Pi^{(1)} \Pi^{(0)}} \= \left(
\begin{smallarray}{rrrrrrrrrrrr}
 \+1 & -1 & 1 & 1 & 1 & 1 & -2 & -12 & 0 & 0 & 0 & 0 \\
 0 & \+1 & 0 & 0 & 0 & 0 & -12 & 24 & 6 & 6 & 6 & 6 \\
 0 & 0 & \+1 & 0 & 0 & 0 & 0 & 6 & 0 & 2 & 2 & 2 \\
 0 & 0 & 0 & \+1 & 0 & 0 & 0 & 6 & 2 & 0 & 2 & 2 \\
 0 & 0 & 0 & 0 & \+1 & 0 & 0 & 6 & 2 & 2 & 0 & 2 \\
 0 & 0 & 0 & 0 & 0 & \+1 & 0 & 6 & 2 & 2 & 2 & 0 \\
 0 & 0 & 0 & 0 & 0 & 0 & 1 & 0 & 0 & 0 & 0 & 0 \\
 0 & 0 & 0 & 0 & 0 & 0 & 1 & 1 & 0 & 0 & 0 & 0 \\
 0 & 0 & 0 & 0 & 0 & 0 & -1 & 0 & \+1 & 0 & 0 & 0 \\
 0 & 0 & 0 & 0 & 0 & 0 & -1 & 0 & 0 & \+1 & 0 & 0 \\
 0 & 0 & 0 & 0 & 0 & 0 & -1 & 0 & 0 & 0 & \+1 & 0 \\
 0 & 0 & 0 & 0 & 0 & 0 & -1 & 0 & 0 & 0 & 0 & \+1 \\
\end{smallarray}
\right)~.
\end{align}
We have used the inverse of the matrix $\text{M}_{E_{0}}$ because the direction of the contour is reversed when changing patches.
\subsection{Recovering monodromies for the quotient manifolds}
\vskip-10pt
Finally, let us briefly comment on the relation of the results presented here to those found for the quotient manifolds in \cite{Candelas:2019llw}. Specialising to the locus $\varphi^i = \varphi$, $\varphi^0 = 1$, the discriminant vanishes for $\varphi \in \{ \frac{1}{25}, \frac{1}{9}, 1\}$. The locus $D_{\{0\}}$ is associated to the first of these points, the loci $D_{\{0,i\}}$ to the second, and the loci $D_{\{0,i,j\}}$ to the last. 

On the locus $\varphi^i = \varphi$, $\varphi^0 = 1$, only four of the elements of $\Pi^{(0)}$ are independent. We collect these into the reduced period vector $\wh \Pi^{(0)}$. This is related to the integral period vector $\Pi_{\IZ_{10/\kappa}}$ of the quotient manifold $\HV/\mathbb{Z}_{10/\kappa}$ by a matrix $\text{T}_{\kappa}$.
\begin{equation} \notag
\Pi^{(0)} \; \defineas \;\left(\begin{matrix}\Pi^{(0);0}_{0}\\[3pt]\Pi^{(0);1}\\[3pt]\Pi^{(0)}_{0}\\[3pt]\Pi^{(0)}_{1} \end{matrix}\right)~,\qquad\Pi_{\IZ_{10/\kappa}}\=\text{T}_{\kappa}\wh\Pi^{(0)}~,\qquad \text{T}_{\kappa}\=\begin{pmatrix}10/\kappa&0&0&\,0\\[3pt]0&2/\kappa&0&\,0\\[3pt]0&0&1&\,0\\[3pt]0&0&0&\,1\end{pmatrix}~.
\end{equation}
We can now give the monodromies $\text{M}_{1},\,\text{M}_{\frac{1}{9}}$, and $\text{M}_{\frac{1}{25}}$ of $\Pi_{\IZ_{10/\kappa}}$. First, take the product of the relevant matrices $\widehat{\text{M}}_{s}$ which give the monodromies of $\wh \Pi^{0}$, and then conjugate by $\text{T}_{\kappa}$ to obtain the monodromies of $\Pi_{\IZ_{10/\kappa}}$. For instance, for the $\mathbb{Z}_{10}$ quotient 
\begin{equation} \notag
\text{M}_{\frac{1}{25}}\=\text{T}_{1}^{-1}\widehat{ \text{M}}_{\{0\}}\text{T}_{1}\=\left(\begin{array}{rrrr}1&\phantom{-1}0&\phantom{-1}0&\phantom{-1}0\\[3pt]0&1&0&0\\[3pt]-10&0&1&0\\[3pt]0&0&0&1\end{array}\right)~.
\end{equation}
\newpage

\section{Counting Curves on the Mirror Hulek-Verrill Manifold} \label{sect:Curve_Counting}
\vskip-10pt
There is an interesting problem in directly counting the numbers of various curves of different degrees on the Hulek-Verrill manifold and its quotients. This serves multiple purposes, such as confirming the predictions of mirror symmetry and counting microstates for some configurations of branes wrapped on various cycles on the manifold. In this section, we will find the rational curves up to degree five, and verify that their number agrees with the instanton numbers of \sref{sect:Mirror_Map}. 

It is good to recognise that the manifolds in $\MHV$ can be realised as blowups of the five singular tetraquadrics $\wh\MHV_{i}$ $i=0,\dots,4$, with 24 nodes, using the procedure of \cite{Candelas:1987kf}. $\wh\MHV_{i}$ are the spaces $\cZ_{5i}$ of the section \sref{sect:webs} and are singular limits of the family corresponding to the configuration
\begin{align} \notag
\cicy{\IP^1\\\IP^1\\\IP^1\\\IP^1}{2 \\ 2 \\ 2 \\ 2}_{\chi\,=-128~.}
\end{align}
Members of the family $\MHV$ are elliptically fibred manifolds, and we are able to compute the discriminant of the fibration using standard methods \cite{Duistermaat}. It turns out that the the first few low-degree rational curves appear as irreducible components of singular fibres of the elliptic fibration as in \fref{fig:Projections_and_Fibres}.
\begin{figure}[H]
	\centering
    \adjustbox{scale=1.2,center}{
	\begin{tikzcd}
		&  
		& L_i \arrow[hookrightarrow]{r}{}
		& \text{H}\Lambda \arrow{d}{\pi_j} 
		& \\		
		& \IP^1 \arrow[hookrightarrow]{r}{\varphi} 
		& F_j \arrow[hookrightarrow]{r}{} \arrow{d}{\pi_{m,n}} 
		& _{\phantom{i}}\widehat{\text{H}\Lambda}_j  \arrow{d}{\pi_{m,n}} 
		& \\		
		& 
		& B\arrow[hookrightarrow]{r}[swap]{} 
		& \IP^1_m \times \IP^1_n
		& 
	\end{tikzcd}}
	\vskip5pt
	\capt{6in}{fig:Projections_and_Fibres}{Structure of the fibrations relevant to counting some rational and elliptic curves. $L_i$ denote the lines on $\MHV$ that are blown down to obtain the singular mirror Hulek-Verrill manifold $\wh \MHV_j$ with the birational map denoted by $\pi_j$. $\wh \MHV_j$ is an elliptically fibred manifold with base $\IP^1 \times \IP^1$, and a generic fibre $F_j$. On the discriminant locus $\Delta = 0$ of the elliptic fibration, the fibre becomes singular. On a special set of points $B$, corresponding to nodes of the discriminant locus, the degenerate fibre is a union of two rational curves.}	
\end{figure}
The explicit embeddings of curves depend non-trivially on the coefficients in the defining polynomials, but the curve counts for generic members of the family of mirror manifolds agree. For this reason we will, in place of explicit expressions, discuss properties of a generic member of the family~$\MHV$.

Parts of this discussion are best framed in terms of various embedding maps with different degrees. Among these appear numerous context-specific rational functions. For this reason we will often use the symbols $r_{k}(z)$, $\widetilde{r}_{k}(z)$, to denote a ratio of two \textit{situation-dependent} polynomials of degree $k$. Two instances of these symbols in this section should not automatically be understood as referring to the same function. In this section the Latin indices generically run from 0 to 4. When two different indices appear in an expression, they are understood to refer to distinct numerical values.
\subsection{Blow-down and elliptic fibration} \label{sect:Blow-Down}
\vskip-10pt
The configuration matrix of $\MHV$ is of the form considered in \cite{Candelas:1987kf}, which means that we can use the contraction procedure to obtain a quadric manifold $\widehat{\MHV}_{i}$ defined by one equation:
\begin{align} \label{eq:MHV_Contraction_Matrices}
\cicy{\IP^1\\\IP^1\\\IP^1\\\IP^1\\\IP^1}{1 & 1\\1 & 1\\1 & 1\\ 1 & 1\\ 1 & 1}_{\chi\,=-80} \longleftrightarrow \qquad\;  \cicy{\IP^1\\\IP^1\\\IP^1\\\IP^1}{2 \\ 2 \\ 2 \\ 2}_{\chi\,=-128~.}
\end{align}
We frequently distinguish the five $\IP^{1}$ factors in the product $(\IP^{1})^{5}$ by subscripts. For example, $\IP^{1}_{i}$ denotes the $i$'th such $\IP^{1}$, and has projective coordinates $Y_{i,0},Y_{i,1}$. Throughout this section, we use affine coordinates $Y_{i}=\frac{Y_{i,1}}{Y_{i,0}}$. This makes the equations simpler, and the instances where projective coordinates are needed for statements to be strictly correct are few. Nonetheless, all polynomials in this section can be homogenised using projective coordinates, and in this way any minor ambiguities relating to points at infinity are cleared up.

To see in detail how the process depicted in \eqref{eq:MHV_Contraction_Matrices} works, let us consider the contraction with respect to $\IP^{1}_{i}$. The equations defining the manifold $\MHV$ can be written as
\begin{align} \label{eq:MHV_Contraction}
\begin{split}
Q^1(\bm{Y}) &\= \alpha_i Y_i + \beta_i~,\\
Q^2(\bm{Y}) &\= \gamma_i Y_i + \delta_i~,
\end{split}
\end{align}
with $\alpha_{i},\beta_{i},\gamma_{i},\delta_{i}$ multilinear functions of the four coordinates $Y_j$, $j \neq i$, and no sum over the repeated $i$ is implied. The pair of conditions \eqref{eq:MHV_Contraction} is equivalent to the single matrix equation
\begin{align} \notag
\left( \begin{matrix}
\alpha_i & \beta_i \\
\gamma_i & \delta_i
\end{matrix}  \right) \left( \begin{matrix}
Y_i\\
1
\end{matrix} \right) \= \mathbf{0}~.
\end{align}
Existence of a solution is equivalent to the vanishing of the determinant of the matrix, that is
\begin{align} \label{eq:Q}
\wh Q^i \;\defineas\;\alpha_i \,\delta_i - \beta_i \, \gamma_i \= 0~,
\end{align}
We denote the variety defined by $\{\wh Q^i=0\} \subset  (\IP^1)^4$ as $\widehat{\MHV}_i$. One can see from \eqref{eq:Q} that $\wh\MHV_{i}$ is a conifold. Since the functions $\alpha_i,\beta_i,\gamma_i,\delta_i$ are multilinear, the corresponding configuration matrix is indeed of the form \eqref{eq:MHV_Contraction_Matrices}.

Note that the varieties $\widehat{\MHV}_i$ are birational to $\MHV$. The projection $\pi_i: \MHV \to \wh{\MHV}_i$ defined by
\begin{align} \notag
\pi_i (Y_i,Y_j,Y_k,Y_m,Y_n) \= (Y_j,Y_k,Y_m,Y_n)
\end{align}
gives a birational map between the varieties. Given a point $(Y_j,Y_k,Y_m,Y_n) \in \wh{\MHV}_i$, with $\alpha_i \neq 0$ or $\gamma_i \neq 0$, the equations $Q^1=Q^2=0$ are solved by the unique point $Y_i = - \frac{\beta_i}{\alpha_i}$ or $Y_i = - \frac{\delta_i}{\gamma_i}$, respectively (when $\alpha_i,\gamma_i \neq 0$, these agree), and the inverse $\pi_i^{-1}$ is well-defined. However, when $\alpha_i = \gamma_i = 0$, the conditions $Q^1 = Q^2 = 0$ are satisfied if and only if $\beta_i = \delta_i = 0$. If this is the case, the equation $Q^1 = Q^2 = 0$ is true for all values of $Y_i$, and the inverse image of the point in $\wh \MHV_i$ is a line $\alpha_i = \beta_i = \gamma_i = \delta_i = 0$ on $\MHV$. For generic values of parameters, including generic $\IZ_5$, $\IZ_5 {\times } \IZ_2$ and $\IZ_5 {\times} \IZ_2 {\times } \IZ_2$ symmetric cases, these equations have 24 solutions. From the definition of $\wh Q^i$,~\eqref{eq:Q}, it is clear that the points satisfying this condition are exactly the singularities of $\wh{\MHV}_i$.

The manifold $\MHV$ is generically a smooth elliptic threefold, while $\wh \MHV_{i}$ is an elliptically fibred singular variety (see \fref{fig:Projections_Elliptic}). To see this explicitly, let us choose the base of the fibration to be $\IP^1_m \times \IP^1_n$. We can view the polynomial $\wh Q^i$ as a biquadratic whose coefficients depend on $Y_m$ and $Y_n$. 
\begin{align} \label{eq:fibrationP0P1}
\wh Q^i(Y_j,Y_k) \= \sum_{a,b=0}^2 A_{a,b}(Y_m,Y_n) \; Y_j^a \, Y_k^b,
\end{align}
where $A_{a,b}$ are functions of the base coordinates $Y_m,\,Y_n$. The exact form of these functions depends on the choice of the Calabi-Yau manifold $\MHV$. This defines a biquadric subvariety $E_{i;m,n}$ of $\IP^1_j \times \IP^1_k$, which is a Calabi-Yau variety of dimension one, and thus an elliptic curve.
\begin{figure}[H]
\hskip155pt
	\begin{tikzcd}		
		&
		& \text{H}\Lambda  \arrow{d}{\pi_{i}} \\			
		& E_{i;m,n} \arrow[hookrightarrow]{r}{} 
		& _{\phantom{i}}\widehat{\text{H}\Lambda}_i  \arrow{d}{\pi_{m,n}} \\		
		& 
		& \IP^1_m \times \IP^1_n. 
	\end{tikzcd}
	\vskip10pt
	\capt{6in}{fig:Projections_Elliptic}{The Elliptic Fibration on $\MHV_i$ with base $\IP^1_m \times \IP^1_n$.}	
\end{figure}
Any biquadratic in $\IP^1_m \times \IP^1_n$ can be transformed into the Weierstrass form \cite{Duistermaat}. To this end, one first computes the quadratic discriminant of \eqref{eq:fibrationP0P1} with respect to $Y_j$.
\begin{align} \notag
\cD_n(Y_k) \= \left(\sum_{a=0}^2 A_{a,1} Y_k^i \right)^2 \!\!- 4 \left( \sum_{a=0}^2 A_{i,2} Y_k^i \right) \!\! \left( \sum_{i=0}^2 A_{i,0} Y_k^i \right) \; \defineas \; b_4 Y_k^4 {+} 4 b_3 Y_k^3 {+} 6 b_2 Y_k^2 {+} 4 b_1 Y_k {+}b_0~. \nonumber
\end{align}
One computes the two ``Eisenstein invariants of plane quartics'' defined in \cite{Duistermaat} for this polynomial:
\begin{align}
\begin{split}
D_{m,n} &\=b_4 b_0 + 3 b_2^2 - 4 b_3 b_1~,\\[3pt]
E_{m,n} &\= b_4 b_1^2 + b_3^2 b_0 - b_4 b_2 b_0 - 2 b_3 b_2 b_1 + b_2^3~,
\end{split}
\end{align}
where each $b_a$ is a function of $Y_{m}$ and $Y_{n}$ These can be used to write the Weierstrass form of the elliptic fibre as
\begin{align} \notag
y^2 &= x^3 -D_{m,n} x + 2E_{m,n}~.
\end{align}
The discriminant of this elliptic curve is
\begin{align} \notag
\Delta_{i;\,m,n} &= -D_{m,n}^{3} + 27 E_{m,n}^{2}~,
\end{align}
where the index $i$ refers to the coordinate with respect to which we have contracted $\MHV$ in order to obtain the singular manifold $\wh \MHV_i$, and the indices $m$ and $n$ refer to the choice of the base of the fibration. It is useful to observe that the discriminants satisfy the relations
\begin{align} \notag
\Delta_{i;\,m,n} = \Delta_{j;\,m,n} = \Delta_{k;\,m,n}~,
\end{align}
In other words, for the purposes of computing the discriminant on the base $\IP_m \times \IP_n$, it does not matter which contraction we choose. We plot the zero loci for three $\Delta_{i;\,m,n}$ in \fref{fig:deltaplot3d}.

In the generic case, $\Delta_{i;\,m,n}$ is an irreducible bidegree $(24,24)$ polynomial.
\begin{align} \notag
\Delta_{i;\,m,n}(Y_m,Y_n) = \sum_{a,b=0}^{24} \alpha_{a,b} Y_m^{a} \, Y_n^{b}~.
\end{align}
In case the manifold is symmetric under $\IZ_2$ or $\IZ_2 {\, \times \,} \IZ_2$, the discriminant satisfies one or both of the following symmetry relations:
\begin{align} \notag
Y_m^{24} \, Y_n^{24} \, \Delta_{i;\,m,n} \left( \frac{1}{Y_m}, \frac{1}{Y_n} \right) &\= \Delta_{i;\,m,n}(Y_m,Y_n)~, \qquad \Delta_{i;\,m,n}(-Y_m,-Y_n) \= \Delta_{i;\,m,n}(Y_m,Y_n)~.
\end{align}
A sketch of $\Delta$ for such a $\IZ_{2} {\, \times \,} \IZ_{2}$ symmetric case is given in \fref{fig:deltaplot}. The vanishing locus of $\Delta$ corresponds to the singular locus of elliptic fibres. The types of singular fibres on elliptic surfaces have been classified by Kodaira \cite{Kodaira1,Kodaira2}. Table \ref{tab:Kodaira_Classification} below contains the cases relevant for us.
\begin{table}[H]
	\begin{center}
	\renewcommand{\arraystretch}{1.1}
	\begin{tabular}{|c|c|c|c|c|l|c|}
		\hline
		Type & Ord$(D)$ & Ord$(E)$ & Ord$(\Delta)$ & Dynkin Label & \hfil Fibre & Number \\ \hline \hline
		\vrule height12pt width0pt depth0pt $I_1$ &  0   &  0   &  1   & $A_1$         &  1 nodal curve     &  Continuum      \\ \hline
		$I_2$ &  0   & 0     & 2     & $A_2$         &  \begin{tabular}[c]{@{}l@{}}2 curves meeting\\ at 2 points\end{tabular}   & 200    \\ \hline
		$II$ &   1  &   1  &   2  & $A_1$        &   1 cuspidal curve   & 192    \\ \hline
	\end{tabular}
	\vskip10pt
	\capt{4.5in}{tab:Kodaira_Classification}{The Kodaira classification of singular fibres that appear in the elliptic fibration over the base $\IP^1_m \times \IP_n^1$.}
	\end{center}	
\end{table}
\vskip-30pt
Generically $\Delta_{i;\,m,n}$ is irreducible, so a generic point on the curve $\Delta_{i;\,m,n} = 0$ corresponds to a singularity of the type $I_1$. In other words, the fibre over a generic point over $\{\Delta_{i;\,m,n}=0\} \subset \IP_m^1 \times \IP^1_n$ is a nodal curve. This is related to the fibration structure of the manifold. Namely, the generic fibre over the projection $\MHV \to \IP_n^1$ is a K3 surface. Furthermore, a K3 surface can be realised as an elliptic fibration over $\IP_m^1$ with exactly 24 nodal curves. As $\Delta_{i;\,m,n}$ is a bidegree 24 polynomial, a generic fibre over $\IP_n^1$ is an elliptically fibred $\IP_m^1$ with 24 nodal fibres.

In addition to these generic points, the discriminant curve $\Delta_{i;\,m,n} = 0$ itself has singularities. We find that on $\wh{\MHV}_i$ these fall into two categories, corresponding to cases $I_2$ and $II$ in the Kodaira classification. In the generic case, there are 200 points of type $I_2$ and 192 of type $II$. These account for all 392 singularities on a generic curve. In accordance with the Kodaira classification, on singularities of type $I_2$, the polynomials $\wh Q^i(Y_m,Y_n)$ factorise, with each factor corresponding to an irreducible rational curve. The two components meet at two points, which are the singularities of the fibre. The only exceptions to this are fibres which contain degree-5 rational curves on $\MHV$ --- the second component of such a fibre is a degree-1 rational curve. When this curve is parallel to $\IP_i$, it is exactly the line which has been blown down to obtain $\wh \MHV_i$, and thus does not appear in the fibres on $\wh \MHV_i$.
\begin{figure}[t]
\begin{center}
\includegraphics[width=8cm, height=8cm]{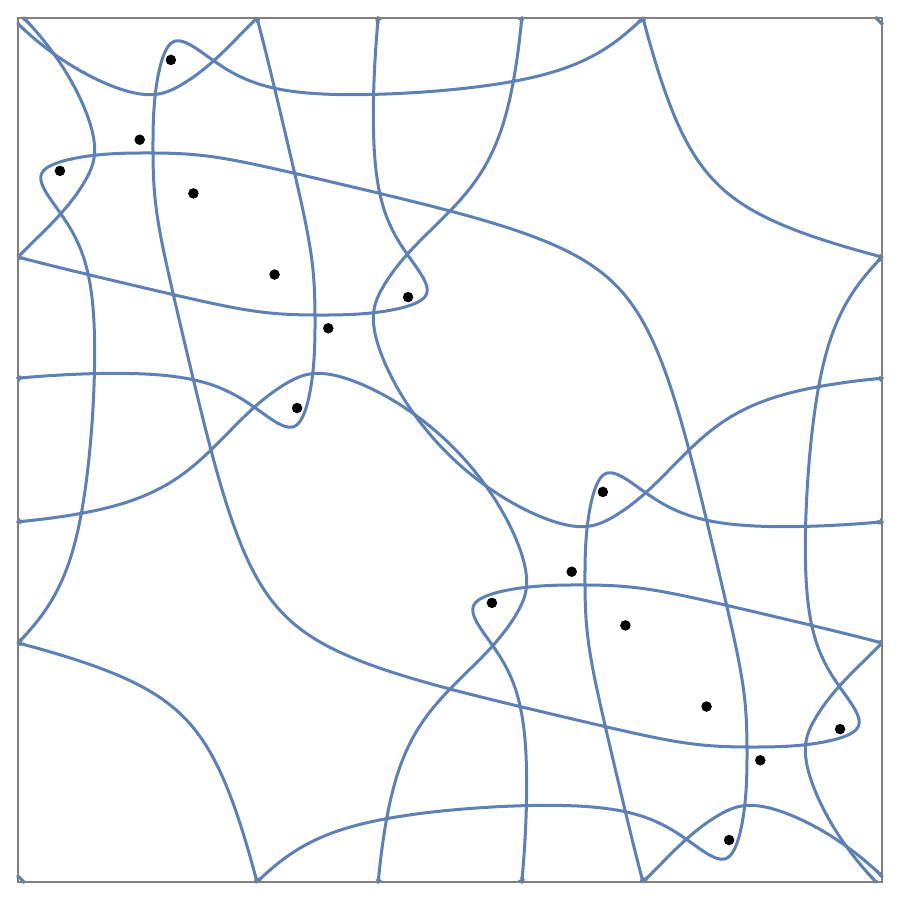}
\vskip8pt
\capt{6in}{fig:deltaplot}{A sketch of the discriminant locus in $\IP^{1}{\times}\IP^{1}$. Opposite edges of the figure are understood to be identified. The real section is drawn. The isolated dots that do not appear to lie on the discriminant locus are `space invaders' that lie on suppressed complex branches of the curve. The sketch is made for a $\IZ_{2}{\times}\IZ_{2}$-symmetric variety as in \eqref{eq:Z5Z2Z2_Variety}, and so the figure is invariant under two reflections. For the values of the parameters for which the sketch is drawn, none of the 192 cusps lie in the real section.}	
\end{center}
\end{figure}
\vspace*{-71pt}
\begin{figure}[H]
\begin{center}
\includegraphics[width=11cm, height=11cm]{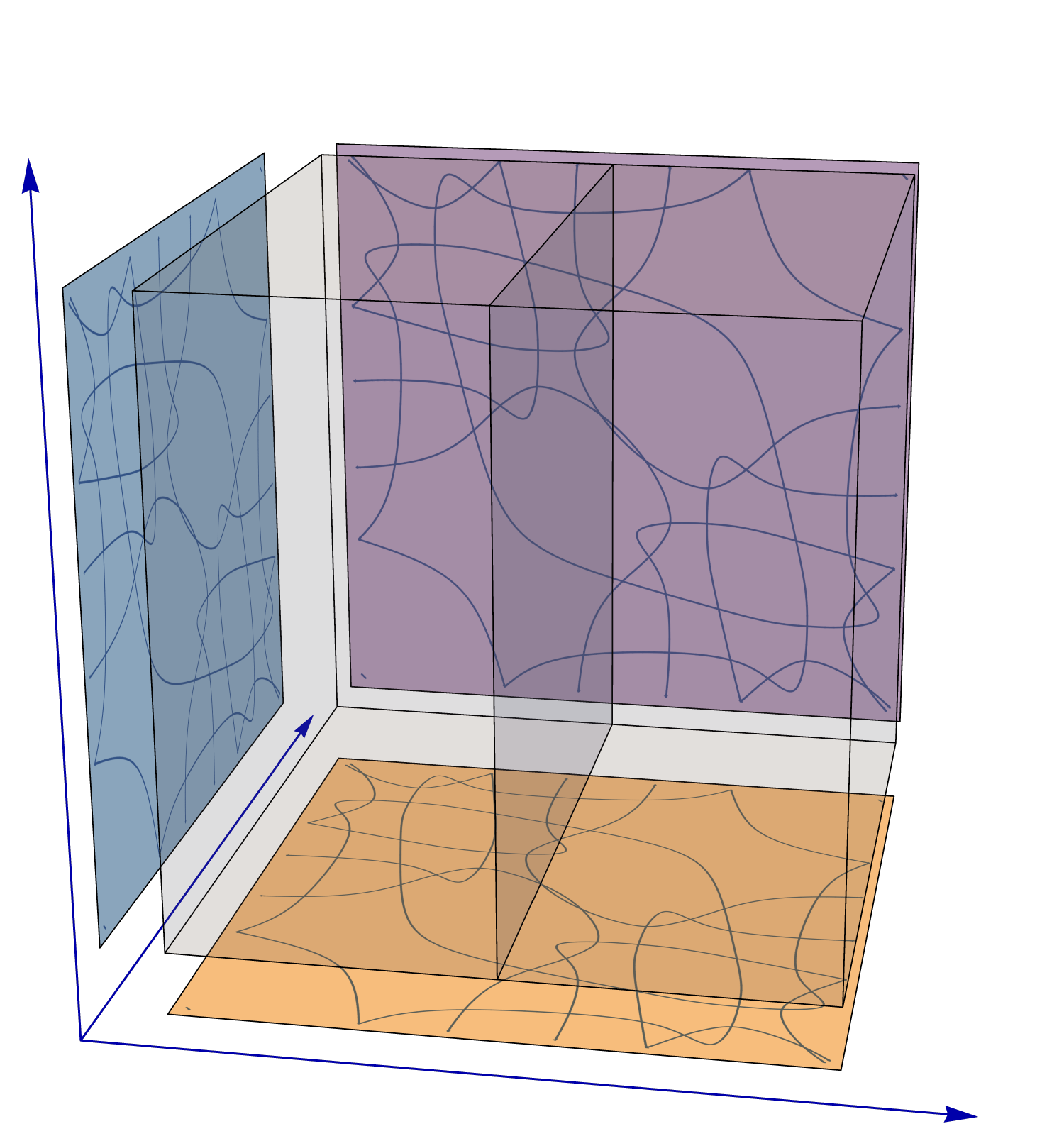}
\vskip8pt
\capt{6in}{fig:deltaplot3d}{A heuristic sketch of the multiple fibrations. Each axis corresponds to a $\IP^{1}$, and the indicated plane corresponds to a K3 fibre of a projection to one of the $\IP^{1}$'s. }
\end{center}
\end{figure}
In what follows, we mostly study the fibres on the singular varieties $\wh{\MHV}_i$. However, using the birational map between $\wh{\MHV}_i$ and $\MHV$ we can lift the curves on $\wh{\MHV}_i$ found this way to curves on $\MHV$. Outside of the exceptional divisors the lift preserves the structure of the fibres. The two-component fibres of Kodaira type $I_2$ are unions of degree 1,2,3,4, and 5 rational curves. In particular, the singular fibres include all lines, quadrics and cubics. We discuss each of these cases in detail in the following subsections \sref{sect:Rational_Lines}, \sref{sect:Rational_Quadrics}, and \sref{sect:Rational_Cubics}. First, however, it is convenient to briefly review some general aspects of curves on $(\IP^1)^5$.

\subsection{Complete intersection curves on \texorpdfstring{$(\IP^1)^5$}{the product of projective spaces}}
\vskip-10pt
It turns out that the curves we consider in the following can be expressed as complete intersections of four polynomials in $(\IP^1)^5$. The degrees and Euler characteristics of such curves are calculable by elementary techniques. Complete intersections on $(\IP^1)^5$ can be systematically searched for, and doing this we obtain some evidence, consistent with the prediction of mirror symmetry, that there are no more curves than those we find here. We consider one-dimensional varieties defined by four equations
\begin{align} \label{eq:four_equations}
p^{1} \= p^{2} \= p^{3} \= p^{4} \= 0~, \qquad \text{with} \qquad \deg_i \left( p^{\alpha} \right) = n_i^{\alpha}.
\end{align}
The two-form dual to the subvariety $p^{\alpha} = 0$ is given by
\begin{align} \notag
\cP^{\alpha} \= \sum_{i=0}^{4} n_i^{\alpha} J_i~,
\end{align}
where $J_i$ is the Kähler, or equivalently volume, form of $\IP_i^1$. Then the dual form of the curve $p^{1} = p^{2} = p^{3} =p^{4}=0$ is
\begin{align} \notag
\cC \; \defineas \; \cP^{1} \wedge \cP^{2} \wedge \cP^{3} \wedge \cP^{4} \= \sum_{\varsigma \in \cS_5} n^{1}_{\varsigma(1)}n^{2}_{\varsigma(2)}n^{3}_{\varsigma(3)}n^{4}_{\varsigma(4)} \, J_{\varsigma(1)} \wedge J_{\varsigma(2)} \wedge J_{\varsigma(3)} \wedge J_{\varsigma(4)}~,
\end{align}
where the sum runs over all permutations of $\{0,\dots,4\}$. The $i$'th degree of a curve dual to $\cC$ is
\begin{align} \label{eq:deg_i_definition}
\deg_i(\cC) &\= \int_{(\IP^1)^5} J_i \wedge \cC \= \sum_{\substack{\varsigma \in \cS_4\\a,b,c,d \neq i}} n^{1}_{\varsigma(a)}n^{2}_{\varsigma(b)}n^{3}_{\varsigma(c)}n^{4}_{\varsigma(d)}~.
\end{align}
The total Chern class of the curve \eqref{eq:four_equations} is given by
\begin{align} \notag
c(\cC) &\= \frac{\prod_{i=0}^4 (1+J_i)^2}{\prod_{\alpha=1}^4 (1 + \sum_{i=0}^4 n_i^{\alpha} J_i)}~.
\end{align}
It is straightforward to compute the Euler characteristic from the third Chern class:
\begin{align} \label{eq:euler_definition}
\chi(\cC) &\= \sum_{\varsigma \in S_5} \left( 2 - \sum_{\alpha=1}^4 n_{\varsigma(0)}^{\alpha} \right) n_{\varsigma(1)}^{1}n_{\varsigma(2)}^{2}n_{\varsigma(3)}^{3}n_{\varsigma(4)}^{4} \= 2 \deg(\cC) - \sum_{i=0}^4 \sum_{\alpha=1}^4 n_i^{\alpha} \deg_i(\cC)~.
\end{align}
These formulae give the degrees and genera of various curves in the following sections. The degrees defined in this way will also agree with the degrees of isomorphisms $\varphi: \IP^1 \to \cC$. 

As we are interested in curves in the Calabi-Yau manifold $\MHV$, we need to make sure that the curve $\cC$ lies completely within this manifold. In the language of algebraic geometry, this is equivalent to requiring that the radical of the ideal generated by the polynomials $p_i$ contains the polynomials $Q^1$ and $Q^2$ which define the $\MHV$ manifold.

It is often convenient to study the lines and other curves on the singular spaces $\wh \MHV_i$, where their connection to the elliptic fibration can be immediately appreciated. Given a curve $\cC$, and a projection $\pi$ to a base $B$, then $\cC$ may project to a curve of $B$, or project to a point. If $\cC$ projects to a curve, it is said to be horizontal in the projection $\pi$, and if $\cC$ projects to a point it is said to be vertical with respect to $\pi$. 

In the following we will study each projection $\pi_{j}$, and we will sometimes say that a curve that projects to a point on the base is \textit{vertical} with respect to the projection while one that projects to a curve on the base is \textit{horizontal}. We will study each case in turn, and finally show that the lines can be associated to a unique degree-5 line and to a node in the discriminant $\Delta_{i;\,m,n}$.
\subsection{Lines} \label{sect:Rational_Lines}
\vskip-10pt
Every degree-one rational curve in $(\IP^{1})^{5}$ is given by a set of four linear equations, each in a single variable. These read, for some $j\in\{0,1,2,3,4\}$ and each $s\in\{0,1,2,3,4\}\setminus{j}$,
\begin{equation} \label{eq:lines}
Y_{s}-y_{s}=0~.
\end{equation}
In this way $\bm{y} = (y_i,y_k,y_m,y_n)$ defines a line $L_{j}$, which is necessarily parallel to $\IP_{j}^1$. Using the data of equations \eqref{eq:lines}, the formulae \eqref{eq:deg_i_definition} and \eqref{eq:euler_definition} tell us that
\begin{align} \notag
\deg_i(L_{j}) \=  \delta_{ij}~, \qquad \chi(L_{j}) \= 2~,
\end{align}
which is exactly as expected for a line. For a line $L_{j}$ to lie on $\MHV$, the solutions to $\eqref{eq:lines}$ must additionally satisfy $Q^{1}=Q^{2}=0$. A substitution reveals that this condition amounts to 
\begin{equation} \notag
\alpha_{j}(\bm{y})+\beta_{j}(\bm{y})Y_{j}\=0~,\qquad\gamma_{j}(\bm{y})+\delta_{j}(\bm{y})Y_{j}\=0~.
\end{equation}
Therefore the $\bm{y}$ must solve $\alpha_{j}=\beta_{j}=\gamma_{j}=\delta_{j}=0$, and so gives a singularity on $\wh\MHV_{j}$. As has already been mentioned, these equations have 24 solutions for each $j$. There are therefore $5\times24=120$ lines. In the $\IZ_{5}$ symmetric case, the permissible values of $\bm{y}$ group into $\IZ_{5}$ orbits and taking the quotient leaves 24 distinct lines. Similarly, in the $\IZ_2$ symmetric cases, the involution $Y_i \mapsto -Y_i$ (or equivalently $Y_{i,0} \leftrightarrow Y_{i,1}$) identifies two lines. On $\MHV / \IZ_5 \times \IZ_2$ there are therefore $12$ lines, each descending from a family of $10$ lines on the covering space. Finally, the generic $\IZ_5 \times \IZ_2 \times \IZ_2$ quotient contains exactly $6$ lines.

\subsubsection*{Horizontal Lines}
\vskip-5pt
For definiteness, let us consider the projection $\pi_4$, the lines $L_2$, and the elliptic fibration $E_{4;0,1}$ with base $\IP^{1}_0 \times \IP^{1}_1$. The lines $L_2$ on $\MHV$ can be understood to arise as blow-ups of singular points $\bm{y}$ on $\wh \MHV_2$, and can be given by the embedding
\begin{equation} \notag
z\mapsto(y_0,y_{1},z,y_{3},y_{4})~.
\end{equation}
The projection $\pi_{4}$ then takes this line to a line in $\wh\MHV_{4}$, given by the embedding
\begin{equation} \notag
z\mapsto(y_0,y_{1},z,y_{3})~.
\end{equation}
Thus $L_2$ forms a part of the fibre of $E_{4;0,1}$ lying over the basepoint $(y_0,y_1)$. Tautologically, this fibre can be realised as the curve defined by the equation
\begin{equation} \notag
\wh Q^{4}(y_0,y_{1},Y_{2},Y_{3})\=0~.
\end{equation}
Reflecting the fact that this fibre contains a line and hence is reducible, the above polynomial factorises into degree-one and degree-three pieces (in homogeneous coordinates). The first factor is of course the equation of the image of the line $L_{2}$ on $\wh\MHV_{4}$. 

The second factor of $\wh Q^4$ has degree (1,2) with respect to $Y_2,Y_3$ and thus is a multidegree $(0,0,1,2)$ curve $\wh\cC_{(0,0,1,2)}$, which meets the line at two points. The map
\begin{align} \label{eq:psi_map}
z\mapsto \left(y_0,y_1,z,r_{2}(z) \right)~
\end{align} 
is a degree $(0,0,1,2)$ isomorphism taking $\IP^1$ to $\wh \cC_{(0,0,1,2)}$.

These curves lift to degree-5 curves $\cC_{(0,0,1,2,2)}$ on $\MHV$. The equations $Q^1 = Q^2 = 0$ are solved by setting $Y_4 = - \frac{\beta_4}{\alpha_4} = - \frac{\delta_4}{\gamma_4}$. Note that $\alpha_4$ and $\beta_4$ are both linear in $Y_0$ and $Y_1$, so substituting in the values of $Y_0$ and $Y_1$ in terms of $z$ from \eqref{eq:psi_map} into the ratio $\frac{\beta_4(z)}{\alpha_4(z)}$ gives a rational function $\wt r_2(z)$ of degree 2, as it can be shown that the quantities $\alpha_{4}(z)$ and $\beta_{4}(z)$ have exactly one linear factor in common. We arrive at a curve $\cC_{(0,0,1,2,2)}$ with an isomorphism $\psi_{4;0,1;2}: \IP^1 \to \cC_{(0,0,1,2,2)}$ given by
\begin{align} \label{eq:psi_MHV}
\psi_{4;0,1;2}(z) \= \left(y_0,y_1,z,r_{2}(z),\widetilde{r}_{2}(z) \right)~.
\end{align}
Therefore, on $\MHV$ the fibre over basepoint $(y_0,y_1)$ consists of two rational curves that meet in two points. According to Kodaira's classification, the point $(y_0,y_1)$ must be a node on the discriminant of this elliptic fibration. Upon projection to $\wh \MHV_4$, this becomes a node of $\Delta_{4;\,0,1}$, which is indeed what we find in the examples we have studied.

Other maps $\psi_{i;m,n;j}$ are defined similarly, with the privileged role of $Y_{4},Y_{0},Y_{1},Y_{2}$ in this construction replaced by $Y_{i},Y_{m},Y_{n},Y_{j}$. We display the interplay between these maps and projections in \fref{fig:Projections_Lines}.
\begin{figure}[h]
	\centering
	\begin{tikzcd}
		&\arrow{d}{\pi_i} \hskip25pt\text{H}\Lambda \supset L_j  &\arrow{l}{\pi_{j}^{-1}} \bm{y} \subset \wh \MHV_j\\			
		\IP^{1} \arrow[r] &\hskip-5pt L_j \subset E_{i;m,n} \arrow[hookrightarrow]{r}{} &  \wh \MHV_i \arrow{d}{\pi_{m,n}}\\
		& & \IP_m^{1} \times \IP_n^{1}
	\end{tikzcd}
	\vskip10pt
	\capt{6in}{fig:Projections_Lines}{A chain of birational maps allows us to see lines $L_i$, corresponding to a singularity of $\wh \MHV_j$ at $\bm{y}$ explicitly as singular fibres on $\wh \MHV_i$ viewed as a fibration over $\IP^{1}_m \times \IP^{1}_n$. The polynomial $\wh Q^i(Y_m,Y_n)$ factorises into two factors, one of degree $(0,1)$, corresponding to the line, and the other of degree $(2,1)$. This latter factor corresponds to a projection of a degree-5 curve down to $\wh \MHV_i$.}	
\end{figure}
\subsubsection*{Vertical Lines}
\vskip-5pt
Let us now focus our attention to the line $L_4$, which is mapped to a point\footnote{This point is not necessarily the same as the $\bm{y}$ in the previous subsection.} $\bm{y}$ by $\pi_4$. By symmetry, over the point $(y_0,y_1)$ on the base $\IP^{1}_0 \times \IP^{1}_1$ in $\MHV$, the fibre is given by the union of the line $L_4$ together with a degree-5 curve $\cC_{(0,0,2,2,1)}$, which meets the line in two points. Projecting this fibre down to $\wh \MHV_4$ maps the line to a point $\bm{y}$, and the degree-5 curve to a degree-4 curve $\wh \cC_{(0,0,2,2)}$, which intersects itself at the point $\bm{y}$. So there exists a birational map $\IP^1 \to \cC_{(0,0,2,2)}$
\begin{align} \label{eq:psi_proj}
z\mapsto \left(y_0^{(a)},y_1^{(a)},r_{2}(z),\widetilde{r}_{2}(z) \right)~,
\end{align}
which is not, however, an isomorphism due to the self-intersection. Such a curve will not fit Kodaira's classification, which can be traced back to the fact that $\wh \MHV_4$ is singular. Indeed, the lift of the fibre is a union of two rational curves meeting at two points, and thus corresponds to a node in the discriminant locus of the fibration $\MHV$. Upon projecting down to $\wh \MHV_4$, this becomes a node of the locus $\Delta_{4;\,0,1} = 0$. An alternative way of arriving at the same conclusion is by noting that, as we have remarked previously, $\Delta_{4;\,0,1} = \Delta_{2;\,0,1}$, and by the previous subsection, $L_4$ corresponds to a node of $\Delta_{2;\,0,1} = 0$.

A straightforward generalisation of the the results of the last two subsections reveals that the 72 lines $L_i$, $L_j$, and $L_k$, together with the degree-5 curves, account for 72 of the nodes of the discriminant locus $\Delta_{i;\,m,n} = 0$. The locus has in total 200 nodes, the rest of which turn out to correspond to curves of degrees 2, 3, and 4, as we show shortly.

\fref{fig:Projections_and_Fibres_curves} sketches the lifts of singular fibres in $\wh \MHV_{4}$ to $\MHV$.
\vskip10pt
\begin{figure}[H]
\centering
\newcommand{\xdownarrow}[1]{%
  {\left\downarrow\vbox to #1{}\right.\kern-\nulldelimiterspace}
}
\begin{center}
\includegraphics[width=16cm, height=8cm]{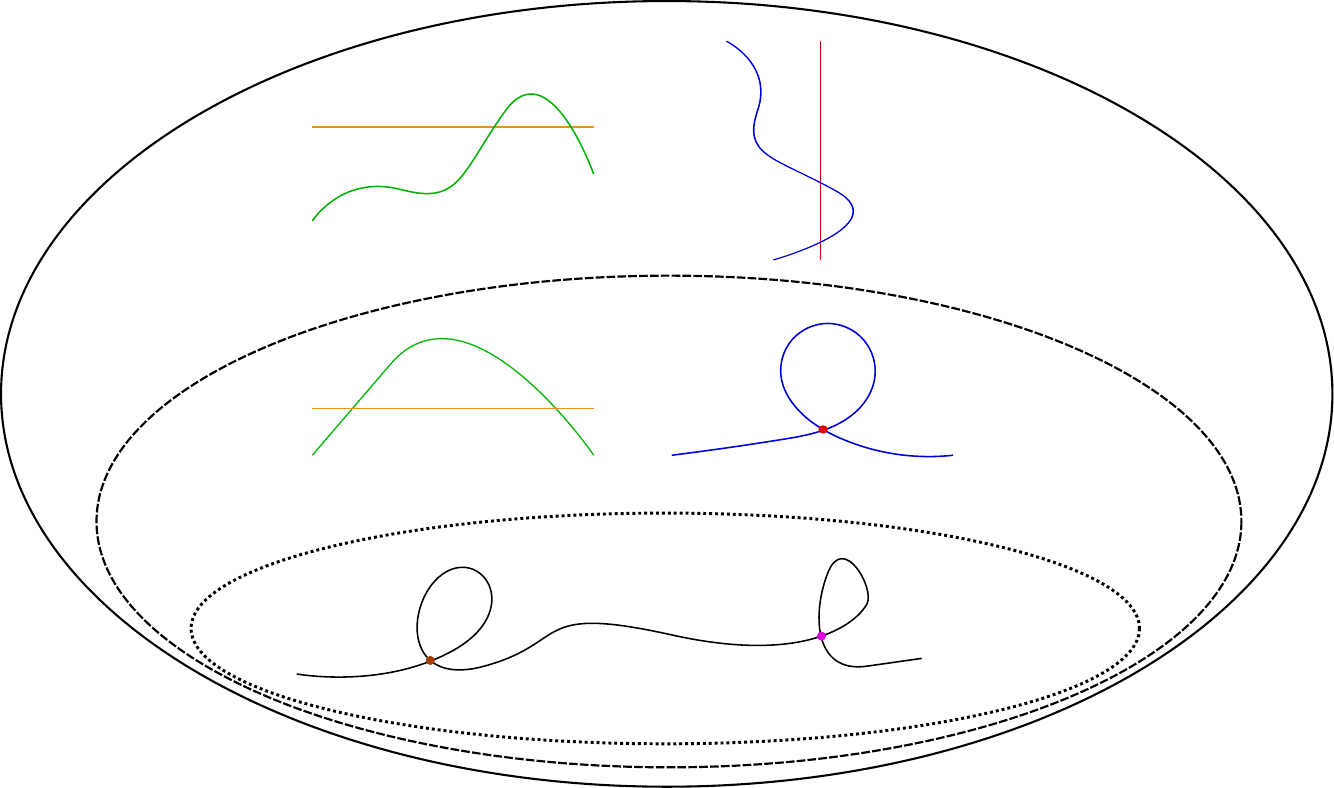}
\end{center}
\place{5}{2.7}{$\MHV$}
\place{5}{2.5}{$\; \, \xdownarrow{0.9cm} \pi_4$}
\place{5}{1.8}{$\wh \MHV_4$}
\place{5}{1.6}{$\; \, \xdownarrow{0.9cm} \pi_{0,1}$}
\place{4.85}{0.95}{$\IP^{1}_0 \times \IP^{1}_1$}
\place{4.05}{3.2}{$L_4^{(a)}$}
\place{3.3}{2.7}{$\cC_{(0,0,2,2,1)}$}
\place{2}{3.2}{$L_2^{(b)}$}
\place{2.5}{2.7}{$\cC_{(0,0,1,2,2)}$}
\place{2.6}{2.15}{$\wh \cC_{(0,0,1,2)}$}
\place{2.2}{1.7}{$\wh L_2^{(b)}$}
\place{4.3}{2.05}{$\wh \cC_{(0,0,2,2)}$}
\place{3.95}{1.6}{$\bm y^{(a)}$}
\place{3}{0.8}{$\Delta_{4;\,0,1}=0$}
\vskip-10pt
\capt{6in}{fig:Projections_and_Fibres_curves}{Schematic representation of elliptic fibres containing lines on $\MHV$. For concreteness, we have chosen here $i=4$, $m=0$, and $n=1$. The largest oval represents the smooth manifold $\MHV$, on which the elliptic fibration over $\IP_0 \times \IP_1$ contains unions of two rational curves. Here we have pictured the fibres which consist of one line and a degree-5 curve. }	
\end{figure}
\newpage
\subsection{Quadrics} \label{sect:Rational_Quadrics}
\vskip-10pt
The analysis of irreducible degree-2 curves proceeds largely along the same lines. Algebraic quadrics on $\MHV$ can be expressed, for a triple $k,m,n$ with constants $q_{k},q_{m},q_{n}$, as the complete intersections 
\begin{align} \label{eq:Quadric_CI_polynomials}
Y_k - q_{k} \= 0~, \quad Y_m - q_{m} \= 0~, \quad Y_n - q_{n} \= 0~, \quad p\left(Y_{0},Y_{1},Y_{2},Y_{3},Y_{4}\right) \=0~.
\end{align}
Here, $p$ is an irreducible multidegree $(1,1,1,1,1)$ polynomial. With $i,j$ denoting the pair in $\{0,1,2,3,4\}\setminus\{k,m,n\}$, the equations \eqref{eq:Quadric_CI_polynomials} define a curve $\cC$ with
\begin{align} \notag
\deg_s(\cC) \= \delta_{s,i} + \delta_{s,j}, \qquad \chi(\cC) \= 2.
\end{align}
While this is not the most general form of a degree-2 curve on $(\IP^1)^5$, it can be shown that the curves which lie in $\MHV$ are of this form. To ensure that a curve defined by \eqref{eq:Quadric_CI_polynomials} lies in $\MHV$, we must have that, specialising to $Y_{i}=q_{i},Y_{j}=q_{j},$ and $Y_{k}=q_{k}$,
\begin{equation} \notag
Q^{1},Q^{2}\in\sqrt{\left\langle p\right\rangle} \= \langle p \rangle~.
\end{equation}
The square root indicates the radical of the ideal $\langle p \rangle$, which in this case is the ideal itself. As $p$ is irreducible and all three polynomials $Q^{1},Q^{2},$ and $p$ are of multidegree (1,1,1,1,1), this requires $p = C Q^1$ or $p=C Q^{2}$, with $C$ a constant. Further, we must have either $Q^1 = Q^2$ or one of the $Q$'s vanishing at $Y_{k}=q_{k},Y_{m}=q_{m},$ and $Y_{n}=q_{n}$. We cannot have both $Q$'s vanishing after this specialisation. In general there are 24 values of $\{q_{k},q_{m},q_{n}\}$ for which these conditions are satisfied. There are 10 ways of choosing the triple $k,m,n$, and so we find 240 curves of degree~2 on $\MHV$. In the $\IZ_{5}$, $\IZ_5 {\times} \IZ_2$, and $\IZ_5 {\times} \IZ_2 {\times} \IZ_2$ symmetric cases, these curves come in families of 5, 10, and 20, respectively, so taking the quotients gives exactly 48 curves on $\MHV/\IZ_5$, 24 on $\MHV/\IZ_5 {\times} \IZ_2$, and 12 on $\MHV/\IZ_5 {\times} \IZ_2 {\times} \IZ_2$. This agrees with the results of \cite{Candelas:2019llw}.

Again, an alternative point-of-view can be obtained by viewing these quadrics as fibres of the elliptic fibrations $\wh \MHV_i \to \IP_m^1 \times \IP_n^1$. Consider the case $(i,j,k,m,n)=(4,2,3,0,1)$. The quadrics $\cC_{(0,0,1,1,0)}$ are isomorphic to $\IP^1$ with the map given by
\begin{align} \notag
z\mapsto \left(q_0, q_1, z, r_{1}(z), q_4 \right).
\end{align}
Upon projection by $\pi_4$, this becomes a quadric on $\wh \MHV_4$ with embedding
\begin{align} \notag
z\mapsto \left(q_0, q_1, z, r_1(z) \right).
\end{align}
The fibre to which this belongs is given by $\wh Q^4(q_0,q_1,Y_2,Y_3)$, which factorises into two degree-$(1,1)$ factors, with the first one corresponding to the quadric $\wh \cC_{(0,0,1,1)}$. The second curve $\wh \cC_{(0,0,1,1)}$ is also a quadric on $\wh \MHV_4$, but can be lifted to $\MHV$. To do this, we again set $Y_4 = - \frac{\beta_4}{\alpha_4}$, to ensure that the lifted curve lies in $\MHV$. Expressing $\alpha_4$ and $\beta_4$ in terms of $z$, the parameter on the curve, this is generically a degree-2 rational function. Thus the lift $\cC_{(0,0,1,1,2)}$ is a degree-4 curve isomorphic to $\IP^{1}$ via
\begin{align} \notag
z \mapsto \left( q_0, q_1, z, r_{1}(z), r_{2}(z) \right).
\end{align}
Similarly, by symmetry we know that there exists a fibre on $\MHV$ which consists of the curves $\cC_{(0,0,0,1,1)}$ and $\cC_{(0,0,2,1,1)}$. Projecting them down to $\wh \MHV_4$ gives a line and cubic, $\wh \cC_{(0,0,0,1)}$ and $\wh \cC_{(0,0,2,1)}$, meeting in two points. By symmetry the curves studied above also meet in two points, in accordance with Kodaira's classification. Thus the 72 quadrics of the form $\cC_{(0,0,1,1,0)}$, $\cC_{(0,0,1,0,1)}$ or $\cC_{(0,0,0,1,1)}$ also each corresponds to a unique node of the discriminant locus $\Delta_{4;\,0,1}$.
\subsection{Cubics}  \label{sect:Rational_Cubics}
\vskip-10pt
Cubic curves whose multidegree is a permutation of $(1,1,1,0,0)$ can also be expressed as complete intersections. The most general cubic curves that can be defined by four multilinear equations are of the form
\begin{equation} \notag
\begin{aligned}
Y_m - c_m&\=0~,\\
Y_n - c_n&\=0~,\\
p \,\defineas\, a_0 + a_1 Y_i + a_2 Y_j + a_3 Y_i Y_j&\=0~,\\
q \,\defineas\, b_0 + b_1 Y_k + b_2 Y_j + b_3 Y_k Y_j&\=0~,
\end{aligned}
\end{equation}
This defines a curve $\cC_{3}$ with
\begin{align} \notag
\deg_i(\cC_{3}) \= \delta_{i,l} + \delta_{i,m} + \delta_{i,n}, \qquad \chi(\cC_{3}) \= 2.
\end{align}
Curves of this form include all cubics lying in $\MHV$. For the curve defined in this way to lie in $\MHV$, the ideal generated by these polynomials must contain the polynomials $Q^1$ and $Q^2$. This condition is equivalent to requiring that there are coefficients $d_{a}$, $e_{b}$ such that when $Y_{m}=c_{m},$ $Y_{n}=c_{n}$
\begin{align} \notag
Q^1 \= d_1 \, p + d_2 \, q + d_3 \, p \, Y_k + d_4 \, q \, Y_i~, \qquad Q^2 \= e_1 \, p + e_2 \, q + e_3 \, p \, Y_k + e_4 \, q \, Y_i~.
\end{align}
For a quintuple $(i,j,k,m,n)$ there are in general exactly 112 solutions to these equations. Summing over the 10 distinct choices of $(i,j,k,m,n)$ gives us 1120 curves of degree 3, which once again come in $\IZ_5$, $\IZ_5 {\times} \IZ_2$, and $\IZ_5 {\times} \IZ_2 {\times} \IZ_2$ invariant families in the symmetric cases. Taking the quotients with respect to $\IZ_5$, $\IZ_5 {\times} \IZ_2$, and $\IZ_5 {\times} \IZ_2 {\times} \IZ_2$ leave $224$, $112$, and $56$ curves of degree 3 respectively, in agreement with \cite{Candelas:2019llw}.

As was the case with the lines and quadrics, the cubics also appear as singular fibres of elliptic fibrations, and in fact account for the remaining 56 nodes of the discriminant locus $\Delta_{i;\,m,n} = 0$. Take again $(i,j,k,m,n)=(4,2,3,0,1)$ to expedite the discussion, and consider the cubic curves $\cC_{(0,0,1,1,1)}$. The projection of this curve to $\wh \MHV_4$ is a quadric $\wh \cC_{(0,0,1,1)}$. As before, this indicates that the polynomial $\wh Q^4(Y_2,Y_3)$ factorises into two components, both of degree $(1,1)$. The isomorphisms with $\IP^1$ are of the form
\begin{align} \notag
z\mapsto \left(c_0, c_1,z,r_{1}(z)\right).
\end{align}
The quantity $\frac{\beta_4}{\alpha_4}$ determining the lift to a curve on $\MHV$ is a priori a ratio of two degree-2 polynomials. However, this is a component of a reducible elliptic fibre inside of which we already have a curve of total degree 3, therefore the two polynomials $\alpha_{4},\beta_{4}$ must share a factor so that the lifts are curves $\cC_{(0,0,1,1,1)}$. The isomorphisms with $\IP^1$ are given by
\begin{align} \notag
z\mapsto \left(c_0, c_1,z,r_{1}(z),\widetilde{r}_{1}(z)\right).
\end{align}
\subsection{Summary}
\vskip-10pt
This completes the classifications of fibres over the nodes of the discriminant curves on singular varieties $\wh \MHV_i$ (over the base $\IP_m \times \IP_n$), and their lifts to $\MHV$. We summarise our findings in \tref{tab:Q_Factorisations} and \tref{tab:Counting_Numbers}, taking $i=4,\,m=0,\,n=1$ for concreteness.
\begin{table}[H] 
	\begin{center}
		\renewcommand{\arraystretch}{1.3}
		\begin{tabular}{|l|l|l|l|l|c|}
			\hline
		\vrule height12pt width0pt depth6pt Type &	Degree~1	& Degree~2	& Curve 1 				& Curve 2					& Number \\\hline \hline
		
         Line	&(0,0,0,0)	& (0,0,2,2)	& $L_4$					& $\cC_{(0,0,2,2,1)}$		& 24 	\\ \hline
		Line	&(0,0,1,0)	& (0,0,1,2)	& $L_2$					& $\cC_{(0,0,1,2,2)}$		& 24    \\ \hline
		Line	&(0,0,0,1)	& (0,0,2,1)	& $L_3$					& $\cC_{(0,0,2,1,2)}$		& 24    \\ \hline
			
		Quadric	&(0,0,1,1)	& (0,0,1,1)	& $\cC_{(0,0,1,1,0)}$	& $\cC_{(0,0,1,1,2)}$		& 24 	\\ \hline
		Quadric	&(0,0,0,1)	& (0,0,2,1)	& $\cC_{(0,0,1,0,1)}$	& $\cC_{(0,0,1,2,1)}$ 		& 24    \\ \hline
		Quadric	&(0,0,1,0)	& (0,0,1,2)	& $\cC_{(0,0,0,1,1)}$	& $\cC_{(0,0,2,1,1)}$		& 24    \\ \hline
		
		Cubic	&(0,0,1,1)	& (0,0,1,1)	& $\cC_{(0,0,1,1,1)}$	&  $\cC_{(0,0,1,1,1)}$		& 56 	\\ \hline	
					
		\end{tabular}
		\vskip10pt
		\capt{6in}{tab:Q_Factorisations}{Factorisations of $\wh Q_4$ over the nodes of the discriminant curve $\Delta_{4;\,0,1}=\Delta_{3;\,0,1}=\Delta_{2;\,0,1}$ and the corresponding curves on the non-singular variety $\MHV$.}
	\end{center}	
\end{table}
\vskip-20pt\begin{table}[H]
	\begin{center}
		\begin{tabular}{|c|c|}
			\hline
		\vrule height12pt width0pt depth6pt $\fp$ &	$n_\fp$ \\
			\hline \hline
			\vrule height12pt width0pt depth0pt(0,0,0,0,1) & 24 \\ \hline
			(0,0,0,1,1) & 24 \\ 
			(0,0,0,0,2) & 0 \\ \hline 	
			(0,0,1,1,1) & 112 \\ 
			(0,0,0,1,2) & 0 \\ 
			(0,0,0,0,3) & 0 \\ \hline		
			(0,0,1,1,2) & 24 \\ 
			(0,0,0,1,3) & 0 \\ 
			(0,0,0,0,4) & 0 \\ 	\hline	
			(0,0,1,2,2) & 24 \\ 
			(0,0,0,1,4) & 0 \\ 
			\vrule height0pt width0pt depth6pt(0,0,0,0,5) & 0 \\ 	\hline							
			\end{tabular}		
		\vskip10pt
		\capt{6.4in}{tab:Counting_Numbers}{The results of this section, giving the curve-counts for some low degrees. The numbers that are related to these by a cyclic permutation are omitted. Note the agreement with the tables in appendix \ref{app:Instanton_Numbers}.}
	\end{center}	
\end{table}
\vspace*{-40pt}

\newpage
\section{Outlook and discussion}
\vskip-10pt
\subsection{Coxeter groups and higher genus invariants}
\vskip-10pt
As the Coxeter group we have described can be seen as acting on the divisors of the manifold, we expect that the identities between instanton numbers obtained from these symmetries persist to all genera for the mirror Hulek-Verrill. For other manifolds admitting such flops, with a similar group appearing at genus 0, one also expects the same group to appear at every genus. The set of such manifolds includes a number of complete intersection Calabi-Yau manifolds. For instance, there is a Coxeter symmetry appearing in the tables of \cite{Hosono:2011np}, appendix B, which gives the instanton numbers at genera 0,1,2 for a two-parameter split of the quintic threefold. The tables at each genus exhibit the same infinite dihedral symmetry.

This raises a tantalising possibility, the focus of ongoing work, to use the symmetry as an aid to computing higher genus instanton numbers. For compact Calabi-Yau threefolds the set of instanton numbers for a given genus is determined by BCOV recursion \cite{Bershadsky:1993ta,Bershadsky:1993cx} \emph{up to} a set of functions $f^{(h)},\,2\leqslant h\leqslant g$, the holomorphic ambiguities. These are rational functions of the moduli, with bounded degrees in the numerator and denominator. In order to fix the ambiguities, one must fix these ambiguities, which can be done, for example, if one can independently compute some instanton numbers, or find relations between instanton numbers.

Coxeter symmetry could conceivably provide a way of establishing such independent relations, making it possible to compute instanton numbers to higher genera for the compact multiparameter geometries possessing these symmetries. However, even if this were true, there would remain the practical matter of implementing the recursion. Due to how the Coxeter symmetries affect the degrees of instanton numbers, it may be required to work to a very high degree in order to fix the ambiguities, and in our present work exceeding degree-29 proved to be challenging. 

Moreover, there exists a simple recipe, obvious to readers of certain schools, for obtaining the instanton numbers on a manifold from those on a split of the manifold. This was not crucial to our main line of discussion, but it is satisfying to see that the instanton numbers $\nu_{i,j,k,l}$ for the family of tetraquadrics, which can be computed by methods similar to those described in this paper, match perfectly with the combinations $\sum_{m}n_{i,j,k,l,m}$. This gives a hope for making progress on geometries with no Coxeter symmetry, such as the quintic, by instead working on a split and then projecting out the last index as above. In fact, the tetraquadric actually possesses its own Coxeter group which is compatible with that of the mirror Hulek-Verrill manifold through the above projection.

The instanton numbers play a role in the microstate counting for 4d $\mathcal{N}=2$ black holes. One perspective on this problem considers the elliptic genus of the Maldacena-Strominger-Witten CFT \cite{Maldacena:1997de}, which describes the microscopic degrees of freedom in an M-theory construction. The Fourier coefficients of this quantity are specific combinations of the instanton numbers at different genera, and it is interesting to consider what Coxeter symmetry tells you about these combinations.

\subsection{Separation of variables, periods, and amplitudes}
\vskip-10pt
The section \sref{sect:Periods} was written with an explicit focus on the Hulek-Verrill manifold, but most of the discussion generalises directly to other multiparameter manifolds. In particular, in spite of lacking a full Picard-Fuchs system, we can use the knowledge of the triple intersection numbers of the mirror, as in \cite{Hosono:1994ax}, to obtain the periods of the Hulek-Verrill manifolds. We formulate this method of finding periods as expanding the periods in the mirror cohomology algebra elements. Additionally, by studying lines in the moduli space, we are able to obtain the full monodromy data and find the periods everywhere in the moduli space. This process is also greatly simplified by symmetry.

%we can appeal to symmetries in order to obtain the periods and a full set of monodromy data. It was already clear from \cite{Hosono:1994ax} how to obtain the series expressions for the periods starting from the holomorphic period and the triple intersection numbers, and expressions very similar to our integral expressions can often be achieved from these alone, although we have not seen this done for a complete set of periods before. 

The periods can be expressed in terms of Bessel function moments. These expressions are particularly interesting where banana Feynman diagrams are concerned, as they facilitate analytic study of the cut diagram in regimes of parameter space where the series expression for the fundamental period does not converge.

\subsection{The Hulek-Verrill Manifolds and modularity}
\vskip-10pt
As the methods of computing local zeta functions for threefolds improve, it becomes possible to consider period evaluations and expressions for these in terms of automorphic L-values. In this paper we have seen that the periods can be expressed as Bessel moments, so modularity considerations make it possible to express these moments in terms of L-values. More generally, the periods of any favourable complete intersection can be written as integrals of products of Meijer-G functions. There is much work to be done in exploiting modularity, or more generally automorphy, in order to obtain interesting identities for such integrals, which may prompt new questions.

We will return to the Hulek-Verrill manifold as a primary example in future work. One project considers the zeta function for the five-parameter family and how modular behaviour of this relates to supersymmetric vacua of IIB supergravity flux compactifications. Our work in this paper serves as crucial input for this future project \cite{SFV}.

There are many interesting relations between the mirror Hulek-Verrill and the tetraquadric that follow from the splitting. The latter does not seem to possess a rank-two attractor on the maximally symmetric locus, i.e. the one-parameter family AESZ16 \cite{CY_Mainz-db}. The former does have rank-two attractors, and we will return elsewhere to studying such pairs of manifolds \cite{GWS}.

Although the zeta function for the one-parameter quotient of the Hulek-Verrill manifold has been extensively tabulated, the same is not true for the full multiparameter family. Progress with this problem is contingent upon possessing a good description of the periods, which we provide in this paper. The results of this paper should make possible a search for rank-two attractors `off the diagonal', where not all moduli are equal.

\section*{Acknowledgements}
\vskip-10pt
We are grateful to Duco van Straten, Alessio Marrani, Sujay Nair, and Dawid Kielak for productive discussions. We also thank Mohamed Elmi for instruction regarding higher-genus instanton numbers. We wish to thank Fabian Ruehle and Andre Lukas for pointing out the connection between birational maps and instanton number symmetries, and for interesting discussions on the topic. We are also grateful to the referees for their helpful comments and suggestions. PK thanks the Osk.~Huttusen S\"a\"ati\"o and the Jenny and Antti Wihuri Foundation for support. JM is supported by EPSRC studentship \#2272658. The authors would like to acknowledge the use of the University of Oxford Advanced Research Computing (ARC) facility in carrying out this work \cite{richards_2015_22558}. PC and XD would like also to thank the Isaac Newton Institute for Mathematical Sciences in Cambrige for support and hospitality during the program New Connections in Number Theory and Physics while completing this project; this workshop was supported by EPSRC grant number EP/R014604/1.

\newpage
\addcontentsline{toc}{section}{Appendices}
\appendix
\section{Toric Geometry Data} \label{app:Toric_Geometry_Data}
\vskip-10pt
Here we gather some data related to the polytopes and toric varieties discussed in \sref{sect:Toric_Geometry}.
\subsection*{The polytope $\wh \Delta$ and the ambient variety $\IP_{\wh \Delta}$}
\vskip-10pt
\begin{table}[H]
\small{\pgfplotstabletypeset[
begin table/.add={}{[t]},
col sep=space,
ignore chars={},
every head row/.style={output empty row,
before row={\hline \multicolumn{4}{|c|}{\vrule height15pt width0pt depth6pt Vertices of $\widehat{\Delta}$ }\\},
after row=\hline\hline},
every first row/.style={before row=\vrule height12pt width0pt depth0pt},
every last row/.style={before row=\vrule height0pt width0pt depth6pt,after row=\hline},
%after row=\hline,    %  Uncomment this to get back lines between all rows.
columns={A,B,A,B},
display columns/0/.style={select equal part entry of={0}{2},column type = {|l}},
display columns/1/.style={select equal part entry of={0}{2},column type = {|l|}},
display columns/2/.style={select equal part entry of={1}{2},column type = {|l}},
display columns/3/.style={select equal part entry of={1}{2},column type = {|l|}},
string type]
{
A B

$u_{1}$ (\,-1,\,\+0,\,\+0,\,\+0)
$u_{2}$ (\,-1,\,\+0,\,\+0,\,\+1)
$u_{3}$ (\,-1,\,\+0,\,\+1,\,\+0)
$u_{4}$ (\,-1,\,\+1,\,\+0,\,\+0)
$u_{5}$ (\,\+0,\,-1,\,\+0,\,\+0)
$u_{6}$ (\,\+0,\,-1,\,\+0,\,\+1)
$u_{7}$ (\,\+0,\,-1,\,\+1,\,\+0)
$u_{8}$ (\,\+0,\,\+0,\,-1,\,\+0)
$u_{9}$ (\,\+0,\,\+0,\,-1,\,\+1)
$u_{10}$ (\,\+0,\,\+0,\,\+0,\,-1)
$u_{11}$ (\,\+0,\,\+0,\,\+0,\,\+1)
$u_{12}$ (\,\+0,\,\+0,\,\+1,\,-1)
$u_{13}$ (\,\+0,\,\+0,\,\+1,\,\+0)
$u_{14}$ (\,\+0,\,\+1,\,-1,\,\+0)
$u_{15}$ (\,\+0,\,\+1,\,\+0,\,-1)
$u_{16}$ (\,\+0,\,\+1,\,\+0,\,\+0)
$u_{17}$ (\,\+1,\,-1,\,\+0,\,\+0)
$u_{18}$ (\,\+1,\,\+0,\,-1,\,\+0)
$u_{19}$ (\,\+1,\,\+0,\,\+0,\,-1)
$u_{20}$ (\,\+1,\,\+0,\,\+0,\,\+0)
}}
\hfill 
\small{\pgfplotstabletypeset[
begin table/.add={}{[t]},
col sep=space,
ignore chars={},
every head row/.style={output empty row,
before row={\hline \multicolumn{4}{|c|}{\vrule height15pt width0pt depth6pt Faces of $\widehat{\Delta}$ }\\},
after row=\hline\hline},
every first row/.style={before row=\vrule height12pt width0pt depth0pt},
every last row/.style={before row=\vrule height0pt width0pt depth6pt,after row=\hline},
%after row=\hline,    %  Uncomment this to get back lines between all rows.
columns={A,B,A,B},
display columns/0/.style={select equal part entry of={0}{2},column type = {|r}},
display columns/1/.style={select equal part entry of={0}{2},column type = {|r|}},
display columns/2/.style={select equal part entry of={1}{2},column type = {|r}},
display columns/3/.style={select equal part entry of={1}{2},column type = {|r|}},
string type]
{
A B

$\rho_{1}$ $y_{1}=1$
$\rho_{2}$ $-y_{1}=1$
$\rho_{3}$ $y_{2}=1$
$\rho_{4}$ $-y_{2}=1$
$\rho_{5}$ $y_{1}+y_{2}=1$
$\rho_{6}$ $-y_{1}-y_{2}=1$
$\rho_{7}$ $y_{3}=1$
$\rho_{8}$ $-y_{3}=1$
$\rho_{9}$ $y_{1}+y_{3}=1$
$\rho_{10}$ $-y_{1}-y_{3}=1$
$\rho_{11}$ $y_{2}+y_{3}=1$
$\rho_{12}$ $-y_{2}-y_{3}=1$
$\rho_{13}$ $y_{1}+y_{2}+y_{3}=1$
$\rho_{14}$ $-y_{1}-y_{2}-y_{3}=1$
$\rho_{15}$ $y_{4}=1$
$\rho_{16}$ $-y_{4}=1$
$\rho_{17}$ $y_{1}+y_{4}=1$
$\rho_{18}$ $-y_{1}-y_{4}=1$
$\rho_{19}$ $y_{2}+y_{4}=1$
$\rho_{20}$ $-y_{2}-y_{4}=1$
$\rho_{21}$ $y_{1}+y_{2}+y_{4}=1$
$\rho_{22}$ $-y_{1}-y_{2}-y_{4}=1$
$\rho_{23}$ $y_{3}+y_{4}=1$
$\rho_{24}$ $-y_{3}-y_{4}=1$
$\rho_{25}$ $y_{1}+y_{3}+y_{4}=1$
$\rho_{26}$ $-y_{1}-y_{3}-y_{4}=1$
$\rho_{27}$ $y_{2}+y_{3}+y_{4}=1$
$\rho_{28}$ $-y_{2}-y_{3}-y_{4}=1$
$\rho_{29}$ $y_{1}+y_{2}+y_{3}+y_{4}=1$
$\rho_{30}$ $-y_{1}-y_{2}-y_{3}-y_{4}=1$
}}
\end{table}
We form a matrix $\widehat{\text{M}}$ out of these vectors,
\begin{align} \notag
\widehat{\text{M}} \= \left(\begin{array}{c}
u_1 \\
u_2 \\
\dots \\
u_{20}
\end{array}\right) \= \left(o_1, o_2, o_3, o_4\right)~.
\end{align}
The nullspace of $\widehat{\text{M}}^T$, expressed in a convenient basis, gives 16 relations between these vectors:
\begin{equation} \notag
\begin{aligned}
u_{i}+u_{21-i}&\=0, & \quad 1\leq i\leq10 ~,\\[3pt]
u_{1}-u_{5}+u_{17}&\=0~,\quad &  u_{1}-u_{8}+u_{18}&\=0~,\quad &  u_{1}-u_{10}+u_{19}&\=0~, \\[3pt]
u_{5}-u_{8}+u_{14}&\=0~, & 
u_{5}-u_{10}+u_{15}&\=0~, & 
u_{8}-u_{10}+u_{12}&\=0~.
\end{aligned}
\end{equation}
Each of these relations corresponds to a scaling symmetry $\IC^* \subset (\IC^*)^{16}$. For example, the relations $u_{1}+u_{20}=0$ and $u_{8}-u_{10}+u_{12}=0$ correspond to scalings
\begin{equation} \notag
\begin{aligned}
\IC_1^* &\; : \; (\eta_1,  \dots,  \eta_{20}) \mapsto (\lambda_1\eta_1, \eta_2,  \dots, \eta_{19}, \lambda_1 \eta_{20})~,\\[3pt]
\IC_{16}^* &\; : \; (\eta_1,  \dots,  \eta_{20}) \mapsto (\eta_1, \eta_2,\dots, \eta_{7}, \lambda_{16}\eta_{8}, \eta_{9}, \lambda_{16}^{-1}\eta_{10},\eta_{11}, \lambda_{16}\eta_{12},\eta_{13}\dots\eta_{19}, \eta_{20})~.
\end{aligned}
\end{equation}
There are four invariant combinations of coordinates that we can identify with the coordinates on the torus $\IT^4 \subset \IP_{\wh \Delta}$. These can be taken to be
\begin{equation} \label{eq:Cox_Invariants_Singular_MHV}
\begin{aligned}
H_1 & \= \eta^{o_1} \= \frac{\eta _{17} \eta _{18} \eta _{19} \eta _{20}}{\eta _1 \eta _{2} \eta _{3} \eta _{4}}~, \quad& H_2 &\= \eta^{o_2} \= \frac{\eta _4 \eta _{14} \eta _{15} \eta _{16}}{\eta _5 \eta _{6} \eta _{7} \eta_{17}}~, \\[6pt]
H_3 & \= \eta^{o_3} \= \frac{\eta _3 \eta _{7} \eta _{12} \eta _{13}}{\eta _8 \eta _9 \eta _{14} \eta _{18}}~, & H_4 &\= \eta^{o_4} \= \frac{\eta _{2} \eta _{6} \eta _{9} \eta _{11}}{\eta_{10} \eta_{12} \eta_{15} \eta_{19}}~.
\end{aligned}
\end{equation} 
\subsection*{The dual polytope $\wh \Delta^*$ and the ambient variety $\IP_{\wh \Delta^*}$}
\vskip-10pt
\begin{table}[H]
\centering
\small{\pgfplotstabletypeset[
begin table/.add={}{[t]},
col sep=space,
ignore chars={},
every head row/.style={output empty row,
before row={\hline \multicolumn{4}{|c|}{\vrule height15pt width0pt depth6pt Vertices of $\widehat{\Delta}^{*}$ }\\},
after row=\hline\hline},
every first row/.style={before row=\vrule height12pt width0pt depth0pt},
every last row/.style={before row=\vrule height0pt width0pt depth6pt,after row=\hline},
%after row=\hline,    %  Uncomment this to get back lines between all rows.
columns={A,B,A,B},
display columns/0/.style={select equal part entry of={0}{2},column type = {|l}},
display columns/1/.style={select equal part entry of={0}{2},column type = {|l|}},
display columns/2/.style={select equal part entry of={1}{2},column type = {|l}},
display columns/3/.style={select equal part entry of={1}{2},column type = {|l|}},
string type]
{
A B

$v_{1}$ (\,\+1,\,\+0,\,\+0,\,\+0)
$v_{2}$ (\,-1,\,\+0,\,\+0,\,\+0)
$v_{3}$ (\,\+0,\,\+1,\,\+0,\,\+0)
$v_{4}$ (\,\+0,\,-1,\,\+0,\,\+0)
$v_{5}$ (\,\+1,\,\+1,\,\+0,\,\+0)
$v_{6}$ (\,-1,\,-1,\,\+0,\,\+0)
$v_{7}$ (\,\+0,\,\+0,\,\+1,\,\+0)
$v_{8}$ (\,\+0,\,\+0,\,-1,\,\+0)
$v_{9}$ (\,\+1,\,\+0,\,\+1,\,\+0)
$v_{10}$ (\,-1,\,\+0,\,-1,\,\+0)
$v_{11}$ (\,\+0,\,\+1,\,\+1,\,\+0)
$v_{12}$ (\,\+0,\,-1,\,-1,\,\+0)
$v_{13}$ (\,\+1,\,\+1,\,\+1,\,\+0)
$v_{14}$ (\,-1,\,-1,\,-1,\,\+0)
$v_{15}$ (\,\+0,\,\+0,\,\+0,\,\+1)
$v_{16}$ (\,\+0,\,\+0,\,\+0,\,-1)
$v_{17}$ (\,\+1,\,\+0,\,\+0,\,\+1)
$v_{18}$ (\,-1,\,\+0,\,\+0,\,-1)
$v_{19}$ (\,\+0,\,\+1,\,\+0,\,\+1)
$v_{20}$ (\,\+0,\,-1,\,\+0,\,-1)
$v_{21}$ (\,\+1,\,\+1,\,\+0,\,\+1)
$v_{22}$ (\,-1,\,-1,\,\+0,\,-1)
$v_{23}$ (\,\+0,\,\+0,\,\+1,\,\+1)
$v_{24}$ (\,\+0,\,\+0,\,-1,\,-1)
$v_{25}$ (\,\+1,\,\+0,\,\+1,\,\+1)
$v_{26}$ (\,-1,\,\+0,\,-1,\,-1)
$v_{27}$ (\,\+0,\,\+1,\,\+1,\,\+1)
$v_{28}$ (\,\+0,\,-1,\,-1,\,-1)
$v_{29}$ (\,\+1,\,\+1,\,\+1,\,\+1)
$v_{30}$ (\,-1,\,-1,\,-1,\,-1)
}}
\qquad 
\small{\pgfplotstabletypeset[
begin table/.add={}{[t]},
col sep=space,
ignore chars={},
every head row/.style={output empty row,
before row={\hline \multicolumn{4}{|c|}{\vrule height15pt width0pt depth6pt Faces of $\widehat{\Delta}^{*}$ }\\},
after row=\hline\hline},
every first row/.style={before row=\vrule height12pt width0pt depth0pt},
every last row/.style={before row=\vrule height0pt width0pt depth6pt,after row=\hline},
%after row=\hline,    %  Uncomment this to get back lines between all rows.
columns={A,B,A,B},
display columns/0/.style={select equal part entry of={0}{2},column type = {|r}},
display columns/1/.style={select equal part entry of={0}{2},column type = {|r|}},
display columns/2/.style={select equal part entry of={1}{2},column type = {|r}},
display columns/3/.style={select equal part entry of={1}{2},column type = {|r|}},
string type]
{
A B

$\tau_{1}$ $-x_{1}=1$
$\tau_{2}$ $-x_{1}+x_{4}=1$
$\tau_{3}$ $-x_{1}+x_{3}=1$
$\tau_{4}$ $-x_{1}+x_{2}=1$
$\tau_{5}$ $-x_{2}=1$
$\tau_{6}$ $-x_{2}+x_{4}=1$
$\tau_{7}$ $-x_{2}+x_{3}=1$
$\tau_{8}$ $-x_{3}=1$
$\tau_{9}$ $-x_{3}+x_{4}=1$
$\tau_{10}$ $-x_{4}=1$
$\tau_{11}$ $x_{4}=1$
$\tau_{12}$ $x_{3}-x_{4}=1$
$\tau_{13}$ $x_{3}=1$
$\tau_{14}$ $x_{2}-x_{3}=1$
$\tau_{15}$ $x_{2}-x_{4}=1$
$\tau_{16}$ $x_{2}=1$
$\tau_{17}$ $x_{1}-x_{2}=1$
$\tau_{18}$ $x_{1}-x_{3}=1$
$\tau_{19}$ $x_{1}-x_{4}=1$
$\tau_{20}$ $x_{1}=1$
}}
\end{table}
We form a matrix $\widehat{\text{W}}$ out of these vectors,
\begin{align} \notag
\widehat{\text{W}} \= \left(\begin{array}{c}
v_1 \\
v_2 \\
\dots \\
v_{30}
\end{array}\right) \= \left(w_1, w_2, w_3, w_4\right)~.
\end{align}
By finding the nullspace of $\widehat{\text{W}}^T$, we find 26 independent relations between the 30 vectors.
\begin{equation} \notag
\quad v_{2i}+v_{2i-1}\=0~,\quad  1\leq i\leq15~, 
\end{equation}
\begin{equation} \notag
\begin{aligned}
v_{7}+v_{15}+v_{24}&\=0~,\qquad&  v_{3}+v_{7}+v_{15}+v_{28}&\=0~,\\[3pt] v_{3}+v_{15}+v_{20}&\=0~,& v_{1}+v_{7}+v_{15}+v_{26}&\=0~,\\[3pt] v_{1}+v_{15}+v_{18}&\=0~,& v_{1}+v_{3}+v_{15}+v_{22}&\=0~,\\[3pt] v_{3}+v_{7}+v_{12}&\=0~,& v_{1}+v_{3}+v_{7}+v_{14}&\=0~,\\[3pt] v_{1}+v_{7}+v_{10}&\=0~,& v_{3}+v_{7}+v_{15}+v_{28}&\=0~,\\[3pt] v_{1}+v_{3}+v_{6}&\=0~.&
\end{aligned}
\end{equation}

Again, each of these relations corresponds to a scaling symmetry $\IC^* \subset (\IC^*)^{26}$. There are four invariant combinations of coordinates that we can identify with the coordinates on the torus ${\IT^4 \subset X_{\wh \Delta^*}}$. These can be taken to be
\begin{equation} \label{eq:Cox_Invariants_Singular_HV}
\begin{aligned}
\Xi_1 & \= \xi^{w_1} \= \frac{\xi _1 \xi _{5} \xi _{9} \xi _{13} \xi _{17} \xi _{21} \xi _{25} \xi _{29}}{\xi _2 \xi _6 \xi _{10} \xi _{14} \xi _{18} \xi _{22} \xi _{26} \xi _{30}}~, \quad& \Xi_2 &\= \xi^{w_2} \= \frac{\xi _3 \xi _{5} \xi _{11} \xi _{13} \xi _{19} \xi _{21} \xi _{27} \xi _{29}}{\xi _4 \xi _6 \xi _{12} \xi _{14} \xi _{20} \xi _{22} \xi _{28} \xi _{30}}~, \\[6pt]
\Xi_3 & \= \xi^{w_3} \= \frac{\xi _7 \xi _{9} \xi _{11} \xi _{13} \xi _{23} \xi _{25} \xi _{27} \xi _{29}}{\xi _8 \xi _{10} \xi _{12} \xi _{14} \xi _{24} \xi _{26} \xi _{28} \xi _{30}}~, & \Xi_4 &\= \xi^{w_4} \= \frac{\xi _{15} \xi _{17} \xi _{19} \xi _{21} \xi _{23} \xi _{25} \xi _{27} \xi _{29}}{\xi _{16} \xi _{18} \xi _{20} \xi _{22} \xi _{24} \xi _{26} \xi _{28} \xi _{30}}~.
\end{aligned}
\end{equation}
\subsection*{The polytope $\nabla^*$ and the ambient variety $\IP_{\nabla^*}$}
\vskip-10pt
\begin{table}[H]
\centering
\small{\pgfplotstabletypeset[
begin table/.add={}{[t]},
col sep=space,
ignore chars={},
every head row/.style={output empty row,
before row={\hline \multicolumn{2}{|c|}{\vrule height15pt width0pt depth6pt Vertices of $\nabla^{*}$ }\\},
after row=\hline\hline},
every first row/.style={before row=\vrule height12pt width0pt depth0pt},
every last row/.style={before row=\vrule height0pt width0pt depth6pt,after row=\hline},
%after row=\hline,    %  Uncomment this to get back lines between all rows.
columns={A,B},
display columns/0/.style={select equal part entry of={0}{1},column type = {|l}},
display columns/1/.style={select equal part entry of={0}{1},column type = {|l|}},
string type]
{
A B

$u_{1}$ (\,\+1,\,\+0,\,\+0,\,\+0,\,\+0)
$u_{2}$ (\,-1,\,\+0,\,\+0,\,\+0,\,\+0)
$u_{3}$ (\,\+0,\,\+1,\,\+0,\,\+0,\,\+0)
$u_{4}$ (\,\+0,\,-1,\,\+0,\,\+0,\,\+0)
$u_{5}$ (\,\+0,\,\+0,\,\+1,\,\+0,\,\+0)
$u_{6}$ (\,\+0,\,\+0,\,-1,\,\+0,\,\+0)
$u_{7}$ (\,\+0,\,\+0,\,\+0,\,\+1,\,\+0)
$u_{8}$ (\,\+0,\,\+0,\,\+0,\,-1,\,\+0)
$u_{9}$ (\,\+0,\,\+0,\,\+0,\,\+0,\,\+1)
$u_{10}$ (\,\+0,\,\+0,\,\+0,\,\+0,\,-1)
}}
\qquad
\small{\pgfplotstabletypeset[
begin table/.add={}{[t]},
col sep=space,
ignore chars={},
every head row/.style={output empty row,
before row={\hline \multicolumn{4}{|c|}{\vrule height15pt width0pt depth6pt Faces of $\nabla^{*}$ }\\},
after row=\hline\hline},
every first row/.style={before row=\vrule height12pt width0pt depth0pt},
every last row/.style={before row=\vrule height0pt width0pt depth6pt,after row=\hline},
%after row=\hline,    %  Uncomment this to get back lines between all rows.
columns={A,B,A,B},
display columns/0/.style={select equal part entry of={0}{2},column type = {|r}},
display columns/1/.style={select equal part entry of={0}{2},column type = {|r|}},
display columns/2/.style={select equal part entry of={1}{2},column type = {|r}},
display columns/3/.style={select equal part entry of={1}{2},column type = {|r|}},
string type]
{
A B

$\tau_{1}$ $-x_{1}-x_{2}-x_{3}-x_{4}-x_{5}=1$
$\tau_{2}$ $-x_{1}-x_{2}-x_{3}-x_{4}+x_{5}=1$
$\tau_{3}$ $-x_{1}-x_{2}-x_{3}+x_{4}-x_{5}=1$
$\tau_{4}$ $-x_{1}-x_{2}-x_{3}+x_{4}+x_{5}=1$
$\tau_{5}$ $-x_{1}-x_{2}+x_{3}-x_{4}-x_{5}=1$
$\tau_{6}$ $-x_{1}-x_{2}+x_{3}-x_{4}+x_{5}=1$
$\tau_{7}$ $-x_{1}-x_{2}+x_{3}+x_{4}-x_{5}=1$
$\tau_{8}$ $-x_{1}-x_{2}+x_{3}+x_{4}+x_{5}=1$
$\tau_{9}$ $-x_{1}+x_{2}-x_{3}-x_{4}-x_{5}=1$
$\tau_{10}$ $-x_{1}+x_{2}-x_{3}-x_{4}+x_{5}=1$
$\tau_{11}$ $-x_{1}+x_{2}-x_{3}+x_{4}-x_{5}=1$
$\tau_{12}$ $-x_{1}+x_{2}-x_{3}+x_{4}+x_{5}=1$
$\tau_{13}$ $-x_{1}+x_{2}+x_{3}-x_{4}-x_{5}=1$
$\tau_{14}$ $-x_{1}+x_{2}+x_{3}-x_{4}+x_{5}=1$
$\tau_{15}$ $-x_{1}+x_{2}+x_{3}+x_{4}-x_{5}=1$
$\tau_{16}$ $-x_{1}+x_{2}+x_{3}+x_{4}+x_{5}=1$
$\tau_{17}$ $x_{1}-x_{2}-x_{3}-x_{4}-x_{5}=1$
$\tau_{18}$ $x_{1}-x_{2}-x_{3}-x_{4}+x_{5}=1$
$\tau_{19}$ $x_{1}-x_{2}-x_{3}+x_{4}-x_{5}=1$
$\tau_{20}$ $x_{1}-x_{2}-x_{3}+x_{4}+x_{5}=1$
$\tau_{21}$ $x_{1}-x_{2}+x_{3}-x_{4}-x_{5}=1$
$\tau_{22}$ $x_{1}-x_{2}+x_{3}-x_{4}+x_{5}=1$
$\tau_{23}$ $x_{1}-x_{2}+x_{3}+x_{4}-x_{5}=1$
$\tau_{24}$ $x_{1}-x_{2}+x_{3}+x_{4}+x_{5}=1$
$\tau_{25}$ $x_{1}+x_{2}-x_{3}-x_{4}-x_{5}=1$
$\tau_{26}$ $x_{1}+x_{2}-x_{3}-x_{4}+x_{5}=1$
$\tau_{27}$ $x_{1}+x_{2}-x_{3}+x_{4}-x_{5}=1$
$\tau_{28}$ $x_{1}+x_{2}-x_{3}+x_{4}+x_{5}=1$
$\tau_{29}$ $x_{1}+x_{2}+x_{3}-x_{4}-x_{5}=1$
$\tau_{30}$ $x_{1}+x_{2}+x_{3}-x_{4}+x_{5}=1$
$\tau_{31}$ $x_{1}+x_{2}+x_{3}+x_{4}-x_{5}=1$
$\tau_{32}$ $x_{1}+x_{2}+x_{3}+x_{4}+x_{5}=1$
}}
\end{table}
A brief inspection reveals that the ten vertices of this polytope share precisely five relations, ${u_{2i}+u_{2i-1}=0}$. Each pair of vertices entering into these relations form a set of homogeneous coordinates for a $\IP^{1}$. This demonstrates that $\IP_{\nabla^{*}}\cong \left(\IP^{1}\right)^{5}$.
\subsection*{The polytope $\Delta^{*}$ and the ambient variety $\IP_{\Delta^{*}}$}
\vskip-10pt
\begin{table}[H]
\centering
\small{\pgfplotstabletypeset[
col sep=space,
ignore chars={},
every head row/.style={output empty row,
before row={\hline \multicolumn{4}{|c|}{\vrule height15pt width0pt depth6pt Vertices of $\Delta^{*}$ }\\},
after row=\hline\hline},
every first row/.style={before row=\vrule height12pt width0pt depth0pt},
every last row/.style={before row=\vrule height0pt width0pt depth6pt,after row=\hline},
%after row=\hline,    %  Uncomment this to get back lines between all rows.
columns={A,B,A,B},
display columns/0/.style={select equal part entry of={0}{2},column type = {|l}},
display columns/1/.style={select equal part entry of={0}{2},column type = {|l|}},
display columns/2/.style={select equal part entry of={1}{2},column type = {|l}},
display columns/3/.style={select equal part entry of={1}{2},column type = {|l|}},
string type]
{
A B

$u_{1}$ (\,\+1,\,\+0,\,\+0,\,\+0,\,\+0)
$u_{2}$ (\,-1,\,\+0,\,\+0,\,\+0,\,\+0)
$u_{3}$ (\,\+0,\,\+1,\,\+0,\,\+0,\,\+0)
$u_{4}$ (\,\+0,\,-1,\,\+0,\,\+0,\,\+0)
$u_{5}$ (\,\+1,\,\+1,\,\+0,\,\+0,\,\+0)
$u_{6}$ (\,-1,\,-1,\,\+0,\,\+0,\,\+0)
$u_{7}$ (\,\+0,\,\+0,\,\+1,\,\+0,\,\+0)
$u_{8}$ (\,\+0,\,\+0,\,-1,\,\+0,\,\+0)
$u_{9}$ (\,\+1,\,\+0,\,\+1,\,\+0,\,\+0)
$u_{10}$ (\,-1,\,\+0,\,-1,\,\+0,\,\+0)
$u_{11}$ (\,\+0,\,\+1,\,\+1,\,\+0,\,\+0)
$u_{12}$ (\,\+0,\,-1,\,-1,\,\+0,\,\+0)
$u_{13}$ (\,\+1,\,\+1,\,\+1,\,\+0,\,\+0)
$u_{14}$ (\,-1,\,-1,\,-1,\,\+0,\,\+0)
$u_{15}$ (\,\+0,\,\+0,\,\+0,\,\+1,\,\+0)
$u_{16}$ (\,\+0,\,\+0,\,\+0,\,-1,\,\+0)
$u_{17}$ (\,\+1,\,\+0,\,\+0,\,\+1,\,\+0)
$u_{18}$ (\,-1,\,\+0,\,\+0,\,-1,\,\+0)
$u_{19}$ (\,\+0,\,\+1,\,\+0,\,\+1,\,\+0)
$u_{20}$ (\,\+0,\,-1,\,\+0,\,-1,\,\+0)
$u_{21}$ (\,\+1,\,\+1,\,\+0,\,\+1,\,\+0)
$u_{22}$ (\,-1,\,-1,\,\+0,\,-1,\,\+0)
$u_{23}$ (\,\+0,\,\+0,\,\+1,\,\+1,\,\+0)
$u_{24}$ (\,\+0,\,\+0,\,-1,\,-1,\,\+0)
$u_{25}$ (\,\+1,\,\+0,\,\+1,\,\+1,\,\+0)
$u_{26}$ (\,-1,\,\+0,\,-1,\,-1,\,\+0)
$u_{27}$ (\,\+0,\,\+1,\,\+1,\,\+1,\,\+0)
$u_{28}$ (\,\+0,\,-1,\,-1,\,-1,\,\+0)
$u_{29}$ (\,\+1,\,\+1,\,\+1,\,\+1,\,\+0)
$u_{30}$ (\,-1,\,-1,\,-1,\,-1,\,\+0)
$u_{31}$ (\,\+0,\,\+0,\,\+0,\,\+0,\,\+1)
$u_{32}$ (\,\+0,\,\+0,\,\+0,\,\+0,\,-1)
$u_{33}$ (\,\+1,\,\+0,\,\+0,\,\+0,\,\+1)
$u_{34}$ (\,-1,\,\+0,\,\+0,\,\+0,\,-1)
$u_{35}$ (\,\+0,\,\+1,\,\+0,\,\+0,\,\+1)
$u_{36}$ (\,\+0,\,-1,\,\+0,\,\+0,\,-1)
$u_{37}$ (\,\+1,\,\+1,\,\+0,\,\+0,\,\+1)
$u_{38}$ (\,-1,\,-1,\,\+0,\,\+0,\,-1)
$u_{39}$ (\,\+0,\,\+0,\,\+1,\,\+0,\,\+1)
$u_{40}$ (\,\+0,\,\+0,\,-1,\,\+0,\,-1)
$u_{41}$ (\,\+1,\,\+0,\,\+1,\,\+0,\,\+1)
$u_{42}$ (\,-1,\,\+0,\,-1,\,\+0,\,-1)
$u_{43}$ (\,\+0,\,\+1,\,\+1,\,\+0,\,\+1)
$u_{44}$ (\,\+0,\,-1,\,-1,\,\+0,\,-1)
$u_{45}$ (\,\+1,\,\+1,\,\+1,\,\+0,\,\+1)
$u_{46}$ (\,-1,\,-1,\,-1,\,\+0,\,-1)
$u_{47}$ (\,\+0,\,\+0,\,\+0,\,\+1,\,\+1)
$u_{48}$ (\,\+0,\,\+0,\,\+0,\,-1,\,-1)
$u_{49}$ (\,\+1,\,\+0,\,\+0,\,\+1,\,\+1)
$u_{50}$ (\,-1,\,\+0,\,\+0,\,-1,\,-1)
$u_{51}$ (\,\+0,\,\+1,\,\+0,\,\+1,\,\+1)
$u_{52}$ (\,\+0,\,-1,\,\+0,\,-1,\,-1)
$u_{53}$ (\,\+1,\,\+1,\,\+0,\,\+1,\,\+1)
$u_{54}$ (\,-1,\,-1,\,\+0,\,-1,\,-1)
$u_{55}$ (\,\+0,\,\+0,\,\+1,\,\+1,\,\+1)
$u_{56}$ (\,\+0,\,\+0,\,-1,\,-1,\,-1)
$u_{57}$ (\,\+1,\,\+0,\,\+1,\,\+1,\,\+1)
$u_{58}$ (\,-1,\,\+0,\,-1,\,-1,\,-1)
$u_{59}$ (\,\+0,\,\+1,\,\+1,\,\+1,\,\+1)
$u_{60}$ (\,\+0,\,-1,\,-1,\,-1,\,-1)
$u_{61}$ (\,\+1,\,\+1,\,\+1,\,\+1,\,\+1)
$u_{62}$ (\,-1,\,-1,\,-1,\,-1,\,-1)
}}
\qquad
\small{\pgfplotstabletypeset[
col sep=space,
ignore chars={},
every head row/.style={output empty row,
before row={\hline \multicolumn{2}{|c|}{\vrule height15pt width0pt depth6pt Faces of $\Delta^{*}$ }\\},
after row=\hline\hline},
every first row/.style={before row=\vrule height12pt width0pt depth0pt},
every last row/.style={before row=\vrule height0pt width0pt depth6pt,after row=\hline},
%after row=\hline,    %  Uncomment this to get back lines between all rows.
columns={A,B},
display columns/0/.style={select equal part entry of={0}{1},column type = {|l}},
display columns/1/.style={select equal part entry of={0}{1},column type = {|r|}},
string type]
{
A B

$\rho_{1}$ $y_{1}=1$
$\rho_{2}$ $y_{2}=1$
$\rho_{3}$ $y_{3}=1$
$\rho_{4}$ $y_{4}=1$
$\rho_{5}$ $y_{5}=1$
$\rho_{6}$ $-y_{1}=1$
$\rho_{7}$ $-y_{2}=1$
$\rho_{8}$ $-y_{3}=1$
$\rho_{9}$ $-y_{4}=1$
$\rho_{10}$ $-y_{5}=1$
$\rho_{11}$ $y_{1}-y_{2}=1$
$\rho_{12}$ $y_{1}-y_{3}=1$
$\rho_{13}$ $y_{1}-y_{4}=1$
$\rho_{14}$ $y_{1}-y_{5}=1$
$\rho_{15}$ $-y_{1}+y_{2}=1$
$\rho_{16}$ $y_{2}-y_{3}=1$
$\rho_{17}$ $y_{2}-y_{4}=1$
$\rho_{18}$ $y_{2}-y_{5}=1$
$\rho_{19}$ $-y_{1}+y_{3}=1$
$\rho_{20}$ $-y_{2}+y_{3}=1$
$\rho_{21}$ $y_{3}-y_{4}=1$
$\rho_{22}$ $y_{3}-y_{5}=1$
$\rho_{23}$ $-y_{1}+y_{4}=1$
$\rho_{24}$ $-y_{2}+y_{4}=1$
$\rho_{25}$ $-y_{3}+y_{4}=1$
$\rho_{26}$ $y_{4}-y_{5}=1$
$\rho_{27}$ $-y_{1}+y_{5}=1$
$\rho_{28}$ $-y_{2}+y_{5}=1$
$\rho_{29}$ $-y_{3}+y_{5}=1$
$\rho_{30}$ $-y_{4}+y_{5}=1$
$\+$ $\+$
}}
\end{table}
As is now familiar, we form a matrix $\text{W}$ out of these vectors,
\begin{align} \notag
\text{W} \= \left(\begin{array}{c}
u_1 \\
u_2 \\
\dots \\
u_{30}
\end{array}\right) \= \left(w_1, w_2, w_3, w_4, w_5\right)~.
\end{align}
By examining the nullspace of $\text{W}^T$, we find 57 independent relations between these 62 vectors.
\newpage
\begin{equation} \notag
u_{2i}+u_{2i-1}\=0,\qquad 1\leq i\leq31~,
\end{equation}\vskip-15pt
\begin{equation} \notag
\begin{aligned}
u_{1}{+}u_{3}{+}u_6&=0~,&  u_{1}{+}u_{7}{+}u_{10}&=0~,& u_{1}{+}u_{15}{+}u_{18}&=0~,\\[3pt]
u_{1}{+}u_{31}{+}u_{34}&=0~,&
u_{3}{+}u_{7}{+}u_{12}&=0~,& u_{3}{+}u_{15}{+}u_{20}&=0~,\\[3pt] u_{3}{+}u_{31}{+}u_{36}&=0~,&
u_{7}{+}u_{15}{+}u_{24}&=0~,&
u_{7}{+}u_{31}{+}u_{40}&=0~,\\[3pt]
u_{15}{+}u_{31}{+}u_{48}&=0~,&
u_{1}{+}u_{3}{+}u_{7}{+}u_{14}&=0~,&
u_{1}{+}u_{3}{+}u_{15}{+}u_{22}&=0~,\\[3pt]
u_{1}{+}u_{3}{+}u_{31}{+}u_{38}&=0~,&
u_{1}{+}u_{7}{+}u_{15}{+}u_{26}&=0~,&
u_{1}{+}u_{7}{+}u_{31}{+}u_{42}&=0~,\\[3pt]
u_{1}{+}u_{15}{+}u_{31}{+}u_{50}&=0~,&
u_{3}{+}u_{7}{+}u_{15}{+}u_{28}&=0~,&
u_{3}{+}u_{7}{+}u_{31}{+}u_{44}&=0~,\\[3pt]
u_{3}{+}u_{15}{+}u_{31}{+}u_{52}&=0~,&
u_{7}{+}u_{15}{+}u_{31}{+}u_{56}&=0~,&
u_{1}{+}u_{3}{+}u_{7}{+}u_{15}{+}u_{30}&=0~,\\[3pt]
u_{1}{+}u_{3}{+}u_{7}{+}u_{31}{+}u_{46}&=0~,&
u_{1}{+}u_{3}{+}u_{15}{+}u_{31}{+}u_{54}&=0~,&
u_{1}{+}u_{7}{+}u_{15}{+}u_{31}{+}u_{58}&=0~,\\[3pt]
u_{3}{+}u_{7}{+}u_{15}{+}u_{31}{+}u_{60}&=0~,&
u_{1}{+}u_{3}{+}u_{7}{+}u_{15}{+}u_{31}{+}u_{62}&=0~.
\end{aligned}
\end{equation}

Each of these relations corresponds to a scaling symmetry $\IC^* \subset (\IC^*)^{57}$. There are five independent invariant combinations of coordinates that we can identify as coordinates on the torus $\IT^5 \subset \IP_{\Delta^{*}}$.\vskip-15pt
\begin{align} \label{eq:Cox_Invariants_HV}
\begin{split}
\Xi_1 &\= \xi^{w_1} \= \frac{\xi _{1} \xi _{5} \xi _{9} \xi _{13} \xi _{17} \xi _{21} \xi _{25} \xi _{29} \xi _{33} \xi _{37} \xi _{41} \xi _{45} \xi _{49} \xi _{53} \xi _{57} \xi _{61}}{\xi_2 \xi _6 \xi _{10} \xi _{14} \xi _{18} \xi _{22} \xi _{26} \xi _{30} \xi _{34} \xi _{38} \xi _{42} \xi _{46} \xi _{50} \xi _{54} \xi _{58} \xi _{62}}~,\\[6pt]
\Xi_2 &\= \xi^{w_2} \=\frac{\xi _{3} \xi _{5} \xi _{11} \xi _{13} \xi _{19} \xi _{21} \xi _{27} \xi _{29} \xi _{35} \xi _{37} \xi _{43} \xi _{45} \xi _{51} \xi _{53} \xi _{59} \xi _{61}}{\xi_4 \xi _6 \xi _{12} \xi _{14} \xi _{20} \xi _{22} \xi _{28} \xi _{30} \xi _{36} \xi _{38} \xi _{44} \xi _{46} \xi _{52} \xi _{54} \xi _{60} \xi _{62}}~, \\[6pt]
\Xi_3 &\= \xi^{w_3} \= \frac{\xi _{7} \xi _{9} \xi _{11} \xi _{13} \xi _{23} \xi _{25} \xi _{27} \xi _{29} \xi _{39} \xi _{41} \xi _{43} \xi _{45} \xi _{55} \xi _{57} \xi _{59} \xi _{61}}{\xi_8 \xi _{10} \xi _{12} \xi _{14} \xi _{24} \xi _{26} \xi _{28} \xi _{30} \xi _{40} \xi _{42} \xi _{44} \xi _{46} \xi _{56} \xi _{58} \xi _{60} \xi _{62}}~, \\[6pt]
\Xi_4 &\= \xi^{w_4} \=\frac{\xi _{15} \xi _{17} \xi _{19} \xi _{21} \xi _{23} \xi _{25} \xi _{27} \xi _{29} \xi _{47} \xi _{49} \xi _{51} \xi _{53} \xi _{55} \xi _{57} \xi _{59} \xi _{61}}{\xi_{16} \xi _{18} \xi _{20} \xi _{22} \xi _{24} \xi _{26} \xi _{28} \xi _{30} \xi _{48} \xi _{50} \xi _{52} \xi _{54} \xi _{56} \xi _{58} \xi _{60} \xi _{62}}~,\\[6pt]
\Xi_5 &\= \xi^{w_5} \= \frac{\xi _{31} \xi _{33} \xi _{35} \xi _{37} \xi _{39} \xi _{41} \xi _{43} \xi _{45} \xi _{47} \xi _{49} \xi _{51} \xi _{53} \xi _{55} \xi _{57} \xi _{59} \xi _{61}}{\xi_{32} \xi _{34} \xi _{36} \xi _{38} \xi _{40} \xi _{42} \xi _{44} \xi _{46} \xi _{48} \xi _{50} \xi _{52} \xi _{54} \xi _{56} \xi _{58} \xi _{60} \xi _{62}}~.
\end{split}
\end{align}

\subsection*{The polytope $\nabla$}
\vskip-10pt
\begin{table}[H]
\centering
\small{\pgfplotstabletypeset[
begin table/.add={}{[t]},
col sep=space,
ignore chars={},
every head row/.style={output empty row,
before row={\hline \multicolumn{6}{|c|}{\vrule height15pt width0pt depth6pt Vertices of $\nabla$ }\\},
after row=\hline\hline},
every first row/.style={before row=\vrule height12pt width0pt depth0pt},
every last row/.style={before row=\vrule height0pt width0pt depth6pt,after row=\hline},
%after row=\hline,    %  Uncomment this to get back lines between all rows.
columns={A,B,A,B,A,B},
display columns/0/.style={select equal part entry of={0}{3},column type = {|l}},
display columns/1/.style={select equal part entry of={0}{3},column type = {|l|}},
display columns/2/.style={select equal part entry of={1}{3},column type = {|l}},
display columns/3/.style={select equal part entry of={1}{3},column type = {|l|}},
display columns/4/.style={select equal part entry of={2}{3},column type = {|l}},
display columns/5/.style={select equal part entry of={2}{3},column type = {|l|}},
string type]
{
A B

$u_{1}$ (\,-1,\,-1,\,-1,\,-1,\,-1)
$u_{2}$ (\,-1,\,-1,\,-1,\,-1,\,\+1)
$u_{3}$ (\,-1,\,-1,\,-1,\,\+1,\,-1)
$u_{4}$ (\,-1,\,-1,\,-1,\,\+1,\,\+1)
$u_{5}$ (\,-1,\,-1,\,\+1,\,-1,\,-1)
$u_{6}$ (\,-1,\,-1,\,\+1,\,-1,\,\+1)
$u_{7}$ (\,-1,\,-1,\,\+1,\,\+1,\,-1)
$u_{8}$ (\,-1,\,-1,\,\+1,\,\+1,\,\+1)
$u_{9}$ (\,-1,\,\+1,\,-1,\,-1,\,-1)
$u_{10}$ (\,-1,\,\+1,\,-1,\,-1,\,\+1)
$u_{11}$ (\,-1,\,\+1,\,-1,\,\+1,\,-1)
$u_{12}$ (\,-1,\,\+1,\,-1,\,\+1,\,\+1)
$u_{13}$ (\,-1,\,\+1,\,\+1,\,-1,\,-1)
$u_{14}$ (\,-1,\,\+1,\,\+1,\,-1,\,\+1)
$u_{15}$ (\,-1,\,\+1,\,\+1,\,\+1,\,-1)
$u_{16}$ (\,-1,\,\+1,\,\+1,\,\+1,\,\+1)
$u_{17}$ (\,\+1,\,-1,\,-1,\,-1,\,-1)
$u_{18}$ (\,\+1,\,-1,\,-1,\,-1,\,\+1)
$u_{19}$ (\,\+1,\,-1,\,-1,\,\+1,\,-1)
$u_{20}$ (\,\+1,\,-1,\,-1,\,\+1,\,\+1)
$u_{21}$ (\,\+1,\,-1,\,\+1,\,-1,\,-1)
$u_{22}$ (\,\+1,\,-1,\,\+1,\,-1,\,\+1)
$u_{23}$ (\,\+1,\,-1,\,\+1,\,\+1,\,-1)
$u_{24}$ (\,\+1,\,-1,\,\+1,\,\+1,\,\+1)
$u_{25}$ (\,\+1,\,\+1,\,-1,\,-1,\,-1)
$u_{26}$ (\,\+1,\,\+1,\,-1,\,-1,\,\+1)
$u_{27}$ (\,\+1,\,\+1,\,-1,\,\+1,\,-1)
$u_{28}$ (\,\+1,\,\+1,\,-1,\,\+1,\,\+1)
$u_{29}$ (\,\+1,\,\+1,\,\+1,\,-1,\,-1)
$u_{30}$ (\,\+1,\,\+1,\,\+1,\,-1,\,\+1)
$u_{31}$ (\,\+1,\,\+1,\,\+1,\,\+1,\,-1)
$u_{32}$ (\,\+1,\,\+1,\,\+1,\,\+1,\,\+1)
}}
\qquad
\small{\pgfplotstabletypeset[
begin table/.add={}{[t]},
col sep=space,
ignore chars={},
every head row/.style={output empty row,
before row={\hline \multicolumn{2}{|c|}{\vrule height15pt width0pt depth6pt Faces of $\nabla$ }\\},
after row=\hline\hline},
every last row/.style={before row=\vrule height0pt width0pt depth9pt,after row=\hline},
%after row=\hline,    %  Uncomment this to get back lines between all rows.
columns={A,B},
display columns/0/.style={select equal part entry of={0}{1},column type = {|l}},
display columns/1/.style={select equal part entry of={0}{1},column type = {|r|}},
string type]
{
A B

$\rho_{1}$ $y_{1}=1$
$\rho_{2}$ $-y_{1}=1$
$\rho_{3}$ $y_{2}=1$
$\rho_{4}$ $-y_{2}=1$
$\rho_{5}$ $y_{3}=1$
$\rho_{6}$ $-y_{3}=1$
$\rho_{7}$ $y_{4}=1$
$\rho_{8}$ $-y_{4}=1$
$\rho_{9}$ $y_{5}=1$
$\rho_{10}$ $-y_{5}=1$
$\+$ $\+$
$\+$ $\+$
}}
\end{table}

\subsection*{The polytope $\Delta$}
\vskip-10pt
\begin{table}[H]
\centering
\small{\pgfplotstabletypeset[
col sep=space,
ignore chars={},
every head row/.style={output empty row,
before row={\hline \multicolumn{2}{|c|}{\vrule height15pt width0pt depth6pt Vertices of $\Delta$ }\\},
after row=\hline\hline},
every first row/.style={before row=\vrule height12pt width0pt depth0pt},
every last row/.style={before row=\vrule height0pt width0pt depth6pt,after row=\hline},
%after row=\hline,    %  Uncomment this to get back lines between all rows.
columns={A,B},
display columns/0/.style={select equal part entry of={0}{1},column type = {|l}},
display columns/1/.style={select equal part entry of={0}{1},column type = {|l|}},
string type]
{
A B

$v_{1}$ (\,\+1,\,\+0,\,\+0,\,\+0,\,\+0)
$v_{2}$ (\,\+0,\,\+1,\,\+0,\,\+0,\,\+0)
$v_{3}$ (\,\+0,\,\+0,\,\+1,\,\+0,\,\+0)
$v_{4}$ (\,\+0,\,\+0,\,\+0,\,\+1,\,\+0)
$v_{5}$ (\,\+0,\,\+0,\,\+0,\,\+0,\,\+1)
$v_{6}$ (\,-1,\,\+0,\,\+0,\,\+0,\,\+0)
$v_{7}$ (\,\+0,\,-1,\,\+0,\,\+0,\,\+0)
$v_{8}$ (\,\+0,\,\+0,\,-1,\,\+0,\,\+0)
$v_{9}$ (\,\+0,\,\+0,\,\+0,\,-1,\,\+0)
$v_{10}$ (\,\+0,\,\+0,\,\+0,\,\+0,\,-1)
$v_{11}$ (\,\+1,\,-1,\,\+0,\,\+0,\,\+0)
$v_{12}$ (\,\+1,\,\+0,\,-1,\,\+0,\,\+0)
$v_{13}$ (\,\+1,\,\+0,\,\+0,\,-1,\,\+0)
$v_{14}$ (\,\+1,\,\+0,\,\+0,\,\+0,\,-1)
$v_{15}$ (\,-1,\,\+1,\,\+0,\,\+0,\,\+0)
$v_{16}$ (\,\+0,\,\+1,\,-1,\,\+0,\,\+0)
$v_{17}$ (\,\+0,\,\+1,\,\+0,\,-1,\,\+0)
$v_{18}$ (\,\+0,\,\+1,\,\+0,\,\+0,\,-1)
$v_{19}$ (\,-1,\,\+0,\,\+1,\,\+0,\,\+0)
$v_{20}$ (\,\+0,\,-1,\,\+1,\,\+0,\,\+0)
$v_{21}$ (\,\+0,\,\+0,\,\+1,\,-1,\,\+0)
$v_{22}$ (\,\+0,\,\+0,\,\+1,\,\+0,\,-1)
$v_{23}$ (\,-1,\,\+0,\,\+0,\,\+1,\,\+0)
$v_{24}$ (\,\+0,\,-1,\,\+0,\,\+1,\,\+0)
$v_{25}$ (\,\+0,\,\+0,\,-1,\,\+1,\,\+0)
$v_{26}$ (\,\+0,\,\+0,\,\+0,\,\+1,\,-1)
$v_{27}$ (\,-1,\,\+0,\,\+0,\,\+0,\,\+1)
$v_{28}$ (\,\+0,\,-1,\,\+0,\,\+0,\,\+1)
$v_{29}$ (\,\+0,\,\+0,\,-1,\,\+0,\,\+1)
$v_{30}$ (\,\+0,\,\+0,\,\+0,\,-1,\,\+1)
$\+$ $\+$
}}
\qquad
\small{\pgfplotstabletypeset[
col sep=space,
ignore chars={},
every head row/.style={output empty row,
before row={\hline \multicolumn{4}{|c|}{\vrule height15pt width0pt depth6pt Faces of $\Delta$ }\\},
after row=\hline\hline},
every first row/.style={before row=\vrule height12pt width0pt depth0pt},
every last row/.style={before row=\vrule height0pt width0pt depth6pt,after row=\hline},
%after row=\hline,    %  Uncomment this to get back lines between all rows.
columns={A,B,A,B},
display columns/0/.style={select equal part entry of={0}{2},column type = {|l}},
display columns/1/.style={select equal part entry of={0}{2},column type = {|r|}},
display columns/2/.style={select equal part entry of={1}{2},column type = {|l}},
display columns/3/.style={select equal part entry of={1}{2},column type = {|r|}},
string type]
{
A B

$\tau_{1}$ $x_{1}=1$
$\tau_{2}$ $-x_{1}=1$
$\tau_{3}$ $x_{2}=1$
$\tau_{4}$ $-x_{2}=1$
$\tau_{5}$ $x_{1}+x_{2}=1$
$\tau_{6}$ $-x_{1}-x_{2}=1$
$\tau_{7}$ $x_{3}=1$
$\tau_{8}$ $-x_{3}=1$
$\tau_{9}$ $x_{1}+x_{3}=1$
$\tau_{10}$ $-x_{1}-x_{3}=1$
$\tau_{11}$ $x_{2}+x_{3}=1$
$\tau_{12}$ $-x_{2}-x_{3}=1$
$\tau_{13}$ $x_{1}+x_{2}+x_{3}=1$
$\tau_{14}$ $-x_{1}-x_{2}-x_{3}=1$
$\tau_{15}$ $x_{4}=1$
$\tau_{16}$ $-x_{4}=1$
$\tau_{17}$ $x_{1}+x_{4}=1$
$\tau_{18}$ $-x_{1}-x_{4}=1$
$\tau_{19}$ $x_{2}+x_{4}=1$
$\tau_{20}$ $-x_{2}-x_{4}=1$
$\tau_{21}$ $x_{1}+x_{2}+x_{4}=1$
$\tau_{22}$ $-x_{1}-x_{2}-x_{4}=1$
$\tau_{23}$ $x_{3}+x_{4}=1$
$\tau_{24}$ $-x_{3}-x_{4}=1$
$\tau_{25}$ $x_{1}+x_{3}+x_{4}=1$
$\tau_{26}$ $-x_{1}-x_{3}-x_{4}=1$
$\tau_{27}$ $x_{2}+x_{3}+x_{4}=1$
$\tau_{28}$ $-x_{2}-x_{3}-x_{4}=1$
$\tau_{29}$ $x_{1}+x_{2}+x_{3}+x_{4}=1$
$\tau_{30}$ $-x_{1}-x_{2}-x_{3}-x_{4}=1$
$\tau_{31}$ $x_{5}=1$
$\tau_{32}$ $-x_{5}=1$
$\tau_{33}$ $x_{1}+x_{5}=1$
$\tau_{34}$ $-x_{1}-x_{5}=1$
$\tau_{35}$ $x_{2}+x_{5}=1$
$\tau_{36}$ $-x_{2}-x_{5}=1$
$\tau_{37}$ $x_{1}+x_{2}+x_{5}=1$
$\tau_{38}$ $-x_{1}-x_{2}-x_{5}=1$
$\tau_{39}$ $x_{3}+x_{5}=1$
$\tau_{40}$ $-x_{3}-x_{5}=1$
$\tau_{41}$ $x_{1}+x_{3}+x_{5}=1$
$\tau_{42}$ $-x_{1}-x_{3}-x_{5}=1$
$\tau_{43}$ $x_{2}+x_{3}+x_{5}=1$
$\tau_{44}$ $-x_{2}-x_{3}-x_{5}=1$
$\tau_{45}$ $x_{1}+x_{2}+x_{3}+x_{5}=1$
$\tau_{46}$ $-x_{1}-x_{2}-x_{3}-x_{5}=1$
$\tau_{47}$ $x_{4}+x_{5}=1$
$\tau_{48}$ $-x_{4}-x_{5}=1$
$\tau_{49}$ $x_{1}+x_{4}+x_{5}=1$
$\tau_{50}$ $-x_{1}-x_{4}-x_{5}=1$
$\tau_{51}$ $x_{2}+x_{4}+x_{5}=1$
$\tau_{52}$ $-x_{2}-x_{4}-x_{5}=1$
$\tau_{53}$ $x_{1}+x_{2}+x_{4}+x_{5}=1$
$\tau_{54}$ $-x_{1}-x_{2}-x_{4}-x_{5}=1$
$\tau_{55}$ $x_{3}+x_{4}+x_{5}=1$
$\tau_{56}$ $-x_{3}-x_{4}-x_{5}=1$
$\tau_{57}$ $x_{1}+x_{3}+x_{4}+x_{5}=1$
$\tau_{58}$ $-x_{1}-x_{3}-x_{4}-x_{5}=1$
$\tau_{59}$ $x_{2}+x_{3}+x_{4}+x_{5}=1$
$\tau_{60}$ $-x_{2}-x_{3}-x_{4}-x_{5}=1$
$\tau_{61}$ $x_{1}+x_{2}+x_{3}+x_{4}+x_{5}=1$
$\tau_{62}$ $-x_{1}-x_{2}-x_{3}-x_{4}-x_{5}=1$
}}
\end{table}

\newpage
\section{Series Expressions for the Bessel Integrals}\label{app:Bessels}
\vskip-10pt
The symbol $\mathbf{p}$ is understood to denote a multi-index $(p_{1},p_{2},p_{3},p_{4},p_{5})$. We adopt a notation $c_{\mathbf{p}}$ for the multinomial coefficients. Recall also the harmonic numbers $H_{n}$ and Polygamma functions $\psi$.
\begin{equation}  \notag
c_{\mathbf{p}} \;\defineas\; \binom{\deg(\bm{p})}{\bm{p}}^2 \= \left(\frac{\left(\sum_{i=1}^{5}p_{i}\right)!}{\prod_{i=1}^{5}p_{i}!}\right)^{2} ~,\qquad H_{n}\=\sum_{k=1}^{n}\frac{1}{k}~, \qquad \psi(z)\=\frac{\dd}{\dd z}\,\log\Gamma(z)~.
\end{equation}
For positive integers $m$ one has the following special values for $\psi$ and its derivatives: 
\begin{equation}  \notag
\begin{aligned}
\psi(m)&\=H_{m-1}-\gamma~,\qquad \psi^{(1)}(m)&\=\frac{\pi^{2}}{6}-\sum_{k=1}^{m-1}\frac{1}{k^{2}}~, \qquad \psi^{(2)}(m)&\=2\left(-\zeta(3)+\sum_{k=1}^{m-1}\frac{1}{k^{3}}\right),
\end{aligned}
\end{equation}
with $\gamma$ the Euler-Mascheroni constant.

With $n$ understood to be a positive integer, we will make frequent use of the following integrals, valid for $\text{Re}[\varphi]>0$.
\begin{equation} \label{eq:useful_integrals}
\begin{aligned}
\int_{0}^{\infty}\!\!\!\dd z \,K_{0}(\sqrt{\varphi}z)z^{2n+1}&\=4^{n}(n!)^{2}\varphi^{-1-n}~,\\[10pt]
\int_{0}^{\infty}\!\!\!\dd z \,K_{0}(\sqrt{\varphi}z)\log\left(\frac{z}{2}\right)z^{2n+1}&\=4^{n}(n!)^{2}\varphi^{-1-n}\left(\psi(n+1)-\frac{1}{2}\log(\varphi)\right),\\[10pt]
\int_{0}^{\infty}\!\!\!\dd z \,K_{0}(\sqrt{\varphi}z)\log\left(\frac{z}{2}\right)^{2}z^{2n+1}&\=4^{n-1}(n!)^{2}\varphi^{-1-n}\left(2\psi^{(1)}(n+1)-2\psi(n+1)+\log(\varphi)\right),\\[10pt]
\int_{0}^{\infty}\!\!\!\dd z \,K_{0}(\sqrt{\varphi}z)\log\left(\frac{z}{2}\right)^{3}z^{2n+1}&\=4^{n-1}(n!)^{2}\varphi^{-1-n}\left(\psi^{(2)}(n+1)\phantom{\frac{1}{2}}\right.\\[5pt]&\hskip-30pt\left.-3\left(\log \varphi-2\psi(n+1)\right)\psi^{(1)}(n+1)-\frac{1}{2}\left(\log \varphi-2\psi(n+1)\right)^{3}\right).
\end{aligned}
\end{equation}
To derive the formulae \eqref{eq:Bessel_pi0} and \eqref{eq:period_series}, recall the following series expressions for the Bessel functions $I_{0}(x)$ and $K_{0}(x)$, which can be substituted in the integrals \eqref{eq:Bessel_pi1}-\eqref{eq:Bessel_pi3}. One should substitute all Bessel functions for their series below, barring one $K_{0}$. Then integrating termwise and applying the above identities allows one obtain the formulae \eqref{eq:Bessel_pi0} and \eqref{eq:period_series}.
\begin{equation} \label{eq:Bessel_series}
\begin{aligned}
I_{0}(x)\=\sum_{n=0}^{\infty}\frac{1}{(n!)^{2}}\left(\frac{x}{2}\right)^{2n}~, \qquad K_{0}(x&)\=-\log\left(\frac{x}{2}\right)I_{0}(x)+\sum_{n=0}^{\infty}\frac{\psi(n+1)}{(n!)^{2}}\left(\frac{x}{2}\right)^{2n}.
\end{aligned}
\end{equation}
\newpage
\section{Parameter Counting} \label{app:Parameters}
\vskip-10pt
The polynomials \eqref{eq:Z_5_Quotient_Polynomials}-\eqref{eq:Z5Z2Z2_Variety} defining the manifolds $\MHV$ and their various quotients contain a number of parameters, which can be viewed as the complex structure parameters of the family $\MHV$. Naïvely it would seem that there are more free parameters in the defining polynomials than there are complex structure parameters. However, a more careful consideration will show that upon correctly accounting for redundancies, the parameter counts indeed agree.

Consider, for concreteness, the varieties in the family $\MHV$ which are symmetric under $\IZ_5 \times \IZ_2$, which we take to be those generated by $S$ and $V$ as in \eqref{eq:Z_5Z_2_generators}. 

We wish to determine the independent parameters in the polynomials $Q^{1}$ and $Q^{2}$ defining this symmetric variety. There are at least two sources of redundancy. The first is that different polynomials can generate the same ideal. The second arises from automorphisms of the ambient variety $\left(\IP^{1}\right)^{5}$. 

We begin by considering the most general $\IZ_5$-invariant polynomials:
\begin{align} \notag
Q^1 &\= A_0 m_{11111} {+} A_1 m_{10000} {+} A_2 m_{11000} {+} A_3 m_{10100} {+} A_4 m_{11100} {+} A_5 m_{11010} {+} A_6 m_{11110} {+} A_7 m_{00000}~,\nonumber\\[3pt]
Q^2 &\= B_0 m_{11111} {+} B_1 m_{10000} {+} B_2 m_{11000} {+} B_3 m_{10100} {+} B_4 m_{11100} {+} B_5 m_{11010} {+} B_6 m_{11110} {+} B_7 m_{00000}~.\nonumber
\end{align}
To have a variety that is invariant under the $\IZ_2$ transformation
\begin{align} \notag
V: \qquad Y_{i,0} \; \leftrightarrow \; Y_{i,1}  \qquad \text{for all }i.
\end{align}
We demand that the ideal $\langle Q^1, Q^2 \rangle$ is invariant under the action of $V$. In this case this reduces to demanding that $VQ^1$ and $V Q^2$ are linear combinations of $Q^1$ and $Q^2$:
\begin{align} \label{eq:V_invariance}
\left(\begin{matrix}
V Q^1\\
V Q^2
\end{matrix}
\right) \= \text{M} \left(\begin{matrix}
Q^1\\
Q^2
\end{matrix}
\right) \qquad \text{for some} \qquad \text{M} \in \text{GL}(2,\IC)~.
\end{align}
Clearly $V^2=\text{Id}$ from which it follows that $\text{M}^2=1$. In the generic case, the matrix $\text{M}$ takes the form
\begin{align} \notag
\text{M} \= \left(\begin{matrix}
a & \+b\\
\frac{1-a^2}{b} & -a 
\end{matrix}
\right).
\end{align} 
This has the Jordan normal form
\begin{align} \notag
\text{M} \= \left(\begin{matrix}
-1 & 0\\
\+0 & 1
\end{matrix}
\right).
\end{align}
Thus, by redefining $Q^1$ and $Q^2$ suitably, the condition \eqref{eq:V_invariance} becomes
\begin{align} \label{eq:V_invariance2}
V Q^1 \= -Q^1 \qquad \text{and} \qquad V Q^2 \= Q^2~.
\end{align}
The only residual redefinitions of $Q^1$ and $Q^2$ are those that keep the diagonalised $M$ fixed, that is rescalings of $Q^1$ and $Q^2$. Leaving these scalings unfixed for the time being, the condition \eqref{eq:V_invariance2} can be solved to give
\begin{align} \label{eq:V_invarianceAB}
A_{7-i} \= A_i~, \qquad B_{7-i} \= B_i~. 
\end{align}
Demanding the condition \eqref{eq:V_invarianceAB} fixes most of the automorphisms of $(\IP^1)^5/\IZ_5$, but there is one remaining family of M\"obius automorphisms of the form
\begin{align} \notag
T: \qquad \frac{Y_{i,0}}{Y_{i,1}} \mapsto \frac{Y_{i,0} + k \, Y_{i,1}}{k \, Y_{i,0} + Y_{i,1}} \qquad \text{with} \qquad k \in \IC\setminus\{1,-1\} \qquad \text{for all } i~.
\end{align}
Transformations of this form preserve the condition \eqref{eq:V_invarianceAB}. The images of $Q^1$ and $Q^2$ can be written down, but the generic form is slightly complicated. We note that
\begin{align} \notag
T (Q^1) \= \frac{(k-1) \left(-A_1 k^3-A_1 k^2-A_2 k^2-A_3 k^2+A_0 \left(k^4+k^3+k^2+k+1\right)-A_1 k\right)}{k^5} m_{00000} + ....
\end{align}
By choosing $k$ suitably, we can force the coefficient of $m_{00000}$ to vanish. Upon redefining the remaining parameters the polynomials $Q^1$ and $Q^2$ become
\begin{align} \notag
Q^1 &\= A_1 m_{10000} {+} A_2 m_{11000} {+} A_3 m_{10100} {-} A_2 m_{11100} {-} A_3 m_{11010} {-} A_1 m_{11110}~, \\[3pt] \notag
Q^2 &\= B_0 m_{11111} {+} B_1 m_{10000} {+} B_2 m_{11000} {+} B_3 m_{10100} {+} B_2 m_{11100} {+} B_3 m_{11010} {+} B_1 m_{11110} {+} B_0 m_{00000}~.
\end{align}
Finally, we can eliminate two parameters by rescaling. This leaves two polynomials with five independent parameters. 
\begin{align} \notag
Q^1 &\= m_{10000} + a_1 m_{11000} + a_2 m_{10100} - a_2 m_{11100} - a_1 m_{11010} - m_{11110}~, \nonumber\\[3pt] \notag
Q^2 &\= m_{11111} + a_3 m_{10000} + a_4 m_{11000} + a_5 m_{10100} + a_4 m_{11100} + a_5 m_{11010} + a_3 m_{11110} + m_{00000}~.
\end{align}

\newpage
\section{Recurrences for Elements of Webs}\label{app:properties_of_W}
\vskip-10pt
Consider the map $\psi_{\bm I}: \bm J \mapsto -2H(\bm J,\bm I)$, which gives a map $\IZ^5 \to \IZ$. Let us denote by $\ell$ the greatest common divisor of the nonzero $\psi_{\bm I}(\bm J)$ over all $\bm J \in \IZ^5$.
\begin{restatable}{claim}{claim:index_component_differences}
Let $\bm J = w \bm I$, then $J_i - I_i = 0 \mod \ell$ for all $i$. Specifically,
\begin{align}
J_i - I_i \= -2 H(\bm n_i (\bm J), \bm I)~,
\end{align}
where $\bm n_i$ satisfies the recurrence relations
\begin{align} \label{eq:n_i_recurrence}
\begin{split}
    \bm n_i(g_i \bm J) &\= \bm n_i(\bm J) + w^{-1} \bme^i~,\\
    \bm n_i(g_j \bm J) &\= \bm n_i(\bm J).   
\end{split}
\end{align}
\end{restatable}
Let us show this by induction. The initial step is straightforward, as we need to only check the component on which the duality acts.
\begin{align} \notag
(g_j \bm I)_j - I_j \= \deg \bm I - 3 I_j \= -2 H(\bme^j, \bm I) \= 0 \mod \ell~.
\end{align}
For the induction step, assume $\bm J = w \bm I$ and that for some vectors $\bm n_i$
\begin{align} \notag
J_i - I_i = -2H(\bm n_i(\bm J), \bm I)~.
\end{align}
Then it is enough to check that $(g_i \bm J)_i - I_i = (g_i \bm J)_i - J_i \mod \ell$ vanishes:
\begin{align} \notag
(g_i \bm J)_i - I_i \= (g_i \bm J)_i - J_i \= \deg \bm J - 3 J_i \= -2H(\bme^i,\bm J) \mod \ell.
\end{align}
Now recalling that the bilinear $H$ is invariant under the action of the Coxeter group, we have that
\begin{align} \notag
-2H(\bme^i,\bm J) \= -2H(\bme^i,w \bm I) \= -2H(w^{-1} \bme^i,\bm I) \= 0 \mod \ell~.
\end{align}
Keeping track of the terms that vanish $\text{modulo } \ell$ gives the recurrence formulae \eqref{eq:n_i_recurrence}. 

An immediate consequence of this is the following claim:
\begin{restatable}{claim}{claim:index_vector_degree_differences}
If $w = g_1 \dots g_l$ is a word in the duality operations and $\bm J = w \bm I$, $\deg \bm J - \deg \bm I \= 0 \mod \ell$. Specifically,
\begin{align} \label{eq:Webs_N_in_terms_of_n}
\deg \bm J - \deg \bm I \= -2H(\bm N(\bm J),\bm I)~, \qquad \text{with} \qquad \bm N(\bm J) = \sum_i \bm n_i(\bm J)~.
\end{align}
\end{restatable}

\newpage
\rightline{\it Look at the book of the Tablets of Heaven, and read what}
\rightline{\it is written upon them, and note every individual fact.}
\vskip5pt
\rightline{Enoch 81.1}
 
\section{Instanton Numbers} \label{app:Instanton_Numbers}
\vskip-10pt
In this appendix we tabulate nonzero instanton numbers up to degree 29 for genera 0 and 1. Our tables only give one multidegree in each $S_{5}$ orbit. For example the number $n_{(0,1,0,0,0)}$ is not explicitly given, but this number equals $n_{(1,0,0,0,0)}=24$ which does appear. Furthermore, instanton numbers that vanish are omitted.

\subsection{Genus-0 instantons}\label{sect:genus0instantons}
\vspace*{-10pt}
\begin{table}[H]
\begin{center}
\capt{5in}{tab:genuszeroinstantons}{The genus-zero instanton numbers $n_{\bm{I}}$ for $\text{deg}(\bm{I})\leqslant 29$}
\end{center}
\end{table}
\vskip-25pt
\newcommand{\tablepreambleInstantonNumbers}[1]{
	\vspace{-1.35cm}
	\begin{center}
		\begin{longtable}{|>{\scriptsize\centering$} p{.8in}<{$} |>{\scriptsize$}p{1.8in}<{$}
				||>{\scriptsize \centering$} p{.8in} <{~$} |>{\scriptsize$}p{1.8in}<{$} |}\hline
			\multicolumn{4}{|c|}{\vrule height 12pt depth7pt width 0pt $\text{deg}(\bm{I})=#1$}
			\tabularnewline[0.5pt] \hline \hline
			\vrule height 12pt depth5pt width 0pt $\normalsize{$\bm{I}$}$ & \hfil $\normalsize{$n_{\bm{I}}$}$ & $\normalsize{$\bm{I}$}$ & \hfil $\normalsize{$n_{\bm{I}}$}$ \tabularnewline[.5pt] \hline
			\endfirsthead
			\hline
			\multicolumn{4}{|l|}{$\text{deg}(\bm{I})=#1$, \sl continued}\tabularnewline[0.5pt] \hline 
			\vrule height 12pt depth5pt width 0pt $\normalsize{$\bm{I}$}$ & \hfil $\normalsize{$n_{\bm{I}}$}$ & $\normalsize{$\bm{I}$}$ &\hfil $\normalsize{$n_{\bm{I}}$}$ \tabularnewline[0.5pt] \hline
			& & & \tabularnewline[-12pt]
			\endhead
			\hline\hline 
			\multicolumn{4}{|r|}{{\footnotesize\sl Continued on the following page}}\tabularnewline[0.5pt] \hline
			\endfoot
			\hline
			\endlastfoot}
		
\newcommand{\tablepostambleInstantonNumbers}{\end{longtable}\end{center}}

\vskip30pt

\tablepreambleInstantonNumbers{1}
(1, 0, 0, 0, 0)	 & 24	 &  	 &  
\tablepostambleInstantonNumbers \vskip-18pt 

\tablepreambleInstantonNumbers{2}
(1, 1, 0, 0, 0)	 & 24	 &  	 &  
\tablepostambleInstantonNumbers \vskip-18pt 

\tablepreambleInstantonNumbers{3}
(1, 1, 1, 0, 0)	 & 112	 &  	 &  
\tablepostambleInstantonNumbers \vskip-18pt 

\tablepreambleInstantonNumbers{4}
 (1, 1, 1, 1, 0)	 & 1104	 & (2, 1, 1, 0, 0)	 & 24
\tablepostambleInstantonNumbers \vskip-18pt 

\tablepreambleInstantonNumbers{5}
 (1, 1, 1, 1, 1)	 & 19200	 & (2, 1, 1, 1, 0)	 & 1104
\tabularnewline[-4pt] 
(2, 2, 1, 0, 0)	 & 24	 &  	 &  
\tablepostambleInstantonNumbers \vskip-18pt 

\tablepreambleInstantonNumbers{6}
 (2, 1, 1, 1, 1)	 & 45408	 & (2, 2, 1, 1, 0)	 & 2800
\tabularnewline[-4pt] 
 (2, 2, 2, 0, 0)	 & 80	 & (3, 1, 1, 1, 0)	 & 112
\tablepostambleInstantonNumbers \vskip-18pt 

\tablepreambleInstantonNumbers{7}
 (2, 2, 1, 1, 1)	 & 212880	 & (2, 2, 2, 1, 0)	 & 14496
\tabularnewline[-4pt] 
 (3, 1, 1, 1, 1)	 & 19200	 & (3, 2, 1, 1, 0)	 & 1104
\tabularnewline[-4pt] 
(3, 2, 2, 0, 0)	 & 24	 &  	 &  
\tablepostambleInstantonNumbers \vskip-18pt 

\tablepreambleInstantonNumbers{8}
 (2, 2, 2, 1, 1)	 & 1691856	 & (2, 2, 2, 2, 0)	 & 122352
\tabularnewline[-4pt] 
 (3, 2, 1, 1, 1)	 & 212880	 & (3, 2, 2, 1, 0)	 & 14496
\tabularnewline[-4pt] 
 (3, 3, 1, 1, 0)	 & 1104	 & (3, 3, 2, 0, 0)	 & 24
\tabularnewline[-4pt] 
 (4, 1, 1, 1, 1)	 & 1104	 & (4, 2, 1, 1, 0)	 & 24
\tablepostambleInstantonNumbers \vskip-18pt 

\tablepreambleInstantonNumbers{9}
 (2, 2, 2, 2, 1)	 & 20299992	 & (3, 2, 2, 1, 1)	 & 3222112
\tabularnewline[-4pt] 
 (3, 2, 2, 2, 0)	 & 234048	 & (3, 3, 1, 1, 1)	 & 434688
\tabularnewline[-4pt] 
 (3, 3, 2, 1, 0)	 & 30624	 & (3, 3, 3, 0, 0)	 & 112
\tabularnewline[-4pt] 
 (4, 2, 1, 1, 1)	 & 45408	 & (4, 2, 2, 1, 0)	 & 2800
\tabularnewline[-4pt] 
(4, 3, 1, 1, 0)	 & 112	 &  	 &  
\tablepostambleInstantonNumbers \vskip-18pt 

\tablepreambleInstantonNumbers{10}
 (2, 2, 2, 2, 2)	 & 341681280	 & (3, 2, 2, 2, 1)	 & 63576576
\tabularnewline[-4pt] 
 (3, 3, 2, 1, 1)	 & 10883712	 & (3, 3, 2, 2, 0)	 & 795936
\tabularnewline[-4pt] 
 (3, 3, 3, 1, 0)	 & 122448	 & (4, 2, 2, 1, 1)	 & 1691856
\tabularnewline[-4pt] 
 (4, 2, 2, 2, 0)	 & 122352	 & (4, 3, 1, 1, 1)	 & 212880
\tabularnewline[-4pt] 
 (4, 3, 2, 1, 0)	 & 14496	 & (4, 3, 3, 0, 0)	 & 24
\tabularnewline[-4pt] 
 (4, 4, 1, 1, 0)	 & 24	 & (5, 2, 1, 1, 1)	 & 1104
\tabularnewline[-4pt] 
(5, 2, 2, 1, 0)	 & 24	 &  	 &  
\tablepostambleInstantonNumbers \vskip-18pt 

\tablepreambleInstantonNumbers{11}
 (3, 2, 2, 2, 2)	 & 1599622824	 & (3, 3, 2, 2, 1)	 & 316997280
\tabularnewline[-4pt] 
 (3, 3, 3, 1, 1)	 & 59097600	 & (3, 3, 3, 2, 0)	 & 4326048
\tabularnewline[-4pt] 
 (4, 2, 2, 2, 1)	 & 63576576	 & (4, 3, 2, 1, 1)	 & 10883712
\tabularnewline[-4pt] 
 (4, 3, 2, 2, 0)	 & 795936	 & (4, 3, 3, 1, 0)	 & 122448
\tabularnewline[-4pt] 
 (4, 4, 1, 1, 1)	 & 212880	 & (4, 4, 2, 1, 0)	 & 14496
\tabularnewline[-4pt] 
 (4, 4, 3, 0, 0)	 & 24	 & (5, 2, 2, 1, 1)	 & 212880
\tabularnewline[-4pt] 
 (5, 2, 2, 2, 0)	 & 14496	 & (5, 3, 1, 1, 1)	 & 19200
\tabularnewline[-4pt] 
(5, 3, 2, 1, 0)	 & 1104	 &  	 &  
\tablepostambleInstantonNumbers \vskip-18pt 

\tablepreambleInstantonNumbers{12}
 (3, 3, 2, 2, 2)	 & 11032046624	 & (3, 3, 3, 2, 1)	 & 2322325968
\tabularnewline[-4pt] 
 (3, 3, 3, 3, 0)	 & 33777312	 & (4, 2, 2, 2, 2)	 & 2624447520
\tabularnewline[-4pt] 
 (4, 3, 2, 2, 1)	 & 529392832	 & (4, 3, 3, 1, 1)	 & 100919904
\tabularnewline[-4pt] 
 (4, 3, 3, 2, 0)	 & 7371792	 & (4, 4, 2, 1, 1)	 & 19420400
\tabularnewline[-4pt] 
 (4, 4, 2, 2, 0)	 & 1423104	 & (4, 4, 3, 1, 0)	 & 234048
\tabularnewline[-4pt] 
 (4, 4, 4, 0, 0)	 & 80	 & (5, 2, 2, 2, 1)	 & 20299992
\tabularnewline[-4pt] 
 (5, 3, 2, 1, 1)	 & 3222112	 & (5, 3, 2, 2, 0)	 & 234048
\tabularnewline[-4pt] 
 (5, 3, 3, 1, 0)	 & 30624	 & (5, 4, 1, 1, 1)	 & 45408
\tabularnewline[-4pt] 
 (5, 4, 2, 1, 0)	 & 2800	 & (6, 2, 2, 1, 1)	 & 2800
\tabularnewline[-4pt] 
 (6, 2, 2, 2, 0)	 & 80	 & (6, 3, 1, 1, 1)	 & 112
\tablepostambleInstantonNumbers \vskip-18pt 

\tablepreambleInstantonNumbers{13}
 (3, 3, 3, 2, 2)	 & 105371446464	 & (3, 3, 3, 3, 1)	 & 23351460864
\tabularnewline[-4pt] 
 (4, 3, 2, 2, 2)	 & 27607031136	 & (4, 3, 3, 2, 1)	 & 5950086192
\tabularnewline[-4pt] 
 (4, 3, 3, 3, 0)	 & 88179456	 & (4, 4, 2, 2, 1)	 & 1426637712
\tabularnewline[-4pt] 
 (4, 4, 3, 1, 1)	 & 282674592	 & (4, 4, 3, 2, 0)	 & 20578560
\tabularnewline[-4pt] 
 (4, 4, 4, 1, 0)	 & 795936	 & (5, 2, 2, 2, 2)	 & 1599622824
\tabularnewline[-4pt] 
 (5, 3, 2, 2, 1)	 & 316997280	 & (5, 3, 3, 1, 1)	 & 59097600
\tabularnewline[-4pt] 
 (5, 3, 3, 2, 0)	 & 4326048	 & (5, 4, 2, 1, 1)	 & 10883712
\tabularnewline[-4pt] 
 (5, 4, 2, 2, 0)	 & 795936	 & (5, 4, 3, 1, 0)	 & 122448
\tabularnewline[-4pt] 
 (5, 4, 4, 0, 0)	 & 24	 & (5, 5, 1, 1, 1)	 & 19200
\tabularnewline[-4pt] 
 (5, 5, 2, 1, 0)	 & 1104	 & (6, 2, 2, 2, 1)	 & 1691856
\tabularnewline[-4pt] 
 (6, 3, 2, 1, 1)	 & 212880	 & (6, 3, 2, 2, 0)	 & 14496
\tabularnewline[-4pt] 
 (6, 3, 3, 1, 0)	 & 1104	 & (6, 4, 1, 1, 1)	 & 1104
\tabularnewline[-4pt] 
(6, 4, 2, 1, 0)	 & 24	 &  	 &  
\tablepostambleInstantonNumbers \vskip-18pt 

\tablepreambleInstantonNumbers{14}
 (3, 3, 3, 3, 2)	 & 1326841710624	 & (4, 3, 3, 2, 2)	 & 377080188864
\tabularnewline[-4pt] 
 (4, 3, 3, 3, 1)	 & 85495746528	 & (4, 4, 2, 2, 2)	 & 103492041648
\tabularnewline[-4pt] 
 (4, 4, 3, 2, 1)	 & 22951602432	 & (4, 4, 3, 3, 0)	 & 347078520
\tabularnewline[-4pt] 
 (4, 4, 4, 1, 1)	 & 1218252960	 & (4, 4, 4, 2, 0)	 & 88177920
\tabularnewline[-4pt] 
 (5, 3, 2, 2, 2)	 & 27607031136	 & (5, 3, 3, 2, 1)	 & 5950086192
\tabularnewline[-4pt] 
 (5, 3, 3, 3, 0)	 & 88179456	 & (5, 4, 2, 2, 1)	 & 1426637712
\tabularnewline[-4pt] 
 (5, 4, 3, 1, 1)	 & 282674592	 & (5, 4, 3, 2, 0)	 & 20578560
\tabularnewline[-4pt] 
 (5, 4, 4, 1, 0)	 & 795936	 & (5, 5, 2, 1, 1)	 & 10883712
\tabularnewline[-4pt] 
 (5, 5, 2, 2, 0)	 & 795936	 & (5, 5, 3, 1, 0)	 & 122448
\tabularnewline[-4pt] 
 (5, 5, 4, 0, 0)	 & 24	 & (6, 2, 2, 2, 2)	 & 341681280
\tabularnewline[-4pt] 
 (6, 3, 2, 2, 1)	 & 63576576	 & (6, 3, 3, 1, 1)	 & 10883712
\tabularnewline[-4pt] 
 (6, 3, 3, 2, 0)	 & 795936	 & (6, 4, 2, 1, 1)	 & 1691856
\tabularnewline[-4pt] 
 (6, 4, 2, 2, 0)	 & 122352	 & (6, 4, 3, 1, 0)	 & 14496
\tabularnewline[-4pt] 
 (6, 5, 1, 1, 1)	 & 1104	 & (6, 5, 2, 1, 0)	 & 24
\tabularnewline[-4pt] 
 (7, 2, 2, 2, 1)	 & 14496	 & (7, 3, 2, 1, 1)	 & 1104
\tabularnewline[-4pt] 
(7, 3, 2, 2, 0)	 & 24	 &  	 &  
\tablepostambleInstantonNumbers \vskip-18pt 

\tablepreambleInstantonNumbers{15}
 (3, 3, 3, 3, 3)	 & 21228933784320	 & (4, 3, 3, 3, 2)	 & 6446376071472
\tabularnewline[-4pt] 
 (4, 4, 3, 2, 2)	 & 1912895782008	 & (4, 4, 3, 3, 1)	 & 443961562528
\tabularnewline[-4pt] 
 (4, 4, 4, 2, 1)	 & 126121309632	 & (4, 4, 4, 3, 0)	 & 1935300720
\tabularnewline[-4pt] 
 (5, 3, 3, 2, 2)	 & 570360079168	 & (5, 3, 3, 3, 1)	 & 130194945024
\tabularnewline[-4pt] 
 (5, 4, 2, 2, 2)	 & 158730945984	 & (5, 4, 3, 2, 1)	 & 35487082592
\tabularnewline[-4pt] 
 (5, 4, 3, 3, 0)	 & 539120544	 & (5, 4, 4, 1, 1)	 & 1944767152
\tabularnewline[-4pt] 
 (5, 4, 4, 2, 0)	 & 140436672	 & (5, 5, 2, 2, 1)	 & 2306418848
\tabularnewline[-4pt] 
 (5, 5, 3, 1, 1)	 & 464696832	 & (5, 5, 3, 2, 0)	 & 33777312
\tabularnewline[-4pt] 
 (5, 5, 4, 1, 0)	 & 1423616	 & (5, 5, 5, 0, 0)	 & 112
\tabularnewline[-4pt] 
 (6, 3, 2, 2, 2)	 & 11032046624	 & (6, 3, 3, 2, 1)	 & 2322325968
\tabularnewline[-4pt] 
 (6, 3, 3, 3, 0)	 & 33777312	 & (6, 4, 2, 2, 1)	 & 529392832
\tabularnewline[-4pt] 
 (6, 4, 3, 1, 1)	 & 100919904	 & (6, 4, 3, 2, 0)	 & 7371792
\tabularnewline[-4pt] 
 (6, 4, 4, 1, 0)	 & 234048	 & (6, 5, 2, 1, 1)	 & 3222112
\tabularnewline[-4pt] 
 (6, 5, 2, 2, 0)	 & 234048	 & (6, 5, 3, 1, 0)	 & 30624
\tabularnewline[-4pt] 
 (6, 6, 1, 1, 1)	 & 112	 & (7, 2, 2, 2, 2)	 & 20299992
\tabularnewline[-4pt] 
 (7, 3, 2, 2, 1)	 & 3222112	 & (7, 3, 3, 1, 1)	 & 434688
\tabularnewline[-4pt] 
 (7, 3, 3, 2, 0)	 & 30624	 & (7, 4, 2, 1, 1)	 & 45408
\tabularnewline[-4pt] 
 (7, 4, 2, 2, 0)	 & 2800	 & (7, 4, 3, 1, 0)	 & 112
\tablepostambleInstantonNumbers \vskip-18pt 

\tablepreambleInstantonNumbers{16}
 (4, 3, 3, 3, 3)	 & 134508124418928	 & (4, 4, 3, 3, 2)	 & 42411173392368
\tabularnewline[-4pt] 
 (4, 4, 4, 2, 2)	 & 13138629854976	 & (4, 4, 4, 3, 1)	 & 3114669545280
\tabularnewline[-4pt] 
 (4, 4, 4, 4, 0)	 & 14386855920	 & (5, 3, 3, 3, 2)	 & 13834674726336
\tabularnewline[-4pt] 
 (5, 4, 3, 2, 2)	 & 4183230238656	 & (5, 4, 3, 3, 1)	 & 980247769056
\tabularnewline[-4pt] 
 (5, 4, 4, 2, 1)	 & 285207114048	 & (5, 4, 4, 3, 0)	 & 4392333792
\tabularnewline[-4pt] 
 (5, 5, 2, 2, 2)	 & 366406656528	 & (5, 5, 3, 2, 1)	 & 83099778720
\tabularnewline[-4pt] 
 (5, 5, 3, 3, 0)	 & 1272585120	 & (5, 5, 4, 1, 1)	 & 4826161680
\tabularnewline[-4pt] 
 (5, 5, 4, 2, 0)	 & 347078520	 & (5, 5, 5, 1, 0)	 & 4326048
\tabularnewline[-4pt] 
 (6, 3, 3, 2, 2)	 & 377080188864	 & (6, 3, 3, 3, 1)	 & 85495746528
\tabularnewline[-4pt] 
 (6, 4, 2, 2, 2)	 & 103492041648	 & (6, 4, 3, 2, 1)	 & 22951602432
\tabularnewline[-4pt] 
 (6, 4, 3, 3, 0)	 & 347078520	 & (6, 4, 4, 1, 1)	 & 1218252960
\tabularnewline[-4pt] 
 (6, 4, 4, 2, 0)	 & 88177920	 & (6, 5, 2, 2, 1)	 & 1426637712
\tabularnewline[-4pt] 
 (6, 5, 3, 1, 1)	 & 282674592	 & (6, 5, 3, 2, 0)	 & 20578560
\tabularnewline[-4pt] 
 (6, 5, 4, 1, 0)	 & 795936	 & (6, 5, 5, 0, 0)	 & 24
\tabularnewline[-4pt] 
 (6, 6, 2, 1, 1)	 & 1691856	 & (6, 6, 2, 2, 0)	 & 122352
\tabularnewline[-4pt] 
 (6, 6, 3, 1, 0)	 & 14496	 & (7, 3, 2, 2, 2)	 & 1599622824
\tabularnewline[-4pt] 
 (7, 3, 3, 2, 1)	 & 316997280	 & (7, 3, 3, 3, 0)	 & 4326048
\tabularnewline[-4pt] 
 (7, 4, 2, 2, 1)	 & 63576576	 & (7, 4, 3, 1, 1)	 & 10883712
\tabularnewline[-4pt] 
 (7, 4, 3, 2, 0)	 & 795936	 & (7, 4, 4, 1, 0)	 & 14496
\tabularnewline[-4pt] 
 (7, 5, 2, 1, 1)	 & 212880	 & (7, 5, 2, 2, 0)	 & 14496
\tabularnewline[-4pt] 
 (7, 5, 3, 1, 0)	 & 1104	 & (8, 2, 2, 2, 2)	 & 122352
\tabularnewline[-4pt] 
 (8, 3, 2, 2, 1)	 & 14496	 & (8, 3, 3, 1, 1)	 & 1104
\tabularnewline[-4pt] 
 (8, 3, 3, 2, 0)	 & 24	 & (8, 4, 2, 1, 1)	 & 24
\tablepostambleInstantonNumbers \vskip-18pt 

\tablepreambleInstantonNumbers{17}
 (4, 4, 3, 3, 3)	 & 1112487680575968	 & (4, 4, 4, 3, 2)	 & 363393804317664
\tabularnewline[-4pt] 
 (4, 4, 4, 4, 1)	 & 28258960027296	 & (5, 3, 3, 3, 3)	 & 389973010495488
\tabularnewline[-4pt] 
 (5, 4, 3, 3, 2)	 & 125365423769760	 & (5, 4, 4, 2, 2)	 & 39692266181304
\tabularnewline[-4pt] 
 (5, 4, 4, 3, 1)	 & 9502910875584	 & (5, 4, 4, 4, 0)	 & 45007048752
\tabularnewline[-4pt] 
 (5, 5, 3, 2, 2)	 & 13073262151968	 & (5, 5, 3, 3, 1)	 & 3100342138368
\tabularnewline[-4pt] 
 (5, 5, 4, 2, 1)	 & 931163905728	 & (5, 5, 4, 3, 0)	 & 14386869840
\tabularnewline[-4pt] 
 (5, 5, 5, 1, 1)	 & 17798444544	 & (5, 5, 5, 2, 0)	 & 1272585120
\tabularnewline[-4pt] 
 (6, 3, 3, 3, 2)	 & 13834674726336	 & (6, 4, 3, 2, 2)	 & 4183230238656
\tabularnewline[-4pt] 
 (6, 4, 3, 3, 1)	 & 980247769056	 & (6, 4, 4, 2, 1)	 & 285207114048
\tabularnewline[-4pt] 
 (6, 4, 4, 3, 0)	 & 4392333792	 & (6, 5, 2, 2, 2)	 & 366406656528
\tabularnewline[-4pt] 
 (6, 5, 3, 2, 1)	 & 83099778720	 & (6, 5, 3, 3, 0)	 & 1272585120
\tabularnewline[-4pt] 
 (6, 5, 4, 1, 1)	 & 4826161680	 & (6, 5, 4, 2, 0)	 & 347078520
\tabularnewline[-4pt] 
 (6, 5, 5, 1, 0)	 & 4326048	 & (6, 6, 2, 2, 1)	 & 1426637712
\tabularnewline[-4pt] 
 (6, 6, 3, 1, 1)	 & 282674592	 & (6, 6, 3, 2, 0)	 & 20578560
\tabularnewline[-4pt] 
 (6, 6, 4, 1, 0)	 & 795936	 & (6, 6, 5, 0, 0)	 & 24
\tabularnewline[-4pt] 
 (7, 3, 3, 2, 2)	 & 105371446464	 & (7, 3, 3, 3, 1)	 & 23351460864
\tabularnewline[-4pt] 
 (7, 4, 2, 2, 2)	 & 27607031136	 & (7, 4, 3, 2, 1)	 & 5950086192
\tabularnewline[-4pt] 
 (7, 4, 3, 3, 0)	 & 88179456	 & (7, 4, 4, 1, 1)	 & 282674592
\tabularnewline[-4pt] 
 (7, 4, 4, 2, 0)	 & 20578560	 & (7, 5, 2, 2, 1)	 & 316997280
\tabularnewline[-4pt] 
 (7, 5, 3, 1, 1)	 & 59097600	 & (7, 5, 3, 2, 0)	 & 4326048
\tabularnewline[-4pt] 
 (7, 5, 4, 1, 0)	 & 122448	 & (7, 6, 2, 1, 1)	 & 212880
\tabularnewline[-4pt] 
 (7, 6, 2, 2, 0)	 & 14496	 & (7, 6, 3, 1, 0)	 & 1104
\tabularnewline[-4pt] 
 (8, 3, 2, 2, 2)	 & 63576576	 & (8, 3, 3, 2, 1)	 & 10883712
\tabularnewline[-4pt] 
 (8, 3, 3, 3, 0)	 & 122448	 & (8, 4, 2, 2, 1)	 & 1691856
\tabularnewline[-4pt] 
 (8, 4, 3, 1, 1)	 & 212880	 & (8, 4, 3, 2, 0)	 & 14496
\tabularnewline[-4pt] 
 (8, 4, 4, 1, 0)	 & 24	 & (8, 5, 2, 1, 1)	 & 1104
\tabularnewline[-4pt] 
(8, 5, 2, 2, 0)	 & 24	 &  	 &  
\tablepostambleInstantonNumbers \vskip-18pt 

\tablepreambleInstantonNumbers{18}
 (4, 4, 4, 3, 3)	 & 11630106886504344	 & (4, 4, 4, 4, 2)	 & 3920585033699328
\tabularnewline[-4pt] 
 (5, 4, 3, 3, 3)	 & 4272828104425920	 & (5, 4, 4, 3, 2)	 & 1423524718242752
\tabularnewline[-4pt] 
 (5, 4, 4, 4, 1)	 & 114110495895360	 & (5, 5, 3, 3, 2)	 & 507096396665312
\tabularnewline[-4pt] 
 (5, 5, 4, 2, 2)	 & 164605655104880	 & (5, 5, 4, 3, 1)	 & 39821013536096
\tabularnewline[-4pt] 
 (5, 5, 4, 4, 0)	 & 193411225936	 & (5, 5, 5, 2, 1)	 & 4217701870608
\tabularnewline[-4pt] 
 (5, 5, 5, 3, 0)	 & 65215603200	 & (6, 3, 3, 3, 3)	 & 552486590320032
\tabularnewline[-4pt] 
 (6, 4, 3, 3, 2)	 & 178677828494464	 & (6, 4, 4, 2, 2)	 & 56949598227232
\tabularnewline[-4pt] 
 (6, 4, 4, 3, 1)	 & 13674852866304	 & (6, 4, 4, 4, 0)	 & 65215569408
\tabularnewline[-4pt] 
 (6, 5, 3, 2, 2)	 & 18954386538304	 & (6, 5, 3, 3, 1)	 & 4510722900128
\tabularnewline[-4pt] 
 (6, 5, 4, 2, 1)	 & 1367836823744	 & (6, 5, 4, 3, 0)	 & 21143067840
\tabularnewline[-4pt] 
 (6, 5, 5, 1, 1)	 & 27120466144	 & (6, 5, 5, 2, 0)	 & 1935300720
\tabularnewline[-4pt] 
 (6, 6, 2, 2, 2)	 & 551803842816	 & (6, 6, 3, 2, 1)	 & 125948336640
\tabularnewline[-4pt] 
 (6, 6, 3, 3, 0)	 & 1935300720	 & (6, 6, 4, 1, 1)	 & 7510615200
\tabularnewline[-4pt] 
 (6, 6, 4, 2, 0)	 & 539115744	 & (6, 6, 5, 1, 0)	 & 7371792
\tabularnewline[-4pt] 
 (6, 6, 6, 0, 0)	 & 80	 & (7, 3, 3, 3, 2)	 & 6446376071472
\tabularnewline[-4pt] 
 (7, 4, 3, 2, 2)	 & 1912895782008	 & (7, 4, 3, 3, 1)	 & 443961562528
\tabularnewline[-4pt] 
 (7, 4, 4, 2, 1)	 & 126121309632	 & (7, 4, 4, 3, 0)	 & 1935300720
\tabularnewline[-4pt] 
 (7, 5, 2, 2, 2)	 & 158730945984	 & (7, 5, 3, 2, 1)	 & 35487082592
\tabularnewline[-4pt] 
 (7, 5, 3, 3, 0)	 & 539120544	 & (7, 5, 4, 1, 1)	 & 1944767152
\tabularnewline[-4pt] 
 (7, 5, 4, 2, 0)	 & 140436672	 & (7, 5, 5, 1, 0)	 & 1423616
\tabularnewline[-4pt] 
 (7, 6, 2, 2, 1)	 & 529392832	 & (7, 6, 3, 1, 1)	 & 100919904
\tabularnewline[-4pt] 
 (7, 6, 3, 2, 0)	 & 7371792	 & (7, 6, 4, 1, 0)	 & 234048
\tabularnewline[-4pt] 
 (7, 7, 2, 1, 1)	 & 45408	 & (7, 7, 2, 2, 0)	 & 2800
\tabularnewline[-4pt] 
 (7, 7, 3, 1, 0)	 & 112	 & (8, 3, 3, 2, 2)	 & 11032046624
\tabularnewline[-4pt] 
 (8, 3, 3, 3, 1)	 & 2322325968	 & (8, 4, 2, 2, 2)	 & 2624447520
\tabularnewline[-4pt] 
 (8, 4, 3, 2, 1)	 & 529392832	 & (8, 4, 3, 3, 0)	 & 7371792
\tabularnewline[-4pt] 
 (8, 4, 4, 1, 1)	 & 19420400	 & (8, 4, 4, 2, 0)	 & 1423104
\tabularnewline[-4pt] 
 (8, 5, 2, 2, 1)	 & 20299992	 & (8, 5, 3, 1, 1)	 & 3222112
\tabularnewline[-4pt] 
 (8, 5, 3, 2, 0)	 & 234048	 & (8, 5, 4, 1, 0)	 & 2800
\tabularnewline[-4pt] 
 (8, 6, 2, 1, 1)	 & 2800	 & (8, 6, 2, 2, 0)	 & 80
\tabularnewline[-4pt] 
 (9, 3, 2, 2, 2)	 & 234048	 & (9, 3, 3, 2, 1)	 & 30624
\tabularnewline[-4pt] 
 (9, 3, 3, 3, 0)	 & 112	 & (9, 4, 2, 2, 1)	 & 2800
\tabularnewline[-4pt] 
(9, 4, 3, 1, 1)	 & 112	 &  	 &  
\tablepostambleInstantonNumbers \vskip-18pt 

\tablepreambleInstantonNumbers{19}
 (4, 4, 4, 4, 3)	 & 149583407202367176	 & (5, 4, 4, 3, 3)	 & 57309129620711136
\tabularnewline[-4pt] 
 (5, 4, 4, 4, 2)	 & 19680157760407104	 & (5, 5, 3, 3, 3)	 & 21671962905320448
\tabularnewline[-4pt] 
 (5, 5, 4, 3, 2)	 & 7371081117191712	 & (5, 5, 4, 4, 1)	 & 609209937409968
\tabularnewline[-4pt] 
 (5, 5, 5, 2, 2)	 & 905275754212800	 & (5, 5, 5, 3, 1)	 & 221145135246336
\tabularnewline[-4pt] 
 (5, 5, 5, 4, 0)	 & 1096632180480	 & (6, 4, 3, 3, 3)	 & 8236673292611808
\tabularnewline[-4pt] 
 (6, 4, 4, 3, 2)	 & 2768640614245200	 & (6, 4, 4, 4, 1)	 & 224917616990784
\tabularnewline[-4pt] 
 (6, 5, 3, 3, 2)	 & 1000740719949936	 & (6, 5, 4, 2, 2)	 & 328447354833120
\tabularnewline[-4pt] 
 (6, 5, 4, 3, 1)	 & 79804026346992	 & (6, 5, 4, 4, 0)	 & 391409808576
\tabularnewline[-4pt] 
 (6, 5, 5, 2, 1)	 & 8748592415904	 & (6, 5, 5, 3, 0)	 & 135171775392
\tabularnewline[-4pt] 
 (6, 6, 3, 2, 2)	 & 39360165257928	 & (6, 6, 3, 3, 1)	 & 9425697295296
\tabularnewline[-4pt] 
 (6, 6, 4, 2, 1)	 & 2910089695872	 & (6, 6, 4, 3, 0)	 & 45007048752
\tabularnewline[-4pt] 
 (6, 6, 5, 1, 1)	 & 61773182400	 & (6, 6, 5, 2, 0)	 & 4392333792
\tabularnewline[-4pt] 
 (6, 6, 6, 1, 0)	 & 20578560	 & (7, 3, 3, 3, 3)	 & 389973010495488
\tabularnewline[-4pt] 
 (7, 4, 3, 3, 2)	 & 125365423769760	 & (7, 4, 4, 2, 2)	 & 39692266181304
\tabularnewline[-4pt] 
 (7, 4, 4, 3, 1)	 & 9502910875584	 & (7, 4, 4, 4, 0)	 & 45007048752
\tabularnewline[-4pt] 
 (7, 5, 3, 2, 2)	 & 13073262151968	 & (7, 5, 3, 3, 1)	 & 3100342138368
\tabularnewline[-4pt] 
 (7, 5, 4, 2, 1)	 & 931163905728	 & (7, 5, 4, 3, 0)	 & 14386869840
\tabularnewline[-4pt] 
 (7, 5, 5, 1, 1)	 & 17798444544	 & (7, 5, 5, 2, 0)	 & 1272585120
\tabularnewline[-4pt] 
 (7, 6, 2, 2, 2)	 & 366406656528	 & (7, 6, 3, 2, 1)	 & 83099778720
\tabularnewline[-4pt] 
 (7, 6, 3, 3, 0)	 & 1272585120	 & (7, 6, 4, 1, 1)	 & 4826161680
\tabularnewline[-4pt] 
 (7, 6, 4, 2, 0)	 & 347078520	 & (7, 6, 5, 1, 0)	 & 4326048
\tabularnewline[-4pt] 
 (7, 6, 6, 0, 0)	 & 24	 & (7, 7, 2, 2, 1)	 & 316997280
\tabularnewline[-4pt] 
 (7, 7, 3, 1, 1)	 & 59097600	 & (7, 7, 3, 2, 0)	 & 4326048
\tabularnewline[-4pt] 
 (7, 7, 4, 1, 0)	 & 122448	 & (8, 3, 3, 3, 2)	 & 1326841710624
\tabularnewline[-4pt] 
 (8, 4, 3, 2, 2)	 & 377080188864	 & (8, 4, 3, 3, 1)	 & 85495746528
\tabularnewline[-4pt] 
 (8, 4, 4, 2, 1)	 & 22951602432	 & (8, 4, 4, 3, 0)	 & 347078520
\tabularnewline[-4pt] 
 (8, 5, 2, 2, 2)	 & 27607031136	 & (8, 5, 3, 2, 1)	 & 5950086192
\tabularnewline[-4pt] 
 (8, 5, 3, 3, 0)	 & 88179456	 & (8, 5, 4, 1, 1)	 & 282674592
\tabularnewline[-4pt] 
 (8, 5, 4, 2, 0)	 & 20578560	 & (8, 5, 5, 1, 0)	 & 122448
\tabularnewline[-4pt] 
 (8, 6, 2, 2, 1)	 & 63576576	 & (8, 6, 3, 1, 1)	 & 10883712
\tabularnewline[-4pt] 
 (8, 6, 3, 2, 0)	 & 795936	 & (8, 6, 4, 1, 0)	 & 14496
\tabularnewline[-4pt] 
 (8, 7, 2, 1, 1)	 & 1104	 & (8, 7, 2, 2, 0)	 & 24
\tabularnewline[-4pt] 
 (9, 3, 3, 2, 2)	 & 316997280	 & (9, 3, 3, 3, 1)	 & 59097600
\tabularnewline[-4pt] 
 (9, 4, 2, 2, 2)	 & 63576576	 & (9, 4, 3, 2, 1)	 & 10883712
\tabularnewline[-4pt] 
 (9, 4, 3, 3, 0)	 & 122448	 & (9, 4, 4, 1, 1)	 & 212880
\tabularnewline[-4pt] 
 (9, 4, 4, 2, 0)	 & 14496	 & (9, 5, 2, 2, 1)	 & 212880
\tabularnewline[-4pt] 
 (9, 5, 3, 1, 1)	 & 19200	 & (9, 5, 3, 2, 0)	 & 1104
\tablepostambleInstantonNumbers \vskip-18pt 

\tablepreambleInstantonNumbers{20}
 (4, 4, 4, 4, 4)	 & 2315758601706011520	 & (5, 4, 4, 4, 3)	 & 920246692052672448
\tabularnewline[-4pt] 
 (5, 5, 4, 3, 3)	 & 362176732991882256	 & (5, 5, 4, 4, 2)	 & 126656377507736616
\tabularnewline[-4pt] 
 (5, 5, 5, 3, 2)	 & 48949713376347552	 & (5, 5, 5, 4, 1)	 & 4162140562025760
\tabularnewline[-4pt] 
 (5, 5, 5, 5, 0)	 & 7888589144400	 & (6, 4, 4, 3, 3)	 & 145074948270672288
\tabularnewline[-4pt] 
 (6, 4, 4, 4, 2)	 & 50310287851264512	 & (6, 5, 3, 3, 3)	 & 55724768553096576
\tabularnewline[-4pt] 
 (6, 5, 4, 3, 2)	 & 19159936729163904	 & (6, 5, 4, 4, 1)	 & 1608297381675072
\tabularnewline[-4pt] 
 (6, 5, 5, 2, 2)	 & 2428815576573408	 & (6, 5, 5, 3, 1)	 & 596073535387056
\tabularnewline[-4pt] 
 (6, 5, 5, 4, 0)	 & 2981800050480	 & (6, 6, 3, 3, 2)	 & 2713101057421728
\tabularnewline[-4pt] 
 (6, 6, 4, 2, 2)	 & 903893653068672	 & (6, 6, 4, 3, 1)	 & 220840621188096
\tabularnewline[-4pt] 
 (6, 6, 4, 4, 0)	 & 1096632086784	 & (6, 6, 5, 2, 1)	 & 25377635878296
\tabularnewline[-4pt] 
 (6, 6, 5, 3, 0)	 & 391409808576	 & (6, 6, 6, 1, 1)	 & 203336907216
\tabularnewline[-4pt] 
 (6, 6, 6, 2, 0)	 & 14386855920	 & (7, 4, 3, 3, 3)	 & 8236673292611808
\tabularnewline[-4pt] 
 (7, 4, 4, 3, 2)	 & 2768640614245200	 & (7, 4, 4, 4, 1)	 & 224917616990784
\tabularnewline[-4pt] 
 (7, 5, 3, 3, 2)	 & 1000740719949936	 & (7, 5, 4, 2, 2)	 & 328447354833120
\tabularnewline[-4pt] 
 (7, 5, 4, 3, 1)	 & 79804026346992	 & (7, 5, 4, 4, 0)	 & 391409808576
\tabularnewline[-4pt] 
 (7, 5, 5, 2, 1)	 & 8748592415904	 & (7, 5, 5, 3, 0)	 & 135171775392
\tabularnewline[-4pt] 
 (7, 6, 3, 2, 2)	 & 39360165257928	 & (7, 6, 3, 3, 1)	 & 9425697295296
\tabularnewline[-4pt] 
 (7, 6, 4, 2, 1)	 & 2910089695872	 & (7, 6, 4, 3, 0)	 & 45007048752
\tabularnewline[-4pt] 
 (7, 6, 5, 1, 1)	 & 61773182400	 & (7, 6, 5, 2, 0)	 & 4392333792
\tabularnewline[-4pt] 
 (7, 6, 6, 1, 0)	 & 20578560	 & (7, 7, 2, 2, 2)	 & 366406656528
\tabularnewline[-4pt] 
 (7, 7, 3, 2, 1)	 & 83099778720	 & (7, 7, 3, 3, 0)	 & 1272585120
\tabularnewline[-4pt] 
 (7, 7, 4, 1, 1)	 & 4826161680	 & (7, 7, 4, 2, 0)	 & 347078520
\tabularnewline[-4pt] 
 (7, 7, 5, 1, 0)	 & 4326048	 & (7, 7, 6, 0, 0)	 & 24
\tabularnewline[-4pt] 
 (8, 3, 3, 3, 3)	 & 134508124418928	 & (8, 4, 3, 3, 2)	 & 42411173392368
\tabularnewline[-4pt] 
 (8, 4, 4, 2, 2)	 & 13138629854976	 & (8, 4, 4, 3, 1)	 & 3114669545280
\tabularnewline[-4pt] 
 (8, 4, 4, 4, 0)	 & 14386855920	 & (8, 5, 3, 2, 2)	 & 4183230238656
\tabularnewline[-4pt] 
 (8, 5, 3, 3, 1)	 & 980247769056	 & (8, 5, 4, 2, 1)	 & 285207114048
\tabularnewline[-4pt] 
 (8, 5, 4, 3, 0)	 & 4392333792	 & (8, 5, 5, 1, 1)	 & 4826161680
\tabularnewline[-4pt] 
 (8, 5, 5, 2, 0)	 & 347078520	 & (8, 6, 2, 2, 2)	 & 103492041648
\tabularnewline[-4pt] 
 (8, 6, 3, 2, 1)	 & 22951602432	 & (8, 6, 3, 3, 0)	 & 347078520
\tabularnewline[-4pt] 
 (8, 6, 4, 1, 1)	 & 1218252960	 & (8, 6, 4, 2, 0)	 & 88177920
\tabularnewline[-4pt] 
 (8, 6, 5, 1, 0)	 & 795936	 & (8, 7, 2, 2, 1)	 & 63576576
\tabularnewline[-4pt] 
 (8, 7, 3, 1, 1)	 & 10883712	 & (8, 7, 3, 2, 0)	 & 795936
\tabularnewline[-4pt] 
 (8, 7, 4, 1, 0)	 & 14496	 & (8, 8, 2, 1, 1)	 & 24
\tabularnewline[-4pt] 
 (9, 3, 3, 3, 2)	 & 105371446464	 & (9, 4, 3, 2, 2)	 & 27607031136
\tabularnewline[-4pt] 
 (9, 4, 3, 3, 1)	 & 5950086192	 & (9, 4, 4, 2, 1)	 & 1426637712
\tabularnewline[-4pt] 
 (9, 4, 4, 3, 0)	 & 20578560	 & (9, 5, 2, 2, 2)	 & 1599622824
\tabularnewline[-4pt] 
 (9, 5, 3, 2, 1)	 & 316997280	 & (9, 5, 3, 3, 0)	 & 4326048
\tabularnewline[-4pt] 
 (9, 5, 4, 1, 1)	 & 10883712	 & (9, 5, 4, 2, 0)	 & 795936
\tabularnewline[-4pt] 
 (9, 5, 5, 1, 0)	 & 1104	 & (9, 6, 2, 2, 1)	 & 1691856
\tabularnewline[-4pt] 
 (9, 6, 3, 1, 1)	 & 212880	 & (9, 6, 3, 2, 0)	 & 14496
\tabularnewline[-4pt] 
 (9, 6, 4, 1, 0)	 & 24	 & (10, 3, 3, 2, 2)	 & 795936
\tabularnewline[-4pt] 
 (10, 3, 3, 3, 1)	 & 122448	 & (10, 4, 2, 2, 2)	 & 122352
\tabularnewline[-4pt] 
 (10, 4, 3, 2, 1)	 & 14496	 & (10, 4, 3, 3, 0)	 & 24
\tabularnewline[-4pt] 
 (10, 4, 4, 1, 1)	 & 24	 & (10, 5, 2, 2, 1)	 & 24
\tablepostambleInstantonNumbers \vskip-18pt 

\tablepreambleInstantonNumbers{21}
 (5, 4, 4, 4, 4)	 & 17389206433621316832	 & (5, 5, 4, 4, 3)	 & 7079567101109436512
\tabularnewline[-4pt] 
 (5, 5, 5, 3, 3)	 & 2860072289627444736	 & (5, 5, 5, 4, 2)	 & 1017289744237857120
\tabularnewline[-4pt] 
 (5, 5, 5, 5, 1)	 & 35306571598392576	 & (6, 4, 4, 4, 3)	 & 2968386268852263168
\tabularnewline[-4pt] 
 (6, 5, 4, 3, 3)	 & 1187054464752608224	 & (6, 5, 4, 4, 2)	 & 419478239436537264
\tabularnewline[-4pt] 
 (6, 5, 5, 3, 2)	 & 165119843412344816	 & (6, 5, 5, 4, 1)	 & 14258867760974432
\tabularnewline[-4pt] 
 (6, 5, 5, 5, 0)	 & 27765085214112	 & (6, 6, 3, 3, 3)	 & 190193228131870512
\tabularnewline[-4pt] 
 (6, 6, 4, 3, 2)	 & 66233922634330080	 & (6, 6, 4, 4, 1)	 & 5658979212554128
\tabularnewline[-4pt] 
 (6, 6, 5, 2, 2)	 & 8718347106041576	 & (6, 6, 5, 3, 1)	 & 2150266975191936
\tabularnewline[-4pt] 
 (6, 6, 5, 4, 0)	 & 10848408360480	 & (6, 6, 6, 2, 1)	 & 99894446151552
\tabularnewline[-4pt] 
 (6, 6, 6, 3, 0)	 & 1535514818112	 & (7, 4, 4, 3, 3)	 & 196866216448867200
\tabularnewline[-4pt] 
 (7, 4, 4, 4, 2)	 & 68481669752665152	 & (7, 5, 3, 3, 3)	 & 75992812385562624
\tabularnewline[-4pt] 
 (7, 5, 4, 3, 2)	 & 26217346711258048	 & (7, 5, 4, 4, 1)	 & 2211223893638272
\tabularnewline[-4pt] 
 (7, 5, 5, 2, 2)	 & 3356453655323136	 & (7, 5, 5, 3, 1)	 & 824874647838720
\tabularnewline[-4pt] 
 (7, 5, 5, 4, 0)	 & 4136092936448	 & (7, 6, 3, 3, 2)	 & 3761948244770304
\tabularnewline[-4pt] 
 (7, 6, 4, 2, 2)	 & 1259132047619264	 & (7, 6, 4, 3, 1)	 & 308134225628128
\tabularnewline[-4pt] 
 (7, 6, 4, 4, 0)	 & 1535514818112	 & (7, 6, 5, 2, 1)	 & 35929933424832
\tabularnewline[-4pt] 
 (7, 6, 5, 3, 0)	 & 553728279360	 & (7, 6, 6, 1, 1)	 & 299302640864
\tabularnewline[-4pt] 
 (7, 6, 6, 2, 0)	 & 21143067840	 & (7, 7, 3, 2, 2)	 & 56389985840000
\tabularnewline[-4pt] 
 (7, 7, 3, 3, 1)	 & 13542066341888	 & (7, 7, 4, 2, 1)	 & 4216297529824
\tabularnewline[-4pt] 
 (7, 7, 4, 3, 0)	 & 65215603200	 & (7, 7, 5, 1, 1)	 & 92396257280
\tabularnewline[-4pt] 
 (7, 7, 5, 2, 0)	 & 6558863360	 & (7, 7, 6, 1, 0)	 & 33777312
\tabularnewline[-4pt] 
 (7, 7, 7, 0, 0)	 & 112	 & (8, 4, 3, 3, 3)	 & 4272828104425920
\tabularnewline[-4pt] 
 (8, 4, 4, 3, 2)	 & 1423524718242752	 & (8, 4, 4, 4, 1)	 & 114110495895360
\tabularnewline[-4pt] 
 (8, 5, 3, 3, 2)	 & 507096396665312	 & (8, 5, 4, 2, 2)	 & 164605655104880
\tabularnewline[-4pt] 
 (8, 5, 4, 3, 1)	 & 39821013536096	 & (8, 5, 4, 4, 0)	 & 193411225936
\tabularnewline[-4pt] 
 (8, 5, 5, 2, 1)	 & 4217701870608	 & (8, 5, 5, 3, 0)	 & 65215603200
\tabularnewline[-4pt] 
 (8, 6, 3, 2, 2)	 & 18954386538304	 & (8, 6, 3, 3, 1)	 & 4510722900128
\tabularnewline[-4pt] 
 (8, 6, 4, 2, 1)	 & 1367836823744	 & (8, 6, 4, 3, 0)	 & 21143067840
\tabularnewline[-4pt] 
 (8, 6, 5, 1, 1)	 & 27120466144	 & (8, 6, 5, 2, 0)	 & 1935300720
\tabularnewline[-4pt] 
 (8, 6, 6, 1, 0)	 & 7371792	 & (8, 7, 2, 2, 2)	 & 158730945984
\tabularnewline[-4pt] 
 (8, 7, 3, 2, 1)	 & 35487082592	 & (8, 7, 3, 3, 0)	 & 539120544
\tabularnewline[-4pt] 
 (8, 7, 4, 1, 1)	 & 1944767152	 & (8, 7, 4, 2, 0)	 & 140436672
\tabularnewline[-4pt] 
 (8, 7, 5, 1, 0)	 & 1423616	 & (8, 8, 2, 2, 1)	 & 20299992
\tabularnewline[-4pt] 
 (8, 8, 3, 1, 1)	 & 3222112	 & (8, 8, 3, 2, 0)	 & 234048
\tabularnewline[-4pt] 
 (8, 8, 4, 1, 0)	 & 2800	 & (9, 3, 3, 3, 3)	 & 21228933784320
\tabularnewline[-4pt] 
 (9, 4, 3, 3, 2)	 & 6446376071472	 & (9, 4, 4, 2, 2)	 & 1912895782008
\tabularnewline[-4pt] 
 (9, 4, 4, 3, 1)	 & 443961562528	 & (9, 4, 4, 4, 0)	 & 1935300720
\tabularnewline[-4pt] 
 (9, 5, 3, 2, 2)	 & 570360079168	 & (9, 5, 3, 3, 1)	 & 130194945024
\tabularnewline[-4pt] 
 (9, 5, 4, 2, 1)	 & 35487082592	 & (9, 5, 4, 3, 0)	 & 539120544
\tabularnewline[-4pt] 
 (9, 5, 5, 1, 1)	 & 464696832	 & (9, 5, 5, 2, 0)	 & 33777312
\tabularnewline[-4pt] 
 (9, 6, 2, 2, 2)	 & 11032046624	 & (9, 6, 3, 2, 1)	 & 2322325968
\tabularnewline[-4pt] 
 (9, 6, 3, 3, 0)	 & 33777312	 & (9, 6, 4, 1, 1)	 & 100919904
\tabularnewline[-4pt] 
 (9, 6, 4, 2, 0)	 & 7371792	 & (9, 6, 5, 1, 0)	 & 30624
\tabularnewline[-4pt] 
 (9, 7, 2, 2, 1)	 & 3222112	 & (9, 7, 3, 1, 1)	 & 434688
\tabularnewline[-4pt] 
 (9, 7, 3, 2, 0)	 & 30624	 & (9, 7, 4, 1, 0)	 & 112
\tabularnewline[-4pt] 
 (10, 3, 3, 3, 2)	 & 2322325968	 & (10, 4, 3, 2, 2)	 & 529392832
\tabularnewline[-4pt] 
 (10, 4, 3, 3, 1)	 & 100919904	 & (10, 4, 4, 2, 1)	 & 19420400
\tabularnewline[-4pt] 
 (10, 4, 4, 3, 0)	 & 234048	 & (10, 5, 2, 2, 2)	 & 20299992
\tabularnewline[-4pt] 
 (10, 5, 3, 2, 1)	 & 3222112	 & (10, 5, 3, 3, 0)	 & 30624
\tabularnewline[-4pt] 
 (10, 5, 4, 1, 1)	 & 45408	 & (10, 5, 4, 2, 0)	 & 2800
\tabularnewline[-4pt] 
 (10, 6, 2, 2, 1)	 & 2800	 & (10, 6, 3, 1, 1)	 & 112
\tablepostambleInstantonNumbers \vskip-18pt 

\tablepreambleInstantonNumbers{22}
 (5, 5, 4, 4, 4)	 & 159832960277398698312	 & (5, 5, 5, 4, 3)	 & 66573482065327669440
\tabularnewline[-4pt] 
 (5, 5, 5, 5, 2)	 & 9952370045915290464	 & (6, 4, 4, 4, 4)	 & 69702170473826178048
\tabularnewline[-4pt] 
 (6, 5, 4, 4, 3)	 & 28814753795787304128	 & (6, 5, 5, 3, 3)	 & 11833136668383611040
\tabularnewline[-4pt] 
 (6, 5, 5, 4, 2)	 & 4252005327651223776	 & (6, 5, 5, 5, 1)	 & 152435152838866176
\tabularnewline[-4pt] 
 (6, 6, 4, 3, 3)	 & 5023740750844977792	 & (6, 6, 4, 4, 2)	 & 1795314514514344416
\tabularnewline[-4pt] 
 (6, 6, 5, 3, 2)	 & 721163569257189312	 & (6, 6, 5, 4, 1)	 & 63276065657309280
\tabularnewline[-4pt] 
 (6, 6, 5, 5, 0)	 & 126532108859856	 & (6, 6, 6, 2, 2)	 & 40767562975883520
\tabularnewline[-4pt] 
 (6, 6, 6, 3, 1)	 & 10102374952223232	 & (6, 6, 6, 4, 0)	 & 51294956593632
\tabularnewline[-4pt] 
 (7, 4, 4, 4, 3)	 & 5273427759409817952	 & (7, 5, 4, 3, 3)	 & 2124595552827372432
\tabularnewline[-4pt] 
 (7, 5, 4, 4, 2)	 & 754387255278771840	 & (7, 5, 5, 3, 2)	 & 299477728365291600
\tabularnewline[-4pt] 
 (7, 5, 5, 4, 1)	 & 26040136828870752	 & (7, 5, 5, 5, 0)	 & 51294957112992
\tabularnewline[-4pt] 
 (7, 6, 3, 3, 3)	 & 346896207708697296	 & (7, 6, 4, 3, 2)	 & 121505012171479176
\tabularnewline[-4pt] 
 (7, 6, 4, 4, 1)	 & 10462960782869952	 & (7, 6, 5, 2, 2)	 & 16270300857476160
\tabularnewline[-4pt] 
 (7, 6, 5, 3, 1)	 & 4021264698687264	 & (7, 6, 5, 4, 0)	 & 20350993239840
\tabularnewline[-4pt] 
 (7, 6, 6, 2, 1)	 & 194378107421760	 & (7, 6, 6, 3, 0)	 & 2981800050480
\tabularnewline[-4pt] 
 (7, 7, 3, 3, 2)	 & 7173870919736064	 & (7, 7, 4, 2, 2)	 & 2422178398686816
\tabularnewline[-4pt] 
 (7, 7, 4, 3, 1)	 & 594508678788528	 & (7, 7, 4, 4, 0)	 & 2981800050480
\tabularnewline[-4pt] 
 (7, 7, 5, 2, 1)	 & 71274491245200	 & (7, 7, 5, 3, 0)	 & 1096632180480
\tabularnewline[-4pt] 
 (7, 7, 6, 1, 1)	 & 639016897824	 & (7, 7, 6, 2, 0)	 & 45007048752
\tabularnewline[-4pt] 
 (7, 7, 7, 1, 0)	 & 88179456	 & (8, 4, 4, 3, 3)	 & 145074948270672288
\tabularnewline[-4pt] 
 (8, 4, 4, 4, 2)	 & 50310287851264512	 & (8, 5, 3, 3, 3)	 & 55724768553096576
\tabularnewline[-4pt] 
 (8, 5, 4, 3, 2)	 & 19159936729163904	 & (8, 5, 4, 4, 1)	 & 1608297381675072
\tabularnewline[-4pt] 
 (8, 5, 5, 2, 2)	 & 2428815576573408	 & (8, 5, 5, 3, 1)	 & 596073535387056
\tabularnewline[-4pt] 
 (8, 5, 5, 4, 0)	 & 2981800050480	 & (8, 6, 3, 3, 2)	 & 2713101057421728
\tabularnewline[-4pt] 
 (8, 6, 4, 2, 2)	 & 903893653068672	 & (8, 6, 4, 3, 1)	 & 220840621188096
\tabularnewline[-4pt] 
 (8, 6, 4, 4, 0)	 & 1096632086784	 & (8, 6, 5, 2, 1)	 & 25377635878296
\tabularnewline[-4pt] 
 (8, 6, 5, 3, 0)	 & 391409808576	 & (8, 6, 6, 1, 1)	 & 203336907216
\tabularnewline[-4pt] 
 (8, 6, 6, 2, 0)	 & 14386855920	 & (8, 7, 3, 2, 2)	 & 39360165257928
\tabularnewline[-4pt] 
 (8, 7, 3, 3, 1)	 & 9425697295296	 & (8, 7, 4, 2, 1)	 & 2910089695872
\tabularnewline[-4pt] 
 (8, 7, 4, 3, 0)	 & 45007048752	 & (8, 7, 5, 1, 1)	 & 61773182400
\tabularnewline[-4pt] 
 (8, 7, 5, 2, 0)	 & 4392333792	 & (8, 7, 6, 1, 0)	 & 20578560
\tabularnewline[-4pt] 
 (8, 7, 7, 0, 0)	 & 24	 & (8, 8, 2, 2, 2)	 & 103492041648
\tabularnewline[-4pt] 
 (8, 8, 3, 2, 1)	 & 22951602432	 & (8, 8, 3, 3, 0)	 & 347078520
\tabularnewline[-4pt] 
 (8, 8, 4, 1, 1)	 & 1218252960	 & (8, 8, 4, 2, 0)	 & 88177920
\tabularnewline[-4pt] 
 (8, 8, 5, 1, 0)	 & 795936	 & (9, 4, 3, 3, 3)	 & 1112487680575968
\tabularnewline[-4pt] 
 (9, 4, 4, 3, 2)	 & 363393804317664	 & (9, 4, 4, 4, 1)	 & 28258960027296
\tabularnewline[-4pt] 
 (9, 5, 3, 3, 2)	 & 125365423769760	 & (9, 5, 4, 2, 2)	 & 39692266181304
\tabularnewline[-4pt] 
 (9, 5, 4, 3, 1)	 & 9502910875584	 & (9, 5, 4, 4, 0)	 & 45007048752
\tabularnewline[-4pt] 
 (9, 5, 5, 2, 1)	 & 931163905728	 & (9, 5, 5, 3, 0)	 & 14386869840
\tabularnewline[-4pt] 
 (9, 6, 3, 2, 2)	 & 4183230238656	 & (9, 6, 3, 3, 1)	 & 980247769056
\tabularnewline[-4pt] 
 (9, 6, 4, 2, 1)	 & 285207114048	 & (9, 6, 4, 3, 0)	 & 4392333792
\tabularnewline[-4pt] 
 (9, 6, 5, 1, 1)	 & 4826161680	 & (9, 6, 5, 2, 0)	 & 347078520
\tabularnewline[-4pt] 
 (9, 6, 6, 1, 0)	 & 795936	 & (9, 7, 2, 2, 2)	 & 27607031136
\tabularnewline[-4pt] 
 (9, 7, 3, 2, 1)	 & 5950086192	 & (9, 7, 3, 3, 0)	 & 88179456
\tabularnewline[-4pt] 
 (9, 7, 4, 1, 1)	 & 282674592	 & (9, 7, 4, 2, 0)	 & 20578560
\tabularnewline[-4pt] 
 (9, 7, 5, 1, 0)	 & 122448	 & (9, 8, 2, 2, 1)	 & 1691856
\tabularnewline[-4pt] 
 (9, 8, 3, 1, 1)	 & 212880	 & (9, 8, 3, 2, 0)	 & 14496
\tabularnewline[-4pt] 
 (9, 8, 4, 1, 0)	 & 24	 & (10, 3, 3, 3, 3)	 & 1326841710624
\tabularnewline[-4pt] 
 (10, 4, 3, 3, 2)	 & 377080188864	 & (10, 4, 4, 2, 2)	 & 103492041648
\tabularnewline[-4pt] 
 (10, 4, 4, 3, 1)	 & 22951602432	 & (10, 4, 4, 4, 0)	 & 88177920
\tabularnewline[-4pt] 
 (10, 5, 3, 2, 2)	 & 27607031136	 & (10, 5, 3, 3, 1)	 & 5950086192
\tabularnewline[-4pt] 
 (10, 5, 4, 2, 1)	 & 1426637712	 & (10, 5, 4, 3, 0)	 & 20578560
\tabularnewline[-4pt] 
 (10, 5, 5, 1, 1)	 & 10883712	 & (10, 5, 5, 2, 0)	 & 795936
\tabularnewline[-4pt] 
 (10, 6, 2, 2, 2)	 & 341681280	 & (10, 6, 3, 2, 1)	 & 63576576
\tabularnewline[-4pt] 
 (10, 6, 3, 3, 0)	 & 795936	 & (10, 6, 4, 1, 1)	 & 1691856
\tabularnewline[-4pt] 
 (10, 6, 4, 2, 0)	 & 122352	 & (10, 6, 5, 1, 0)	 & 24
\tabularnewline[-4pt] 
 (10, 7, 2, 2, 1)	 & 14496	 & (10, 7, 3, 1, 1)	 & 1104
\tabularnewline[-4pt] 
 (10, 7, 3, 2, 0)	 & 24	 & (11, 3, 3, 3, 2)	 & 4326048
\tabularnewline[-4pt] 
 (11, 4, 3, 2, 2)	 & 795936	 & (11, 4, 3, 3, 1)	 & 122448
\tabularnewline[-4pt] 
 (11, 4, 4, 2, 1)	 & 14496	 & (11, 4, 4, 3, 0)	 & 24
\tabularnewline[-4pt] 
 (11, 5, 2, 2, 2)	 & 14496	 & (11, 5, 3, 2, 1)	 & 1104
\tablepostambleInstantonNumbers \vskip-18pt 

\tablepreambleInstantonNumbers{23}
 (5, 5, 5, 4, 4)	 & 1763388068567027864736	 & (5, 5, 5, 5, 3)	 & 749974117128947225088
\tabularnewline[-4pt] 
 (6, 5, 4, 4, 4)	 & 790863904443723569376	 & (6, 5, 5, 4, 3)	 & 334370838391810248432
\tabularnewline[-4pt] 
 (6, 5, 5, 5, 2)	 & 51265779665018065536	 & (6, 6, 4, 4, 3)	 & 147828049737997120632
\tabularnewline[-4pt] 
 (6, 6, 5, 3, 3)	 & 61757539943858380704	 & (6, 6, 5, 4, 2)	 & 22420496977021999680
\tabularnewline[-4pt] 
 (6, 6, 5, 5, 1)	 & 830212985215356432	 & (6, 6, 6, 3, 2)	 & 3982038442779651984
\tabularnewline[-4pt] 
 (6, 6, 6, 4, 1)	 & 354725791310991552	 & (6, 6, 6, 5, 0)	 & 725912434085952
\tabularnewline[-4pt] 
 (7, 4, 4, 4, 4)	 & 157391952685989052728	 & (7, 5, 4, 4, 3)	 & 65617907601886711296
\tabularnewline[-4pt] 
 (7, 5, 5, 3, 3)	 & 27190620887571766272	 & (7, 5, 5, 4, 2)	 & 9824161857371476896
\tabularnewline[-4pt] 
 (7, 5, 5, 5, 1)	 & 358330187751266304	 & (7, 6, 4, 3, 3)	 & 11689593863624674656
\tabularnewline[-4pt] 
 (7, 6, 4, 4, 2)	 & 4202606608677077184	 & (7, 6, 5, 3, 2)	 & 1706830027589928192
\tabularnewline[-4pt] 
 (7, 6, 5, 4, 1)	 & 150993571342096992	 & (7, 6, 5, 5, 0)	 & 305922925426848
\tabularnewline[-4pt] 
 (7, 6, 6, 2, 2)	 & 100108346194477248	 & (7, 6, 6, 3, 1)	 & 24863416450991904
\tabularnewline[-4pt] 
 (7, 6, 6, 4, 0)	 & 126532108859856	 & (7, 7, 3, 3, 3)	 & 841539378868429824
\tabularnewline[-4pt] 
 (7, 7, 4, 3, 2)	 & 297115911452589936	 & (7, 7, 4, 4, 1)	 & 25857038420140320
\tabularnewline[-4pt] 
 (7, 7, 5, 2, 2)	 & 40746789567213888	 & (7, 7, 5, 3, 1)	 & 10097809547695104
\tabularnewline[-4pt] 
 (7, 7, 5, 4, 0)	 & 51294957112992	 & (7, 7, 6, 2, 1)	 & 515881389602064
\tabularnewline[-4pt] 
 (7, 7, 6, 3, 0)	 & 7888589144400	 & (7, 7, 7, 1, 1)	 & 1927069671936
\tabularnewline[-4pt] 
 (7, 7, 7, 2, 0)	 & 135171775392	 & (8, 4, 4, 4, 3)	 & 5273427759409817952
\tabularnewline[-4pt] 
 (8, 5, 4, 3, 3)	 & 2124595552827372432	 & (8, 5, 4, 4, 2)	 & 754387255278771840
\tabularnewline[-4pt] 
 (8, 5, 5, 3, 2)	 & 299477728365291600	 & (8, 5, 5, 4, 1)	 & 26040136828870752
\tabularnewline[-4pt] 
 (8, 5, 5, 5, 0)	 & 51294957112992	 & (8, 6, 3, 3, 3)	 & 346896207708697296
\tabularnewline[-4pt] 
 (8, 6, 4, 3, 2)	 & 121505012171479176	 & (8, 6, 4, 4, 1)	 & 10462960782869952
\tabularnewline[-4pt] 
 (8, 6, 5, 2, 2)	 & 16270300857476160	 & (8, 6, 5, 3, 1)	 & 4021264698687264
\tabularnewline[-4pt] 
 (8, 6, 5, 4, 0)	 & 20350993239840	 & (8, 6, 6, 2, 1)	 & 194378107421760
\tabularnewline[-4pt] 
 (8, 6, 6, 3, 0)	 & 2981800050480	 & (8, 7, 3, 3, 2)	 & 7173870919736064
\tabularnewline[-4pt] 
 (8, 7, 4, 2, 2)	 & 2422178398686816	 & (8, 7, 4, 3, 1)	 & 594508678788528
\tabularnewline[-4pt] 
 (8, 7, 4, 4, 0)	 & 2981800050480	 & (8, 7, 5, 2, 1)	 & 71274491245200
\tabularnewline[-4pt] 
 (8, 7, 5, 3, 0)	 & 1096632180480	 & (8, 7, 6, 1, 1)	 & 639016897824
\tabularnewline[-4pt] 
 (8, 7, 6, 2, 0)	 & 45007048752	 & (8, 7, 7, 1, 0)	 & 88179456
\tabularnewline[-4pt] 
 (8, 8, 3, 2, 2)	 & 39360165257928	 & (8, 8, 3, 3, 1)	 & 9425697295296
\tabularnewline[-4pt] 
 (8, 8, 4, 2, 1)	 & 2910089695872	 & (8, 8, 4, 3, 0)	 & 45007048752
\tabularnewline[-4pt] 
 (8, 8, 5, 1, 1)	 & 61773182400	 & (8, 8, 5, 2, 0)	 & 4392333792
\tabularnewline[-4pt] 
 (8, 8, 6, 1, 0)	 & 20578560	 & (8, 8, 7, 0, 0)	 & 24
\tabularnewline[-4pt] 
 (9, 4, 4, 3, 3)	 & 57309129620711136	 & (9, 4, 4, 4, 2)	 & 19680157760407104
\tabularnewline[-4pt] 
 (9, 5, 3, 3, 3)	 & 21671962905320448	 & (9, 5, 4, 3, 2)	 & 7371081117191712
\tabularnewline[-4pt] 
 (9, 5, 4, 4, 1)	 & 609209937409968	 & (9, 5, 5, 2, 2)	 & 905275754212800
\tabularnewline[-4pt] 
 (9, 5, 5, 3, 1)	 & 221145135246336	 & (9, 5, 5, 4, 0)	 & 1096632180480
\tabularnewline[-4pt] 
 (9, 6, 3, 3, 2)	 & 1000740719949936	 & (9, 6, 4, 2, 2)	 & 328447354833120
\tabularnewline[-4pt] 
 (9, 6, 4, 3, 1)	 & 79804026346992	 & (9, 6, 4, 4, 0)	 & 391409808576
\tabularnewline[-4pt] 
 (9, 6, 5, 2, 1)	 & 8748592415904	 & (9, 6, 5, 3, 0)	 & 135171775392
\tabularnewline[-4pt] 
 (9, 6, 6, 1, 1)	 & 61773182400	 & (9, 6, 6, 2, 0)	 & 4392333792
\tabularnewline[-4pt] 
 (9, 7, 3, 2, 2)	 & 13073262151968	 & (9, 7, 3, 3, 1)	 & 3100342138368
\tabularnewline[-4pt] 
 (9, 7, 4, 2, 1)	 & 931163905728	 & (9, 7, 4, 3, 0)	 & 14386869840
\tabularnewline[-4pt] 
 (9, 7, 5, 1, 1)	 & 17798444544	 & (9, 7, 5, 2, 0)	 & 1272585120
\tabularnewline[-4pt] 
 (9, 7, 6, 1, 0)	 & 4326048	 & (9, 8, 2, 2, 2)	 & 27607031136
\tabularnewline[-4pt] 
 (9, 8, 3, 2, 1)	 & 5950086192	 & (9, 8, 3, 3, 0)	 & 88179456
\tabularnewline[-4pt] 
 (9, 8, 4, 1, 1)	 & 282674592	 & (9, 8, 4, 2, 0)	 & 20578560
\tabularnewline[-4pt] 
 (9, 8, 5, 1, 0)	 & 122448	 & (9, 9, 2, 2, 1)	 & 212880
\tabularnewline[-4pt] 
 (9, 9, 3, 1, 1)	 & 19200	 & (9, 9, 3, 2, 0)	 & 1104
\tabularnewline[-4pt] 
 (10, 4, 3, 3, 3)	 & 134508124418928	 & (10, 4, 4, 3, 2)	 & 42411173392368
\tabularnewline[-4pt] 
 (10, 4, 4, 4, 1)	 & 3114669545280	 & (10, 5, 3, 3, 2)	 & 13834674726336
\tabularnewline[-4pt] 
 (10, 5, 4, 2, 2)	 & 4183230238656	 & (10, 5, 4, 3, 1)	 & 980247769056
\tabularnewline[-4pt] 
 (10, 5, 4, 4, 0)	 & 4392333792	 & (10, 5, 5, 2, 1)	 & 83099778720
\tabularnewline[-4pt] 
 (10, 5, 5, 3, 0)	 & 1272585120	 & (10, 6, 3, 2, 2)	 & 377080188864
\tabularnewline[-4pt] 
 (10, 6, 3, 3, 1)	 & 85495746528	 & (10, 6, 4, 2, 1)	 & 22951602432
\tabularnewline[-4pt] 
 (10, 6, 4, 3, 0)	 & 347078520	 & (10, 6, 5, 1, 1)	 & 282674592
\tabularnewline[-4pt] 
 (10, 6, 5, 2, 0)	 & 20578560	 & (10, 6, 6, 1, 0)	 & 14496
\tabularnewline[-4pt] 
 (10, 7, 2, 2, 2)	 & 1599622824	 & (10, 7, 3, 2, 1)	 & 316997280
\tabularnewline[-4pt] 
 (10, 7, 3, 3, 0)	 & 4326048	 & (10, 7, 4, 1, 1)	 & 10883712
\tabularnewline[-4pt] 
 (10, 7, 4, 2, 0)	 & 795936	 & (10, 7, 5, 1, 0)	 & 1104
\tabularnewline[-4pt] 
 (10, 8, 2, 2, 1)	 & 14496	 & (10, 8, 3, 1, 1)	 & 1104
\tabularnewline[-4pt] 
 (10, 8, 3, 2, 0)	 & 24	 & (11, 3, 3, 3, 3)	 & 23351460864
\tabularnewline[-4pt] 
 (11, 4, 3, 3, 2)	 & 5950086192	 & (11, 4, 4, 2, 2)	 & 1426637712
\tabularnewline[-4pt] 
 (11, 4, 4, 3, 1)	 & 282674592	 & (11, 4, 4, 4, 0)	 & 795936
\tabularnewline[-4pt] 
 (11, 5, 3, 2, 2)	 & 316997280	 & (11, 5, 3, 3, 1)	 & 59097600
\tabularnewline[-4pt] 
 (11, 5, 4, 2, 1)	 & 10883712	 & (11, 5, 4, 3, 0)	 & 122448
\tabularnewline[-4pt] 
 (11, 5, 5, 1, 1)	 & 19200	 & (11, 5, 5, 2, 0)	 & 1104
\tabularnewline[-4pt] 
 (11, 6, 2, 2, 2)	 & 1691856	 & (11, 6, 3, 2, 1)	 & 212880
\tabularnewline[-4pt] 
 (11, 6, 3, 3, 0)	 & 1104	 & (11, 6, 4, 1, 1)	 & 1104
\tabularnewline[-4pt] 
(11, 6, 4, 2, 0)	 & 24	 &  	 &  
\tablepostambleInstantonNumbers \vskip-18pt 

\tablepreambleInstantonNumbers{24}
 (5, 5, 5, 5, 4)	 & 22958958469178899286112	 & (6, 5, 5, 4, 4)	 & 10565118218002014469248
\tabularnewline[-4pt] 
 (6, 5, 5, 5, 3)	 & 4556958521329222612288	 & (6, 6, 4, 4, 4)	 & 4831190355131709036288
\tabularnewline[-4pt] 
 (6, 6, 5, 4, 3)	 & 2073506802039240412736	 & (6, 6, 5, 5, 2)	 & 325919053252224299168
\tabularnewline[-4pt] 
 (6, 6, 6, 3, 3)	 & 398117530652334602832	 & (6, 6, 6, 4, 2)	 & 145958872012447992432
\tabularnewline[-4pt] 
 (6, 6, 6, 5, 1)	 & 5571858588504821712	 & (6, 6, 6, 6, 0)	 & 5101035241706976
\tabularnewline[-4pt] 
 (7, 5, 4, 4, 4)	 & 2234583126440197477248	 & (7, 5, 5, 4, 3)	 & 953357306203227960000
\tabularnewline[-4pt] 
 (7, 5, 5, 5, 2)	 & 148388879628408287328	 & (7, 6, 4, 4, 3)	 & 426992405813739053760
\tabularnewline[-4pt] 
 (7, 6, 5, 3, 3)	 & 180227194372605425904	 & (7, 6, 5, 4, 2)	 & 65822715717348500960
\tabularnewline[-4pt] 
 (7, 6, 5, 5, 1)	 & 2483294974158049312	 & (7, 6, 6, 3, 2)	 & 12017787465197578008
\tabularnewline[-4pt] 
 (7, 6, 6, 4, 1)	 & 1079565550915867008	 & (7, 6, 6, 5, 0)	 & 2235977596096128
\tabularnewline[-4pt] 
 (7, 7, 4, 3, 3)	 & 35230204567671156080	 & (7, 7, 4, 4, 2)	 & 12756877670900976952
\tabularnewline[-4pt] 
 (7, 7, 5, 3, 2)	 & 5251183397613765152	 & (7, 7, 5, 4, 1)	 & 468967501173905952
\tabularnewline[-4pt] 
 (7, 7, 5, 5, 0)	 & 964399018545152	 & (7, 7, 6, 2, 2)	 & 322136481160659232
\tabularnewline[-4pt] 
 (7, 7, 6, 3, 1)	 & 80203179581197904	 & (7, 7, 6, 4, 0)	 & 408865565088240
\tabularnewline[-4pt] 
 (7, 7, 7, 2, 1)	 & 1823829689450016	 & (7, 7, 7, 3, 0)	 & 27765085214112
\tabularnewline[-4pt] 
 (8, 4, 4, 4, 4)	 & 205889395932163617312	 & (8, 5, 4, 4, 3)	 & 86069413996832124352
\tabularnewline[-4pt] 
 (8, 5, 5, 3, 3)	 & 35767925041629127584	 & (8, 5, 5, 4, 2)	 & 12945658352928829152
\tabularnewline[-4pt] 
 (8, 5, 5, 5, 1)	 & 474753531842519136	 & (8, 6, 4, 3, 3)	 & 15439042985968145360
\tabularnewline[-4pt] 
 (8, 6, 4, 4, 2)	 & 5561099164846002688	 & (8, 6, 5, 3, 2)	 & 2266457718090172160
\tabularnewline[-4pt] 
 (8, 6, 5, 4, 1)	 & 201010148340707968	 & (8, 6, 5, 5, 0)	 & 408865565088240
\tabularnewline[-4pt] 
 (8, 6, 6, 2, 2)	 & 134490191192202528	 & (8, 6, 6, 3, 1)	 & 33425310221788864
\tabularnewline[-4pt] 
 (8, 6, 6, 4, 0)	 & 170193514498560	 & (8, 7, 3, 3, 3)	 & 1126346577851592960
\tabularnewline[-4pt] 
 (8, 7, 4, 3, 2)	 & 398660145480045856	 & (8, 7, 4, 4, 1)	 & 34807080531177792
\tabularnewline[-4pt] 
 (8, 7, 5, 2, 2)	 & 55084399995750144	 & (8, 7, 5, 3, 1)	 & 13661798641742976
\tabularnewline[-4pt] 
 (8, 7, 5, 4, 0)	 & 69468841810240	 & (8, 7, 6, 2, 1)	 & 710229601026304
\tabularnewline[-4pt] 
 (8, 7, 6, 3, 0)	 & 10848408360480	 & (8, 7, 7, 1, 1)	 & 2760956499680
\tabularnewline[-4pt] 
 (8, 7, 7, 2, 0)	 & 193411225936	 & (8, 8, 3, 3, 2)	 & 9867511282052976
\tabularnewline[-4pt] 
 (8, 8, 4, 2, 2)	 & 3345447617629184	 & (8, 8, 4, 3, 1)	 & 822222499846400
\tabularnewline[-4pt] 
 (8, 8, 4, 4, 0)	 & 4136092740352	 & (8, 8, 5, 2, 1)	 & 99885127862240
\tabularnewline[-4pt] 
 (8, 8, 5, 3, 0)	 & 1535514818112	 & (8, 8, 6, 1, 1)	 & 927231904320
\tabularnewline[-4pt] 
 (8, 8, 6, 2, 0)	 & 65215569408	 & (8, 8, 7, 1, 0)	 & 140436672
\tabularnewline[-4pt] 
 (8, 8, 8, 0, 0)	 & 80	 & (9, 4, 4, 4, 3)	 & 2968386268852263168
\tabularnewline[-4pt] 
 (9, 5, 4, 3, 3)	 & 1187054464752608224	 & (9, 5, 4, 4, 2)	 & 419478239436537264
\tabularnewline[-4pt] 
 (9, 5, 5, 3, 2)	 & 165119843412344816	 & (9, 5, 5, 4, 1)	 & 14258867760974432
\tabularnewline[-4pt] 
 (9, 5, 5, 5, 0)	 & 27765085214112	 & (9, 6, 3, 3, 3)	 & 190193228131870512
\tabularnewline[-4pt] 
 (9, 6, 4, 3, 2)	 & 66233922634330080	 & (9, 6, 4, 4, 1)	 & 5658979212554128
\tabularnewline[-4pt] 
 (9, 6, 5, 2, 2)	 & 8718347106041576	 & (9, 6, 5, 3, 1)	 & 2150266975191936
\tabularnewline[-4pt] 
 (9, 6, 5, 4, 0)	 & 10848408360480	 & (9, 6, 6, 2, 1)	 & 99894446151552
\tabularnewline[-4pt] 
 (9, 6, 6, 3, 0)	 & 1535514818112	 & (9, 7, 3, 3, 2)	 & 3761948244770304
\tabularnewline[-4pt] 
 (9, 7, 4, 2, 2)	 & 1259132047619264	 & (9, 7, 4, 3, 1)	 & 308134225628128
\tabularnewline[-4pt] 
 (9, 7, 4, 4, 0)	 & 1535514818112	 & (9, 7, 5, 2, 1)	 & 35929933424832
\tabularnewline[-4pt] 
 (9, 7, 5, 3, 0)	 & 553728279360	 & (9, 7, 6, 1, 1)	 & 299302640864
\tabularnewline[-4pt] 
 (9, 7, 6, 2, 0)	 & 21143067840	 & (9, 7, 7, 1, 0)	 & 33777312
\tabularnewline[-4pt] 
 (9, 8, 3, 2, 2)	 & 18954386538304	 & (9, 8, 3, 3, 1)	 & 4510722900128
\tabularnewline[-4pt] 
 (9, 8, 4, 2, 1)	 & 1367836823744	 & (9, 8, 4, 3, 0)	 & 21143067840
\tabularnewline[-4pt] 
 (9, 8, 5, 1, 1)	 & 27120466144	 & (9, 8, 5, 2, 0)	 & 1935300720
\tabularnewline[-4pt] 
 (9, 8, 6, 1, 0)	 & 7371792	 & (9, 9, 2, 2, 2)	 & 11032046624
\tabularnewline[-4pt] 
 (9, 9, 3, 2, 1)	 & 2322325968	 & (9, 9, 3, 3, 0)	 & 33777312
\tabularnewline[-4pt] 
 (9, 9, 4, 1, 1)	 & 100919904	 & (9, 9, 4, 2, 0)	 & 7371792
\tabularnewline[-4pt] 
 (9, 9, 5, 1, 0)	 & 30624	 & (10, 4, 4, 3, 3)	 & 11630106886504344
\tabularnewline[-4pt] 
 (10, 4, 4, 4, 2)	 & 3920585033699328	 & (10, 5, 3, 3, 3)	 & 4272828104425920
\tabularnewline[-4pt] 
 (10, 5, 4, 3, 2)	 & 1423524718242752	 & (10, 5, 4, 4, 1)	 & 114110495895360
\tabularnewline[-4pt] 
 (10, 5, 5, 2, 2)	 & 164605655104880	 & (10, 5, 5, 3, 1)	 & 39821013536096
\tabularnewline[-4pt] 
 (10, 5, 5, 4, 0)	 & 193411225936	 & (10, 6, 3, 3, 2)	 & 178677828494464
\tabularnewline[-4pt] 
 (10, 6, 4, 2, 2)	 & 56949598227232	 & (10, 6, 4, 3, 1)	 & 13674852866304
\tabularnewline[-4pt] 
 (10, 6, 4, 4, 0)	 & 65215569408	 & (10, 6, 5, 2, 1)	 & 1367836823744
\tabularnewline[-4pt] 
 (10, 6, 5, 3, 0)	 & 21143067840	 & (10, 6, 6, 1, 1)	 & 7510615200
\tabularnewline[-4pt] 
 (10, 6, 6, 2, 0)	 & 539115744	 & (10, 7, 3, 2, 2)	 & 1912895782008
\tabularnewline[-4pt] 
 (10, 7, 3, 3, 1)	 & 443961562528	 & (10, 7, 4, 2, 1)	 & 126121309632
\tabularnewline[-4pt] 
 (10, 7, 4, 3, 0)	 & 1935300720	 & (10, 7, 5, 1, 1)	 & 1944767152
\tabularnewline[-4pt] 
 (10, 7, 5, 2, 0)	 & 140436672	 & (10, 7, 6, 1, 0)	 & 234048
\tabularnewline[-4pt] 
 (10, 8, 2, 2, 2)	 & 2624447520	 & (10, 8, 3, 2, 1)	 & 529392832
\tabularnewline[-4pt] 
 (10, 8, 3, 3, 0)	 & 7371792	 & (10, 8, 4, 1, 1)	 & 19420400
\tabularnewline[-4pt] 
 (10, 8, 4, 2, 0)	 & 1423104	 & (10, 8, 5, 1, 0)	 & 2800
\tabularnewline[-4pt] 
 (10, 9, 2, 2, 1)	 & 2800	 & (10, 9, 3, 1, 1)	 & 112
\tabularnewline[-4pt] 
 (11, 4, 3, 3, 3)	 & 6446376071472	 & (11, 4, 4, 3, 2)	 & 1912895782008
\tabularnewline[-4pt] 
 (11, 4, 4, 4, 1)	 & 126121309632	 & (11, 5, 3, 3, 2)	 & 570360079168
\tabularnewline[-4pt] 
 (11, 5, 4, 2, 2)	 & 158730945984	 & (11, 5, 4, 3, 1)	 & 35487082592
\tabularnewline[-4pt] 
 (11, 5, 4, 4, 0)	 & 140436672	 & (11, 5, 5, 2, 1)	 & 2306418848
\tabularnewline[-4pt] 
 (11, 5, 5, 3, 0)	 & 33777312	 & (11, 6, 3, 2, 2)	 & 11032046624
\tabularnewline[-4pt] 
 (11, 6, 3, 3, 1)	 & 2322325968	 & (11, 6, 4, 2, 1)	 & 529392832
\tabularnewline[-4pt] 
 (11, 6, 4, 3, 0)	 & 7371792	 & (11, 6, 5, 1, 1)	 & 3222112
\tabularnewline[-4pt] 
 (11, 6, 5, 2, 0)	 & 234048	 & (11, 7, 2, 2, 2)	 & 20299992
\tabularnewline[-4pt] 
 (11, 7, 3, 2, 1)	 & 3222112	 & (11, 7, 3, 3, 0)	 & 30624
\tabularnewline[-4pt] 
 (11, 7, 4, 1, 1)	 & 45408	 & (11, 7, 4, 2, 0)	 & 2800
\tabularnewline[-4pt] 
 (12, 3, 3, 3, 3)	 & 33777312	 & (12, 4, 3, 3, 2)	 & 7371792
\tabularnewline[-4pt] 
 (12, 4, 4, 2, 2)	 & 1423104	 & (12, 4, 4, 3, 1)	 & 234048
\tabularnewline[-4pt] 
 (12, 4, 4, 4, 0)	 & 80	 & (12, 5, 3, 2, 2)	 & 234048
\tabularnewline[-4pt] 
 (12, 5, 3, 3, 1)	 & 30624	 & (12, 5, 4, 2, 1)	 & 2800
\tabularnewline[-4pt] 
(12, 6, 2, 2, 2)	 & 80	 &  	 &  
\tablepostambleInstantonNumbers \vskip-18pt 

\tablepreambleInstantonNumbers{25}
 (5, 5, 5, 5, 5)	 & 347718598088041789328640	 & (6, 5, 5, 5, 4)	 & 163766423699355653551056
\tabularnewline[-4pt] 
 (6, 6, 5, 4, 4)	 & 76746430278444036385392	 & (6, 6, 5, 5, 3)	 & 33557221088952835248384
\tabularnewline[-4pt] 
 (6, 6, 6, 4, 3)	 & 15579801166584314831616	 & (6, 6, 6, 5, 2)	 & 2507158978553441682912
\tabularnewline[-4pt] 
 (6, 6, 6, 6, 1)	 & 45075021198059982144	 & (7, 5, 5, 4, 4)	 & 36602428260502812573792
\tabularnewline[-4pt] 
 (7, 5, 5, 5, 3)	 & 15930480413967177684480	 & (7, 6, 4, 4, 4)	 & 16950676810888359150336
\tabularnewline[-4pt] 
 (7, 6, 5, 4, 3)	 & 7345251761305389562560	 & (7, 6, 5, 5, 2)	 & 1172715223879828113648
\tabularnewline[-4pt] 
 (7, 6, 6, 3, 3)	 & 1445782834458789325920	 & (7, 6, 6, 4, 2)	 & 533243466879375407808
\tabularnewline[-4pt] 
 (7, 6, 6, 5, 1)	 & 20734174826253969312	 & (7, 6, 6, 6, 0)	 & 19503820669876800
\tabularnewline[-4pt] 
 (7, 7, 4, 4, 3)	 & 1560763765722117846528	 & (7, 7, 5, 3, 3)	 & 666467844013257615360
\tabularnewline[-4pt] 
 (7, 7, 5, 4, 2)	 & 245004909605415502560	 & (7, 7, 5, 5, 1)	 & 9433084896265973760
\tabularnewline[-4pt] 
 (7, 7, 6, 3, 2)	 & 46154206945493038080	 & (7, 7, 6, 4, 1)	 & 4182469007721935136
\tabularnewline[-4pt] 
 (7, 7, 6, 5, 0)	 & 8765016259161504	 & (7, 7, 7, 2, 2)	 & 1329872815417735680
\tabularnewline[-4pt] 
 (7, 7, 7, 3, 1)	 & 331877990439469056	 & (7, 7, 7, 4, 0)	 & 1692511362069504
\tabularnewline[-4pt] 
 (8, 5, 4, 4, 4)	 & 3727698169135125498096	 & (8, 5, 5, 4, 3)	 & 1597193840680108612560
\tabularnewline[-4pt] 
 (8, 5, 5, 5, 2)	 & 250363560088071437904	 & (8, 6, 4, 4, 3)	 & 719772958474534446912
\tabularnewline[-4pt] 
 (8, 6, 5, 3, 3)	 & 305275969074406465728	 & (8, 6, 5, 4, 2)	 & 111802499676622459032
\tabularnewline[-4pt] 
 (8, 6, 5, 5, 1)	 & 4254301857683952288	 & (8, 6, 6, 3, 2)	 & 20678681485694660736
\tabularnewline[-4pt] 
 (8, 6, 6, 4, 1)	 & 1864566688856423904	 & (8, 6, 6, 5, 0)	 & 3881643757375656
\tabularnewline[-4pt] 
 (8, 7, 4, 3, 3)	 & 60590920000179493056	 & (8, 7, 4, 4, 2)	 & 22012186784542835520
\tabularnewline[-4pt] 
 (8, 7, 5, 3, 2)	 & 9117897040377080832	 & (8, 7, 5, 4, 1)	 & 817746667654917168
\tabularnewline[-4pt] 
 (8, 7, 5, 5, 0)	 & 1692511362069504	 & (8, 7, 6, 2, 2)	 & 571128202199454336
\tabularnewline[-4pt] 
 (8, 7, 6, 3, 1)	 & 142342287006477504	 & (8, 7, 6, 4, 0)	 & 725912434085952
\tabularnewline[-4pt] 
 (8, 7, 7, 2, 1)	 & 3377194221012096	 & (8, 7, 7, 3, 0)	 & 51294957112992
\tabularnewline[-4pt] 
 (8, 8, 3, 3, 3)	 & 2006276928131711424	 & (8, 8, 4, 3, 2)	 & 713471511849776160
\tabularnewline[-4pt] 
 (8, 8, 4, 4, 1)	 & 62675569121448240	 & (8, 8, 5, 2, 2)	 & 100010833402440120
\tabularnewline[-4pt] 
 (8, 8, 5, 3, 1)	 & 24840263013168672	 & (8, 8, 5, 4, 0)	 & 126532108859856
\tabularnewline[-4pt] 
 (8, 8, 6, 2, 1)	 & 1335301022489328	 & (8, 8, 6, 3, 0)	 & 20350993239840
\tabularnewline[-4pt] 
 (8, 8, 7, 1, 1)	 & 5601159429504	 & (8, 8, 7, 2, 0)	 & 391409808576
\tabularnewline[-4pt] 
 (8, 8, 8, 1, 0)	 & 347078520	 & (9, 4, 4, 4, 4)	 & 157391952685989052728
\tabularnewline[-4pt] 
 (9, 5, 4, 4, 3)	 & 65617907601886711296	 & (9, 5, 5, 3, 3)	 & 27190620887571766272
\tabularnewline[-4pt] 
 (9, 5, 5, 4, 2)	 & 9824161857371476896	 & (9, 5, 5, 5, 1)	 & 358330187751266304
\tabularnewline[-4pt] 
 (9, 6, 4, 3, 3)	 & 11689593863624674656	 & (9, 6, 4, 4, 2)	 & 4202606608677077184
\tabularnewline[-4pt] 
 (9, 6, 5, 3, 2)	 & 1706830027589928192	 & (9, 6, 5, 4, 1)	 & 150993571342096992
\tabularnewline[-4pt] 
 (9, 6, 5, 5, 0)	 & 305922925426848	 & (9, 6, 6, 2, 2)	 & 100108346194477248
\tabularnewline[-4pt] 
 (9, 6, 6, 3, 1)	 & 24863416450991904	 & (9, 6, 6, 4, 0)	 & 126532108859856
\tabularnewline[-4pt] 
 (9, 7, 3, 3, 3)	 & 841539378868429824	 & (9, 7, 4, 3, 2)	 & 297115911452589936
\tabularnewline[-4pt] 
 (9, 7, 4, 4, 1)	 & 25857038420140320	 & (9, 7, 5, 2, 2)	 & 40746789567213888
\tabularnewline[-4pt] 
 (9, 7, 5, 3, 1)	 & 10097809547695104	 & (9, 7, 5, 4, 0)	 & 51294957112992
\tabularnewline[-4pt] 
 (9, 7, 6, 2, 1)	 & 515881389602064	 & (9, 7, 6, 3, 0)	 & 7888589144400
\tabularnewline[-4pt] 
 (9, 7, 7, 1, 1)	 & 1927069671936	 & (9, 7, 7, 2, 0)	 & 135171775392
\tabularnewline[-4pt] 
 (9, 8, 3, 3, 2)	 & 7173870919736064	 & (9, 8, 4, 2, 2)	 & 2422178398686816
\tabularnewline[-4pt] 
 (9, 8, 4, 3, 1)	 & 594508678788528	 & (9, 8, 4, 4, 0)	 & 2981800050480
\tabularnewline[-4pt] 
 (9, 8, 5, 2, 1)	 & 71274491245200	 & (9, 8, 5, 3, 0)	 & 1096632180480
\tabularnewline[-4pt] 
 (9, 8, 6, 1, 1)	 & 639016897824	 & (9, 8, 6, 2, 0)	 & 45007048752
\tabularnewline[-4pt] 
 (9, 8, 7, 1, 0)	 & 88179456	 & (9, 8, 8, 0, 0)	 & 24
\tabularnewline[-4pt] 
 (9, 9, 3, 2, 2)	 & 13073262151968	 & (9, 9, 3, 3, 1)	 & 3100342138368
\tabularnewline[-4pt] 
 (9, 9, 4, 2, 1)	 & 931163905728	 & (9, 9, 4, 3, 0)	 & 14386869840
\tabularnewline[-4pt] 
 (9, 9, 5, 1, 1)	 & 17798444544	 & (9, 9, 5, 2, 0)	 & 1272585120
\tabularnewline[-4pt] 
 (9, 9, 6, 1, 0)	 & 4326048	 & (10, 4, 4, 4, 3)	 & 920246692052672448
\tabularnewline[-4pt] 
 (10, 5, 4, 3, 3)	 & 362176732991882256	 & (10, 5, 4, 4, 2)	 & 126656377507736616
\tabularnewline[-4pt] 
 (10, 5, 5, 3, 2)	 & 48949713376347552	 & (10, 5, 5, 4, 1)	 & 4162140562025760
\tabularnewline[-4pt] 
 (10, 5, 5, 5, 0)	 & 7888589144400	 & (10, 6, 3, 3, 3)	 & 55724768553096576
\tabularnewline[-4pt] 
 (10, 6, 4, 3, 2)	 & 19159936729163904	 & (10, 6, 4, 4, 1)	 & 1608297381675072
\tabularnewline[-4pt] 
 (10, 6, 5, 2, 2)	 & 2428815576573408	 & (10, 6, 5, 3, 1)	 & 596073535387056
\tabularnewline[-4pt] 
 (10, 6, 5, 4, 0)	 & 2981800050480	 & (10, 6, 6, 2, 1)	 & 25377635878296
\tabularnewline[-4pt] 
 (10, 6, 6, 3, 0)	 & 391409808576	 & (10, 7, 3, 3, 2)	 & 1000740719949936
\tabularnewline[-4pt] 
 (10, 7, 4, 2, 2)	 & 328447354833120	 & (10, 7, 4, 3, 1)	 & 79804026346992
\tabularnewline[-4pt] 
 (10, 7, 4, 4, 0)	 & 391409808576	 & (10, 7, 5, 2, 1)	 & 8748592415904
\tabularnewline[-4pt] 
 (10, 7, 5, 3, 0)	 & 135171775392	 & (10, 7, 6, 1, 1)	 & 61773182400
\tabularnewline[-4pt] 
 (10, 7, 6, 2, 0)	 & 4392333792	 & (10, 7, 7, 1, 0)	 & 4326048
\tabularnewline[-4pt] 
 (10, 8, 3, 2, 2)	 & 4183230238656	 & (10, 8, 3, 3, 1)	 & 980247769056
\tabularnewline[-4pt] 
 (10, 8, 4, 2, 1)	 & 285207114048	 & (10, 8, 4, 3, 0)	 & 4392333792
\tabularnewline[-4pt] 
 (10, 8, 5, 1, 1)	 & 4826161680	 & (10, 8, 5, 2, 0)	 & 347078520
\tabularnewline[-4pt] 
 (10, 8, 6, 1, 0)	 & 795936	 & (10, 9, 2, 2, 2)	 & 1599622824
\tabularnewline[-4pt] 
 (10, 9, 3, 2, 1)	 & 316997280	 & (10, 9, 3, 3, 0)	 & 4326048
\tabularnewline[-4pt] 
 (10, 9, 4, 1, 1)	 & 10883712	 & (10, 9, 4, 2, 0)	 & 795936
\tabularnewline[-4pt] 
 (10, 9, 5, 1, 0)	 & 1104	 & (10, 10, 2, 2, 1)	 & 24
\tabularnewline[-4pt] 
 (11, 4, 4, 3, 3)	 & 1112487680575968	 & (11, 4, 4, 4, 2)	 & 363393804317664
\tabularnewline[-4pt] 
 (11, 5, 3, 3, 3)	 & 389973010495488	 & (11, 5, 4, 3, 2)	 & 125365423769760
\tabularnewline[-4pt] 
 (11, 5, 4, 4, 1)	 & 9502910875584	 & (11, 5, 5, 2, 2)	 & 13073262151968
\tabularnewline[-4pt] 
 (11, 5, 5, 3, 1)	 & 3100342138368	 & (11, 5, 5, 4, 0)	 & 14386869840
\tabularnewline[-4pt] 
 (11, 6, 3, 3, 2)	 & 13834674726336	 & (11, 6, 4, 2, 2)	 & 4183230238656
\tabularnewline[-4pt] 
 (11, 6, 4, 3, 1)	 & 980247769056	 & (11, 6, 4, 4, 0)	 & 4392333792
\tabularnewline[-4pt] 
 (11, 6, 5, 2, 1)	 & 83099778720	 & (11, 6, 5, 3, 0)	 & 1272585120
\tabularnewline[-4pt] 
 (11, 6, 6, 1, 1)	 & 282674592	 & (11, 6, 6, 2, 0)	 & 20578560
\tabularnewline[-4pt] 
 (11, 7, 3, 2, 2)	 & 105371446464	 & (11, 7, 3, 3, 1)	 & 23351460864
\tabularnewline[-4pt] 
 (11, 7, 4, 2, 1)	 & 5950086192	 & (11, 7, 4, 3, 0)	 & 88179456
\tabularnewline[-4pt] 
 (11, 7, 5, 1, 1)	 & 59097600	 & (11, 7, 5, 2, 0)	 & 4326048
\tabularnewline[-4pt] 
 (11, 7, 6, 1, 0)	 & 1104	 & (11, 8, 2, 2, 2)	 & 63576576
\tabularnewline[-4pt] 
 (11, 8, 3, 2, 1)	 & 10883712	 & (11, 8, 3, 3, 0)	 & 122448
\tabularnewline[-4pt] 
 (11, 8, 4, 1, 1)	 & 212880	 & (11, 8, 4, 2, 0)	 & 14496
\tabularnewline[-4pt] 
 (12, 4, 3, 3, 3)	 & 85495746528	 & (12, 4, 4, 3, 2)	 & 22951602432
\tabularnewline[-4pt] 
 (12, 4, 4, 4, 1)	 & 1218252960	 & (12, 5, 3, 3, 2)	 & 5950086192
\tabularnewline[-4pt] 
 (12, 5, 4, 2, 2)	 & 1426637712	 & (12, 5, 4, 3, 1)	 & 282674592
\tabularnewline[-4pt] 
 (12, 5, 4, 4, 0)	 & 795936	 & (12, 5, 5, 2, 1)	 & 10883712
\tabularnewline[-4pt] 
 (12, 5, 5, 3, 0)	 & 122448	 & (12, 6, 3, 2, 2)	 & 63576576
\tabularnewline[-4pt] 
 (12, 6, 3, 3, 1)	 & 10883712	 & (12, 6, 4, 2, 1)	 & 1691856
\tabularnewline[-4pt] 
 (12, 6, 4, 3, 0)	 & 14496	 & (12, 6, 5, 1, 1)	 & 1104
\tabularnewline[-4pt] 
 (12, 6, 5, 2, 0)	 & 24	 & (12, 7, 2, 2, 2)	 & 14496
\tabularnewline[-4pt] 
(12, 7, 3, 2, 1)	 & 1104	 &  	 &  
\tablepostambleInstantonNumbers \vskip-18pt 

\tablepreambleInstantonNumbers{26}
 (6, 5, 5, 5, 5)	 & 2910174233830401416162688	 & (6, 6, 5, 5, 4)	 & 1393776642755701910391504
\tabularnewline[-4pt] 
 (6, 6, 6, 4, 4)	 & 664776534906643820467776	 & (6, 6, 6, 5, 3)	 & 294433074567120966718080
\tabularnewline[-4pt] 
 (6, 6, 6, 6, 2)	 & 22914149837439123291648	 & (7, 5, 5, 5, 4)	 & 683358195482651060173200
\tabularnewline[-4pt] 
 (7, 6, 5, 4, 4)	 & 324239422338282700223616	 & (7, 6, 5, 5, 3)	 & 143065987059929651882064
\tabularnewline[-4pt] 
 (7, 6, 6, 4, 3)	 & 67330722644161744497600	 & (7, 6, 6, 5, 2)	 & 11001712604766772877568
\tabularnewline[-4pt] 
 (7, 6, 6, 6, 1)	 & 204289830851585811840	 & (7, 7, 4, 4, 4)	 & 73760592549207990341160
\tabularnewline[-4pt] 
 (7, 7, 5, 4, 3)	 & 32292308644789613776992	 & (7, 7, 5, 5, 2)	 & 5240983985031424336512
\tabularnewline[-4pt] 
 (7, 7, 6, 3, 3)	 & 6528373563454253739936	 & (7, 7, 6, 4, 2)	 & 2422666442072438912352
\tabularnewline[-4pt] 
 (7, 7, 6, 5, 1)	 & 95977233617823957552	 & (7, 7, 6, 6, 0)	 & 92700939550359360
\tabularnewline[-4pt] 
 (7, 7, 7, 3, 2)	 & 221273208968851435344	 & (7, 7, 7, 4, 1)	 & 20221514209776438144
\tabularnewline[-4pt] 
 (7, 7, 7, 5, 0)	 & 42801528146793216	 & (8, 5, 5, 4, 4)	 & 76121903698269498879600
\tabularnewline[-4pt] 
 (8, 5, 5, 5, 3)	 & 33298242026156998722144	 & (8, 6, 4, 4, 4)	 & 35504946591945154063104
\tabularnewline[-4pt] 
 (8, 6, 5, 4, 3)	 & 15467950978366663032576	 & (8, 6, 5, 5, 2)	 & 2490930240945503131824
\tabularnewline[-4pt] 
 (8, 6, 6, 3, 3)	 & 3087105410543684178144	 & (8, 6, 6, 4, 2)	 & 1142319760025546317200
\tabularnewline[-4pt] 
 (8, 6, 6, 5, 1)	 & 44858935490060258472	 & (8, 6, 6, 6, 0)	 & 42801528135993600
\tabularnewline[-4pt] 
 (8, 7, 4, 4, 3)	 & 3345260435684206623648	 & (8, 7, 5, 3, 3)	 & 1437664544143525963632
\tabularnewline[-4pt] 
 (8, 7, 5, 4, 2)	 & 530380950844598802240	 & (8, 7, 5, 5, 1)	 & 20644575230653895136
\tabularnewline[-4pt] 
 (8, 7, 6, 3, 2)	 & 101657485900434092424	 & (8, 7, 6, 4, 1)	 & 9254014284061822464
\tabularnewline[-4pt] 
 (8, 7, 6, 5, 0)	 & 19503820669876800	 & (8, 7, 7, 2, 2)	 & 3047920567708923264
\tabularnewline[-4pt] 
 (8, 7, 7, 3, 1)	 & 761479183438470384	 & (8, 7, 7, 4, 0)	 & 3881643757375656
\tabularnewline[-4pt] 
 (8, 8, 4, 3, 3)	 & 135135632721772486224	 & (8, 8, 4, 4, 2)	 & 49318322079952346112
\tabularnewline[-4pt] 
 (8, 8, 5, 3, 2)	 & 20610476713078747200	 & (8, 8, 5, 4, 1)	 & 1859138760210276768
\tabularnewline[-4pt] 
 (8, 8, 5, 5, 0)	 & 3881643757375656	 & (8, 8, 6, 2, 2)	 & 1329629977546611936
\tabularnewline[-4pt] 
 (8, 8, 6, 3, 1)	 & 331824982853181696	 & (8, 8, 6, 4, 0)	 & 1692511359568896
\tabularnewline[-4pt] 
 (8, 8, 7, 2, 1)	 & 8359186921934400	 & (8, 8, 7, 3, 0)	 & 126532108859856
\tabularnewline[-4pt] 
 (8, 8, 8, 1, 1)	 & 15746747463456	 & (8, 8, 8, 2, 0)	 & 1096632086784
\tabularnewline[-4pt] 
 (9, 5, 4, 4, 4)	 & 3727698169135125498096	 & (9, 5, 5, 4, 3)	 & 1597193840680108612560
\tabularnewline[-4pt] 
 (9, 5, 5, 5, 2)	 & 250363560088071437904	 & (9, 6, 4, 4, 3)	 & 719772958474534446912
\tabularnewline[-4pt] 
 (9, 6, 5, 3, 3)	 & 305275969074406465728	 & (9, 6, 5, 4, 2)	 & 111802499676622459032
\tabularnewline[-4pt] 
 (9, 6, 5, 5, 1)	 & 4254301857683952288	 & (9, 6, 6, 3, 2)	 & 20678681485694660736
\tabularnewline[-4pt] 
 (9, 6, 6, 4, 1)	 & 1864566688856423904	 & (9, 6, 6, 5, 0)	 & 3881643757375656
\tabularnewline[-4pt] 
 (9, 7, 4, 3, 3)	 & 60590920000179493056	 & (9, 7, 4, 4, 2)	 & 22012186784542835520
\tabularnewline[-4pt] 
 (9, 7, 5, 3, 2)	 & 9117897040377080832	 & (9, 7, 5, 4, 1)	 & 817746667654917168
\tabularnewline[-4pt] 
 (9, 7, 5, 5, 0)	 & 1692511362069504	 & (9, 7, 6, 2, 2)	 & 571128202199454336
\tabularnewline[-4pt] 
 (9, 7, 6, 3, 1)	 & 142342287006477504	 & (9, 7, 6, 4, 0)	 & 725912434085952
\tabularnewline[-4pt] 
 (9, 7, 7, 2, 1)	 & 3377194221012096	 & (9, 7, 7, 3, 0)	 & 51294957112992
\tabularnewline[-4pt] 
 (9, 8, 3, 3, 3)	 & 2006276928131711424	 & (9, 8, 4, 3, 2)	 & 713471511849776160
\tabularnewline[-4pt] 
 (9, 8, 4, 4, 1)	 & 62675569121448240	 & (9, 8, 5, 2, 2)	 & 100010833402440120
\tabularnewline[-4pt] 
 (9, 8, 5, 3, 1)	 & 24840263013168672	 & (9, 8, 5, 4, 0)	 & 126532108859856
\tabularnewline[-4pt] 
 (9, 8, 6, 2, 1)	 & 1335301022489328	 & (9, 8, 6, 3, 0)	 & 20350993239840
\tabularnewline[-4pt] 
 (9, 8, 7, 1, 1)	 & 5601159429504	 & (9, 8, 7, 2, 0)	 & 391409808576
\tabularnewline[-4pt] 
 (9, 8, 8, 1, 0)	 & 347078520	 & (9, 9, 3, 3, 2)	 & 7173870919736064
\tabularnewline[-4pt] 
 (9, 9, 4, 2, 2)	 & 2422178398686816	 & (9, 9, 4, 3, 1)	 & 594508678788528
\tabularnewline[-4pt] 
 (9, 9, 4, 4, 0)	 & 2981800050480	 & (9, 9, 5, 2, 1)	 & 71274491245200
\tabularnewline[-4pt] 
 (9, 9, 5, 3, 0)	 & 1096632180480	 & (9, 9, 6, 1, 1)	 & 639016897824
\tabularnewline[-4pt] 
 (9, 9, 6, 2, 0)	 & 45007048752	 & (9, 9, 7, 1, 0)	 & 88179456
\tabularnewline[-4pt] 
 (9, 9, 8, 0, 0)	 & 24	 & (10, 4, 4, 4, 4)	 & 69702170473826178048
\tabularnewline[-4pt] 
 (10, 5, 4, 4, 3)	 & 28814753795787304128	 & (10, 5, 5, 3, 3)	 & 11833136668383611040
\tabularnewline[-4pt] 
 (10, 5, 5, 4, 2)	 & 4252005327651223776	 & (10, 5, 5, 5, 1)	 & 152435152838866176
\tabularnewline[-4pt] 
 (10, 6, 4, 3, 3)	 & 5023740750844977792	 & (10, 6, 4, 4, 2)	 & 1795314514514344416
\tabularnewline[-4pt] 
 (10, 6, 5, 3, 2)	 & 721163569257189312	 & (10, 6, 5, 4, 1)	 & 63276065657309280
\tabularnewline[-4pt] 
 (10, 6, 5, 5, 0)	 & 126532108859856	 & (10, 6, 6, 2, 2)	 & 40767562975883520
\tabularnewline[-4pt] 
 (10, 6, 6, 3, 1)	 & 10102374952223232	 & (10, 6, 6, 4, 0)	 & 51294956593632
\tabularnewline[-4pt] 
 (10, 7, 3, 3, 3)	 & 346896207708697296	 & (10, 7, 4, 3, 2)	 & 121505012171479176
\tabularnewline[-4pt] 
 (10, 7, 4, 4, 1)	 & 10462960782869952	 & (10, 7, 5, 2, 2)	 & 16270300857476160
\tabularnewline[-4pt] 
 (10, 7, 5, 3, 1)	 & 4021264698687264	 & (10, 7, 5, 4, 0)	 & 20350993239840
\tabularnewline[-4pt] 
 (10, 7, 6, 2, 1)	 & 194378107421760	 & (10, 7, 6, 3, 0)	 & 2981800050480
\tabularnewline[-4pt] 
 (10, 7, 7, 1, 1)	 & 639016897824	 & (10, 7, 7, 2, 0)	 & 45007048752
\tabularnewline[-4pt] 
 (10, 8, 3, 3, 2)	 & 2713101057421728	 & (10, 8, 4, 2, 2)	 & 903893653068672
\tabularnewline[-4pt] 
 (10, 8, 4, 3, 1)	 & 220840621188096	 & (10, 8, 4, 4, 0)	 & 1096632086784
\tabularnewline[-4pt] 
 (10, 8, 5, 2, 1)	 & 25377635878296	 & (10, 8, 5, 3, 0)	 & 391409808576
\tabularnewline[-4pt] 
 (10, 8, 6, 1, 1)	 & 203336907216	 & (10, 8, 6, 2, 0)	 & 14386855920
\tabularnewline[-4pt] 
 (10, 8, 7, 1, 0)	 & 20578560	 & (10, 9, 3, 2, 2)	 & 4183230238656
\tabularnewline[-4pt] 
 (10, 9, 3, 3, 1)	 & 980247769056	 & (10, 9, 4, 2, 1)	 & 285207114048
\tabularnewline[-4pt] 
 (10, 9, 4, 3, 0)	 & 4392333792	 & (10, 9, 5, 1, 1)	 & 4826161680
\tabularnewline[-4pt] 
 (10, 9, 5, 2, 0)	 & 347078520	 & (10, 9, 6, 1, 0)	 & 795936
\tabularnewline[-4pt] 
 (10, 10, 2, 2, 2)	 & 341681280	 & (10, 10, 3, 2, 1)	 & 63576576
\tabularnewline[-4pt] 
 (10, 10, 3, 3, 0)	 & 795936	 & (10, 10, 4, 1, 1)	 & 1691856
\tabularnewline[-4pt] 
 (10, 10, 4, 2, 0)	 & 122352	 & (10, 10, 5, 1, 0)	 & 24
\tabularnewline[-4pt] 
 (11, 4, 4, 4, 3)	 & 149583407202367176	 & (11, 5, 4, 3, 3)	 & 57309129620711136
\tabularnewline[-4pt] 
 (11, 5, 4, 4, 2)	 & 19680157760407104	 & (11, 5, 5, 3, 2)	 & 7371081117191712
\tabularnewline[-4pt] 
 (11, 5, 5, 4, 1)	 & 609209937409968	 & (11, 5, 5, 5, 0)	 & 1096632180480
\tabularnewline[-4pt] 
 (11, 6, 3, 3, 3)	 & 8236673292611808	 & (11, 6, 4, 3, 2)	 & 2768640614245200
\tabularnewline[-4pt] 
 (11, 6, 4, 4, 1)	 & 224917616990784	 & (11, 6, 5, 2, 2)	 & 328447354833120
\tabularnewline[-4pt] 
 (11, 6, 5, 3, 1)	 & 79804026346992	 & (11, 6, 5, 4, 0)	 & 391409808576
\tabularnewline[-4pt] 
 (11, 6, 6, 2, 1)	 & 2910089695872	 & (11, 6, 6, 3, 0)	 & 45007048752
\tabularnewline[-4pt] 
 (11, 7, 3, 3, 2)	 & 125365423769760	 & (11, 7, 4, 2, 2)	 & 39692266181304
\tabularnewline[-4pt] 
 (11, 7, 4, 3, 1)	 & 9502910875584	 & (11, 7, 4, 4, 0)	 & 45007048752
\tabularnewline[-4pt] 
 (11, 7, 5, 2, 1)	 & 931163905728	 & (11, 7, 5, 3, 0)	 & 14386869840
\tabularnewline[-4pt] 
 (11, 7, 6, 1, 1)	 & 4826161680	 & (11, 7, 6, 2, 0)	 & 347078520
\tabularnewline[-4pt] 
 (11, 7, 7, 1, 0)	 & 122448	 & (11, 8, 3, 2, 2)	 & 377080188864
\tabularnewline[-4pt] 
 (11, 8, 3, 3, 1)	 & 85495746528	 & (11, 8, 4, 2, 1)	 & 22951602432
\tabularnewline[-4pt] 
 (11, 8, 4, 3, 0)	 & 347078520	 & (11, 8, 5, 1, 1)	 & 282674592
\tabularnewline[-4pt] 
 (11, 8, 5, 2, 0)	 & 20578560	 & (11, 8, 6, 1, 0)	 & 14496
\tabularnewline[-4pt] 
 (11, 9, 2, 2, 2)	 & 63576576	 & (11, 9, 3, 2, 1)	 & 10883712
\tabularnewline[-4pt] 
 (11, 9, 3, 3, 0)	 & 122448	 & (11, 9, 4, 1, 1)	 & 212880
\tabularnewline[-4pt] 
 (11, 9, 4, 2, 0)	 & 14496	 & (12, 4, 4, 3, 3)	 & 42411173392368
\tabularnewline[-4pt] 
 (12, 4, 4, 4, 2)	 & 13138629854976	 & (12, 5, 3, 3, 3)	 & 13834674726336
\tabularnewline[-4pt] 
 (12, 5, 4, 3, 2)	 & 4183230238656	 & (12, 5, 4, 4, 1)	 & 285207114048
\tabularnewline[-4pt] 
 (12, 5, 5, 2, 2)	 & 366406656528	 & (12, 5, 5, 3, 1)	 & 83099778720
\tabularnewline[-4pt] 
 (12, 5, 5, 4, 0)	 & 347078520	 & (12, 6, 3, 3, 2)	 & 377080188864
\tabularnewline[-4pt] 
 (12, 6, 4, 2, 2)	 & 103492041648	 & (12, 6, 4, 3, 1)	 & 22951602432
\tabularnewline[-4pt] 
 (12, 6, 4, 4, 0)	 & 88177920	 & (12, 6, 5, 2, 1)	 & 1426637712
\tabularnewline[-4pt] 
 (12, 6, 5, 3, 0)	 & 20578560	 & (12, 6, 6, 1, 1)	 & 1691856
\tabularnewline[-4pt] 
 (12, 6, 6, 2, 0)	 & 122352	 & (12, 7, 3, 2, 2)	 & 1599622824
\tabularnewline[-4pt] 
 (12, 7, 3, 3, 1)	 & 316997280	 & (12, 7, 4, 2, 1)	 & 63576576
\tabularnewline[-4pt] 
 (12, 7, 4, 3, 0)	 & 795936	 & (12, 7, 5, 1, 1)	 & 212880
\tabularnewline[-4pt] 
 (12, 7, 5, 2, 0)	 & 14496	 & (12, 8, 2, 2, 2)	 & 122352
\tabularnewline[-4pt] 
 (12, 8, 3, 2, 1)	 & 14496	 & (12, 8, 3, 3, 0)	 & 24
\tabularnewline[-4pt] 
 (12, 8, 4, 1, 1)	 & 24	 & (13, 4, 3, 3, 3)	 & 88179456
\tabularnewline[-4pt] 
 (13, 4, 4, 3, 2)	 & 20578560	 & (13, 4, 4, 4, 1)	 & 795936
\tabularnewline[-4pt] 
 (13, 5, 3, 3, 2)	 & 4326048	 & (13, 5, 4, 2, 2)	 & 795936
\tabularnewline[-4pt] 
 (13, 5, 4, 3, 1)	 & 122448	 & (13, 5, 4, 4, 0)	 & 24
\tabularnewline[-4pt] 
 (13, 5, 5, 2, 1)	 & 1104	 & (13, 6, 3, 2, 2)	 & 14496
\tabularnewline[-4pt] 
 (13, 6, 3, 3, 1)	 & 1104	 & (13, 6, 4, 2, 1)	 & 24
\tablepostambleInstantonNumbers \vskip-18pt 

\tablepreambleInstantonNumbers{27}
 (6, 6, 5, 5, 5)	 & 28656849112544426796718608	 & (6, 6, 6, 5, 4)	 & 13944411721206459640109952
\tabularnewline[-4pt] 
 (6, 6, 6, 6, 3)	 & 3030705830464261116958752	 & (7, 5, 5, 5, 5)	 & 14401495635309838652737536
\tabularnewline[-4pt] 
 (7, 6, 5, 5, 4)	 & 6979017914791123565948416	 & (7, 6, 6, 4, 4)	 & 3369981367793558156370720
\tabularnewline[-4pt] 
 (7, 6, 6, 5, 3)	 & 1505729217469676504230592	 & (7, 6, 6, 6, 2)	 & 120468636234042768002112
\tabularnewline[-4pt] 
 (7, 7, 5, 4, 4)	 & 1670002775792759585584480	 & (7, 7, 5, 5, 3)	 & 743754012075104160058368
\tabularnewline[-4pt] 
 (7, 7, 6, 4, 3)	 & 355021932713338577724288	 & (7, 7, 6, 5, 2)	 & 58904390293772710703920
\tabularnewline[-4pt] 
 (7, 7, 6, 6, 1)	 & 1129438555365292906784	 & (7, 7, 7, 3, 3)	 & 36027026438881932128256
\tabularnewline[-4pt] 
 (7, 7, 7, 4, 2)	 & 13449577349429667122112	 & (7, 7, 7, 5, 1)	 & 542448112625738749440
\tabularnewline[-4pt] 
 (7, 7, 7, 6, 0)	 & 536474722655969280	 & (8, 5, 5, 5, 4)	 & 1737004389084229283537184
\tabularnewline[-4pt] 
 (8, 6, 5, 4, 4)	 & 830547124981920578846912	 & (8, 6, 5, 5, 3)	 & 368512581418692060268192
\tabularnewline[-4pt] 
 (8, 6, 6, 4, 3)	 & 174894056440451268167776	 & (8, 6, 6, 5, 2)	 & 28841649917004887522080
\tabularnewline[-4pt] 
 (8, 6, 6, 6, 1)	 & 546003716872720702848	 & (8, 7, 4, 4, 4)	 & 192443720708788001680608
\tabularnewline[-4pt] 
 (8, 7, 5, 4, 3)	 & 84779574976667980526368	 & (8, 7, 5, 5, 2)	 & 13895992208920106396544
\tabularnewline[-4pt] 
 (8, 7, 6, 3, 3)	 & 17421547466353577240096	 & (8, 7, 6, 4, 2)	 & 6488505552963384035984
\tabularnewline[-4pt] 
 (8, 7, 6, 5, 1)	 & 259867472483449630240	 & (8, 7, 6, 6, 0)	 & 254791938658803840
\tabularnewline[-4pt] 
 (8, 7, 7, 3, 2)	 & 612363334253114849568	 & (8, 7, 7, 4, 1)	 & 56225444916409818816
\tabularnewline[-4pt] 
 (8, 7, 7, 5, 0)	 & 119602242975339008	 & (8, 8, 4, 4, 3)	 & 9090329485714422369312
\tabularnewline[-4pt] 
 (8, 8, 5, 3, 3)	 & 3937816074403396325984	 & (8, 8, 5, 4, 2)	 & 1458934381540619175680
\tabularnewline[-4pt] 
 (8, 8, 5, 5, 1)	 & 57539715512784775920	 & (8, 8, 6, 3, 2)	 & 285713812377367900976
\tabularnewline[-4pt] 
 (8, 8, 6, 4, 1)	 & 26146648523244093888	 & (8, 8, 6, 5, 0)	 & 55456767284050560
\tabularnewline[-4pt] 
 (8, 8, 7, 2, 2)	 & 8999069221638485760	 & (8, 8, 7, 3, 1)	 & 2251004857405139040
\tabularnewline[-4pt] 
 (8, 8, 7, 4, 0)	 & 11461248223336400	 & (8, 8, 8, 2, 1)	 & 27132798580132512
\tabularnewline[-4pt] 
 (8, 8, 8, 3, 0)	 & 408865565088240	 & (9, 5, 5, 4, 4)	 & 96959783616963943030208
\tabularnewline[-4pt] 
 (9, 5, 5, 5, 3)	 & 42482450279496628079616	 & (9, 6, 4, 4, 4)	 & 45328721421594382226880
\tabularnewline[-4pt] 
 (9, 6, 5, 4, 3)	 & 19781742107681508906368	 & (9, 6, 5, 5, 2)	 & 3194412266903770307840
\tabularnewline[-4pt] 
 (9, 6, 6, 3, 3)	 & 3965695431311102575520	 & (9, 6, 6, 4, 2)	 & 1468942204326701835456
\tabularnewline[-4pt] 
 (9, 6, 6, 5, 1)	 & 57866098737278532384	 & (9, 6, 6, 6, 0)	 & 55456767284050560
\tabularnewline[-4pt] 
 (9, 7, 4, 4, 3)	 & 4302653199079323937920	 & (9, 7, 5, 3, 3)	 & 1852914051293922601984
\tabularnewline[-4pt] 
 (9, 7, 5, 4, 2)	 & 684339544738107833984	 & (9, 7, 5, 5, 1)	 & 26729153888914424832
\tabularnewline[-4pt] 
 (9, 7, 6, 3, 2)	 & 131895336423525168320	 & (9, 7, 6, 4, 1)	 & 12023512502480326944
\tabularnewline[-4pt] 
 (9, 7, 6, 5, 0)	 & 25383601647232320	 & (9, 7, 7, 2, 2)	 & 4005072998835530048
\tabularnewline[-4pt] 
 (9, 7, 7, 3, 1)	 & 1000944581364139008	 & (9, 7, 7, 4, 0)	 & 5101035246140064
\tabularnewline[-4pt] 
 (9, 8, 4, 3, 3)	 & 176048749037882981792	 & (9, 8, 4, 4, 2)	 & 64342058578990962432
\tabularnewline[-4pt] 
 (9, 8, 5, 3, 2)	 & 26964980823333578944	 & (9, 8, 5, 4, 1)	 & 2436667741602374528
\tabularnewline[-4pt] 
 (9, 8, 5, 5, 0)	 & 5101035246140064	 & (9, 8, 6, 2, 2)	 & 1755969430220426496
\tabularnewline[-4pt] 
 (9, 8, 6, 3, 1)	 & 438395891787545984	 & (9, 8, 6, 4, 0)	 & 2235977596096128
\tabularnewline[-4pt] 
 (9, 8, 7, 2, 1)	 & 11256418126498304	 & (9, 8, 7, 3, 0)	 & 170193515484672
\tabularnewline[-4pt] 
 (9, 8, 8, 1, 1)	 & 22072657897776	 & (9, 8, 8, 2, 0)	 & 1535514818112
\tabularnewline[-4pt] 
 (9, 9, 3, 3, 3)	 & 2670197361402514944	 & (9, 9, 4, 3, 2)	 & 951716992566363648
\tabularnewline[-4pt] 
 (9, 9, 4, 4, 1)	 & 83845604993916000	 & (9, 9, 5, 2, 2)	 & 134329574207275104
\tabularnewline[-4pt] 
 (9, 9, 5, 3, 1)	 & 33386278590988800	 & (9, 9, 5, 4, 0)	 & 170193515484672
\tabularnewline[-4pt] 
 (9, 9, 6, 2, 1)	 & 1823776449179136	 & (9, 9, 6, 3, 0)	 & 27765085214112
\tabularnewline[-4pt] 
 (9, 9, 7, 1, 1)	 & 7933211814912	 & (9, 9, 7, 2, 0)	 & 553728279360
\tabularnewline[-4pt] 
 (9, 9, 8, 1, 0)	 & 539120544	 & (9, 9, 9, 0, 0)	 & 112
\tabularnewline[-4pt] 
 (10, 5, 4, 4, 4)	 & 2234583126440197477248	 & (10, 5, 5, 4, 3)	 & 953357306203227960000
\tabularnewline[-4pt] 
 (10, 5, 5, 5, 2)	 & 148388879628408287328	 & (10, 6, 4, 4, 3)	 & 426992405813739053760
\tabularnewline[-4pt] 
 (10, 6, 5, 3, 3)	 & 180227194372605425904	 & (10, 6, 5, 4, 2)	 & 65822715717348500960
\tabularnewline[-4pt] 
 (10, 6, 5, 5, 1)	 & 2483294974158049312	 & (10, 6, 6, 3, 2)	 & 12017787465197578008
\tabularnewline[-4pt] 
 (10, 6, 6, 4, 1)	 & 1079565550915867008	 & (10, 6, 6, 5, 0)	 & 2235977596096128
\tabularnewline[-4pt] 
 (10, 7, 4, 3, 3)	 & 35230204567671156080	 & (10, 7, 4, 4, 2)	 & 12756877670900976952
\tabularnewline[-4pt] 
 (10, 7, 5, 3, 2)	 & 5251183397613765152	 & (10, 7, 5, 4, 1)	 & 468967501173905952
\tabularnewline[-4pt] 
 (10, 7, 5, 5, 0)	 & 964399018545152	 & (10, 7, 6, 2, 2)	 & 322136481160659232
\tabularnewline[-4pt] 
 (10, 7, 6, 3, 1)	 & 80203179581197904	 & (10, 7, 6, 4, 0)	 & 408865565088240
\tabularnewline[-4pt] 
 (10, 7, 7, 2, 1)	 & 1823829689450016	 & (10, 7, 7, 3, 0)	 & 27765085214112
\tabularnewline[-4pt] 
 (10, 8, 3, 3, 3)	 & 1126346577851592960	 & (10, 8, 4, 3, 2)	 & 398660145480045856
\tabularnewline[-4pt] 
 (10, 8, 4, 4, 1)	 & 34807080531177792	 & (10, 8, 5, 2, 2)	 & 55084399995750144
\tabularnewline[-4pt] 
 (10, 8, 5, 3, 1)	 & 13661798641742976	 & (10, 8, 5, 4, 0)	 & 69468841810240
\tabularnewline[-4pt] 
 (10, 8, 6, 2, 1)	 & 710229601026304	 & (10, 8, 6, 3, 0)	 & 10848408360480
\tabularnewline[-4pt] 
 (10, 8, 7, 1, 1)	 & 2760956499680	 & (10, 8, 7, 2, 0)	 & 193411225936
\tabularnewline[-4pt] 
 (10, 8, 8, 1, 0)	 & 140436672	 & (10, 9, 3, 3, 2)	 & 3761948244770304
\tabularnewline[-4pt] 
 (10, 9, 4, 2, 2)	 & 1259132047619264	 & (10, 9, 4, 3, 1)	 & 308134225628128
\tabularnewline[-4pt] 
 (10, 9, 4, 4, 0)	 & 1535514818112	 & (10, 9, 5, 2, 1)	 & 35929933424832
\tabularnewline[-4pt] 
 (10, 9, 5, 3, 0)	 & 553728279360	 & (10, 9, 6, 1, 1)	 & 299302640864
\tabularnewline[-4pt] 
 (10, 9, 6, 2, 0)	 & 21143067840	 & (10, 9, 7, 1, 0)	 & 33777312
\tabularnewline[-4pt] 
 (10, 10, 3, 2, 2)	 & 1912895782008	 & (10, 10, 3, 3, 1)	 & 443961562528
\tabularnewline[-4pt] 
 (10, 10, 4, 2, 1)	 & 126121309632	 & (10, 10, 4, 3, 0)	 & 1935300720
\tabularnewline[-4pt] 
 (10, 10, 5, 1, 1)	 & 1944767152	 & (10, 10, 5, 2, 0)	 & 140436672
\tabularnewline[-4pt] 
 (10, 10, 6, 1, 0)	 & 234048	 & (11, 4, 4, 4, 4)	 & 17389206433621316832
\tabularnewline[-4pt] 
 (11, 5, 4, 4, 3)	 & 7079567101109436512	 & (11, 5, 5, 3, 3)	 & 2860072289627444736
\tabularnewline[-4pt] 
 (11, 5, 5, 4, 2)	 & 1017289744237857120	 & (11, 5, 5, 5, 1)	 & 35306571598392576
\tabularnewline[-4pt] 
 (11, 6, 4, 3, 3)	 & 1187054464752608224	 & (11, 6, 4, 4, 2)	 & 419478239436537264
\tabularnewline[-4pt] 
 (11, 6, 5, 3, 2)	 & 165119843412344816	 & (11, 6, 5, 4, 1)	 & 14258867760974432
\tabularnewline[-4pt] 
 (11, 6, 5, 5, 0)	 & 27765085214112	 & (11, 6, 6, 2, 2)	 & 8718347106041576
\tabularnewline[-4pt] 
 (11, 6, 6, 3, 1)	 & 2150266975191936	 & (11, 6, 6, 4, 0)	 & 10848408360480
\tabularnewline[-4pt] 
 (11, 7, 3, 3, 3)	 & 75992812385562624	 & (11, 7, 4, 3, 2)	 & 26217346711258048
\tabularnewline[-4pt] 
 (11, 7, 4, 4, 1)	 & 2211223893638272	 & (11, 7, 5, 2, 2)	 & 3356453655323136
\tabularnewline[-4pt] 
 (11, 7, 5, 3, 1)	 & 824874647838720	 & (11, 7, 5, 4, 0)	 & 4136092936448
\tabularnewline[-4pt] 
 (11, 7, 6, 2, 1)	 & 35929933424832	 & (11, 7, 6, 3, 0)	 & 553728279360
\tabularnewline[-4pt] 
 (11, 7, 7, 1, 1)	 & 92396257280	 & (11, 7, 7, 2, 0)	 & 6558863360
\tabularnewline[-4pt] 
 (11, 8, 3, 3, 2)	 & 507096396665312	 & (11, 8, 4, 2, 2)	 & 164605655104880
\tabularnewline[-4pt] 
 (11, 8, 4, 3, 1)	 & 39821013536096	 & (11, 8, 4, 4, 0)	 & 193411225936
\tabularnewline[-4pt] 
 (11, 8, 5, 2, 1)	 & 4217701870608	 & (11, 8, 5, 3, 0)	 & 65215603200
\tabularnewline[-4pt] 
 (11, 8, 6, 1, 1)	 & 27120466144	 & (11, 8, 6, 2, 0)	 & 1935300720
\tabularnewline[-4pt] 
 (11, 8, 7, 1, 0)	 & 1423616	 & (11, 9, 3, 2, 2)	 & 570360079168
\tabularnewline[-4pt] 
 (11, 9, 3, 3, 1)	 & 130194945024	 & (11, 9, 4, 2, 1)	 & 35487082592
\tabularnewline[-4pt] 
 (11, 9, 4, 3, 0)	 & 539120544	 & (11, 9, 5, 1, 1)	 & 464696832
\tabularnewline[-4pt] 
 (11, 9, 5, 2, 0)	 & 33777312	 & (11, 9, 6, 1, 0)	 & 30624
\tabularnewline[-4pt] 
 (11, 10, 2, 2, 2)	 & 20299992	 & (11, 10, 3, 2, 1)	 & 3222112
\tabularnewline[-4pt] 
 (11, 10, 3, 3, 0)	 & 30624	 & (11, 10, 4, 1, 1)	 & 45408
\tabularnewline[-4pt] 
 (11, 10, 4, 2, 0)	 & 2800	 & (12, 4, 4, 4, 3)	 & 11630106886504344
\tabularnewline[-4pt] 
 (12, 5, 4, 3, 3)	 & 4272828104425920	 & (12, 5, 4, 4, 2)	 & 1423524718242752
\tabularnewline[-4pt] 
 (12, 5, 5, 3, 2)	 & 507096396665312	 & (12, 5, 5, 4, 1)	 & 39821013536096
\tabularnewline[-4pt] 
 (12, 5, 5, 5, 0)	 & 65215603200	 & (12, 6, 3, 3, 3)	 & 552486590320032
\tabularnewline[-4pt] 
 (12, 6, 4, 3, 2)	 & 178677828494464	 & (12, 6, 4, 4, 1)	 & 13674852866304
\tabularnewline[-4pt] 
 (12, 6, 5, 2, 2)	 & 18954386538304	 & (12, 6, 5, 3, 1)	 & 4510722900128
\tabularnewline[-4pt] 
 (12, 6, 5, 4, 0)	 & 21143067840	 & (12, 6, 6, 2, 1)	 & 125948336640
\tabularnewline[-4pt] 
 (12, 6, 6, 3, 0)	 & 1935300720	 & (12, 7, 3, 3, 2)	 & 6446376071472
\tabularnewline[-4pt] 
 (12, 7, 4, 2, 2)	 & 1912895782008	 & (12, 7, 4, 3, 1)	 & 443961562528
\tabularnewline[-4pt] 
 (12, 7, 4, 4, 0)	 & 1935300720	 & (12, 7, 5, 2, 1)	 & 35487082592
\tabularnewline[-4pt] 
 (12, 7, 5, 3, 0)	 & 539120544	 & (12, 7, 6, 1, 1)	 & 100919904
\tabularnewline[-4pt] 
 (12, 7, 6, 2, 0)	 & 7371792	 & (12, 7, 7, 1, 0)	 & 112
\tabularnewline[-4pt] 
 (12, 8, 3, 2, 2)	 & 11032046624	 & (12, 8, 3, 3, 1)	 & 2322325968
\tabularnewline[-4pt] 
 (12, 8, 4, 2, 1)	 & 529392832	 & (12, 8, 4, 3, 0)	 & 7371792
\tabularnewline[-4pt] 
 (12, 8, 5, 1, 1)	 & 3222112	 & (12, 8, 5, 2, 0)	 & 234048
\tabularnewline[-4pt] 
 (12, 9, 2, 2, 2)	 & 234048	 & (12, 9, 3, 2, 1)	 & 30624
\tabularnewline[-4pt] 
 (12, 9, 3, 3, 0)	 & 112	 & (12, 9, 4, 1, 1)	 & 112
\tabularnewline[-4pt] 
 (13, 4, 4, 3, 3)	 & 443961562528	 & (13, 4, 4, 4, 2)	 & 126121309632
\tabularnewline[-4pt] 
 (13, 5, 3, 3, 3)	 & 130194945024	 & (13, 5, 4, 3, 2)	 & 35487082592
\tabularnewline[-4pt] 
 (13, 5, 4, 4, 1)	 & 1944767152	 & (13, 5, 5, 2, 2)	 & 2306418848
\tabularnewline[-4pt] 
 (13, 5, 5, 3, 1)	 & 464696832	 & (13, 5, 5, 4, 0)	 & 1423616
\tabularnewline[-4pt] 
 (13, 6, 3, 3, 2)	 & 2322325968	 & (13, 6, 4, 2, 2)	 & 529392832
\tabularnewline[-4pt] 
 (13, 6, 4, 3, 1)	 & 100919904	 & (13, 6, 4, 4, 0)	 & 234048
\tabularnewline[-4pt] 
 (13, 6, 5, 2, 1)	 & 3222112	 & (13, 6, 5, 3, 0)	 & 30624
\tabularnewline[-4pt] 
 (13, 6, 6, 1, 1)	 & 112	 & (13, 7, 3, 2, 2)	 & 3222112
\tabularnewline[-4pt] 
 (13, 7, 3, 3, 1)	 & 434688	 & (13, 7, 4, 2, 1)	 & 45408
\tabularnewline[-4pt] 
(13, 7, 4, 3, 0)	 & 112	 &  	 &  
\tablepostambleInstantonNumbers \vskip-18pt 

\tablepreambleInstantonNumbers{28}
 (6, 6, 6, 5, 5)	 & 327684614387349299961738768	 & (6, 6, 6, 6, 4)	 & 161823659616827892042946656
\tabularnewline[-4pt] 
 (7, 6, 5, 5, 5)	 & 167783614906668761262716784	 & (7, 6, 6, 5, 4)	 & 82579048510474932784060128
\tabularnewline[-4pt] 
 (7, 6, 6, 6, 3)	 & 18312063250016785426332456	 & (7, 7, 5, 5, 4)	 & 41952996407118579100732512
\tabularnewline[-4pt] 
 (7, 7, 6, 4, 4)	 & 20512359802778934105982824	 & (7, 7, 6, 5, 3)	 & 9244971964576432583359680
\tabularnewline[-4pt] 
 (7, 7, 6, 6, 2)	 & 760089538306826431748976	 & (7, 7, 7, 4, 3)	 & 2248511154703665686900736
\tabularnewline[-4pt] 
 (7, 7, 7, 5, 2)	 & 378607952063724321532320	 & (7, 7, 7, 6, 1)	 & 7483863739103052384864
\tabularnewline[-4pt] 
 (7, 7, 7, 7, 0)	 & 3704581973944705776	 & (8, 5, 5, 5, 5)	 & 43981223391578028025767312
\tabularnewline[-4pt] 
 (8, 6, 5, 5, 4)	 & 21480065424682369924330608	 & (8, 6, 6, 4, 4)	 & 10456756549205905359458304
\tabularnewline[-4pt] 
 (8, 6, 6, 5, 3)	 & 4698818465684165361643776	 & (8, 6, 6, 6, 2)	 & 382713191877285462148512
\tabularnewline[-4pt] 
 (8, 7, 5, 4, 4)	 & 5236870677368742358091328	 & (8, 7, 5, 5, 3)	 & 2346380263735039996887072
\tabularnewline[-4pt] 
 (8, 7, 6, 4, 3)	 & 1130432594965350429319440	 & (8, 7, 6, 5, 2)	 & 189389511540055226332992
\tabularnewline[-4pt] 
 (8, 7, 6, 6, 1)	 & 3705422029800182188416	 & (8, 7, 7, 3, 3)	 & 118167437729899245359856
\tabularnewline[-4pt] 
 (8, 7, 7, 4, 2)	 & 44276794310053780437840	 & (8, 7, 7, 5, 1)	 & 1805398290592460769984
\tabularnewline[-4pt] 
 (8, 7, 7, 6, 0)	 & 1810611871504105272	 & (8, 8, 4, 4, 4)	 & 624198973773146079716784
\tabularnewline[-4pt] 
 (8, 8, 5, 4, 3)	 & 276957561140740817672064	 & (8, 8, 5, 5, 2)	 & 45903620873883847141776
\tabularnewline[-4pt] 
 (8, 8, 6, 3, 3)	 & 57992334650190556570488	 & (8, 8, 6, 4, 2)	 & 21685078061596939766784
\tabularnewline[-4pt] 
 (8, 8, 6, 5, 1)	 & 878968309413423252864	 & (8, 8, 6, 6, 0)	 & 875827020273329664
\tabularnewline[-4pt] 
 (8, 8, 7, 3, 2)	 & 2125395706871678176416	 & (8, 8, 7, 4, 1)	 & 196113949182715052160
\tabularnewline[-4pt] 
 (8, 8, 7, 5, 0)	 & 419093788958668992	 & (8, 8, 8, 2, 2)	 & 33641794610965862400
\tabularnewline[-4pt] 
 (8, 8, 8, 3, 1)	 & 8424735061429730304	 & (8, 8, 8, 4, 0)	 & 42801528135993600
\tabularnewline[-4pt] 
 (9, 5, 5, 5, 4)	 & 2754129399126265116578688	 & (9, 6, 5, 4, 4)	 & 1321756163879149610446896
\tabularnewline[-4pt] 
 (9, 6, 5, 5, 3)	 & 588017185467849919723392	 & (9, 6, 6, 4, 3)	 & 280190865205345515202656
\tabularnewline[-4pt] 
 (9, 6, 6, 5, 2)	 & 46407141063611686053552	 & (9, 6, 6, 6, 1)	 & 886526887354737645072
\tabularnewline[-4pt] 
 (9, 7, 4, 4, 4)	 & 308970407090480886905664	 & (9, 7, 5, 4, 3)	 & 136517982401519160854256
\tabularnewline[-4pt] 
 (9, 7, 5, 5, 2)	 & 22480410038381810846784	 & (9, 7, 6, 3, 3)	 & 28271565861485256037152
\tabularnewline[-4pt] 
 (9, 7, 6, 4, 2)	 & 10547216287172185866240	 & (9, 7, 6, 5, 1)	 & 424552799703766464912
\tabularnewline[-4pt] 
 (9, 7, 6, 6, 0)	 & 419093788958668992	 & (9, 7, 7, 3, 2)	 & 1010995377658903433280
\tabularnewline[-4pt] 
 (9, 7, 7, 4, 1)	 & 93022849378461001968	 & (9, 7, 7, 5, 0)	 & 198280729061595552
\tabularnewline[-4pt] 
 (9, 8, 4, 4, 3)	 & 14880953729756521482240	 & (9, 8, 5, 3, 3)	 & 6470374366380782830464
\tabularnewline[-4pt] 
 (9, 8, 5, 4, 2)	 & 2401953482064409297584	 & (9, 8, 5, 5, 1)	 & 95308266595738550640
\tabularnewline[-4pt] 
 (9, 8, 6, 3, 2)	 & 475191508986206197632	 & (9, 8, 6, 4, 1)	 & 43589536992230477208
\tabularnewline[-4pt] 
 (9, 8, 6, 5, 0)	 & 92700939550359360	 & (9, 8, 7, 2, 2)	 & 15318198529776992064
\tabularnewline[-4pt] 
 (9, 8, 7, 3, 1)	 & 3833600055583272480	 & (9, 8, 7, 4, 0)	 & 19503820669876800
\tabularnewline[-4pt] 
 (9, 8, 8, 2, 1)	 & 48279338403693048	 & (9, 8, 8, 3, 0)	 & 725912434085952
\tabularnewline[-4pt] 
 (9, 9, 4, 3, 3)	 & 297520326374626846896	 & (9, 9, 4, 4, 2)	 & 109036439304691948608
\tabularnewline[-4pt] 
 (9, 9, 5, 3, 2)	 & 45945931585469072928	 & (9, 9, 5, 4, 1)	 & 4165834565805729216
\tabularnewline[-4pt] 
 (9, 9, 5, 5, 0)	 & 8765016259161504	 & (9, 9, 6, 2, 2)	 & 3046779784226901072
\tabularnewline[-4pt] 
 (9, 9, 6, 3, 1)	 & 761207182511922096	 & (9, 9, 6, 4, 0)	 & 3881643757375656
\tabularnewline[-4pt] 
 (9, 9, 7, 2, 1)	 & 20278800720533664	 & (9, 9, 7, 3, 0)	 & 305922925426848
\tabularnewline[-4pt] 
 (9, 9, 8, 1, 1)	 & 42951164308896	 & (9, 9, 8, 2, 0)	 & 2981800050480
\tabularnewline[-4pt] 
 (9, 9, 9, 1, 0)	 & 1272585120	 & (10, 5, 5, 4, 4)	 & 76121903698269498879600
\tabularnewline[-4pt] 
 (10, 5, 5, 5, 3)	 & 33298242026156998722144	 & (10, 6, 4, 4, 4)	 & 35504946591945154063104
\tabularnewline[-4pt] 
 (10, 6, 5, 4, 3)	 & 15467950978366663032576	 & (10, 6, 5, 5, 2)	 & 2490930240945503131824
\tabularnewline[-4pt] 
 (10, 6, 6, 3, 3)	 & 3087105410543684178144	 & (10, 6, 6, 4, 2)	 & 1142319760025546317200
\tabularnewline[-4pt] 
 (10, 6, 6, 5, 1)	 & 44858935490060258472	 & (10, 6, 6, 6, 0)	 & 42801528135993600
\tabularnewline[-4pt] 
 (10, 7, 4, 4, 3)	 & 3345260435684206623648	 & (10, 7, 5, 3, 3)	 & 1437664544143525963632
\tabularnewline[-4pt] 
 (10, 7, 5, 4, 2)	 & 530380950844598802240	 & (10, 7, 5, 5, 1)	 & 20644575230653895136
\tabularnewline[-4pt] 
 (10, 7, 6, 3, 2)	 & 101657485900434092424	 & (10, 7, 6, 4, 1)	 & 9254014284061822464
\tabularnewline[-4pt] 
 (10, 7, 6, 5, 0)	 & 19503820669876800	 & (10, 7, 7, 2, 2)	 & 3047920567708923264
\tabularnewline[-4pt] 
 (10, 7, 7, 3, 1)	 & 761479183438470384	 & (10, 7, 7, 4, 0)	 & 3881643757375656
\tabularnewline[-4pt] 
 (10, 8, 4, 3, 3)	 & 135135632721772486224	 & (10, 8, 4, 4, 2)	 & 49318322079952346112
\tabularnewline[-4pt] 
 (10, 8, 5, 3, 2)	 & 20610476713078747200	 & (10, 8, 5, 4, 1)	 & 1859138760210276768
\tabularnewline[-4pt] 
 (10, 8, 5, 5, 0)	 & 3881643757375656	 & (10, 8, 6, 2, 2)	 & 1329629977546611936
\tabularnewline[-4pt] 
 (10, 8, 6, 3, 1)	 & 331824982853181696	 & (10, 8, 6, 4, 0)	 & 1692511359568896
\tabularnewline[-4pt] 
 (10, 8, 7, 2, 1)	 & 8359186921934400	 & (10, 8, 7, 3, 0)	 & 126532108859856
\tabularnewline[-4pt] 
 (10, 8, 8, 1, 1)	 & 15746747463456	 & (10, 8, 8, 2, 0)	 & 1096632086784
\tabularnewline[-4pt] 
 (10, 9, 3, 3, 3)	 & 2006276928131711424	 & (10, 9, 4, 3, 2)	 & 713471511849776160
\tabularnewline[-4pt] 
 (10, 9, 4, 4, 1)	 & 62675569121448240	 & (10, 9, 5, 2, 2)	 & 100010833402440120
\tabularnewline[-4pt] 
 (10, 9, 5, 3, 1)	 & 24840263013168672	 & (10, 9, 5, 4, 0)	 & 126532108859856
\tabularnewline[-4pt] 
 (10, 9, 6, 2, 1)	 & 1335301022489328	 & (10, 9, 6, 3, 0)	 & 20350993239840
\tabularnewline[-4pt] 
 (10, 9, 7, 1, 1)	 & 5601159429504	 & (10, 9, 7, 2, 0)	 & 391409808576
\tabularnewline[-4pt] 
 (10, 9, 8, 1, 0)	 & 347078520	 & (10, 9, 9, 0, 0)	 & 24
\tabularnewline[-4pt] 
 (10, 10, 3, 3, 2)	 & 2713101057421728	 & (10, 10, 4, 2, 2)	 & 903893653068672
\tabularnewline[-4pt] 
 (10, 10, 4, 3, 1)	 & 220840621188096	 & (10, 10, 4, 4, 0)	 & 1096632086784
\tabularnewline[-4pt] 
 (10, 10, 5, 2, 1)	 & 25377635878296	 & (10, 10, 5, 3, 0)	 & 391409808576
\tabularnewline[-4pt] 
 (10, 10, 6, 1, 1)	 & 203336907216	 & (10, 10, 6, 2, 0)	 & 14386855920
\tabularnewline[-4pt] 
 (10, 10, 7, 1, 0)	 & 20578560	 & (11, 5, 4, 4, 4)	 & 790863904443723569376
\tabularnewline[-4pt] 
 (11, 5, 5, 4, 3)	 & 334370838391810248432	 & (11, 5, 5, 5, 2)	 & 51265779665018065536
\tabularnewline[-4pt] 
 (11, 6, 4, 4, 3)	 & 147828049737997120632	 & (11, 6, 5, 3, 3)	 & 61757539943858380704
\tabularnewline[-4pt] 
 (11, 6, 5, 4, 2)	 & 22420496977021999680	 & (11, 6, 5, 5, 1)	 & 830212985215356432
\tabularnewline[-4pt] 
 (11, 6, 6, 3, 2)	 & 3982038442779651984	 & (11, 6, 6, 4, 1)	 & 354725791310991552
\tabularnewline[-4pt] 
 (11, 6, 6, 5, 0)	 & 725912434085952	 & (11, 7, 4, 3, 3)	 & 11689593863624674656
\tabularnewline[-4pt] 
 (11, 7, 4, 4, 2)	 & 4202606608677077184	 & (11, 7, 5, 3, 2)	 & 1706830027589928192
\tabularnewline[-4pt] 
 (11, 7, 5, 4, 1)	 & 150993571342096992	 & (11, 7, 5, 5, 0)	 & 305922925426848
\tabularnewline[-4pt] 
 (11, 7, 6, 2, 2)	 & 100108346194477248	 & (11, 7, 6, 3, 1)	 & 24863416450991904
\tabularnewline[-4pt] 
 (11, 7, 6, 4, 0)	 & 126532108859856	 & (11, 7, 7, 2, 1)	 & 515881389602064
\tabularnewline[-4pt] 
 (11, 7, 7, 3, 0)	 & 7888589144400	 & (11, 8, 3, 3, 3)	 & 346896207708697296
\tabularnewline[-4pt] 
 (11, 8, 4, 3, 2)	 & 121505012171479176	 & (11, 8, 4, 4, 1)	 & 10462960782869952
\tabularnewline[-4pt] 
 (11, 8, 5, 2, 2)	 & 16270300857476160	 & (11, 8, 5, 3, 1)	 & 4021264698687264
\tabularnewline[-4pt] 
 (11, 8, 5, 4, 0)	 & 20350993239840	 & (11, 8, 6, 2, 1)	 & 194378107421760
\tabularnewline[-4pt] 
 (11, 8, 6, 3, 0)	 & 2981800050480	 & (11, 8, 7, 1, 1)	 & 639016897824
\tabularnewline[-4pt] 
 (11, 8, 7, 2, 0)	 & 45007048752	 & (11, 8, 8, 1, 0)	 & 20578560
\tabularnewline[-4pt] 
 (11, 9, 3, 3, 2)	 & 1000740719949936	 & (11, 9, 4, 2, 2)	 & 328447354833120
\tabularnewline[-4pt] 
 (11, 9, 4, 3, 1)	 & 79804026346992	 & (11, 9, 4, 4, 0)	 & 391409808576
\tabularnewline[-4pt] 
 (11, 9, 5, 2, 1)	 & 8748592415904	 & (11, 9, 5, 3, 0)	 & 135171775392
\tabularnewline[-4pt] 
 (11, 9, 6, 1, 1)	 & 61773182400	 & (11, 9, 6, 2, 0)	 & 4392333792
\tabularnewline[-4pt] 
 (11, 9, 7, 1, 0)	 & 4326048	 & (11, 10, 3, 2, 2)	 & 377080188864
\tabularnewline[-4pt] 
 (11, 10, 3, 3, 1)	 & 85495746528	 & (11, 10, 4, 2, 1)	 & 22951602432
\tabularnewline[-4pt] 
 (11, 10, 4, 3, 0)	 & 347078520	 & (11, 10, 5, 1, 1)	 & 282674592
\tabularnewline[-4pt] 
 (11, 10, 5, 2, 0)	 & 20578560	 & (11, 10, 6, 1, 0)	 & 14496
\tabularnewline[-4pt] 
 (11, 11, 2, 2, 2)	 & 1691856	 & (11, 11, 3, 2, 1)	 & 212880
\tabularnewline[-4pt] 
 (11, 11, 3, 3, 0)	 & 1104	 & (11, 11, 4, 1, 1)	 & 1104
\tabularnewline[-4pt] 
 (11, 11, 4, 2, 0)	 & 24	 & (12, 4, 4, 4, 4)	 & 2315758601706011520
\tabularnewline[-4pt] 
 (12, 5, 4, 4, 3)	 & 920246692052672448	 & (12, 5, 5, 3, 3)	 & 362176732991882256
\tabularnewline[-4pt] 
 (12, 5, 5, 4, 2)	 & 126656377507736616	 & (12, 5, 5, 5, 1)	 & 4162140562025760
\tabularnewline[-4pt] 
 (12, 6, 4, 3, 3)	 & 145074948270672288	 & (12, 6, 4, 4, 2)	 & 50310287851264512
\tabularnewline[-4pt] 
 (12, 6, 5, 3, 2)	 & 19159936729163904	 & (12, 6, 5, 4, 1)	 & 1608297381675072
\tabularnewline[-4pt] 
 (12, 6, 5, 5, 0)	 & 2981800050480	 & (12, 6, 6, 2, 2)	 & 903893653068672
\tabularnewline[-4pt] 
 (12, 6, 6, 3, 1)	 & 220840621188096	 & (12, 6, 6, 4, 0)	 & 1096632086784
\tabularnewline[-4pt] 
 (12, 7, 3, 3, 3)	 & 8236673292611808	 & (12, 7, 4, 3, 2)	 & 2768640614245200
\tabularnewline[-4pt] 
 (12, 7, 4, 4, 1)	 & 224917616990784	 & (12, 7, 5, 2, 2)	 & 328447354833120
\tabularnewline[-4pt] 
 (12, 7, 5, 3, 1)	 & 79804026346992	 & (12, 7, 5, 4, 0)	 & 391409808576
\tabularnewline[-4pt] 
 (12, 7, 6, 2, 1)	 & 2910089695872	 & (12, 7, 6, 3, 0)	 & 45007048752
\tabularnewline[-4pt] 
 (12, 7, 7, 1, 1)	 & 4826161680	 & (12, 7, 7, 2, 0)	 & 347078520
\tabularnewline[-4pt] 
 (12, 8, 3, 3, 2)	 & 42411173392368	 & (12, 8, 4, 2, 2)	 & 13138629854976
\tabularnewline[-4pt] 
 (12, 8, 4, 3, 1)	 & 3114669545280	 & (12, 8, 4, 4, 0)	 & 14386855920
\tabularnewline[-4pt] 
 (12, 8, 5, 2, 1)	 & 285207114048	 & (12, 8, 5, 3, 0)	 & 4392333792
\tabularnewline[-4pt] 
 (12, 8, 6, 1, 1)	 & 1218252960	 & (12, 8, 6, 2, 0)	 & 88177920
\tabularnewline[-4pt] 
 (12, 8, 7, 1, 0)	 & 14496	 & (12, 9, 3, 2, 2)	 & 27607031136
\tabularnewline[-4pt] 
 (12, 9, 3, 3, 1)	 & 5950086192	 & (12, 9, 4, 2, 1)	 & 1426637712
\tabularnewline[-4pt] 
 (12, 9, 4, 3, 0)	 & 20578560	 & (12, 9, 5, 1, 1)	 & 10883712
\tabularnewline[-4pt] 
 (12, 9, 5, 2, 0)	 & 795936	 & (12, 9, 6, 1, 0)	 & 24
\tabularnewline[-4pt] 
 (12, 10, 2, 2, 2)	 & 122352	 & (12, 10, 3, 2, 1)	 & 14496
\tabularnewline[-4pt] 
 (12, 10, 3, 3, 0)	 & 24	 & (12, 10, 4, 1, 1)	 & 24
\tabularnewline[-4pt] 
 (13, 4, 4, 4, 3)	 & 363393804317664	 & (13, 5, 4, 3, 3)	 & 125365423769760
\tabularnewline[-4pt] 
 (13, 5, 4, 4, 2)	 & 39692266181304	 & (13, 5, 5, 3, 2)	 & 13073262151968
\tabularnewline[-4pt] 
 (13, 5, 5, 4, 1)	 & 931163905728	 & (13, 5, 5, 5, 0)	 & 1272585120
\tabularnewline[-4pt] 
 (13, 6, 3, 3, 3)	 & 13834674726336	 & (13, 6, 4, 3, 2)	 & 4183230238656
\tabularnewline[-4pt] 
 (13, 6, 4, 4, 1)	 & 285207114048	 & (13, 6, 5, 2, 2)	 & 366406656528
\tabularnewline[-4pt] 
 (13, 6, 5, 3, 1)	 & 83099778720	 & (13, 6, 5, 4, 0)	 & 347078520
\tabularnewline[-4pt] 
 (13, 6, 6, 2, 1)	 & 1426637712	 & (13, 6, 6, 3, 0)	 & 20578560
\tabularnewline[-4pt] 
 (13, 7, 3, 3, 2)	 & 105371446464	 & (13, 7, 4, 2, 2)	 & 27607031136
\tabularnewline[-4pt] 
 (13, 7, 4, 3, 1)	 & 5950086192	 & (13, 7, 4, 4, 0)	 & 20578560
\tabularnewline[-4pt] 
 (13, 7, 5, 2, 1)	 & 316997280	 & (13, 7, 5, 3, 0)	 & 4326048
\tabularnewline[-4pt] 
 (13, 7, 6, 1, 1)	 & 212880	 & (13, 7, 6, 2, 0)	 & 14496
\tabularnewline[-4pt] 
 (13, 8, 3, 2, 2)	 & 63576576	 & (13, 8, 3, 3, 1)	 & 10883712
\tabularnewline[-4pt] 
 (13, 8, 4, 2, 1)	 & 1691856	 & (13, 8, 4, 3, 0)	 & 14496
\tabularnewline[-4pt] 
 (13, 8, 5, 1, 1)	 & 1104	 & (13, 8, 5, 2, 0)	 & 24
\tabularnewline[-4pt] 
 (14, 4, 4, 3, 3)	 & 347078520	 & (14, 4, 4, 4, 2)	 & 88177920
\tabularnewline[-4pt] 
 (14, 5, 3, 3, 3)	 & 88179456	 & (14, 5, 4, 3, 2)	 & 20578560
\tabularnewline[-4pt] 
 (14, 5, 4, 4, 1)	 & 795936	 & (14, 5, 5, 2, 2)	 & 795936
\tabularnewline[-4pt] 
 (14, 5, 5, 3, 1)	 & 122448	 & (14, 5, 5, 4, 0)	 & 24
\tabularnewline[-4pt] 
 (14, 6, 3, 3, 2)	 & 795936	 & (14, 6, 4, 2, 2)	 & 122352
\tabularnewline[-4pt] 
 (14, 6, 4, 3, 1)	 & 14496	 & (14, 6, 5, 2, 1)	 & 24
\tabularnewline[-4pt] 
(14, 7, 3, 2, 2)	 & 24	 &  	 &  
\tablepostambleInstantonNumbers \vskip-18pt 

\tablepreambleInstantonNumbers{29}
 (6, 6, 6, 6, 5)	 & 4300779721074151241480884704	 & (7, 6, 6, 5, 5)	 & 2240812589775895583844156576
\tabularnewline[-4pt] 
 (7, 6, 6, 6, 4)	 & 1118592262447208419494700224	 & (7, 7, 5, 5, 5)	 & 1163393252471836868861786112
\tabularnewline[-4pt] 
 (7, 7, 6, 5, 4)	 & 579083249672629145095030368	 & (7, 7, 6, 6, 3)	 & 130956575256689626362575040
\tabularnewline[-4pt] 
 (7, 7, 7, 4, 4)	 & 147818363038488716192722368	 & (7, 7, 7, 5, 3)	 & 67179875933573491930042368
\tabularnewline[-4pt] 
 (7, 7, 7, 6, 2)	 & 5668780293104872727438208	 & (7, 7, 7, 7, 1)	 & 58444196223515692468224
\tabularnewline[-4pt] 
 (8, 6, 5, 5, 5)	 & 611210312904590985126412992	 & (8, 6, 6, 5, 4)	 & 303193957695295574167972368
\tabularnewline[-4pt] 
 (8, 6, 6, 6, 3)	 & 68161114972295875241706048	 & (8, 7, 5, 5, 4)	 & 155642883145653607449683376
\tabularnewline[-4pt] 
 (8, 7, 6, 4, 4)	 & 76752707872006994882755104	 & (8, 7, 6, 5, 3)	 & 34795170486545493534538224
\tabularnewline[-4pt] 
 (8, 7, 6, 6, 2)	 & 2913373653270128828358720	 & (8, 7, 7, 4, 3)	 & 8644490285282936739547776
\tabularnewline[-4pt] 
 (8, 7, 7, 5, 2)	 & 1469714544745529210370240	 & (8, 7, 7, 6, 1)	 & 29632320884127789040512
\tabularnewline[-4pt] 
 (8, 7, 7, 7, 0)	 & 15060587910821007264	 & (8, 8, 5, 4, 4)	 & 20081002987269237834527544
\tabularnewline[-4pt] 
 (8, 8, 5, 5, 3)	 & 9056138605664527122727872	 & (8, 8, 6, 4, 3)	 & 4407669387277685669560512
\tabularnewline[-4pt] 
 (8, 8, 6, 5, 2)	 & 746119098679547758175328	 & (8, 8, 6, 6, 1)	 & 14913192160928967502416
\tabularnewline[-4pt] 
 (8, 8, 7, 3, 3)	 & 475983733946283470184480	 & (8, 8, 7, 4, 2)	 & 179029959278599331183904
\tabularnewline[-4pt] 
 (8, 8, 7, 5, 1)	 & 7382355219180836928000	 & (8, 8, 7, 6, 0)	 & 7505020393627384992
\tabularnewline[-4pt] 
 (8, 8, 8, 3, 2)	 & 9106787701513392933312	 & (8, 8, 8, 4, 1)	 & 844343421475958442384
\tabularnewline[-4pt] 
 (8, 8, 8, 5, 0)	 & 1810611871504105272	 & (9, 5, 5, 5, 5)	 & 85113274584443674676815872
\tabularnewline[-4pt] 
 (9, 6, 5, 5, 4)	 & 41753519869402757126310048	 & (9, 6, 6, 4, 4)	 & 20420265705357860466725736
\tabularnewline[-4pt] 
 (9, 6, 6, 5, 3)	 & 9205492921536158089178400	 & (9, 6, 6, 6, 2)	 & 757338998385448742259408
\tabularnewline[-4pt] 
 (9, 7, 5, 4, 4)	 & 10288587378544353628959936	 & (9, 7, 5, 5, 3)	 & 4625435952934481006642688
\tabularnewline[-4pt] 
 (9, 7, 6, 4, 3)	 & 2240131001946976373134464	 & (9, 7, 6, 5, 2)	 & 377337473299364349892128
\tabularnewline[-4pt] 
 (9, 7, 6, 6, 1)	 & 7465381920406413655872	 & (9, 7, 7, 3, 3)	 & 238119849054264183598080
\tabularnewline[-4pt] 
 (9, 7, 7, 4, 2)	 & 89401881734473077280272	 & (9, 7, 7, 5, 1)	 & 3667019460175374709248
\tabularnewline[-4pt] 
 (9, 7, 7, 6, 0)	 & 3704581973944705776	 & (9, 8, 4, 4, 4)	 & 1250148843664649970662352
\tabularnewline[-4pt] 
 (9, 8, 5, 4, 3)	 & 556910170480793670482400	 & (9, 8, 5, 5, 2)	 & 92873867581256305971840
\tabularnewline[-4pt] 
 (9, 8, 6, 3, 3)	 & 117846752322559629255984	 & (9, 8, 6, 4, 2)	 & 44162182140827544636144
\tabularnewline[-4pt] 
 (9, 8, 6, 5, 1)	 & 1801675431551746250880	 & (9, 8, 6, 6, 0)	 & 1810611871504105272
\tabularnewline[-4pt] 
 (9, 8, 7, 3, 2)	 & 4421119882261082749248	 & (9, 8, 7, 4, 1)	 & 408994002687679224816
\tabularnewline[-4pt] 
 (9, 8, 7, 5, 0)	 & 875827020316064256	 & (9, 8, 8, 2, 2)	 & 72939895256309497680
\tabularnewline[-4pt] 
 (9, 8, 8, 3, 1)	 & 18275815076572138848	 & (9, 8, 8, 4, 0)	 & 92700939550359360
\tabularnewline[-4pt] 
 (9, 9, 4, 4, 3)	 & 30907191740331009622272	 & (9, 9, 5, 3, 3)	 & 13510524682548173641728
\tabularnewline[-4pt] 
 (9, 9, 5, 4, 2)	 & 5029248618061980310752	 & (9, 9, 5, 5, 1)	 & 201260631020776796160
\tabularnewline[-4pt] 
 (9, 9, 6, 3, 2)	 & 1009523697408684687360	 & (9, 9, 6, 4, 1)	 & 92904071198229763248
\tabularnewline[-4pt] 
 (9, 9, 6, 5, 0)	 & 198280729061595552	 & (9, 9, 7, 2, 2)	 & 33639461017770972672
\tabularnewline[-4pt] 
 (9, 9, 7, 3, 1)	 & 8424230964009345024	 & (9, 9, 7, 4, 0)	 & 42801528146793216
\tabularnewline[-4pt] 
 (9, 9, 8, 2, 1)	 & 112932475851555408	 & (9, 9, 8, 3, 0)	 & 1692511362069504
\tabularnewline[-4pt] 
 (9, 9, 9, 1, 1)	 & 113958894140160	 & (9, 9, 9, 2, 0)	 & 7888589144400
\tabularnewline[-4pt] 
 (10, 5, 5, 5, 4)	 & 2754129399126265116578688	 & (10, 6, 5, 4, 4)	 & 1321756163879149610446896
\tabularnewline[-4pt] 
 (10, 6, 5, 5, 3)	 & 588017185467849919723392	 & (10, 6, 6, 4, 3)	 & 280190865205345515202656
\tabularnewline[-4pt] 
 (10, 6, 6, 5, 2)	 & 46407141063611686053552	 & (10, 6, 6, 6, 1)	 & 886526887354737645072
\tabularnewline[-4pt] 
 (10, 7, 4, 4, 4)	 & 308970407090480886905664	 & (10, 7, 5, 4, 3)	 & 136517982401519160854256
\tabularnewline[-4pt] 
 (10, 7, 5, 5, 2)	 & 22480410038381810846784	 & (10, 7, 6, 3, 3)	 & 28271565861485256037152
\tabularnewline[-4pt] 
 (10, 7, 6, 4, 2)	 & 10547216287172185866240	 & (10, 7, 6, 5, 1)	 & 424552799703766464912
\tabularnewline[-4pt] 
 (10, 7, 6, 6, 0)	 & 419093788958668992	 & (10, 7, 7, 3, 2)	 & 1010995377658903433280
\tabularnewline[-4pt] 
 (10, 7, 7, 4, 1)	 & 93022849378461001968	 & (10, 7, 7, 5, 0)	 & 198280729061595552
\tabularnewline[-4pt] 
 (10, 8, 4, 4, 3)	 & 14880953729756521482240	 & (10, 8, 5, 3, 3)	 & 6470374366380782830464
\tabularnewline[-4pt] 
 (10, 8, 5, 4, 2)	 & 2401953482064409297584	 & (10, 8, 5, 5, 1)	 & 95308266595738550640
\tabularnewline[-4pt] 
 (10, 8, 6, 3, 2)	 & 475191508986206197632	 & (10, 8, 6, 4, 1)	 & 43589536992230477208
\tabularnewline[-4pt] 
 (10, 8, 6, 5, 0)	 & 92700939550359360	 & (10, 8, 7, 2, 2)	 & 15318198529776992064
\tabularnewline[-4pt] 
 (10, 8, 7, 3, 1)	 & 3833600055583272480	 & (10, 8, 7, 4, 0)	 & 19503820669876800
\tabularnewline[-4pt] 
 (10, 8, 8, 2, 1)	 & 48279338403693048	 & (10, 8, 8, 3, 0)	 & 725912434085952
\tabularnewline[-4pt] 
 (10, 9, 4, 3, 3)	 & 297520326374626846896	 & (10, 9, 4, 4, 2)	 & 109036439304691948608
\tabularnewline[-4pt] 
 (10, 9, 5, 3, 2)	 & 45945931585469072928	 & (10, 9, 5, 4, 1)	 & 4165834565805729216
\tabularnewline[-4pt] 
 (10, 9, 5, 5, 0)	 & 8765016259161504	 & (10, 9, 6, 2, 2)	 & 3046779784226901072
\tabularnewline[-4pt] 
 (10, 9, 6, 3, 1)	 & 761207182511922096	 & (10, 9, 6, 4, 0)	 & 3881643757375656
\tabularnewline[-4pt] 
 (10, 9, 7, 2, 1)	 & 20278800720533664	 & (10, 9, 7, 3, 0)	 & 305922925426848
\tabularnewline[-4pt] 
 (10, 9, 8, 1, 1)	 & 42951164308896	 & (10, 9, 8, 2, 0)	 & 2981800050480
\tabularnewline[-4pt] 
 (10, 9, 9, 1, 0)	 & 1272585120	 & (10, 10, 3, 3, 3)	 & 2006276928131711424
\tabularnewline[-4pt] 
 (10, 10, 4, 3, 2)	 & 713471511849776160	 & (10, 10, 4, 4, 1)	 & 62675569121448240
\tabularnewline[-4pt] 
 (10, 10, 5, 2, 2)	 & 100010833402440120	 & (10, 10, 5, 3, 1)	 & 24840263013168672
\tabularnewline[-4pt] 
 (10, 10, 5, 4, 0)	 & 126532108859856	 & (10, 10, 6, 2, 1)	 & 1335301022489328
\tabularnewline[-4pt] 
 (10, 10, 6, 3, 0)	 & 20350993239840	 & (10, 10, 7, 1, 1)	 & 5601159429504
\tabularnewline[-4pt] 
 (10, 10, 7, 2, 0)	 & 391409808576	 & (10, 10, 8, 1, 0)	 & 347078520
\tabularnewline[-4pt] 
 (10, 10, 9, 0, 0)	 & 24	 & (11, 5, 5, 4, 4)	 & 36602428260502812573792
\tabularnewline[-4pt] 
 (11, 5, 5, 5, 3)	 & 15930480413967177684480	 & (11, 6, 4, 4, 4)	 & 16950676810888359150336
\tabularnewline[-4pt] 
 (11, 6, 5, 4, 3)	 & 7345251761305389562560	 & (11, 6, 5, 5, 2)	 & 1172715223879828113648
\tabularnewline[-4pt] 
 (11, 6, 6, 3, 3)	 & 1445782834458789325920	 & (11, 6, 6, 4, 2)	 & 533243466879375407808
\tabularnewline[-4pt] 
 (11, 6, 6, 5, 1)	 & 20734174826253969312	 & (11, 6, 6, 6, 0)	 & 19503820669876800
\tabularnewline[-4pt] 
 (11, 7, 4, 4, 3)	 & 1560763765722117846528	 & (11, 7, 5, 3, 3)	 & 666467844013257615360
\tabularnewline[-4pt] 
 (11, 7, 5, 4, 2)	 & 245004909605415502560	 & (11, 7, 5, 5, 1)	 & 9433084896265973760
\tabularnewline[-4pt] 
 (11, 7, 6, 3, 2)	 & 46154206945493038080	 & (11, 7, 6, 4, 1)	 & 4182469007721935136
\tabularnewline[-4pt] 
 (11, 7, 6, 5, 0)	 & 8765016259161504	 & (11, 7, 7, 2, 2)	 & 1329872815417735680
\tabularnewline[-4pt] 
 (11, 7, 7, 3, 1)	 & 331877990439469056	 & (11, 7, 7, 4, 0)	 & 1692511362069504
\tabularnewline[-4pt] 
 (11, 8, 4, 3, 3)	 & 60590920000179493056	 & (11, 8, 4, 4, 2)	 & 22012186784542835520
\tabularnewline[-4pt] 
 (11, 8, 5, 3, 2)	 & 9117897040377080832	 & (11, 8, 5, 4, 1)	 & 817746667654917168
\tabularnewline[-4pt] 
 (11, 8, 5, 5, 0)	 & 1692511362069504	 & (11, 8, 6, 2, 2)	 & 571128202199454336
\tabularnewline[-4pt] 
 (11, 8, 6, 3, 1)	 & 142342287006477504	 & (11, 8, 6, 4, 0)	 & 725912434085952
\tabularnewline[-4pt] 
 (11, 8, 7, 2, 1)	 & 3377194221012096	 & (11, 8, 7, 3, 0)	 & 51294957112992
\tabularnewline[-4pt] 
 (11, 8, 8, 1, 1)	 & 5601159429504	 & (11, 8, 8, 2, 0)	 & 391409808576
\tabularnewline[-4pt] 
 (11, 9, 3, 3, 3)	 & 841539378868429824	 & (11, 9, 4, 3, 2)	 & 297115911452589936
\tabularnewline[-4pt] 
 (11, 9, 4, 4, 1)	 & 25857038420140320	 & (11, 9, 5, 2, 2)	 & 40746789567213888
\tabularnewline[-4pt] 
 (11, 9, 5, 3, 1)	 & 10097809547695104	 & (11, 9, 5, 4, 0)	 & 51294957112992
\tabularnewline[-4pt] 
 (11, 9, 6, 2, 1)	 & 515881389602064	 & (11, 9, 6, 3, 0)	 & 7888589144400
\tabularnewline[-4pt] 
 (11, 9, 7, 1, 1)	 & 1927069671936	 & (11, 9, 7, 2, 0)	 & 135171775392
\tabularnewline[-4pt] 
 (11, 9, 8, 1, 0)	 & 88179456	 & (11, 10, 3, 3, 2)	 & 1000740719949936
\tabularnewline[-4pt] 
 (11, 10, 4, 2, 2)	 & 328447354833120	 & (11, 10, 4, 3, 1)	 & 79804026346992
\tabularnewline[-4pt] 
 (11, 10, 4, 4, 0)	 & 391409808576	 & (11, 10, 5, 2, 1)	 & 8748592415904
\tabularnewline[-4pt] 
 (11, 10, 5, 3, 0)	 & 135171775392	 & (11, 10, 6, 1, 1)	 & 61773182400
\tabularnewline[-4pt] 
 (11, 10, 6, 2, 0)	 & 4392333792	 & (11, 10, 7, 1, 0)	 & 4326048
\tabularnewline[-4pt] 
 (11, 11, 3, 2, 2)	 & 105371446464	 & (11, 11, 3, 3, 1)	 & 23351460864
\tabularnewline[-4pt] 
 (11, 11, 4, 2, 1)	 & 5950086192	 & (11, 11, 4, 3, 0)	 & 88179456
\tabularnewline[-4pt] 
 (11, 11, 5, 1, 1)	 & 59097600	 & (11, 11, 5, 2, 0)	 & 4326048
\tabularnewline[-4pt] 
 (11, 11, 6, 1, 0)	 & 1104	 & (12, 5, 4, 4, 4)	 & 159832960277398698312
\tabularnewline[-4pt] 
 (12, 5, 5, 4, 3)	 & 66573482065327669440	 & (12, 5, 5, 5, 2)	 & 9952370045915290464
\tabularnewline[-4pt] 
 (12, 6, 4, 4, 3)	 & 28814753795787304128	 & (12, 6, 5, 3, 3)	 & 11833136668383611040
\tabularnewline[-4pt] 
 (12, 6, 5, 4, 2)	 & 4252005327651223776	 & (12, 6, 5, 5, 1)	 & 152435152838866176
\tabularnewline[-4pt] 
 (12, 6, 6, 3, 2)	 & 721163569257189312	 & (12, 6, 6, 4, 1)	 & 63276065657309280
\tabularnewline[-4pt] 
 (12, 6, 6, 5, 0)	 & 126532108859856	 & (12, 7, 4, 3, 3)	 & 2124595552827372432
\tabularnewline[-4pt] 
 (12, 7, 4, 4, 2)	 & 754387255278771840	 & (12, 7, 5, 3, 2)	 & 299477728365291600
\tabularnewline[-4pt] 
 (12, 7, 5, 4, 1)	 & 26040136828870752	 & (12, 7, 5, 5, 0)	 & 51294957112992
\tabularnewline[-4pt] 
 (12, 7, 6, 2, 2)	 & 16270300857476160	 & (12, 7, 6, 3, 1)	 & 4021264698687264
\tabularnewline[-4pt] 
 (12, 7, 6, 4, 0)	 & 20350993239840	 & (12, 7, 7, 2, 1)	 & 71274491245200
\tabularnewline[-4pt] 
 (12, 7, 7, 3, 0)	 & 1096632180480	 & (12, 8, 3, 3, 3)	 & 55724768553096576
\tabularnewline[-4pt] 
 (12, 8, 4, 3, 2)	 & 19159936729163904	 & (12, 8, 4, 4, 1)	 & 1608297381675072
\tabularnewline[-4pt] 
 (12, 8, 5, 2, 2)	 & 2428815576573408	 & (12, 8, 5, 3, 1)	 & 596073535387056
\tabularnewline[-4pt] 
 (12, 8, 5, 4, 0)	 & 2981800050480	 & (12, 8, 6, 2, 1)	 & 25377635878296
\tabularnewline[-4pt] 
 (12, 8, 6, 3, 0)	 & 391409808576	 & (12, 8, 7, 1, 1)	 & 61773182400
\tabularnewline[-4pt] 
 (12, 8, 7, 2, 0)	 & 4392333792	 & (12, 8, 8, 1, 0)	 & 795936
\tabularnewline[-4pt] 
 (12, 9, 3, 3, 2)	 & 125365423769760	 & (12, 9, 4, 2, 2)	 & 39692266181304
\tabularnewline[-4pt] 
 (12, 9, 4, 3, 1)	 & 9502910875584	 & (12, 9, 4, 4, 0)	 & 45007048752
\tabularnewline[-4pt] 
 (12, 9, 5, 2, 1)	 & 931163905728	 & (12, 9, 5, 3, 0)	 & 14386869840
\tabularnewline[-4pt] 
 (12, 9, 6, 1, 1)	 & 4826161680	 & (12, 9, 6, 2, 0)	 & 347078520
\tabularnewline[-4pt] 
 (12, 9, 7, 1, 0)	 & 122448	 & (12, 10, 3, 2, 2)	 & 27607031136
\tabularnewline[-4pt] 
 (12, 10, 3, 3, 1)	 & 5950086192	 & (12, 10, 4, 2, 1)	 & 1426637712
\tabularnewline[-4pt] 
 (12, 10, 4, 3, 0)	 & 20578560	 & (12, 10, 5, 1, 1)	 & 10883712
\tabularnewline[-4pt] 
 (12, 10, 5, 2, 0)	 & 795936	 & (12, 10, 6, 1, 0)	 & 24
\tabularnewline[-4pt] 
 (12, 11, 2, 2, 2)	 & 14496	 & (12, 11, 3, 2, 1)	 & 1104
\tabularnewline[-4pt] 
 (13, 4, 4, 4, 4)	 & 149583407202367176	 & (13, 5, 4, 4, 3)	 & 57309129620711136
\tabularnewline[-4pt] 
 (13, 5, 5, 3, 3)	 & 21671962905320448	 & (13, 5, 5, 4, 2)	 & 7371081117191712
\tabularnewline[-4pt] 
 (13, 5, 5, 5, 1)	 & 221145135246336	 & (13, 6, 4, 3, 3)	 & 8236673292611808
\tabularnewline[-4pt] 
 (13, 6, 4, 4, 2)	 & 2768640614245200	 & (13, 6, 5, 3, 2)	 & 1000740719949936
\tabularnewline[-4pt] 
 (13, 6, 5, 4, 1)	 & 79804026346992	 & (13, 6, 5, 5, 0)	 & 135171775392
\tabularnewline[-4pt] 
 (13, 6, 6, 2, 2)	 & 39360165257928	 & (13, 6, 6, 3, 1)	 & 9425697295296
\tabularnewline[-4pt] 
 (13, 6, 6, 4, 0)	 & 45007048752	 & (13, 7, 3, 3, 3)	 & 389973010495488
\tabularnewline[-4pt] 
 (13, 7, 4, 3, 2)	 & 125365423769760	 & (13, 7, 4, 4, 1)	 & 9502910875584
\tabularnewline[-4pt] 
 (13, 7, 5, 2, 2)	 & 13073262151968	 & (13, 7, 5, 3, 1)	 & 3100342138368
\tabularnewline[-4pt] 
 (13, 7, 5, 4, 0)	 & 14386869840	 & (13, 7, 6, 2, 1)	 & 83099778720
\tabularnewline[-4pt] 
 (13, 7, 6, 3, 0)	 & 1272585120	 & (13, 7, 7, 1, 1)	 & 59097600
\tabularnewline[-4pt] 
 (13, 7, 7, 2, 0)	 & 4326048	 & (13, 8, 3, 3, 2)	 & 1326841710624
\tabularnewline[-4pt] 
 (13, 8, 4, 2, 2)	 & 377080188864	 & (13, 8, 4, 3, 1)	 & 85495746528
\tabularnewline[-4pt] 
 (13, 8, 4, 4, 0)	 & 347078520	 & (13, 8, 5, 2, 1)	 & 5950086192
\tabularnewline[-4pt] 
 (13, 8, 5, 3, 0)	 & 88179456	 & (13, 8, 6, 1, 1)	 & 10883712
\tabularnewline[-4pt] 
 (13, 8, 6, 2, 0)	 & 795936	 & (13, 9, 3, 2, 2)	 & 316997280
\tabularnewline[-4pt] 
 (13, 9, 3, 3, 1)	 & 59097600	 & (13, 9, 4, 2, 1)	 & 10883712
\tabularnewline[-4pt] 
 (13, 9, 4, 3, 0)	 & 122448	 & (13, 9, 5, 1, 1)	 & 19200
\tabularnewline[-4pt] 
 (13, 9, 5, 2, 0)	 & 1104	 & (14, 4, 4, 4, 3)	 & 3114669545280
\tabularnewline[-4pt] 
 (14, 5, 4, 3, 3)	 & 980247769056	 & (14, 5, 4, 4, 2)	 & 285207114048
\tabularnewline[-4pt] 
 (14, 5, 5, 3, 2)	 & 83099778720	 & (14, 5, 5, 4, 1)	 & 4826161680
\tabularnewline[-4pt] 
 (14, 5, 5, 5, 0)	 & 4326048	 & (14, 6, 3, 3, 3)	 & 85495746528
\tabularnewline[-4pt] 
 (14, 6, 4, 3, 2)	 & 22951602432	 & (14, 6, 4, 4, 1)	 & 1218252960
\tabularnewline[-4pt] 
 (14, 6, 5, 2, 2)	 & 1426637712	 & (14, 6, 5, 3, 1)	 & 282674592
\tabularnewline[-4pt] 
 (14, 6, 5, 4, 0)	 & 795936	 & (14, 6, 6, 2, 1)	 & 1691856
\tabularnewline[-4pt] 
 (14, 6, 6, 3, 0)	 & 14496	 & (14, 7, 3, 3, 2)	 & 316997280
\tabularnewline[-4pt] 
 (14, 7, 4, 2, 2)	 & 63576576	 & (14, 7, 4, 3, 1)	 & 10883712
\tabularnewline[-4pt] 
 (14, 7, 4, 4, 0)	 & 14496	 & (14, 7, 5, 2, 1)	 & 212880
\tabularnewline[-4pt] 
 (14, 7, 5, 3, 0)	 & 1104	 & (14, 8, 3, 2, 2)	 & 14496
\tabularnewline[-4pt] 
 (14, 8, 3, 3, 1)	 & 1104	 & (14, 8, 4, 2, 1)	 & 24
\tablepostambleInstantonNumbers \vskip-18pt 

\newcommand{\tablepreambleInstantonNumbersGone}[1]{
	\vspace{-1.35cm}
	\begin{center}
		\begin{longtable}{|>{\scriptsize\centering$} p{.8in}<{$} |>{\scriptsize$}p{1.9in}<{$}
				||>{\scriptsize\centering$} p{.8in} <{~$} |>{\scriptsize$}p{1.9in}<{$} |}\hline
			\multicolumn{4}{|c|}{\vrule height 13pt depth8pt width 0pt $\text{deg}(\bm{I})=#1$}
			\tabularnewline[0.5pt] \hline \hline
			\vrule height 12pt depth5pt width 0pt $\normalsize{$\bm{I}$}$ & \hfil $\normalsize{$d_{\bm{I}}$}$ & $\normalsize{$\bm{I}$}$ & \hfil $\normalsize{$d_{\bm{I}}$}$ \tabularnewline[.5pt] \hline
			\endfirsthead
			\hline
			\multicolumn{4}{|l|}{$\text{deg}(\bm{I})=#1$, \sl continued}\tabularnewline[0.5pt] \hline
			\vrule height 12pt depth5pt width 0pt $\normalsize{$\bm{I}$}$ & \hfil $\normalsize{$d_{\bm{I}}$}$ & $\normalsize{$\bm{I}$}$ & \hfil $\normalsize{$d_{\bm{I}}$}$ \tabularnewline[0.5pt] \hline
			& & & \tabularnewline[-12pt]
			\endhead
			\hline\hline 
			\multicolumn{4}{|r|}{{\footnotesize\sl Continued on the following page}}\tabularnewline[0.5pt] \hline
			\endfoot
			\hline
			\endlastfoot}
\newpage
\subsection{Genus-1 instantons}\label{sect:genus1instantons}
\vspace*{-10pt}
\addtocounter{table}{-29}
\begin{table}[H]
\begin{center}
\capt{5in}{tab:genusoneinstantons}{The genus-one instanton numbers $d_{\bm{I}}$ for $\text{deg}(\bm{I})\leqslant 29$}
\end{center}
\end{table}
\tablepreambleInstantonNumbersGone{6}
(2, 2, 2, 0, 0)	 & $4$	 &  	 &  
\tablepostambleInstantonNumbers \vskip-18pt 

\tablepreambleInstantonNumbersGone{7}
(2, 2, 2, 1, 0)	 & $\llap{-}48$	 &  	 &  
\tablepostambleInstantonNumbers \vskip-18pt 

\tablepreambleInstantonNumbersGone{8}
 (2, 2, 2, 1, 1)	 & $528$ 	 & (2, 2, 2, 2, 0)	 & $\llap{-}2292$
\tabularnewline[-4pt] 
(3, 2, 2, 1, 0)	 & $\llap{-}48$	 &  	 &  
\tablepostambleInstantonNumbers \vskip-18pt 

\tablepreambleInstantonNumbersGone{9}
 (2, 2, 2, 2, 1)	 & $29808$ 	 & (3, 2, 2, 1, 1)	 & $928$
\tabularnewline[-4pt] 
 (3, 2, 2, 2, 0)	 & $\llap{-}5600$ 	 & (3, 3, 2, 1, 0)	 & $\llap{-}224$
\tablepostambleInstantonNumbers \vskip-18pt 

\tablepreambleInstantonNumbersGone{10}
 (2, 2, 2, 2, 2)	 & $3666312$ 	 & (3, 2, 2, 2, 1)	 & $104352$
\tabularnewline[-4pt] 
 (3, 3, 2, 1, 1)	 & $4320$ 	 & (3, 3, 2, 2, 0)	 & $\llap{-}29136$
\tabularnewline[-4pt] 
 (3, 3, 3, 1, 0)	 & $\llap{-}2208$ 	 & (4, 2, 2, 1, 1)	 & $528$
\tabularnewline[-4pt] 
 (4, 2, 2, 2, 0)	 & $\llap{-}2292$ 	 & (4, 3, 2, 1, 0)	 & $\llap{-}48$
\tablepostambleInstantonNumbers \vskip-18pt 

\tablepreambleInstantonNumbersGone{11}
 (3, 2, 2, 2, 2)	 & $22958688$ 	 & (3, 3, 2, 2, 1)	 & $679968$
\tabularnewline[-4pt] 
 (3, 3, 3, 1, 1)	 & $30720$ 	 & (3, 3, 3, 2, 0)	 & $\llap{-}251520$
\tabularnewline[-4pt] 
 (4, 2, 2, 2, 1)	 & $104352$ 	 & (4, 3, 2, 1, 1)	 & $4320$
\tabularnewline[-4pt] 
 (4, 3, 2, 2, 0)	 & $\llap{-}29136$ 	 & (4, 3, 3, 1, 0)	 & $\llap{-}2208$
\tabularnewline[-4pt] 
 (4, 4, 2, 1, 0)	 & $\llap{-}48$ 	 & (5, 2, 2, 2, 0)	 & $\llap{-}48$
\tablepostambleInstantonNumbers \vskip-18pt 

\tablepreambleInstantonNumbersGone{12}
 (3, 3, 2, 2, 2)	 & $230549312$ 	 & (3, 3, 3, 2, 1)	 & $6953664$
\tabularnewline[-4pt] 
 (3, 3, 3, 3, 0)	 & $\llap{-}3031872$ 	 & (4, 2, 2, 2, 2)	 & $40083960$
\tabularnewline[-4pt] 
 (4, 3, 2, 2, 1)	 & $1194656$ 	 & (4, 3, 3, 1, 1)	 & $42560$
\tabularnewline[-4pt] 
 (4, 3, 3, 2, 0)	 & $\llap{-}484896$ 	 & (4, 4, 2, 1, 1)	 & $10400$
\tabularnewline[-4pt] 
 (4, 4, 2, 2, 0)	 & $\llap{-}61760$ 	 & (4, 4, 3, 1, 0)	 & $\llap{-}5600$
\tabularnewline[-4pt] 
 (4, 4, 4, 0, 0)	 & $4$ 	 & (5, 2, 2, 2, 1)	 & $29808$
\tabularnewline[-4pt] 
 (5, 3, 2, 1, 1)	 & $928$ 	 & (5, 3, 2, 2, 0)	 & $\llap{-}5600$
\tabularnewline[-4pt] 
 (5, 3, 3, 1, 0)	 & $\llap{-}224$ 	 & (6, 2, 2, 2, 0)	 & $4$
\tablepostambleInstantonNumbers \vskip-18pt 

\tablepreambleInstantonNumbersGone{13}
 (3, 3, 3, 2, 2)	 & $3347625888$ 	 & (3, 3, 3, 3, 1)	 & $99761664$
\tabularnewline[-4pt] 
 (4, 3, 2, 2, 2)	 & $652777584$ 	 & (4, 3, 3, 2, 1)	 & $19494816$
\tabularnewline[-4pt] 
 (4, 3, 3, 3, 0)	 & $\llap{-}9395616$ 	 & (4, 4, 2, 2, 1)	 & $3692400$
\tabularnewline[-4pt] 
 (4, 4, 3, 1, 1)	 & $73824$ 	 & (4, 4, 3, 2, 0)	 & $\llap{-}1679040$
\tabularnewline[-4pt] 
 (4, 4, 4, 1, 0)	 & $\llap{-}29136$ 	 & (5, 2, 2, 2, 2)	 & $22958688$
\tabularnewline[-4pt] 
 (5, 3, 2, 2, 1)	 & $679968$ 	 & (5, 3, 3, 1, 1)	 & $30720$
\tabularnewline[-4pt] 
 (5, 3, 3, 2, 0)	 & $\llap{-}251520$ 	 & (5, 4, 2, 1, 1)	 & $4320$
\tabularnewline[-4pt] 
 (5, 4, 2, 2, 0)	 & $\llap{-}29136$ 	 & (5, 4, 3, 1, 0)	 & $\llap{-}2208$
\tabularnewline[-4pt] 
 (6, 2, 2, 2, 1)	 & $528$ 	 & (6, 3, 2, 2, 0)	 & $\llap{-}48$
\tablepostambleInstantonNumbers \vskip-18pt 

\tablepreambleInstantonNumbersGone{14}
 (3, 3, 3, 3, 2)	 & $65707393920$ 	 & (4, 3, 3, 2, 2)	 & $14105356368$
\tabularnewline[-4pt] 
 (4, 3, 3, 3, 1)	 & $411633120$ 	 & (4, 4, 2, 2, 2)	 & $2937953580$
\tabularnewline[-4pt] 
 (4, 4, 3, 2, 1)	 & $86694528$ 	 & (4, 4, 3, 3, 0)	 & $\llap{-}46049040$
\tabularnewline[-4pt] 
 (4, 4, 4, 1, 1)	 & $\llap{-}317232$ 	 & (4, 4, 4, 2, 0)	 & $\llap{-}9396672$
\tabularnewline[-4pt] 
 (5, 3, 2, 2, 2)	 & $652777584$ 	 & (5, 3, 3, 2, 1)	 & $19494816$
\tabularnewline[-4pt] 
 (5, 3, 3, 3, 0)	 & $\llap{-}9395616$ 	 & (5, 4, 2, 2, 1)	 & $3692400$
\tabularnewline[-4pt] 
 (5, 4, 3, 1, 1)	 & $73824$ 	 & (5, 4, 3, 2, 0)	 & $\llap{-}1679040$
\tabularnewline[-4pt] 
 (5, 4, 4, 1, 0)	 & $\llap{-}29136$ 	 & (5, 5, 2, 1, 1)	 & $4320$
\tabularnewline[-4pt] 
 (5, 5, 2, 2, 0)	 & $\llap{-}29136$ 	 & (5, 5, 3, 1, 0)	 & $\llap{-}2208$
\tabularnewline[-4pt] 
 (6, 2, 2, 2, 2)	 & $3666312$ 	 & (6, 3, 2, 2, 1)	 & $104352$
\tabularnewline[-4pt] 
 (6, 3, 3, 1, 1)	 & $4320$ 	 & (6, 3, 3, 2, 0)	 & $\llap{-}29136$
\tabularnewline[-4pt] 
 (6, 4, 2, 1, 1)	 & $528$ 	 & (6, 4, 2, 2, 0)	 & $\llap{-}2292$
\tabularnewline[-4pt] 
 (6, 4, 3, 1, 0)	 & $\llap{-}48$ 	 & (7, 2, 2, 2, 1)	 & $\llap{-}48$
\tablepostambleInstantonNumbers \vskip-18pt 

\tablepreambleInstantonNumbersGone{15}
 (3, 3, 3, 3, 3)	 & $1668835805184$ 	 & (4, 3, 3, 3, 2)	 & $385951211712$
\tabularnewline[-4pt] 
 (4, 4, 3, 2, 2)	 & $87650018048$ 	 & (4, 4, 3, 3, 1)	 & $2496782816$
\tabularnewline[-4pt] 
 (4, 4, 4, 2, 1)	 & $561090816$ 	 & (4, 4, 4, 3, 0)	 & $\llap{-}327015680$
\tabularnewline[-4pt] 
 (5, 3, 3, 2, 2)	 & $22327107072$ 	 & (5, 3, 3, 3, 1)	 & $646886400$
\tabularnewline[-4pt] 
 (5, 4, 2, 2, 2)	 & $4750051104$ 	 & (5, 4, 3, 2, 1)	 & $138982240$
\tabularnewline[-4pt] 
 (5, 4, 3, 3, 0)	 & $\llap{-}76342880$ 	 & (5, 4, 4, 1, 1)	 & $\llap{-}1114976$
\tabularnewline[-4pt] 
 (5, 4, 4, 2, 0)	 & $\llap{-}16170272$ 	 & (5, 5, 2, 2, 1)	 & $6414464$
\tabularnewline[-4pt] 
 (5, 5, 3, 1, 1)	 & $75776$ 	 & (5, 5, 3, 2, 0)	 & $\llap{-}3031872$
\tabularnewline[-4pt] 
 (5, 5, 4, 1, 0)	 & $\llap{-}61920$ 	 & (6, 3, 2, 2, 2)	 & $230549312$
\tabularnewline[-4pt] 
 (6, 3, 3, 2, 1)	 & $6953664$ 	 & (6, 3, 3, 3, 0)	 & $\llap{-}3031872$
\tabularnewline[-4pt] 
 (6, 4, 2, 2, 1)	 & $1194656$ 	 & (6, 4, 3, 1, 1)	 & $42560$
\tabularnewline[-4pt] 
 (6, 4, 3, 2, 0)	 & $\llap{-}484896$ 	 & (6, 4, 4, 1, 0)	 & $\llap{-}5600$
\tabularnewline[-4pt] 
 (6, 5, 2, 1, 1)	 & $928$ 	 & (6, 5, 2, 2, 0)	 & $\llap{-}5600$
\tabularnewline[-4pt] 
 (6, 5, 3, 1, 0)	 & $\llap{-}224$ 	 & (7, 2, 2, 2, 2)	 & $29808$
\tabularnewline[-4pt] 
 (7, 3, 2, 2, 1)	 & $928$ 	 & (7, 3, 3, 2, 0)	 & $\llap{-}224$
\tablepostambleInstantonNumbers \vskip-18pt 

\tablepreambleInstantonNumbersGone{16}
 (4, 3, 3, 3, 3)	 & $13029814091424$ 	 & (4, 4, 3, 3, 2)	 & $3154648420512$
\tabularnewline[-4pt] 
 (4, 4, 4, 2, 2)	 & $755118268080$ 	 & (4, 4, 4, 3, 1)	 & $20875131744$
\tabularnewline[-4pt] 
 (4, 4, 4, 4, 0)	 & $\llap{-}3110590260$ 	 & (5, 3, 3, 3, 2)	 & $894337855968$
\tabularnewline[-4pt] 
 (5, 4, 3, 2, 2)	 & $208350582720$ 	 & (5, 4, 3, 3, 1)	 & $5848333440$
\tabularnewline[-4pt] 
 (5, 4, 4, 2, 1)	 & $1342319904$ 	 & (5, 4, 4, 3, 0)	 & $\llap{-}824199120$
\tabularnewline[-4pt] 
 (5, 5, 2, 2, 2)	 & $12168742800$ 	 & (5, 5, 3, 2, 1)	 & $351706176$
\tabularnewline[-4pt] 
 (5, 5, 3, 3, 0)	 & $\llap{-}203310240$ 	 & (5, 5, 4, 1, 1)	 & $\llap{-}6126048$
\tabularnewline[-4pt] 
 (5, 5, 4, 2, 0)	 & $\llap{-}46049040$ 	 & (5, 5, 5, 1, 0)	 & $\llap{-}251520$
\tabularnewline[-4pt] 
 (6, 3, 3, 2, 2)	 & $14105356368$ 	 & (6, 3, 3, 3, 1)	 & $411633120$
\tabularnewline[-4pt] 
 (6, 4, 2, 2, 2)	 & $2937953580$ 	 & (6, 4, 3, 2, 1)	 & $86694528$
\tabularnewline[-4pt] 
 (6, 4, 3, 3, 0)	 & $\llap{-}46049040$ 	 & (6, 4, 4, 1, 1)	 & $\llap{-}317232$
\tabularnewline[-4pt] 
 (6, 4, 4, 2, 0)	 & $\llap{-}9396672$ 	 & (6, 5, 2, 2, 1)	 & $3692400$
\tabularnewline[-4pt] 
 (6, 5, 3, 1, 1)	 & $73824$ 	 & (6, 5, 3, 2, 0)	 & $\llap{-}1679040$
\tabularnewline[-4pt] 
 (6, 5, 4, 1, 0)	 & $\llap{-}29136$ 	 & (6, 6, 2, 1, 1)	 & $528$
\tabularnewline[-4pt] 
 (6, 6, 2, 2, 0)	 & $\llap{-}2292$ 	 & (6, 6, 3, 1, 0)	 & $\llap{-}48$
\tabularnewline[-4pt] 
 (7, 3, 2, 2, 2)	 & $22958688$ 	 & (7, 3, 3, 2, 1)	 & $679968$
\tabularnewline[-4pt] 
 (7, 3, 3, 3, 0)	 & $\llap{-}251520$ 	 & (7, 4, 2, 2, 1)	 & $104352$
\tabularnewline[-4pt] 
 (7, 4, 3, 1, 1)	 & $4320$ 	 & (7, 4, 3, 2, 0)	 & $\llap{-}29136$
\tabularnewline[-4pt] 
 (7, 4, 4, 1, 0)	 & $\llap{-}48$ 	 & (7, 5, 2, 2, 0)	 & $\llap{-}48$
\tabularnewline[-4pt] 
 (8, 2, 2, 2, 2)	 & $\llap{-}2292$ 	 & (8, 3, 2, 2, 1)	 & $\llap{-}48$
\tablepostambleInstantonNumbers \vskip-18pt 

\tablepreambleInstantonNumbersGone{17}
 (4, 4, 3, 3, 3)	 & $135453779066496$ 	 & (4, 4, 4, 3, 2)	 & $34155140507184$
\tabularnewline[-4pt] 
 (4, 4, 4, 4, 1)	 & $228415121472$ 	 & (5, 3, 3, 3, 3)	 & $41704406393856$
\tabularnewline[-4pt] 
 (5, 4, 3, 3, 2)	 & $10348372749216$ 	 & (5, 4, 4, 2, 2)	 & $2545705442112$
\tabularnewline[-4pt] 
 (5, 4, 4, 3, 1)	 & $68863079616$ 	 & (5, 4, 4, 4, 0)	 & $\llap{-}11043084816$
\tabularnewline[-4pt] 
 (5, 5, 3, 2, 2)	 & $733831612704$ 	 & (5, 5, 3, 3, 1)	 & $20194851840$
\tabularnewline[-4pt] 
 (5, 5, 4, 2, 1)	 & $4741984896$ 	 & (5, 5, 4, 3, 0)	 & $\llap{-}3110582880$
\tabularnewline[-4pt] 
 (5, 5, 5, 1, 1)	 & $\llap{-}46978560$ 	 & (5, 5, 5, 2, 0)	 & $\llap{-}203310240$
\tabularnewline[-4pt] 
 (6, 3, 3, 3, 2)	 & $894337855968$ 	 & (6, 4, 3, 2, 2)	 & $208350582720$
\tabularnewline[-4pt] 
 (6, 4, 3, 3, 1)	 & $5848333440$ 	 & (6, 4, 4, 2, 1)	 & $1342319904$
\tabularnewline[-4pt] 
 (6, 4, 4, 3, 0)	 & $\llap{-}824199120$ 	 & (6, 5, 2, 2, 2)	 & $12168742800$
\tabularnewline[-4pt] 
 (6, 5, 3, 2, 1)	 & $351706176$ 	 & (6, 5, 3, 3, 0)	 & $\llap{-}203310240$
\tabularnewline[-4pt] 
 (6, 5, 4, 1, 1)	 & $\llap{-}6126048$ 	 & (6, 5, 4, 2, 0)	 & $\llap{-}46049040$
\tabularnewline[-4pt] 
 (6, 5, 5, 1, 0)	 & $\llap{-}251520$ 	 & (6, 6, 2, 2, 1)	 & $3692400$
\tabularnewline[-4pt] 
 (6, 6, 3, 1, 1)	 & $73824$ 	 & (6, 6, 3, 2, 0)	 & $\llap{-}1679040$
\tabularnewline[-4pt] 
 (6, 6, 4, 1, 0)	 & $\llap{-}29136$ 	 & (7, 3, 3, 2, 2)	 & $3347625888$
\tabularnewline[-4pt] 
 (7, 3, 3, 3, 1)	 & $99761664$ 	 & (7, 4, 2, 2, 2)	 & $652777584$
\tabularnewline[-4pt] 
 (7, 4, 3, 2, 1)	 & $19494816$ 	 & (7, 4, 3, 3, 0)	 & $\llap{-}9395616$
\tabularnewline[-4pt] 
 (7, 4, 4, 1, 1)	 & $73824$ 	 & (7, 4, 4, 2, 0)	 & $\llap{-}1679040$
\tabularnewline[-4pt] 
 (7, 5, 2, 2, 1)	 & $679968$ 	 & (7, 5, 3, 1, 1)	 & $30720$
\tabularnewline[-4pt] 
 (7, 5, 3, 2, 0)	 & $\llap{-}251520$ 	 & (7, 5, 4, 1, 0)	 & $\llap{-}2208$
\tabularnewline[-4pt] 
 (7, 6, 2, 2, 0)	 & $\llap{-}48$ 	 & (8, 3, 2, 2, 2)	 & $104352$
\tabularnewline[-4pt] 
 (8, 3, 3, 2, 1)	 & $4320$ 	 & (8, 3, 3, 3, 0)	 & $\llap{-}2208$
\tabularnewline[-4pt] 
 (8, 4, 2, 2, 1)	 & $528$ 	 & (8, 4, 3, 2, 0)	 & $\llap{-}48$
\tablepostambleInstantonNumbers \vskip-18pt 

\tablepreambleInstantonNumbersGone{18}
 (4, 4, 4, 3, 3)	 & $1803381971700144$ 	 & (4, 4, 4, 4, 2)	 & $470537427014352$
\tabularnewline[-4pt] 
 (5, 4, 3, 3, 3)	 & $586171325733792$ 	 & (5, 4, 4, 3, 2)	 & $151342528026688$
\tabularnewline[-4pt] 
 (5, 4, 4, 4, 1)	 & $1011188967744$ 	 & (5, 5, 3, 3, 2)	 & $47693058783296$
\tabularnewline[-4pt] 
 (5, 5, 4, 2, 2)	 & $12091316695232$ 	 & (5, 5, 4, 3, 1)	 & $318330381792$
\tabularnewline[-4pt] 
 (5, 5, 4, 4, 0)	 & $\llap{-}55127514240$ 	 & (5, 5, 5, 2, 1)	 & $23350187616$
\tabularnewline[-4pt] 
 (5, 5, 5, 3, 0)	 & $\llap{-}16642969280$ 	 & (6, 3, 3, 3, 3)	 & $60862991224384$
\tabularnewline[-4pt] 
 (6, 4, 3, 3, 2)	 & $15219924472416$ 	 & (6, 4, 4, 2, 2)	 & $3775716012840$
\tabularnewline[-4pt] 
 (6, 4, 4, 3, 1)	 & $101366312448$ 	 & (6, 4, 4, 4, 0)	 & $\llap{-}16642956928$
\tabularnewline[-4pt] 
 (6, 5, 3, 2, 2)	 & $1103727042528$ 	 & (6, 5, 3, 3, 1)	 & $30154035584$
\tabularnewline[-4pt] 
 (6, 5, 4, 2, 1)	 & $7112117856$ 	 & (6, 5, 4, 3, 0)	 & $\llap{-}4775506080$
\tabularnewline[-4pt] 
 (6, 5, 5, 1, 1)	 & $\llap{-}87015936$ 	 & (6, 5, 5, 2, 0)	 & $\llap{-}327015680$
\tabularnewline[-4pt] 
 (6, 6, 2, 2, 2)	 & $19270744144$ 	 & (6, 6, 3, 2, 1)	 & $554248640$
\tabularnewline[-4pt] 
 (6, 6, 3, 3, 0)	 & $\llap{-}327015680$ 	 & (6, 6, 4, 1, 1)	 & $\llap{-}12495360$
\tabularnewline[-4pt] 
 (6, 6, 4, 2, 0)	 & $\llap{-}76341160$ 	 & (6, 6, 5, 1, 0)	 & $\llap{-}484896$
\tabularnewline[-4pt] 
 (6, 6, 6, 0, 0)	 & $4$ 	 & (7, 3, 3, 3, 2)	 & $385951211712$
\tabularnewline[-4pt] 
 (7, 4, 3, 2, 2)	 & $87650018048$ 	 & (7, 4, 3, 3, 1)	 & $2496782816$
\tabularnewline[-4pt] 
 (7, 4, 4, 2, 1)	 & $561090816$ 	 & (7, 4, 4, 3, 0)	 & $\llap{-}327015680$
\tabularnewline[-4pt] 
 (7, 5, 2, 2, 2)	 & $4750051104$ 	 & (7, 5, 3, 2, 1)	 & $138982240$
\tabularnewline[-4pt] 
 (7, 5, 3, 3, 0)	 & $\llap{-}76342880$ 	 & (7, 5, 4, 1, 1)	 & $\llap{-}1114976$
\tabularnewline[-4pt] 
 (7, 5, 4, 2, 0)	 & $\llap{-}16170272$ 	 & (7, 5, 5, 1, 0)	 & $\llap{-}61920$
\tabularnewline[-4pt] 
 (7, 6, 2, 2, 1)	 & $1194656$ 	 & (7, 6, 3, 1, 1)	 & $42560$
\tabularnewline[-4pt] 
 (7, 6, 3, 2, 0)	 & $\llap{-}484896$ 	 & (7, 6, 4, 1, 0)	 & $\llap{-}5600$
\tabularnewline[-4pt] 
 (8, 3, 3, 2, 2)	 & $230549312$ 	 & (8, 3, 3, 3, 1)	 & $6953664$
\tabularnewline[-4pt] 
 (8, 4, 2, 2, 2)	 & $40083960$ 	 & (8, 4, 3, 2, 1)	 & $1194656$
\tabularnewline[-4pt] 
 (8, 4, 3, 3, 0)	 & $\llap{-}484896$ 	 & (8, 4, 4, 1, 1)	 & $10400$
\tabularnewline[-4pt] 
 (8, 4, 4, 2, 0)	 & $\llap{-}61760$ 	 & (8, 5, 2, 2, 1)	 & $29808$
\tabularnewline[-4pt] 
 (8, 5, 3, 1, 1)	 & $928$ 	 & (8, 5, 3, 2, 0)	 & $\llap{-}5600$
\tabularnewline[-4pt] 
 (8, 6, 2, 2, 0)	 & $4$ 	 & (9, 3, 2, 2, 2)	 & $\llap{-}5600$
\tabularnewline[-4pt] 
(9, 3, 3, 2, 1)	 & $\llap{-}224$	 &  	 &  
\tablepostambleInstantonNumbers \vskip-18pt 

\tablepreambleInstantonNumbersGone{19}
 (4, 4, 4, 4, 3)	 & $29809312235610960$ 	 & (5, 4, 4, 3, 3)	 & $10159668608774304$
\tabularnewline[-4pt] 
 (5, 4, 4, 4, 2)	 & $2707370108500416$ 	 & (5, 5, 3, 3, 3)	 & $3417190702574592$
\tabularnewline[-4pt] 
 (5, 5, 4, 3, 2)	 & $903625742797728$ 	 & (5, 5, 4, 4, 1)	 & $6007581031968$
\tabularnewline[-4pt] 
 (5, 5, 5, 2, 2)	 & $77522333436960$ 	 & (5, 5, 5, 3, 1)	 & $1974181959168$
\tabularnewline[-4pt] 
 (5, 5, 5, 4, 0)	 & $\llap{-}368134832160$ 	 & (6, 4, 3, 3, 3)	 & $1190848151512512$
\tabularnewline[-4pt] 
 (6, 4, 4, 3, 2)	 & $310831260169488$ 	 & (6, 4, 4, 4, 1)	 & $2072265197088$
\tabularnewline[-4pt] 
 (6, 5, 3, 3, 2)	 & $99761061359136$ 	 & (6, 5, 4, 2, 2)	 & $25630803734064$
\tabularnewline[-4pt] 
 (6, 5, 4, 3, 1)	 & $665062141248$ 	 & (6, 5, 4, 4, 0)	 & $\llap{-}119442727776$
\tabularnewline[-4pt] 
 (6, 5, 5, 2, 1)	 & $49806889344$ 	 & (6, 5, 5, 3, 0)	 & $\llap{-}37176746592$
\tabularnewline[-4pt] 
 (6, 6, 3, 2, 2)	 & $2461712752416$ 	 & (6, 6, 3, 3, 1)	 & $66382892544$
\tabularnewline[-4pt] 
 (6, 6, 4, 2, 1)	 & $15766948032$ 	 & (6, 6, 4, 3, 0)	 & $\llap{-}11043084816$
\tabularnewline[-4pt] 
 (6, 6, 5, 1, 1)	 & $\llap{-}273996960$ 	 & (6, 6, 5, 2, 0)	 & $\llap{-}824199120$
\tabularnewline[-4pt] 
 (6, 6, 6, 1, 0)	 & $\llap{-}1679040$ 	 & (7, 3, 3, 3, 3)	 & $41704406393856$
\tabularnewline[-4pt] 
 (7, 4, 3, 3, 2)	 & $10348372749216$ 	 & (7, 4, 4, 2, 2)	 & $2545705442112$
\tabularnewline[-4pt] 
 (7, 4, 4, 3, 1)	 & $68863079616$ 	 & (7, 4, 4, 4, 0)	 & $\llap{-}11043084816$
\tabularnewline[-4pt] 
 (7, 5, 3, 2, 2)	 & $733831612704$ 	 & (7, 5, 3, 3, 1)	 & $20194851840$
\tabularnewline[-4pt] 
 (7, 5, 4, 2, 1)	 & $4741984896$ 	 & (7, 5, 4, 3, 0)	 & $\llap{-}3110582880$
\tabularnewline[-4pt] 
 (7, 5, 5, 1, 1)	 & $\llap{-}46978560$ 	 & (7, 5, 5, 2, 0)	 & $\llap{-}203310240$
\tabularnewline[-4pt] 
 (7, 6, 2, 2, 2)	 & $12168742800$ 	 & (7, 6, 3, 2, 1)	 & $351706176$
\tabularnewline[-4pt] 
 (7, 6, 3, 3, 0)	 & $\llap{-}203310240$ 	 & (7, 6, 4, 1, 1)	 & $\llap{-}6126048$
\tabularnewline[-4pt] 
 (7, 6, 4, 2, 0)	 & $\llap{-}46049040$ 	 & (7, 6, 5, 1, 0)	 & $\llap{-}251520$
\tabularnewline[-4pt] 
 (7, 7, 2, 2, 1)	 & $679968$ 	 & (7, 7, 3, 1, 1)	 & $30720$
\tabularnewline[-4pt] 
 (7, 7, 3, 2, 0)	 & $\llap{-}251520$ 	 & (7, 7, 4, 1, 0)	 & $\llap{-}2208$
\tabularnewline[-4pt] 
 (8, 3, 3, 3, 2)	 & $65707393920$ 	 & (8, 4, 3, 2, 2)	 & $14105356368$
\tabularnewline[-4pt] 
 (8, 4, 3, 3, 1)	 & $411633120$ 	 & (8, 4, 4, 2, 1)	 & $86694528$
\tabularnewline[-4pt] 
 (8, 4, 4, 3, 0)	 & $\llap{-}46049040$ 	 & (8, 5, 2, 2, 2)	 & $652777584$
\tabularnewline[-4pt] 
 (8, 5, 3, 2, 1)	 & $19494816$ 	 & (8, 5, 3, 3, 0)	 & $\llap{-}9395616$
\tabularnewline[-4pt] 
 (8, 5, 4, 1, 1)	 & $73824$ 	 & (8, 5, 4, 2, 0)	 & $\llap{-}1679040$
\tabularnewline[-4pt] 
 (8, 5, 5, 1, 0)	 & $\llap{-}2208$ 	 & (8, 6, 2, 2, 1)	 & $104352$
\tabularnewline[-4pt] 
 (8, 6, 3, 1, 1)	 & $4320$ 	 & (8, 6, 3, 2, 0)	 & $\llap{-}29136$
\tabularnewline[-4pt] 
 (8, 6, 4, 1, 0)	 & $\llap{-}48$ 	 & (9, 3, 3, 2, 2)	 & $679968$
\tabularnewline[-4pt] 
 (9, 3, 3, 3, 1)	 & $30720$ 	 & (9, 4, 2, 2, 2)	 & $104352$
\tabularnewline[-4pt] 
 (9, 4, 3, 2, 1)	 & $4320$ 	 & (9, 4, 3, 3, 0)	 & $\llap{-}2208$
\tabularnewline[-4pt] 
(9, 4, 4, 2, 0)	 & $\llap{-}48$	 &  	 &  
\tablepostambleInstantonNumbers \vskip-18pt 

\tablepreambleInstantonNumbersGone{20}
 (4, 4, 4, 4, 4)	 & $597237294763420872$ 	 & (5, 4, 4, 4, 3)	 & $211913083229294304$
\tabularnewline[-4pt] 
 (5, 5, 4, 3, 3)	 & $74503089764268384$ 	 & (5, 5, 4, 4, 2)	 & $20254395759934128$
\tabularnewline[-4pt] 
 (5, 5, 5, 3, 2)	 & $7011987726247008$ 	 & (5, 5, 5, 4, 1)	 & $46131775979616$
\tabularnewline[-4pt] 
 (5, 5, 5, 5, 0)	 & $\llap{-}3138370134624$ 	 & (6, 4, 4, 3, 3)	 & $27569906261747088$
\tabularnewline[-4pt] 
 (6, 4, 4, 4, 2)	 & $7431936473155680$ 	 & (6, 5, 3, 3, 3)	 & $9449678610817056$
\tabularnewline[-4pt] 
 (6, 5, 4, 3, 2)	 & $2530955656217280$ 	 & (6, 5, 4, 4, 1)	 & $16731712682064$
\tabularnewline[-4pt] 
 (6, 5, 5, 2, 2)	 & $225463832566752$ 	 & (6, 5, 5, 3, 1)	 & $5619611350656$
\tabularnewline[-4pt] 
 (6, 5, 5, 4, 0)	 & $\llap{-}1093125957120$ 	 & (6, 6, 3, 3, 2)	 & $294058742512224$
\tabularnewline[-4pt] 
 (6, 6, 4, 2, 2)	 & $76906534439280$ 	 & (6, 6, 4, 3, 1)	 & $1954674541824$
\tabularnewline[-4pt] 
 (6, 6, 4, 4, 0)	 & $\llap{-}368134868160$ 	 & (6, 6, 5, 2, 1)	 & $149497953456$
\tabularnewline[-4pt] 
 (6, 6, 5, 3, 0)	 & $\llap{-}119442727776$ 	 & (6, 6, 6, 1, 1)	 & $\llap{-}1334560224$
\tabularnewline[-4pt] 
 (6, 6, 6, 2, 0)	 & $\llap{-}3110590260$ 	 & (7, 4, 3, 3, 3)	 & $1190848151512512$
\tabularnewline[-4pt] 
 (7, 4, 4, 3, 2)	 & $310831260169488$ 	 & (7, 4, 4, 4, 1)	 & $2072265197088$
\tabularnewline[-4pt] 
 (7, 5, 3, 3, 2)	 & $99761061359136$ 	 & (7, 5, 4, 2, 2)	 & $25630803734064$
\tabularnewline[-4pt] 
 (7, 5, 4, 3, 1)	 & $665062141248$ 	 & (7, 5, 4, 4, 0)	 & $\llap{-}119442727776$
\tabularnewline[-4pt] 
 (7, 5, 5, 2, 1)	 & $49806889344$ 	 & (7, 5, 5, 3, 0)	 & $\llap{-}37176746592$
\tabularnewline[-4pt] 
 (7, 6, 3, 2, 2)	 & $2461712752416$ 	 & (7, 6, 3, 3, 1)	 & $66382892544$
\tabularnewline[-4pt] 
 (7, 6, 4, 2, 1)	 & $15766948032$ 	 & (7, 6, 4, 3, 0)	 & $\llap{-}11043084816$
\tabularnewline[-4pt] 
 (7, 6, 5, 1, 1)	 & $\llap{-}273996960$ 	 & (7, 6, 5, 2, 0)	 & $\llap{-}824199120$
\tabularnewline[-4pt] 
 (7, 6, 6, 1, 0)	 & $\llap{-}1679040$ 	 & (7, 7, 2, 2, 2)	 & $12168742800$
\tabularnewline[-4pt] 
 (7, 7, 3, 2, 1)	 & $351706176$ 	 & (7, 7, 3, 3, 0)	 & $\llap{-}203310240$
\tabularnewline[-4pt] 
 (7, 7, 4, 1, 1)	 & $\llap{-}6126048$ 	 & (7, 7, 4, 2, 0)	 & $\llap{-}46049040$
\tabularnewline[-4pt] 
 (7, 7, 5, 1, 0)	 & $\llap{-}251520$ 	 & (8, 3, 3, 3, 3)	 & $13029814091424$
\tabularnewline[-4pt] 
 (8, 4, 3, 3, 2)	 & $3154648420512$ 	 & (8, 4, 4, 2, 2)	 & $755118268080$
\tabularnewline[-4pt] 
 (8, 4, 4, 3, 1)	 & $20875131744$ 	 & (8, 4, 4, 4, 0)	 & $\llap{-}3110590260$
\tabularnewline[-4pt] 
 (8, 5, 3, 2, 2)	 & $208350582720$ 	 & (8, 5, 3, 3, 1)	 & $5848333440$
\tabularnewline[-4pt] 
 (8, 5, 4, 2, 1)	 & $1342319904$ 	 & (8, 5, 4, 3, 0)	 & $\llap{-}824199120$
\tabularnewline[-4pt] 
 (8, 5, 5, 1, 1)	 & $\llap{-}6126048$ 	 & (8, 5, 5, 2, 0)	 & $\llap{-}46049040$
\tabularnewline[-4pt] 
 (8, 6, 2, 2, 2)	 & $2937953580$ 	 & (8, 6, 3, 2, 1)	 & $86694528$
\tabularnewline[-4pt] 
 (8, 6, 3, 3, 0)	 & $\llap{-}46049040$ 	 & (8, 6, 4, 1, 1)	 & $\llap{-}317232$
\tabularnewline[-4pt] 
 (8, 6, 4, 2, 0)	 & $\llap{-}9396672$ 	 & (8, 6, 5, 1, 0)	 & $\llap{-}29136$
\tabularnewline[-4pt] 
 (8, 7, 2, 2, 1)	 & $104352$ 	 & (8, 7, 3, 1, 1)	 & $4320$
\tabularnewline[-4pt] 
 (8, 7, 3, 2, 0)	 & $\llap{-}29136$ 	 & (8, 7, 4, 1, 0)	 & $\llap{-}48$
\tabularnewline[-4pt] 
 (9, 3, 3, 3, 2)	 & $3347625888$ 	 & (9, 4, 3, 2, 2)	 & $652777584$
\tabularnewline[-4pt] 
 (9, 4, 3, 3, 1)	 & $19494816$ 	 & (9, 4, 4, 2, 1)	 & $3692400$
\tabularnewline[-4pt] 
 (9, 4, 4, 3, 0)	 & $\llap{-}1679040$ 	 & (9, 5, 2, 2, 2)	 & $22958688$
\tabularnewline[-4pt] 
 (9, 5, 3, 2, 1)	 & $679968$ 	 & (9, 5, 3, 3, 0)	 & $\llap{-}251520$
\tabularnewline[-4pt] 
 (9, 5, 4, 1, 1)	 & $4320$ 	 & (9, 5, 4, 2, 0)	 & $\llap{-}29136$
\tabularnewline[-4pt] 
 (9, 6, 2, 2, 1)	 & $528$ 	 & (9, 6, 3, 2, 0)	 & $\llap{-}48$
\tabularnewline[-4pt] 
 (10, 3, 3, 2, 2)	 & $\llap{-}29136$ 	 & (10, 3, 3, 3, 1)	 & $\llap{-}2208$
\tabularnewline[-4pt] 
 (10, 4, 2, 2, 2)	 & $\llap{-}2292$ 	 & (10, 4, 3, 2, 1)	 & $\llap{-}48$
\tablepostambleInstantonNumbers \vskip-18pt 

\tablepreambleInstantonNumbersGone{21}
 (5, 4, 4, 4, 4)	 & $5224733955268106112$ 	 & (5, 5, 4, 4, 3)	 & $1905129808949968992$
\tabularnewline[-4pt] 
 (5, 5, 5, 3, 3)	 & $689915910456635392$ 	 & (5, 5, 5, 4, 2)	 & $190939831236687552$
\tabularnewline[-4pt] 
 (5, 5, 5, 5, 1)	 & $442883019280896$ 	 & (6, 4, 4, 4, 3)	 & $742174782726416480$
\tabularnewline[-4pt] 
 (6, 5, 4, 3, 3)	 & $265869597857942752$ 	 & (6, 5, 4, 4, 2)	 & $73138223025414832$
\tabularnewline[-4pt] 
 (6, 5, 5, 3, 2)	 & $25872896752378400$ 	 & (6, 5, 5, 4, 1)	 & $168513518883456$
\tabularnewline[-4pt] 
 (6, 5, 5, 5, 0)	 & $\llap{-}12215408263200$ 	 & (6, 6, 3, 3, 3)	 & $35367733329831456$
\tabularnewline[-4pt] 
 (6, 6, 4, 3, 2)	 & $9613082109166896$ 	 & (6, 6, 4, 4, 1)	 & $63004921310816$
\tabularnewline[-4pt] 
 (6, 6, 5, 2, 2)	 & $894838196834976$ 	 & (6, 6, 5, 3, 1)	 & $21688029832256$
\tabularnewline[-4pt] 
 (6, 6, 5, 4, 0)	 & $\llap{-}4429601736480$ 	 & (6, 6, 6, 2, 1)	 & $604322817696$
\tabularnewline[-4pt] 
 (6, 6, 6, 3, 0)	 & $\llap{-}531223501536$ 	 & (7, 4, 4, 3, 3)	 & $38231916995852064$
\tabularnewline[-4pt] 
 (7, 4, 4, 4, 2)	 & $10343926869883680$ 	 & (7, 5, 3, 3, 3)	 & $13183353406838784$
\tabularnewline[-4pt] 
 (7, 5, 4, 3, 2)	 & $3545177906512576$ 	 & (7, 5, 4, 4, 1)	 & $23383915823040$
\tabularnewline[-4pt] 
 (7, 5, 5, 2, 2)	 & $319546789488192$ 	 & (7, 5, 5, 3, 1)	 & $7905805097984$
\tabularnewline[-4pt] 
 (7, 5, 5, 4, 0)	 & $\llap{-}1559208760288$ 	 & (7, 6, 3, 3, 2)	 & $418651558115072$
\tabularnewline[-4pt] 
 (7, 6, 4, 2, 2)	 & $110090172437152$ 	 & (7, 6, 4, 3, 1)	 & $2778335397408$
\tabularnewline[-4pt] 
 (7, 6, 4, 4, 0)	 & $\llap{-}531223501536$ 	 & (7, 6, 5, 2, 1)	 & $213298398784$
\tabularnewline[-4pt] 
 (7, 6, 5, 3, 0)	 & $\llap{-}174588053440$ 	 & (7, 6, 6, 1, 1)	 & $\llap{-}2202805408$
\tabularnewline[-4pt] 
 (7, 6, 6, 2, 0)	 & $\llap{-}4775506080$ 	 & (7, 7, 3, 2, 2)	 & $3650629114944$
\tabularnewline[-4pt] 
 (7, 7, 3, 3, 1)	 & $97849986048$ 	 & (7, 7, 4, 2, 1)	 & $23296907136$
\tabularnewline[-4pt] 
 (7, 7, 4, 3, 0)	 & $\llap{-}16642969280$ 	 & (7, 7, 5, 1, 1)	 & $\llap{-}471020544$
\tabularnewline[-4pt] 
 (7, 7, 5, 2, 0)	 & $\llap{-}1292723968$ 	 & (7, 7, 6, 1, 0)	 & $\llap{-}3031872$
\tabularnewline[-4pt] 
 (8, 4, 3, 3, 3)	 & $586171325733792$ 	 & (8, 4, 4, 3, 2)	 & $151342528026688$
\tabularnewline[-4pt] 
 (8, 4, 4, 4, 1)	 & $1011188967744$ 	 & (8, 5, 3, 3, 2)	 & $47693058783296$
\tabularnewline[-4pt] 
 (8, 5, 4, 2, 2)	 & $12091316695232$ 	 & (8, 5, 4, 3, 1)	 & $318330381792$
\tabularnewline[-4pt] 
 (8, 5, 4, 4, 0)	 & $\llap{-}55127514240$ 	 & (8, 5, 5, 2, 1)	 & $23350187616$
\tabularnewline[-4pt] 
 (8, 5, 5, 3, 0)	 & $\llap{-}16642969280$ 	 & (8, 6, 3, 2, 2)	 & $1103727042528$
\tabularnewline[-4pt] 
 (8, 6, 3, 3, 1)	 & $30154035584$ 	 & (8, 6, 4, 2, 1)	 & $7112117856$
\tabularnewline[-4pt] 
 (8, 6, 4, 3, 0)	 & $\llap{-}4775506080$ 	 & (8, 6, 5, 1, 1)	 & $\llap{-}87015936$
\tabularnewline[-4pt] 
 (8, 6, 5, 2, 0)	 & $\llap{-}327015680$ 	 & (8, 6, 6, 1, 0)	 & $\llap{-}484896$
\tabularnewline[-4pt] 
 (8, 7, 2, 2, 2)	 & $4750051104$ 	 & (8, 7, 3, 2, 1)	 & $138982240$
\tabularnewline[-4pt] 
 (8, 7, 3, 3, 0)	 & $\llap{-}76342880$ 	 & (8, 7, 4, 1, 1)	 & $\llap{-}1114976$
\tabularnewline[-4pt] 
 (8, 7, 4, 2, 0)	 & $\llap{-}16170272$ 	 & (8, 7, 5, 1, 0)	 & $\llap{-}61920$
\tabularnewline[-4pt] 
 (8, 8, 2, 2, 1)	 & $29808$ 	 & (8, 8, 3, 1, 1)	 & $928$
\tabularnewline[-4pt] 
 (8, 8, 3, 2, 0)	 & $\llap{-}5600$ 	 & (9, 3, 3, 3, 3)	 & $1668835805184$
\tabularnewline[-4pt] 
 (9, 4, 3, 3, 2)	 & $385951211712$ 	 & (9, 4, 4, 2, 2)	 & $87650018048$
\tabularnewline[-4pt] 
 (9, 4, 4, 3, 1)	 & $2496782816$ 	 & (9, 4, 4, 4, 0)	 & $\llap{-}327015680$
\tabularnewline[-4pt] 
 (9, 5, 3, 2, 2)	 & $22327107072$ 	 & (9, 5, 3, 3, 1)	 & $646886400$
\tabularnewline[-4pt] 
 (9, 5, 4, 2, 1)	 & $138982240$ 	 & (9, 5, 4, 3, 0)	 & $\llap{-}76342880$
\tabularnewline[-4pt] 
 (9, 5, 5, 1, 1)	 & $75776$ 	 & (9, 5, 5, 2, 0)	 & $\llap{-}3031872$
\tabularnewline[-4pt] 
 (9, 6, 2, 2, 2)	 & $230549312$ 	 & (9, 6, 3, 2, 1)	 & $6953664$
\tabularnewline[-4pt] 
 (9, 6, 3, 3, 0)	 & $\llap{-}3031872$ 	 & (9, 6, 4, 1, 1)	 & $42560$
\tabularnewline[-4pt] 
 (9, 6, 4, 2, 0)	 & $\llap{-}484896$ 	 & (9, 6, 5, 1, 0)	 & $\llap{-}224$
\tabularnewline[-4pt] 
 (9, 7, 2, 2, 1)	 & $928$ 	 & (9, 7, 3, 2, 0)	 & $\llap{-}224$
\tabularnewline[-4pt] 
 (10, 3, 3, 3, 2)	 & $6953664$ 	 & (10, 4, 3, 2, 2)	 & $1194656$
\tabularnewline[-4pt] 
 (10, 4, 3, 3, 1)	 & $42560$ 	 & (10, 4, 4, 2, 1)	 & $10400$
\tabularnewline[-4pt] 
 (10, 4, 4, 3, 0)	 & $\llap{-}5600$ 	 & (10, 5, 2, 2, 2)	 & $29808$
\tabularnewline[-4pt] 
 (10, 5, 3, 2, 1)	 & $928$ 	 & (10, 5, 3, 3, 0)	 & $\llap{-}224$
\tablepostambleInstantonNumbers \vskip-18pt 

\tablepreambleInstantonNumbersGone{22}
 (5, 5, 4, 4, 4)	 & $56439747241501122192$ 	 & (5, 5, 5, 4, 3)	 & $21104661093843211008$
\tabularnewline[-4pt] 
 (5, 5, 5, 5, 2)	 & $2206547301748229184$ 	 & (6, 4, 4, 4, 4)	 & $22962634839334473072$
\tabularnewline[-4pt] 
 (6, 5, 4, 4, 3)	 & $8521174156401735360$ 	 & (6, 5, 5, 3, 3)	 & $3144478089605250720$
\tabularnewline[-4pt] 
 (6, 5, 5, 4, 2)	 & $879997438952410320$ 	 & (6, 5, 5, 5, 1)	 & $2055320972401920$
\tabularnewline[-4pt] 
 (6, 6, 4, 3, 3)	 & $1243841228095131744$ 	 & (6, 6, 4, 4, 2)	 & $346443574026127704$
\tabularnewline[-4pt] 
 (6, 6, 5, 3, 2)	 & $125469245902996512$ 	 & (6, 6, 5, 4, 1)	 & $805991883499728$
\tabularnewline[-4pt] 
 (6, 6, 5, 5, 0)	 & $\llap{-}62415555336480$ 	 & (6, 6, 6, 2, 2)	 & $4694772196282128$
\tabularnewline[-4pt] 
 (6, 6, 6, 3, 1)	 & $109867633989312$ 	 & (6, 6, 6, 4, 0)	 & $\llap{-}23657222126952$
\tabularnewline[-4pt] 
 (7, 4, 4, 4, 3)	 & $1369210879561818480$ 	 & (7, 5, 4, 3, 3)	 & $494845153306899264$
\tabularnewline[-4pt] 
 (7, 5, 4, 4, 2)	 & $136879891485939456$ 	 & (7, 5, 5, 3, 2)	 & $48909204818311680$
\tabularnewline[-4pt] 
 (7, 5, 5, 4, 1)	 & $316717033197408$ 	 & (7, 5, 5, 5, 0)	 & $\llap{-}23657221999872$
\tabularnewline[-4pt] 
 (7, 6, 3, 3, 3)	 & $67323025901888832$ 	 & (7, 6, 4, 3, 2)	 & $18421518301369920$
\tabularnewline[-4pt] 
 (7, 6, 4, 4, 1)	 & $120114507109632$ 	 & (7, 6, 5, 2, 2)	 & $1749317561578944$
\tabularnewline[-4pt] 
 (7, 6, 5, 3, 1)	 & $41786034116160$ 	 & (7, 6, 5, 4, 0)	 & $\llap{-}8738280013680$
\tabularnewline[-4pt] 
 (7, 6, 6, 2, 1)	 & $1179552933984$ 	 & (7, 6, 6, 3, 0)	 & $\llap{-}1093125957120$
\tabularnewline[-4pt] 
 (7, 7, 3, 3, 2)	 & $840536568752160$ 	 & (7, 7, 4, 2, 2)	 & $223306712255904$
\tabularnewline[-4pt] 
 (7, 7, 4, 3, 1)	 & $5559537648384$ 	 & (7, 7, 4, 4, 0)	 & $\llap{-}1093125957120$
\tabularnewline[-4pt] 
 (7, 7, 5, 2, 1)	 & $428753909184$ 	 & (7, 7, 5, 3, 0)	 & $\llap{-}368134832160$
\tabularnewline[-4pt] 
 (7, 7, 6, 1, 1)	 & $\llap{-}5776067616$ 	 & (7, 7, 6, 2, 0)	 & $\llap{-}11043084816$
\tabularnewline[-4pt] 
 (7, 7, 7, 1, 0)	 & $\llap{-}9395616$ 	 & (8, 4, 4, 3, 3)	 & $27569906261747088$
\tabularnewline[-4pt] 
 (8, 4, 4, 4, 2)	 & $7431936473155680$ 	 & (8, 5, 3, 3, 3)	 & $9449678610817056$
\tabularnewline[-4pt] 
 (8, 5, 4, 3, 2)	 & $2530955656217280$ 	 & (8, 5, 4, 4, 1)	 & $16731712682064$
\tabularnewline[-4pt] 
 (8, 5, 5, 2, 2)	 & $225463832566752$ 	 & (8, 5, 5, 3, 1)	 & $5619611350656$
\tabularnewline[-4pt] 
 (8, 5, 5, 4, 0)	 & $\llap{-}1093125957120$ 	 & (8, 6, 3, 3, 2)	 & $294058742512224$
\tabularnewline[-4pt] 
 (8, 6, 4, 2, 2)	 & $76906534439280$ 	 & (8, 6, 4, 3, 1)	 & $1954674541824$
\tabularnewline[-4pt] 
 (8, 6, 4, 4, 0)	 & $\llap{-}368134868160$ 	 & (8, 6, 5, 2, 1)	 & $149497953456$
\tabularnewline[-4pt] 
 (8, 6, 5, 3, 0)	 & $\llap{-}119442727776$ 	 & (8, 6, 6, 1, 1)	 & $\llap{-}1334560224$
\tabularnewline[-4pt] 
 (8, 6, 6, 2, 0)	 & $\llap{-}3110590260$ 	 & (8, 7, 3, 2, 2)	 & $2461712752416$
\tabularnewline[-4pt] 
 (8, 7, 3, 3, 1)	 & $66382892544$ 	 & (8, 7, 4, 2, 1)	 & $15766948032$
\tabularnewline[-4pt] 
 (8, 7, 4, 3, 0)	 & $\llap{-}11043084816$ 	 & (8, 7, 5, 1, 1)	 & $\llap{-}273996960$
\tabularnewline[-4pt] 
 (8, 7, 5, 2, 0)	 & $\llap{-}824199120$ 	 & (8, 7, 6, 1, 0)	 & $\llap{-}1679040$
\tabularnewline[-4pt] 
 (8, 8, 2, 2, 2)	 & $2937953580$ 	 & (8, 8, 3, 2, 1)	 & $86694528$
\tabularnewline[-4pt] 
 (8, 8, 3, 3, 0)	 & $\llap{-}46049040$ 	 & (8, 8, 4, 1, 1)	 & $\llap{-}317232$
\tabularnewline[-4pt] 
 (8, 8, 4, 2, 0)	 & $\llap{-}9396672$ 	 & (8, 8, 5, 1, 0)	 & $\llap{-}29136$
\tabularnewline[-4pt] 
 (9, 4, 3, 3, 3)	 & $135453779066496$ 	 & (9, 4, 4, 3, 2)	 & $34155140507184$
\tabularnewline[-4pt] 
 (9, 4, 4, 4, 1)	 & $228415121472$ 	 & (9, 5, 3, 3, 2)	 & $10348372749216$
\tabularnewline[-4pt] 
 (9, 5, 4, 2, 2)	 & $2545705442112$ 	 & (9, 5, 4, 3, 1)	 & $68863079616$
\tabularnewline[-4pt] 
 (9, 5, 4, 4, 0)	 & $\llap{-}11043084816$ 	 & (9, 5, 5, 2, 1)	 & $4741984896$
\tabularnewline[-4pt] 
 (9, 5, 5, 3, 0)	 & $\llap{-}3110582880$ 	 & (9, 6, 3, 2, 2)	 & $208350582720$
\tabularnewline[-4pt] 
 (9, 6, 3, 3, 1)	 & $5848333440$ 	 & (9, 6, 4, 2, 1)	 & $1342319904$
\tabularnewline[-4pt] 
 (9, 6, 4, 3, 0)	 & $\llap{-}824199120$ 	 & (9, 6, 5, 1, 1)	 & $\llap{-}6126048$
\tabularnewline[-4pt] 
 (9, 6, 5, 2, 0)	 & $\llap{-}46049040$ 	 & (9, 6, 6, 1, 0)	 & $\llap{-}29136$
\tabularnewline[-4pt] 
 (9, 7, 2, 2, 2)	 & $652777584$ 	 & (9, 7, 3, 2, 1)	 & $19494816$
\tabularnewline[-4pt] 
 (9, 7, 3, 3, 0)	 & $\llap{-}9395616$ 	 & (9, 7, 4, 1, 1)	 & $73824$
\tabularnewline[-4pt] 
 (9, 7, 4, 2, 0)	 & $\llap{-}1679040$ 	 & (9, 7, 5, 1, 0)	 & $\llap{-}2208$
\tabularnewline[-4pt] 
 (9, 8, 2, 2, 1)	 & $528$ 	 & (9, 8, 3, 2, 0)	 & $\llap{-}48$
\tabularnewline[-4pt] 
 (10, 3, 3, 3, 3)	 & $65707393920$ 	 & (10, 4, 3, 3, 2)	 & $14105356368$
\tabularnewline[-4pt] 
 (10, 4, 4, 2, 2)	 & $2937953580$ 	 & (10, 4, 4, 3, 1)	 & $86694528$
\tabularnewline[-4pt] 
 (10, 4, 4, 4, 0)	 & $\llap{-}9396672$ 	 & (10, 5, 3, 2, 2)	 & $652777584$
\tabularnewline[-4pt] 
 (10, 5, 3, 3, 1)	 & $19494816$ 	 & (10, 5, 4, 2, 1)	 & $3692400$
\tabularnewline[-4pt] 
 (10, 5, 4, 3, 0)	 & $\llap{-}1679040$ 	 & (10, 5, 5, 1, 1)	 & $4320$
\tabularnewline[-4pt] 
 (10, 5, 5, 2, 0)	 & $\llap{-}29136$ 	 & (10, 6, 2, 2, 2)	 & $3666312$
\tabularnewline[-4pt] 
 (10, 6, 3, 2, 1)	 & $104352$ 	 & (10, 6, 3, 3, 0)	 & $\llap{-}29136$
\tabularnewline[-4pt] 
 (10, 6, 4, 1, 1)	 & $528$ 	 & (10, 6, 4, 2, 0)	 & $\llap{-}2292$
\tabularnewline[-4pt] 
 (10, 7, 2, 2, 1)	 & $\llap{-}48$ 	 & (11, 3, 3, 3, 2)	 & $\llap{-}251520$
\tabularnewline[-4pt] 
 (11, 4, 3, 2, 2)	 & $\llap{-}29136$ 	 & (11, 4, 3, 3, 1)	 & $\llap{-}2208$
\tabularnewline[-4pt] 
 (11, 4, 4, 2, 1)	 & $\llap{-}48$ 	 & (11, 5, 2, 2, 2)	 & $\llap{-}48$
\tablepostambleInstantonNumbers \vskip-18pt 

\tablepreambleInstantonNumbersGone{23}
 (5, 5, 5, 4, 4)	 & $736557199694836925664$ 	 & (5, 5, 5, 5, 3)	 & $281693674984303028736$
\tabularnewline[-4pt] 
 (6, 5, 4, 4, 4)	 & $309132463141878510000$ 	 & (6, 5, 5, 4, 3)	 & $117552717975524482368$
\tabularnewline[-4pt] 
 (6, 5, 5, 5, 2)	 & $12636043517074729152$ 	 & (6, 6, 4, 4, 3)	 & $48619672420408782672$
\tabularnewline[-4pt] 
 (6, 6, 5, 3, 3)	 & $18293419124197908384$ 	 & (6, 6, 5, 4, 2)	 & $5175740190127316160$
\tabularnewline[-4pt] 
 (6, 6, 5, 5, 1)	 & $12132995150259168$ 	 & (6, 6, 6, 3, 2)	 & $777895843534310448$
\tabularnewline[-4pt] 
 (6, 6, 6, 4, 1)	 & $4906985498846880$ 	 & (6, 6, 6, 5, 0)	 & $\llap{-}405156007308576$
\tabularnewline[-4pt] 
 (7, 4, 4, 4, 4)	 & $54505767240269122368$ 	 & (7, 5, 4, 4, 3)	 & $20425869092209172544$
\tabularnewline[-4pt] 
 (7, 5, 5, 3, 3)	 & $7616709702249560064$ 	 & (7, 5, 5, 4, 2)	 & $2144674430360819712$
\tabularnewline[-4pt] 
 (7, 5, 5, 5, 1)	 & $5019686189816832$ 	 & (7, 6, 4, 3, 3)	 & $3057298493530985568$
\tabularnewline[-4pt] 
 (7, 6, 4, 4, 2)	 & $857299960689895392$ 	 & (7, 6, 5, 3, 2)	 & $314475153857119104$
\tabularnewline[-4pt] 
 (7, 6, 5, 4, 1)	 & $2001252354617280$ 	 & (7, 6, 5, 5, 0)	 & $\llap{-}160791639748800$
\tabularnewline[-4pt] 
 (7, 6, 6, 2, 2)	 & $12276819589743504$ 	 & (7, 6, 6, 3, 1)	 & $281059260470880$
\tabularnewline[-4pt] 
 (7, 6, 6, 4, 0)	 & $\llap{-}62415555336480$ 	 & (7, 7, 3, 3, 3)	 & $173735334568551936$
\tabularnewline[-4pt] 
 (7, 7, 4, 3, 2)	 & $47974716780673152$ 	 & (7, 7, 4, 4, 1)	 & $310320577356576$
\tabularnewline[-4pt] 
 (7, 7, 5, 2, 2)	 & $4682747289739200$ 	 & (7, 7, 5, 3, 1)	 & $109488862752768$
\tabularnewline[-4pt] 
 (7, 7, 5, 4, 0)	 & $\llap{-}23657221999872$ 	 & (7, 7, 6, 2, 1)	 & $3117411980160$
\tabularnewline[-4pt] 
 (7, 7, 6, 3, 0)	 & $\llap{-}3138370134624$ 	 & (7, 7, 7, 1, 1)	 & $\llap{-}22739284992$
\tabularnewline[-4pt] 
 (7, 7, 7, 2, 0)	 & $\llap{-}37176746592$ 	 & (8, 4, 4, 4, 3)	 & $1369210879561818480$
\tabularnewline[-4pt] 
 (8, 5, 4, 3, 3)	 & $494845153306899264$ 	 & (8, 5, 4, 4, 2)	 & $136879891485939456$
\tabularnewline[-4pt] 
 (8, 5, 5, 3, 2)	 & $48909204818311680$ 	 & (8, 5, 5, 4, 1)	 & $316717033197408$
\tabularnewline[-4pt] 
 (8, 5, 5, 5, 0)	 & $\llap{-}23657221999872$ 	 & (8, 6, 3, 3, 3)	 & $67323025901888832$
\tabularnewline[-4pt] 
 (8, 6, 4, 3, 2)	 & $18421518301369920$ 	 & (8, 6, 4, 4, 1)	 & $120114507109632$
\tabularnewline[-4pt] 
 (8, 6, 5, 2, 2)	 & $1749317561578944$ 	 & (8, 6, 5, 3, 1)	 & $41786034116160$
\tabularnewline[-4pt] 
 (8, 6, 5, 4, 0)	 & $\llap{-}8738280013680$ 	 & (8, 6, 6, 2, 1)	 & $1179552933984$
\tabularnewline[-4pt] 
 (8, 6, 6, 3, 0)	 & $\llap{-}1093125957120$ 	 & (8, 7, 3, 3, 2)	 & $840536568752160$
\tabularnewline[-4pt] 
 (8, 7, 4, 2, 2)	 & $223306712255904$ 	 & (8, 7, 4, 3, 1)	 & $5559537648384$
\tabularnewline[-4pt] 
 (8, 7, 4, 4, 0)	 & $\llap{-}1093125957120$ 	 & (8, 7, 5, 2, 1)	 & $428753909184$
\tabularnewline[-4pt] 
 (8, 7, 5, 3, 0)	 & $\llap{-}368134832160$ 	 & (8, 7, 6, 1, 1)	 & $\llap{-}5776067616$
\tabularnewline[-4pt] 
 (8, 7, 6, 2, 0)	 & $\llap{-}11043084816$ 	 & (8, 7, 7, 1, 0)	 & $\llap{-}9395616$
\tabularnewline[-4pt] 
 (8, 8, 3, 2, 2)	 & $2461712752416$ 	 & (8, 8, 3, 3, 1)	 & $66382892544$
\tabularnewline[-4pt] 
 (8, 8, 4, 2, 1)	 & $15766948032$ 	 & (8, 8, 4, 3, 0)	 & $\llap{-}11043084816$
\tabularnewline[-4pt] 
 (8, 8, 5, 1, 1)	 & $\llap{-}273996960$ 	 & (8, 8, 5, 2, 0)	 & $\llap{-}824199120$
\tabularnewline[-4pt] 
 (8, 8, 6, 1, 0)	 & $\llap{-}1679040$ 	 & (9, 4, 4, 3, 3)	 & $10159668608774304$
\tabularnewline[-4pt] 
 (9, 4, 4, 4, 2)	 & $2707370108500416$ 	 & (9, 5, 3, 3, 3)	 & $3417190702574592$
\tabularnewline[-4pt] 
 (9, 5, 4, 3, 2)	 & $903625742797728$ 	 & (9, 5, 4, 4, 1)	 & $6007581031968$
\tabularnewline[-4pt] 
 (9, 5, 5, 2, 2)	 & $77522333436960$ 	 & (9, 5, 5, 3, 1)	 & $1974181959168$
\tabularnewline[-4pt] 
 (9, 5, 5, 4, 0)	 & $\llap{-}368134832160$ 	 & (9, 6, 3, 3, 2)	 & $99761061359136$
\tabularnewline[-4pt] 
 (9, 6, 4, 2, 2)	 & $25630803734064$ 	 & (9, 6, 4, 3, 1)	 & $665062141248$
\tabularnewline[-4pt] 
 (9, 6, 4, 4, 0)	 & $\llap{-}119442727776$ 	 & (9, 6, 5, 2, 1)	 & $49806889344$
\tabularnewline[-4pt] 
 (9, 6, 5, 3, 0)	 & $\llap{-}37176746592$ 	 & (9, 6, 6, 1, 1)	 & $\llap{-}273996960$
\tabularnewline[-4pt] 
 (9, 6, 6, 2, 0)	 & $\llap{-}824199120$ 	 & (9, 7, 3, 2, 2)	 & $733831612704$
\tabularnewline[-4pt] 
 (9, 7, 3, 3, 1)	 & $20194851840$ 	 & (9, 7, 4, 2, 1)	 & $4741984896$
\tabularnewline[-4pt] 
 (9, 7, 4, 3, 0)	 & $\llap{-}3110582880$ 	 & (9, 7, 5, 1, 1)	 & $\llap{-}46978560$
\tabularnewline[-4pt] 
 (9, 7, 5, 2, 0)	 & $\llap{-}203310240$ 	 & (9, 7, 6, 1, 0)	 & $\llap{-}251520$
\tabularnewline[-4pt] 
 (9, 8, 2, 2, 2)	 & $652777584$ 	 & (9, 8, 3, 2, 1)	 & $19494816$
\tabularnewline[-4pt] 
 (9, 8, 3, 3, 0)	 & $\llap{-}9395616$ 	 & (9, 8, 4, 1, 1)	 & $73824$
\tabularnewline[-4pt] 
 (9, 8, 4, 2, 0)	 & $\llap{-}1679040$ 	 & (9, 8, 5, 1, 0)	 & $\llap{-}2208$
\tabularnewline[-4pt] 
 (10, 4, 3, 3, 3)	 & $13029814091424$ 	 & (10, 4, 4, 3, 2)	 & $3154648420512$
\tabularnewline[-4pt] 
 (10, 4, 4, 4, 1)	 & $20875131744$ 	 & (10, 5, 3, 3, 2)	 & $894337855968$
\tabularnewline[-4pt] 
 (10, 5, 4, 2, 2)	 & $208350582720$ 	 & (10, 5, 4, 3, 1)	 & $5848333440$
\tabularnewline[-4pt] 
 (10, 5, 4, 4, 0)	 & $\llap{-}824199120$ 	 & (10, 5, 5, 2, 1)	 & $351706176$
\tabularnewline[-4pt] 
 (10, 5, 5, 3, 0)	 & $\llap{-}203310240$ 	 & (10, 6, 3, 2, 2)	 & $14105356368$
\tabularnewline[-4pt] 
 (10, 6, 3, 3, 1)	 & $411633120$ 	 & (10, 6, 4, 2, 1)	 & $86694528$
\tabularnewline[-4pt] 
 (10, 6, 4, 3, 0)	 & $\llap{-}46049040$ 	 & (10, 6, 5, 1, 1)	 & $73824$
\tabularnewline[-4pt] 
 (10, 6, 5, 2, 0)	 & $\llap{-}1679040$ 	 & (10, 6, 6, 1, 0)	 & $\llap{-}48$
\tabularnewline[-4pt] 
 (10, 7, 2, 2, 2)	 & $22958688$ 	 & (10, 7, 3, 2, 1)	 & $679968$
\tabularnewline[-4pt] 
 (10, 7, 3, 3, 0)	 & $\llap{-}251520$ 	 & (10, 7, 4, 1, 1)	 & $4320$
\tabularnewline[-4pt] 
 (10, 7, 4, 2, 0)	 & $\llap{-}29136$ 	 & (10, 8, 2, 2, 1)	 & $\llap{-}48$
\tabularnewline[-4pt] 
 (11, 3, 3, 3, 3)	 & $99761664$ 	 & (11, 4, 3, 3, 2)	 & $19494816$
\tabularnewline[-4pt] 
 (11, 4, 4, 2, 2)	 & $3692400$ 	 & (11, 4, 4, 3, 1)	 & $73824$
\tabularnewline[-4pt] 
 (11, 4, 4, 4, 0)	 & $\llap{-}29136$ 	 & (11, 5, 3, 2, 2)	 & $679968$
\tabularnewline[-4pt] 
 (11, 5, 3, 3, 1)	 & $30720$ 	 & (11, 5, 4, 2, 1)	 & $4320$
\tabularnewline[-4pt] 
 (11, 5, 4, 3, 0)	 & $\llap{-}2208$ 	 & (11, 6, 2, 2, 2)	 & $528$
\tablepostambleInstantonNumbers \vskip-18pt 

\tablepreambleInstantonNumbersGone{24}
 (5, 5, 5, 5, 4)	 & $11398902454359592613184$ 	 & (6, 5, 5, 4, 4)	 & $4920814265932180005216$
\tabularnewline[-4pt] 
 (6, 5, 5, 5, 3)	 & $1911288345346635589568$ 	 & (6, 6, 4, 4, 4)	 & $2110546919582910040272$
\tabularnewline[-4pt] 
 (6, 6, 5, 4, 3)	 & $816025913316248333120$ 	 & (6, 6, 5, 5, 2)	 & $90095498612893458272$
\tabularnewline[-4pt] 
 (6, 6, 6, 3, 3)	 & $132628671085592097024$ 	 & (6, 6, 6, 4, 2)	 & $37906436889576198396$
\tabularnewline[-4pt] 
 (6, 6, 6, 5, 1)	 & $88798104825584352$ 	 & (6, 6, 6, 6, 0)	 & $\llap{-}3238317070364520$
\tabularnewline[-4pt] 
 (7, 5, 4, 4, 4)	 & $928035242813679386880$ 	 & (7, 5, 5, 4, 3)	 & $356555435715410957376$
\tabularnewline[-4pt] 
 (7, 5, 5, 5, 2)	 & $38978014973683179264$ 	 & (7, 6, 4, 4, 3)	 & $149677043087457971504$
\tabularnewline[-4pt] 
 (7, 6, 5, 3, 3)	 & $56981868551941098752$ 	 & (7, 6, 5, 4, 2)	 & $16227227828645916976$
\tabularnewline[-4pt] 
 (7, 6, 5, 5, 1)	 & $38035145900721056$ 	 & (7, 6, 6, 3, 2)	 & $2517326301790872576$
\tabularnewline[-4pt] 
 (7, 6, 6, 4, 1)	 & $15666901809220000$ 	 & (7, 6, 6, 5, 0)	 & $\llap{-}1345692401785920$
\tabularnewline[-4pt] 
 (7, 7, 4, 3, 3)	 & $9881551554602229408$ 	 & (7, 7, 4, 4, 2)	 & $2792965386085320880$
\tabularnewline[-4pt] 
 (7, 7, 5, 3, 2)	 & $1040501263302552608$ 	 & (7, 7, 5, 4, 1)	 & $6536958768444736$
\tabularnewline[-4pt] 
 (7, 7, 5, 5, 0)	 & $\llap{-}548768158982912$ 	 & (7, 7, 6, 2, 2)	 & $42751365570060704$
\tabularnewline[-4pt] 
 (7, 7, 6, 3, 1)	 & $950472340606656$ 	 & (7, 7, 6, 4, 0)	 & $\llap{-}219322647849280$
\tabularnewline[-4pt] 
 (7, 7, 7, 2, 1)	 & $10748457996288$ 	 & (7, 7, 7, 3, 0)	 & $\llap{-}12215408263200$
\tabularnewline[-4pt] 
 (8, 4, 4, 4, 4)	 & $72441168290052535416$ 	 & (8, 5, 4, 4, 3)	 & $27232968315848722400$
\tabularnewline[-4pt] 
 (8, 5, 5, 3, 3)	 & $10189105935322611904$ 	 & (8, 5, 5, 4, 2)	 & $2874629165830386272$
\tabularnewline[-4pt] 
 (8, 5, 5, 5, 1)	 & $6730910671190400$ 	 & (8, 6, 4, 3, 3)	 & $4109159918573469824$
\tabularnewline[-4pt] 
 (8, 6, 4, 4, 2)	 & $1154720984430059840$ 	 & (8, 6, 5, 3, 2)	 & $425294962053433344$
\tabularnewline[-4pt] 
 (8, 6, 5, 4, 1)	 & $2697580082412768$ 	 & (8, 6, 5, 5, 0)	 & $\llap{-}219322647849280$
\tabularnewline[-4pt] 
 (8, 6, 6, 2, 2)	 & $16827059517736904$ 	 & (8, 6, 6, 3, 1)	 & $382345702776672$
\tabularnewline[-4pt] 
 (8, 6, 6, 4, 0)	 & $\llap{-}85781801535360$ 	 & (8, 7, 3, 3, 3)	 & $237188563748902272$
\tabularnewline[-4pt] 
 (8, 7, 4, 3, 2)	 & $65682988009064448$ 	 & (8, 7, 4, 4, 1)	 & $423655401709600$
\tabularnewline[-4pt] 
 (8, 7, 5, 2, 2)	 & $6466615047655104$ 	 & (8, 7, 5, 3, 1)	 & $150108480756288$
\tabularnewline[-4pt] 
 (8, 7, 5, 4, 0)	 & $\llap{-}32779917518560$ 	 & (8, 7, 6, 2, 1)	 & $4272120540288$
\tabularnewline[-4pt] 
 (8, 7, 6, 3, 0)	 & $\llap{-}4429601736480$ 	 & (8, 7, 7, 1, 1)	 & $\llap{-}35309984768$
\tabularnewline[-4pt] 
 (8, 7, 7, 2, 0)	 & $\llap{-}55127514240$ 	 & (8, 8, 3, 3, 2)	 & $1185430621063936$
\tabularnewline[-4pt] 
 (8, 8, 4, 2, 2)	 & $316457581635168$ 	 & (8, 8, 4, 3, 1)	 & $7826902863232$
\tabularnewline[-4pt] 
 (8, 8, 4, 4, 0)	 & $\llap{-}1559208686912$ 	 & (8, 8, 5, 2, 1)	 & $604016711936$
\tabularnewline[-4pt] 
 (8, 8, 5, 3, 0)	 & $\llap{-}531223501536$ 	 & (8, 8, 6, 1, 1)	 & $\llap{-}9200679520$
\tabularnewline[-4pt] 
 (8, 8, 6, 2, 0)	 & $\llap{-}16642956928$ 	 & (8, 8, 7, 1, 0)	 & $\llap{-}16170272$
\tabularnewline[-4pt] 
 (8, 8, 8, 0, 0)	 & $4$ 	 & (9, 4, 4, 4, 3)	 & $742174782726416480$
\tabularnewline[-4pt] 
 (9, 5, 4, 3, 3)	 & $265869597857942752$ 	 & (9, 5, 4, 4, 2)	 & $73138223025414832$
\tabularnewline[-4pt] 
 (9, 5, 5, 3, 2)	 & $25872896752378400$ 	 & (9, 5, 5, 4, 1)	 & $168513518883456$
\tabularnewline[-4pt] 
 (9, 5, 5, 5, 0)	 & $\llap{-}12215408263200$ 	 & (9, 6, 3, 3, 3)	 & $35367733329831456$
\tabularnewline[-4pt] 
 (9, 6, 4, 3, 2)	 & $9613082109166896$ 	 & (9, 6, 4, 4, 1)	 & $63004921310816$
\tabularnewline[-4pt] 
 (9, 6, 5, 2, 2)	 & $894838196834976$ 	 & (9, 6, 5, 3, 1)	 & $21688029832256$
\tabularnewline[-4pt] 
 (9, 6, 5, 4, 0)	 & $\llap{-}4429601736480$ 	 & (9, 6, 6, 2, 1)	 & $604322817696$
\tabularnewline[-4pt] 
 (9, 6, 6, 3, 0)	 & $\llap{-}531223501536$ 	 & (9, 7, 3, 3, 2)	 & $418651558115072$
\tabularnewline[-4pt] 
 (9, 7, 4, 2, 2)	 & $110090172437152$ 	 & (9, 7, 4, 3, 1)	 & $2778335397408$
\tabularnewline[-4pt] 
 (9, 7, 4, 4, 0)	 & $\llap{-}531223501536$ 	 & (9, 7, 5, 2, 1)	 & $213298398784$
\tabularnewline[-4pt] 
 (9, 7, 5, 3, 0)	 & $\llap{-}174588053440$ 	 & (9, 7, 6, 1, 1)	 & $\llap{-}2202805408$
\tabularnewline[-4pt] 
 (9, 7, 6, 2, 0)	 & $\llap{-}4775506080$ 	 & (9, 7, 7, 1, 0)	 & $\llap{-}3031872$
\tabularnewline[-4pt] 
 (9, 8, 3, 2, 2)	 & $1103727042528$ 	 & (9, 8, 3, 3, 1)	 & $30154035584$
\tabularnewline[-4pt] 
 (9, 8, 4, 2, 1)	 & $7112117856$ 	 & (9, 8, 4, 3, 0)	 & $\llap{-}4775506080$
\tabularnewline[-4pt] 
 (9, 8, 5, 1, 1)	 & $\llap{-}87015936$ 	 & (9, 8, 5, 2, 0)	 & $\llap{-}327015680$
\tabularnewline[-4pt] 
 (9, 8, 6, 1, 0)	 & $\llap{-}484896$ 	 & (9, 9, 2, 2, 2)	 & $230549312$
\tabularnewline[-4pt] 
 (9, 9, 3, 2, 1)	 & $6953664$ 	 & (9, 9, 3, 3, 0)	 & $\llap{-}3031872$
\tabularnewline[-4pt] 
 (9, 9, 4, 1, 1)	 & $42560$ 	 & (9, 9, 4, 2, 0)	 & $\llap{-}484896$
\tabularnewline[-4pt] 
 (9, 9, 5, 1, 0)	 & $\llap{-}224$ 	 & (10, 4, 4, 3, 3)	 & $1803381971700144$
\tabularnewline[-4pt] 
 (10, 4, 4, 4, 2)	 & $470537427014352$ 	 & (10, 5, 3, 3, 3)	 & $586171325733792$
\tabularnewline[-4pt] 
 (10, 5, 4, 3, 2)	 & $151342528026688$ 	 & (10, 5, 4, 4, 1)	 & $1011188967744$
\tabularnewline[-4pt] 
 (10, 5, 5, 2, 2)	 & $12091316695232$ 	 & (10, 5, 5, 3, 1)	 & $318330381792$
\tabularnewline[-4pt] 
 (10, 5, 5, 4, 0)	 & $\llap{-}55127514240$ 	 & (10, 6, 3, 3, 2)	 & $15219924472416$
\tabularnewline[-4pt] 
 (10, 6, 4, 2, 2)	 & $3775716012840$ 	 & (10, 6, 4, 3, 1)	 & $101366312448$
\tabularnewline[-4pt] 
 (10, 6, 4, 4, 0)	 & $\llap{-}16642956928$ 	 & (10, 6, 5, 2, 1)	 & $7112117856$
\tabularnewline[-4pt] 
 (10, 6, 5, 3, 0)	 & $\llap{-}4775506080$ 	 & (10, 6, 6, 1, 1)	 & $\llap{-}12495360$
\tabularnewline[-4pt] 
 (10, 6, 6, 2, 0)	 & $\llap{-}76341160$ 	 & (10, 7, 3, 2, 2)	 & $87650018048$
\tabularnewline[-4pt] 
 (10, 7, 3, 3, 1)	 & $2496782816$ 	 & (10, 7, 4, 2, 1)	 & $561090816$
\tabularnewline[-4pt] 
 (10, 7, 4, 3, 0)	 & $\llap{-}327015680$ 	 & (10, 7, 5, 1, 1)	 & $\llap{-}1114976$
\tabularnewline[-4pt] 
 (10, 7, 5, 2, 0)	 & $\llap{-}16170272$ 	 & (10, 7, 6, 1, 0)	 & $\llap{-}5600$
\tabularnewline[-4pt] 
 (10, 8, 2, 2, 2)	 & $40083960$ 	 & (10, 8, 3, 2, 1)	 & $1194656$
\tabularnewline[-4pt] 
 (10, 8, 3, 3, 0)	 & $\llap{-}484896$ 	 & (10, 8, 4, 1, 1)	 & $10400$
\tabularnewline[-4pt] 
 (10, 8, 4, 2, 0)	 & $\llap{-}61760$ 	 & (11, 4, 3, 3, 3)	 & $385951211712$
\tabularnewline[-4pt] 
 (11, 4, 4, 3, 2)	 & $87650018048$ 	 & (11, 4, 4, 4, 1)	 & $561090816$
\tabularnewline[-4pt] 
 (11, 5, 3, 3, 2)	 & $22327107072$ 	 & (11, 5, 4, 2, 2)	 & $4750051104$
\tabularnewline[-4pt] 
 (11, 5, 4, 3, 1)	 & $138982240$ 	 & (11, 5, 4, 4, 0)	 & $\llap{-}16170272$
\tabularnewline[-4pt] 
 (11, 5, 5, 2, 1)	 & $6414464$ 	 & (11, 5, 5, 3, 0)	 & $\llap{-}3031872$
\tabularnewline[-4pt] 
 (11, 6, 3, 2, 2)	 & $230549312$ 	 & (11, 6, 3, 3, 1)	 & $6953664$
\tabularnewline[-4pt] 
 (11, 6, 4, 2, 1)	 & $1194656$ 	 & (11, 6, 4, 3, 0)	 & $\llap{-}484896$
\tabularnewline[-4pt] 
 (11, 6, 5, 1, 1)	 & $928$ 	 & (11, 6, 5, 2, 0)	 & $\llap{-}5600$
\tabularnewline[-4pt] 
 (11, 7, 2, 2, 2)	 & $29808$ 	 & (11, 7, 3, 2, 1)	 & $928$
\tabularnewline[-4pt] 
 (11, 7, 3, 3, 0)	 & $\llap{-}224$ 	 & (12, 3, 3, 3, 3)	 & $\llap{-}3031872$
\tabularnewline[-4pt] 
 (12, 4, 3, 3, 2)	 & $\llap{-}484896$ 	 & (12, 4, 4, 2, 2)	 & $\llap{-}61760$
\tabularnewline[-4pt] 
 (12, 4, 4, 3, 1)	 & $\llap{-}5600$ 	 & (12, 4, 4, 4, 0)	 & $4$
\tabularnewline[-4pt] 
 (12, 5, 3, 2, 2)	 & $\llap{-}5600$ 	 & (12, 5, 3, 3, 1)	 & $\llap{-}224$
\tabularnewline[-4pt] 
(12, 6, 2, 2, 2)	 & $4$	 &  	 &  
\tablepostambleInstantonNumbers \vskip-18pt 

\tablepreambleInstantonNumbersGone{25}
 (5, 5, 5, 5, 5)	 & $205994740015586336392704$ 	 & (6, 5, 5, 5, 4)	 & $91183351722137909496000$
\tabularnewline[-4pt] 
 (6, 6, 5, 4, 4)	 & $40163166616613493587568$ 	 & (6, 6, 5, 5, 3)	 & $15832086999292901742336$
\tabularnewline[-4pt] 
 (6, 6, 6, 4, 3)	 & $6912402731596987607520$ 	 & (6, 6, 6, 5, 2)	 & $782286842060773523040$
\tabularnewline[-4pt] 
 (6, 6, 6, 6, 1)	 & $786907500135989952$ 	 & (7, 5, 5, 4, 4)	 & $18266693407737182177184$
\tabularnewline[-4pt] 
 (7, 5, 5, 5, 3)	 & $7166819934960635255808$ 	 & (7, 6, 4, 4, 4)	 & $7947181852777331886432$
\tabularnewline[-4pt] 
 (7, 6, 5, 4, 3)	 & $3105932030883925870176$ 	 & (7, 6, 5, 5, 2)	 & $348836588057363657472$
\tabularnewline[-4pt] 
 (7, 6, 6, 3, 3)	 & $519156967290582783648$ 	 & (7, 6, 6, 4, 2)	 & $149323799470083244320$
\tabularnewline[-4pt] 
 (7, 6, 6, 5, 1)	 & $348635537576206944$ 	 & (7, 6, 6, 6, 0)	 & $\llap{-}13461969999093600$
\tabularnewline[-4pt] 
 (7, 7, 4, 4, 3)	 & $590236080316778614944$ 	 & (7, 7, 5, 3, 3)	 & $227671001933190288384$
\tabularnewline[-4pt] 
 (7, 7, 5, 4, 2)	 & $65291193554279696544$ 	 & (7, 7, 5, 5, 1)	 & $152724917794335744$
\tabularnewline[-4pt] 
 (7, 7, 6, 3, 2)	 & $10495212981639920256$ 	 & (7, 7, 6, 4, 1)	 & $64196301394125984$
\tabularnewline[-4pt] 
 (7, 7, 6, 5, 0)	 & $\llap{-}5758034709276000$ 	 & (7, 7, 7, 2, 2)	 & $193409949629545344$
\tabularnewline[-4pt] 
 (7, 7, 7, 3, 1)	 & $4142934152742912$ 	 & (7, 7, 7, 4, 0)	 & $\llap{-}1000067627051904$
\tabularnewline[-4pt] 
 (8, 5, 4, 4, 4)	 & $1592870480807489129616$ 	 & (8, 5, 5, 4, 3)	 & $614995894410428607648$
\tabularnewline[-4pt] 
 (8, 5, 5, 5, 2)	 & $67768604147499442944$ 	 & (8, 6, 4, 4, 3)	 & $260001810225111467808$
\tabularnewline[-4pt] 
 (8, 6, 5, 3, 3)	 & $99531632175997196352$ 	 & (8, 6, 5, 4, 2)	 & $28431306071376793872$
\tabularnewline[-4pt] 
 (8, 6, 5, 5, 1)	 & $66591254167125120$ 	 & (8, 6, 6, 3, 2)	 & $4476560442767249184$
\tabularnewline[-4pt] 
 (8, 6, 6, 4, 1)	 & $27663244132912416$ 	 & (8, 6, 6, 5, 0)	 & $\llap{-}2421429008571216$
\tabularnewline[-4pt] 
 (8, 7, 4, 3, 3)	 & $17566723688872737408$ 	 & (8, 7, 4, 4, 2)	 & $4983325664622060480$
\tabularnewline[-4pt] 
 (8, 7, 5, 3, 2)	 & $1869909171436016640$ 	 & (8, 7, 5, 4, 1)	 & $11668379661795360$
\tabularnewline[-4pt] 
 (8, 7, 5, 5, 0)	 & $\llap{-}1000067627051904$ 	 & (8, 7, 6, 2, 2)	 & $78667271480859264$
\tabularnewline[-4pt] 
 (8, 7, 6, 3, 1)	 & $1722815728418688$ 	 & (8, 7, 6, 4, 0)	 & $\llap{-}405156007308576$
\tabularnewline[-4pt] 
 (8, 7, 7, 2, 1)	 & $19460158766688$ 	 & (8, 7, 7, 3, 0)	 & $\llap{-}23657221999872$
\tabularnewline[-4pt] 
 (8, 8, 3, 3, 3)	 & $439198792163267520$ 	 & (8, 8, 4, 3, 2)	 & $122281608111146832$
\tabularnewline[-4pt] 
 (8, 8, 4, 4, 1)	 & $784225690196688$ 	 & (8, 8, 5, 2, 2)	 & $12238194838809408$
\tabularnewline[-4pt] 
 (8, 8, 5, 3, 1)	 & $280061917894368$ 	 & (8, 8, 5, 4, 0)	 & $\llap{-}62415555336480$
\tabularnewline[-4pt] 
 (8, 8, 6, 2, 1)	 & $7932337626384$ 	 & (8, 8, 6, 3, 0)	 & $\llap{-}8738280013680$
\tabularnewline[-4pt] 
 (8, 8, 7, 1, 1)	 & $\llap{-}83145266112$ 	 & (8, 8, 7, 2, 0)	 & $\llap{-}119442727776$
\tabularnewline[-4pt] 
 (8, 8, 8, 1, 0)	 & $\llap{-}46049040$ 	 & (9, 4, 4, 4, 4)	 & $54505767240269122368$
\tabularnewline[-4pt] 
 (9, 5, 4, 4, 3)	 & $20425869092209172544$ 	 & (9, 5, 5, 3, 3)	 & $7616709702249560064$
\tabularnewline[-4pt] 
 (9, 5, 5, 4, 2)	 & $2144674430360819712$ 	 & (9, 5, 5, 5, 1)	 & $5019686189816832$
\tabularnewline[-4pt] 
 (9, 6, 4, 3, 3)	 & $3057298493530985568$ 	 & (9, 6, 4, 4, 2)	 & $857299960689895392$
\tabularnewline[-4pt] 
 (9, 6, 5, 3, 2)	 & $314475153857119104$ 	 & (9, 6, 5, 4, 1)	 & $2001252354617280$
\tabularnewline[-4pt] 
 (9, 6, 5, 5, 0)	 & $\llap{-}160791639748800$ 	 & (9, 6, 6, 2, 2)	 & $12276819589743504$
\tabularnewline[-4pt] 
 (9, 6, 6, 3, 1)	 & $281059260470880$ 	 & (9, 6, 6, 4, 0)	 & $\llap{-}62415555336480$
\tabularnewline[-4pt] 
 (9, 7, 3, 3, 3)	 & $173735334568551936$ 	 & (9, 7, 4, 3, 2)	 & $47974716780673152$
\tabularnewline[-4pt] 
 (9, 7, 4, 4, 1)	 & $310320577356576$ 	 & (9, 7, 5, 2, 2)	 & $4682747289739200$
\tabularnewline[-4pt] 
 (9, 7, 5, 3, 1)	 & $109488862752768$ 	 & (9, 7, 5, 4, 0)	 & $\llap{-}23657221999872$
\tabularnewline[-4pt] 
 (9, 7, 6, 2, 1)	 & $3117411980160$ 	 & (9, 7, 6, 3, 0)	 & $\llap{-}3138370134624$
\tabularnewline[-4pt] 
 (9, 7, 7, 1, 1)	 & $\llap{-}22739284992$ 	 & (9, 7, 7, 2, 0)	 & $\llap{-}37176746592$
\tabularnewline[-4pt] 
 (9, 8, 3, 3, 2)	 & $840536568752160$ 	 & (9, 8, 4, 2, 2)	 & $223306712255904$
\tabularnewline[-4pt] 
 (9, 8, 4, 3, 1)	 & $5559537648384$ 	 & (9, 8, 4, 4, 0)	 & $\llap{-}1093125957120$
\tabularnewline[-4pt] 
 (9, 8, 5, 2, 1)	 & $428753909184$ 	 & (9, 8, 5, 3, 0)	 & $\llap{-}368134832160$
\tabularnewline[-4pt] 
 (9, 8, 6, 1, 1)	 & $\llap{-}5776067616$ 	 & (9, 8, 6, 2, 0)	 & $\llap{-}11043084816$
\tabularnewline[-4pt] 
 (9, 8, 7, 1, 0)	 & $\llap{-}9395616$ 	 & (9, 9, 3, 2, 2)	 & $733831612704$
\tabularnewline[-4pt] 
 (9, 9, 3, 3, 1)	 & $20194851840$ 	 & (9, 9, 4, 2, 1)	 & $4741984896$
\tabularnewline[-4pt] 
 (9, 9, 4, 3, 0)	 & $\llap{-}3110582880$ 	 & (9, 9, 5, 1, 1)	 & $\llap{-}46978560$
\tabularnewline[-4pt] 
 (9, 9, 5, 2, 0)	 & $\llap{-}203310240$ 	 & (9, 9, 6, 1, 0)	 & $\llap{-}251520$
\tabularnewline[-4pt] 
 (10, 4, 4, 4, 3)	 & $211913083229294304$ 	 & (10, 5, 4, 3, 3)	 & $74503089764268384$
\tabularnewline[-4pt] 
 (10, 5, 4, 4, 2)	 & $20254395759934128$ 	 & (10, 5, 5, 3, 2)	 & $7011987726247008$
\tabularnewline[-4pt] 
 (10, 5, 5, 4, 1)	 & $46131775979616$ 	 & (10, 5, 5, 5, 0)	 & $\llap{-}3138370134624$
\tabularnewline[-4pt] 
 (10, 6, 3, 3, 3)	 & $9449678610817056$ 	 & (10, 6, 4, 3, 2)	 & $2530955656217280$
\tabularnewline[-4pt] 
 (10, 6, 4, 4, 1)	 & $16731712682064$ 	 & (10, 6, 5, 2, 2)	 & $225463832566752$
\tabularnewline[-4pt] 
 (10, 6, 5, 3, 1)	 & $5619611350656$ 	 & (10, 6, 5, 4, 0)	 & $\llap{-}1093125957120$
\tabularnewline[-4pt] 
 (10, 6, 6, 2, 1)	 & $149497953456$ 	 & (10, 6, 6, 3, 0)	 & $\llap{-}119442727776$
\tabularnewline[-4pt] 
 (10, 7, 3, 3, 2)	 & $99761061359136$ 	 & (10, 7, 4, 2, 2)	 & $25630803734064$
\tabularnewline[-4pt] 
 (10, 7, 4, 3, 1)	 & $665062141248$ 	 & (10, 7, 4, 4, 0)	 & $\llap{-}119442727776$
\tabularnewline[-4pt] 
 (10, 7, 5, 2, 1)	 & $49806889344$ 	 & (10, 7, 5, 3, 0)	 & $\llap{-}37176746592$
\tabularnewline[-4pt] 
 (10, 7, 6, 1, 1)	 & $\llap{-}273996960$ 	 & (10, 7, 6, 2, 0)	 & $\llap{-}824199120$
\tabularnewline[-4pt] 
 (10, 7, 7, 1, 0)	 & $\llap{-}251520$ 	 & (10, 8, 3, 2, 2)	 & $208350582720$
\tabularnewline[-4pt] 
 (10, 8, 3, 3, 1)	 & $5848333440$ 	 & (10, 8, 4, 2, 1)	 & $1342319904$
\tabularnewline[-4pt] 
 (10, 8, 4, 3, 0)	 & $\llap{-}824199120$ 	 & (10, 8, 5, 1, 1)	 & $\llap{-}6126048$
\tabularnewline[-4pt] 
 (10, 8, 5, 2, 0)	 & $\llap{-}46049040$ 	 & (10, 8, 6, 1, 0)	 & $\llap{-}29136$
\tabularnewline[-4pt] 
 (10, 9, 2, 2, 2)	 & $22958688$ 	 & (10, 9, 3, 2, 1)	 & $679968$
\tabularnewline[-4pt] 
 (10, 9, 3, 3, 0)	 & $\llap{-}251520$ 	 & (10, 9, 4, 1, 1)	 & $4320$
\tabularnewline[-4pt] 
 (10, 9, 4, 2, 0)	 & $\llap{-}29136$ 	 & (11, 4, 4, 3, 3)	 & $135453779066496$
\tabularnewline[-4pt] 
 (11, 4, 4, 4, 2)	 & $34155140507184$ 	 & (11, 5, 3, 3, 3)	 & $41704406393856$
\tabularnewline[-4pt] 
 (11, 5, 4, 3, 2)	 & $10348372749216$ 	 & (11, 5, 4, 4, 1)	 & $68863079616$
\tabularnewline[-4pt] 
 (11, 5, 5, 2, 2)	 & $733831612704$ 	 & (11, 5, 5, 3, 1)	 & $20194851840$
\tabularnewline[-4pt] 
 (11, 5, 5, 4, 0)	 & $\llap{-}3110582880$ 	 & (11, 6, 3, 3, 2)	 & $894337855968$
\tabularnewline[-4pt] 
 (11, 6, 4, 2, 2)	 & $208350582720$ 	 & (11, 6, 4, 3, 1)	 & $5848333440$
\tabularnewline[-4pt] 
 (11, 6, 4, 4, 0)	 & $\llap{-}824199120$ 	 & (11, 6, 5, 2, 1)	 & $351706176$
\tabularnewline[-4pt] 
 (11, 6, 5, 3, 0)	 & $\llap{-}203310240$ 	 & (11, 6, 6, 1, 1)	 & $73824$
\tabularnewline[-4pt] 
 (11, 6, 6, 2, 0)	 & $\llap{-}1679040$ 	 & (11, 7, 3, 2, 2)	 & $3347625888$
\tabularnewline[-4pt] 
 (11, 7, 3, 3, 1)	 & $99761664$ 	 & (11, 7, 4, 2, 1)	 & $19494816$
\tabularnewline[-4pt] 
 (11, 7, 4, 3, 0)	 & $\llap{-}9395616$ 	 & (11, 7, 5, 1, 1)	 & $30720$
\tabularnewline[-4pt] 
 (11, 7, 5, 2, 0)	 & $\llap{-}251520$ 	 & (11, 8, 2, 2, 2)	 & $104352$
\tabularnewline[-4pt] 
 (11, 8, 3, 2, 1)	 & $4320$ 	 & (11, 8, 3, 3, 0)	 & $\llap{-}2208$
\tabularnewline[-4pt] 
 (11, 8, 4, 2, 0)	 & $\llap{-}48$ 	 & (12, 4, 3, 3, 3)	 & $411633120$
\tabularnewline[-4pt] 
 (12, 4, 4, 3, 2)	 & $86694528$ 	 & (12, 4, 4, 4, 1)	 & $\llap{-}317232$
\tabularnewline[-4pt] 
 (12, 5, 3, 3, 2)	 & $19494816$ 	 & (12, 5, 4, 2, 2)	 & $3692400$
\tabularnewline[-4pt] 
 (12, 5, 4, 3, 1)	 & $73824$ 	 & (12, 5, 4, 4, 0)	 & $\llap{-}29136$
\tabularnewline[-4pt] 
 (12, 5, 5, 2, 1)	 & $4320$ 	 & (12, 5, 5, 3, 0)	 & $\llap{-}2208$
\tabularnewline[-4pt] 
 (12, 6, 3, 2, 2)	 & $104352$ 	 & (12, 6, 3, 3, 1)	 & $4320$
\tabularnewline[-4pt] 
 (12, 6, 4, 2, 1)	 & $528$ 	 & (12, 6, 4, 3, 0)	 & $\llap{-}48$
\tabularnewline[-4pt] 
(12, 7, 2, 2, 2)	 & $\llap{-}48$	 &  	 &  
\tablepostambleInstantonNumbers \vskip-18pt 

\tablepreambleInstantonNumbersGone{26}
 (6, 5, 5, 5, 5)	 & $1942248611676865375886400$ 	 & (6, 6, 5, 5, 4)	 & $875590499820813091496688$
\tabularnewline[-4pt] 
 (6, 6, 6, 4, 4)	 & $393169006303652095072272$ 	 & (6, 6, 6, 5, 3)	 & $157120006208930268178944$
\tabularnewline[-4pt] 
 (6, 6, 6, 6, 2)	 & $8106453008262631136544$ 	 & (7, 5, 5, 5, 4)	 & $410426378940650223011136$
\tabularnewline[-4pt] 
 (7, 6, 5, 4, 4)	 & $183290892867377049167040$ 	 & (7, 6, 5, 5, 3)	 & $72977409472143740469600$
\tabularnewline[-4pt] 
 (7, 6, 6, 4, 3)	 & $32349736654103967569424$ 	 & (7, 6, 6, 5, 2)	 & $3721525089274735889952$
\tabularnewline[-4pt] 
 (7, 6, 6, 6, 1)	 & $3782526785435078400$ 	 & (7, 7, 4, 4, 4)	 & $37482832972502804145504$
\tabularnewline[-4pt] 
 (7, 7, 5, 4, 3)	 & $14816698735976435302944$ 	 & (7, 7, 5, 5, 2)	 & $1693905042344566617888$
\tabularnewline[-4pt] 
 (7, 7, 6, 3, 3)	 & $2551778870797810769088$ 	 & (7, 7, 6, 4, 2)	 & $738648123808652479056$
\tabularnewline[-4pt] 
 (7, 7, 6, 5, 1)	 & $1714200026193282432$ 	 & (7, 7, 6, 6, 0)	 & $\llap{-}70199768003592720$
\tabularnewline[-4pt] 
 (7, 7, 7, 3, 2)	 & $55156658249311147008$ 	 & (7, 7, 7, 4, 1)	 & $330197068437575904$
\tabularnewline[-4pt] 
 (7, 7, 7, 5, 0)	 & $\llap{-}30974226462689184$ 	 & (8, 5, 5, 4, 4)	 & $39480418106981466533520$
\tabularnewline[-4pt] 
 (8, 5, 5, 5, 3)	 & $15578338962013218600864$ 	 & (8, 6, 4, 4, 4)	 & $17316166253014236259776$
\tabularnewline[-4pt] 
 (8, 6, 5, 4, 3)	 & $6808548519814415880000$ 	 & (8, 6, 5, 5, 2)	 & $772032475082783116752$
\tabularnewline[-4pt] 
 (8, 6, 6, 3, 3)	 & $1156070611877661822288$ 	 & (8, 6, 6, 4, 2)	 & $333685276507832498868$
\tabularnewline[-4pt] 
 (8, 6, 6, 5, 1)	 & $776764446454911264$ 	 & (8, 6, 6, 6, 0)	 & $\llap{-}30974226462442944$
\tabularnewline[-4pt] 
 (8, 7, 4, 4, 3)	 & $1320063534010559077776$ 	 & (8, 7, 5, 3, 3)	 & $512903625458504140800$
\tabularnewline[-4pt] 
 (8, 7, 5, 4, 2)	 & $147653937102100467360$ 	 & (8, 7, 5, 5, 1)	 & $344594558995733856$
\tabularnewline[-4pt] 
 (8, 7, 6, 3, 2)	 & $24205677459521479872$ 	 & (8, 7, 6, 4, 1)	 & $146449762016579712$
\tabularnewline[-4pt] 
 (8, 7, 6, 5, 0)	 & $\llap{-}13461969999093600$ 	 & (8, 7, 7, 2, 2)	 & $466497292363869360$
\tabularnewline[-4pt] 
 (8, 7, 7, 3, 1)	 & $9764259633947136$ 	 & (8, 7, 7, 4, 0)	 & $\llap{-}2421429008571216$
\tabularnewline[-4pt] 
 (8, 8, 4, 3, 3)	 & $41111092665529983696$ 	 & (8, 8, 4, 4, 2)	 & $11721366876475799760$
\tabularnewline[-4pt] 
 (8, 8, 5, 3, 2)	 & $4443062729668252800$ 	 & (8, 8, 5, 4, 1)	 & $27442949686074432$
\tabularnewline[-4pt] 
 (8, 8, 5, 5, 0)	 & $\llap{-}2421429008571216$ 	 & (8, 8, 6, 2, 2)	 & $193232919723748536$
\tabularnewline[-4pt] 
 (8, 8, 6, 3, 1)	 & $4137353643774720$ 	 & (8, 8, 6, 4, 0)	 & $\llap{-}1000067627325696$
\tabularnewline[-4pt] 
 (8, 8, 7, 2, 1)	 & $46040552189760$ 	 & (8, 8, 7, 3, 0)	 & $\llap{-}62415555336480$
\tabularnewline[-4pt] 
 (8, 8, 8, 1, 1)	 & $\llap{-}285485916336$ 	 & (8, 8, 8, 2, 0)	 & $\llap{-}368134868160$
\tabularnewline[-4pt] 
 (9, 5, 4, 4, 4)	 & $1592870480807489129616$ 	 & (9, 5, 5, 4, 3)	 & $614995894410428607648$
\tabularnewline[-4pt] 
 (9, 5, 5, 5, 2)	 & $67768604147499442944$ 	 & (9, 6, 4, 4, 3)	 & $260001810225111467808$
\tabularnewline[-4pt] 
 (9, 6, 5, 3, 3)	 & $99531632175997196352$ 	 & (9, 6, 5, 4, 2)	 & $28431306071376793872$
\tabularnewline[-4pt] 
 (9, 6, 5, 5, 1)	 & $66591254167125120$ 	 & (9, 6, 6, 3, 2)	 & $4476560442767249184$
\tabularnewline[-4pt] 
 (9, 6, 6, 4, 1)	 & $27663244132912416$ 	 & (9, 6, 6, 5, 0)	 & $\llap{-}2421429008571216$
\tabularnewline[-4pt] 
 (9, 7, 4, 3, 3)	 & $17566723688872737408$ 	 & (9, 7, 4, 4, 2)	 & $4983325664622060480$
\tabularnewline[-4pt] 
 (9, 7, 5, 3, 2)	 & $1869909171436016640$ 	 & (9, 7, 5, 4, 1)	 & $11668379661795360$
\tabularnewline[-4pt] 
 (9, 7, 5, 5, 0)	 & $\llap{-}1000067627051904$ 	 & (9, 7, 6, 2, 2)	 & $78667271480859264$
\tabularnewline[-4pt] 
 (9, 7, 6, 3, 1)	 & $1722815728418688$ 	 & (9, 7, 6, 4, 0)	 & $\llap{-}405156007308576$
\tabularnewline[-4pt] 
 (9, 7, 7, 2, 1)	 & $19460158766688$ 	 & (9, 7, 7, 3, 0)	 & $\llap{-}23657221999872$
\tabularnewline[-4pt] 
 (9, 8, 3, 3, 3)	 & $439198792163267520$ 	 & (9, 8, 4, 3, 2)	 & $122281608111146832$
\tabularnewline[-4pt] 
 (9, 8, 4, 4, 1)	 & $784225690196688$ 	 & (9, 8, 5, 2, 2)	 & $12238194838809408$
\tabularnewline[-4pt] 
 (9, 8, 5, 3, 1)	 & $280061917894368$ 	 & (9, 8, 5, 4, 0)	 & $\llap{-}62415555336480$
\tabularnewline[-4pt] 
 (9, 8, 6, 2, 1)	 & $7932337626384$ 	 & (9, 8, 6, 3, 0)	 & $\llap{-}8738280013680$
\tabularnewline[-4pt] 
 (9, 8, 7, 1, 1)	 & $\llap{-}83145266112$ 	 & (9, 8, 7, 2, 0)	 & $\llap{-}119442727776$
\tabularnewline[-4pt] 
 (9, 8, 8, 1, 0)	 & $\llap{-}46049040$ 	 & (9, 9, 3, 3, 2)	 & $840536568752160$
\tabularnewline[-4pt] 
 (9, 9, 4, 2, 2)	 & $223306712255904$ 	 & (9, 9, 4, 3, 1)	 & $5559537648384$
\tabularnewline[-4pt] 
 (9, 9, 4, 4, 0)	 & $\llap{-}1093125957120$ 	 & (9, 9, 5, 2, 1)	 & $428753909184$
\tabularnewline[-4pt] 
 (9, 9, 5, 3, 0)	 & $\llap{-}368134832160$ 	 & (9, 9, 6, 1, 1)	 & $\llap{-}5776067616$
\tabularnewline[-4pt] 
 (9, 9, 6, 2, 0)	 & $\llap{-}11043084816$ 	 & (9, 9, 7, 1, 0)	 & $\llap{-}9395616$
\tabularnewline[-4pt] 
 (10, 4, 4, 4, 4)	 & $22962634839334473072$ 	 & (10, 5, 4, 4, 3)	 & $8521174156401735360$
\tabularnewline[-4pt] 
 (10, 5, 5, 3, 3)	 & $3144478089605250720$ 	 & (10, 5, 5, 4, 2)	 & $879997438952410320$
\tabularnewline[-4pt] 
 (10, 5, 5, 5, 1)	 & $2055320972401920$ 	 & (10, 6, 4, 3, 3)	 & $1243841228095131744$
\tabularnewline[-4pt] 
 (10, 6, 4, 4, 2)	 & $346443574026127704$ 	 & (10, 6, 5, 3, 2)	 & $125469245902996512$
\tabularnewline[-4pt] 
 (10, 6, 5, 4, 1)	 & $805991883499728$ 	 & (10, 6, 5, 5, 0)	 & $\llap{-}62415555336480$
\tabularnewline[-4pt] 
 (10, 6, 6, 2, 2)	 & $4694772196282128$ 	 & (10, 6, 6, 3, 1)	 & $109867633989312$
\tabularnewline[-4pt] 
 (10, 6, 6, 4, 0)	 & $\llap{-}23657222126952$ 	 & (10, 7, 3, 3, 3)	 & $67323025901888832$
\tabularnewline[-4pt] 
 (10, 7, 4, 3, 2)	 & $18421518301369920$ 	 & (10, 7, 4, 4, 1)	 & $120114507109632$
\tabularnewline[-4pt] 
 (10, 7, 5, 2, 2)	 & $1749317561578944$ 	 & (10, 7, 5, 3, 1)	 & $41786034116160$
\tabularnewline[-4pt] 
 (10, 7, 5, 4, 0)	 & $\llap{-}8738280013680$ 	 & (10, 7, 6, 2, 1)	 & $1179552933984$
\tabularnewline[-4pt] 
 (10, 7, 6, 3, 0)	 & $\llap{-}1093125957120$ 	 & (10, 7, 7, 1, 1)	 & $\llap{-}5776067616$
\tabularnewline[-4pt] 
 (10, 7, 7, 2, 0)	 & $\llap{-}11043084816$ 	 & (10, 8, 3, 3, 2)	 & $294058742512224$
\tabularnewline[-4pt] 
 (10, 8, 4, 2, 2)	 & $76906534439280$ 	 & (10, 8, 4, 3, 1)	 & $1954674541824$
\tabularnewline[-4pt] 
 (10, 8, 4, 4, 0)	 & $\llap{-}368134868160$ 	 & (10, 8, 5, 2, 1)	 & $149497953456$
\tabularnewline[-4pt] 
 (10, 8, 5, 3, 0)	 & $\llap{-}119442727776$ 	 & (10, 8, 6, 1, 1)	 & $\llap{-}1334560224$
\tabularnewline[-4pt] 
 (10, 8, 6, 2, 0)	 & $\llap{-}3110590260$ 	 & (10, 8, 7, 1, 0)	 & $\llap{-}1679040$
\tabularnewline[-4pt] 
 (10, 9, 3, 2, 2)	 & $208350582720$ 	 & (10, 9, 3, 3, 1)	 & $5848333440$
\tabularnewline[-4pt] 
 (10, 9, 4, 2, 1)	 & $1342319904$ 	 & (10, 9, 4, 3, 0)	 & $\llap{-}824199120$
\tabularnewline[-4pt] 
 (10, 9, 5, 1, 1)	 & $\llap{-}6126048$ 	 & (10, 9, 5, 2, 0)	 & $\llap{-}46049040$
\tabularnewline[-4pt] 
 (10, 9, 6, 1, 0)	 & $\llap{-}29136$ 	 & (10, 10, 2, 2, 2)	 & $3666312$
\tabularnewline[-4pt] 
 (10, 10, 3, 2, 1)	 & $104352$ 	 & (10, 10, 3, 3, 0)	 & $\llap{-}29136$
\tabularnewline[-4pt] 
 (10, 10, 4, 1, 1)	 & $528$ 	 & (10, 10, 4, 2, 0)	 & $\llap{-}2292$
\tabularnewline[-4pt] 
 (11, 4, 4, 4, 3)	 & $29809312235610960$ 	 & (11, 5, 4, 3, 3)	 & $10159668608774304$
\tabularnewline[-4pt] 
 (11, 5, 4, 4, 2)	 & $2707370108500416$ 	 & (11, 5, 5, 3, 2)	 & $903625742797728$
\tabularnewline[-4pt] 
 (11, 5, 5, 4, 1)	 & $6007581031968$ 	 & (11, 5, 5, 5, 0)	 & $\llap{-}368134832160$
\tabularnewline[-4pt] 
 (11, 6, 3, 3, 3)	 & $1190848151512512$ 	 & (11, 6, 4, 3, 2)	 & $310831260169488$
\tabularnewline[-4pt] 
 (11, 6, 4, 4, 1)	 & $2072265197088$ 	 & (11, 6, 5, 2, 2)	 & $25630803734064$
\tabularnewline[-4pt] 
 (11, 6, 5, 3, 1)	 & $665062141248$ 	 & (11, 6, 5, 4, 0)	 & $\llap{-}119442727776$
\tabularnewline[-4pt] 
 (11, 6, 6, 2, 1)	 & $15766948032$ 	 & (11, 6, 6, 3, 0)	 & $\llap{-}11043084816$
\tabularnewline[-4pt] 
 (11, 7, 3, 3, 2)	 & $10348372749216$ 	 & (11, 7, 4, 2, 2)	 & $2545705442112$
\tabularnewline[-4pt] 
 (11, 7, 4, 3, 1)	 & $68863079616$ 	 & (11, 7, 4, 4, 0)	 & $\llap{-}11043084816$
\tabularnewline[-4pt] 
 (11, 7, 5, 2, 1)	 & $4741984896$ 	 & (11, 7, 5, 3, 0)	 & $\llap{-}3110582880$
\tabularnewline[-4pt] 
 (11, 7, 6, 1, 1)	 & $\llap{-}6126048$ 	 & (11, 7, 6, 2, 0)	 & $\llap{-}46049040$
\tabularnewline[-4pt] 
 (11, 7, 7, 1, 0)	 & $\llap{-}2208$ 	 & (11, 8, 3, 2, 2)	 & $14105356368$
\tabularnewline[-4pt] 
 (11, 8, 3, 3, 1)	 & $411633120$ 	 & (11, 8, 4, 2, 1)	 & $86694528$
\tabularnewline[-4pt] 
 (11, 8, 4, 3, 0)	 & $\llap{-}46049040$ 	 & (11, 8, 5, 1, 1)	 & $73824$
\tabularnewline[-4pt] 
 (11, 8, 5, 2, 0)	 & $\llap{-}1679040$ 	 & (11, 8, 6, 1, 0)	 & $\llap{-}48$
\tabularnewline[-4pt] 
 (11, 9, 2, 2, 2)	 & $104352$ 	 & (11, 9, 3, 2, 1)	 & $4320$
\tabularnewline[-4pt] 
 (11, 9, 3, 3, 0)	 & $\llap{-}2208$ 	 & (11, 9, 4, 2, 0)	 & $\llap{-}48$
\tabularnewline[-4pt] 
 (12, 4, 4, 3, 3)	 & $3154648420512$ 	 & (12, 4, 4, 4, 2)	 & $755118268080$
\tabularnewline[-4pt] 
 (12, 5, 3, 3, 3)	 & $894337855968$ 	 & (12, 5, 4, 3, 2)	 & $208350582720$
\tabularnewline[-4pt] 
 (12, 5, 4, 4, 1)	 & $1342319904$ 	 & (12, 5, 5, 2, 2)	 & $12168742800$
\tabularnewline[-4pt] 
 (12, 5, 5, 3, 1)	 & $351706176$ 	 & (12, 5, 5, 4, 0)	 & $\llap{-}46049040$
\tabularnewline[-4pt] 
 (12, 6, 3, 3, 2)	 & $14105356368$ 	 & (12, 6, 4, 2, 2)	 & $2937953580$
\tabularnewline[-4pt] 
 (12, 6, 4, 3, 1)	 & $86694528$ 	 & (12, 6, 4, 4, 0)	 & $\llap{-}9396672$
\tabularnewline[-4pt] 
 (12, 6, 5, 2, 1)	 & $3692400$ 	 & (12, 6, 5, 3, 0)	 & $\llap{-}1679040$
\tabularnewline[-4pt] 
 (12, 6, 6, 1, 1)	 & $528$ 	 & (12, 6, 6, 2, 0)	 & $\llap{-}2292$
\tabularnewline[-4pt] 
 (12, 7, 3, 2, 2)	 & $22958688$ 	 & (12, 7, 3, 3, 1)	 & $679968$
\tabularnewline[-4pt] 
 (12, 7, 4, 2, 1)	 & $104352$ 	 & (12, 7, 4, 3, 0)	 & $\llap{-}29136$
\tabularnewline[-4pt] 
 (12, 7, 5, 2, 0)	 & $\llap{-}48$ 	 & (12, 8, 2, 2, 2)	 & $\llap{-}2292$
\tabularnewline[-4pt] 
 (12, 8, 3, 2, 1)	 & $\llap{-}48$ 	 & (13, 4, 3, 3, 3)	 & $\llap{-}9395616$
\tabularnewline[-4pt] 
 (13, 4, 4, 3, 2)	 & $\llap{-}1679040$ 	 & (13, 4, 4, 4, 1)	 & $\llap{-}29136$
\tabularnewline[-4pt] 
 (13, 5, 3, 3, 2)	 & $\llap{-}251520$ 	 & (13, 5, 4, 2, 2)	 & $\llap{-}29136$
\tabularnewline[-4pt] 
 (13, 5, 4, 3, 1)	 & $\llap{-}2208$ 	 & (13, 6, 3, 2, 2)	 & $\llap{-}48$
\tablepostambleInstantonNumbers \vskip-18pt 

\tablepreambleInstantonNumbersGone{27}
 (6, 6, 5, 5, 5)	 & $21652833697465345825473216$ 	 & (6, 6, 6, 5, 4)	 & $9930103217664737498362752$
\tabularnewline[-4pt] 
 (6, 6, 6, 6, 3)	 & $1836744316255728093377504$ 	 & (7, 5, 5, 5, 5)	 & $10424016378695267249799168$
\tabularnewline[-4pt] 
 (7, 6, 5, 5, 4)	 & $4760519298349779631098496$ 	 & (7, 6, 6, 4, 4)	 & $2166849385502855883360624$
\tabularnewline[-4pt] 
 (7, 6, 6, 5, 3)	 & $874186334599890564570432$ 	 & (7, 6, 6, 6, 2)	 & $46465382392154451939072$
\tabularnewline[-4pt] 
 (7, 7, 5, 4, 4)	 & $1028056808615702374843808$ 	 & (7, 7, 5, 5, 3)	 & $413486815490189461567488$
\tabularnewline[-4pt] 
 (7, 7, 6, 4, 3)	 & $186191937088342721822720$ 	 & (7, 7, 6, 5, 2)	 & $21768872364464612982848$
\tabularnewline[-4pt] 
 (7, 7, 6, 6, 1)	 & $22298898608487186016$ 	 & (7, 7, 7, 3, 3)	 & $15448516510046832328704$
\tabularnewline[-4pt] 
 (7, 7, 7, 4, 2)	 & $4498860807385583308224$ 	 & (7, 7, 7, 5, 1)	 & $10343645286248254464$
\tabularnewline[-4pt] 
 (7, 7, 7, 6, 0)	 & $\llap{-}448609384225666560$ 	 & (8, 5, 5, 5, 4)	 & $1092905783425309905030912$
\tabularnewline[-4pt] 
 (8, 6, 5, 4, 4)	 & $492306078671771466820064$ 	 & (8, 6, 5, 5, 3)	 & $197228805113267715996416$
\tabularnewline[-4pt] 
 (8, 6, 6, 4, 3)	 & $88255562432100365628448$ 	 & (8, 6, 6, 5, 2)	 & $10255257716817787654800$
\tabularnewline[-4pt] 
 (8, 6, 6, 6, 1)	 & $10474506505204294080$ 	 & (8, 7, 4, 4, 4)	 & $102771309648644620264992$
\tabularnewline[-4pt] 
 (8, 7, 5, 4, 3)	 & $40909136655697997550496$ 	 & (8, 7, 5, 5, 2)	 & $4727639000530609605248$
\tabularnewline[-4pt] 
 (8, 7, 6, 3, 3)	 & $7175765870665787614560$ 	 & (8, 7, 6, 4, 2)	 & $2085054959697204491664$
\tabularnewline[-4pt] 
 (8, 7, 6, 5, 1)	 & $4813453384531950048$ 	 & (8, 7, 6, 6, 0)	 & $\llap{-}204417944244944160$
\tabularnewline[-4pt] 
 (8, 7, 7, 3, 2)	 & $161547529688023582240$ 	 & (8, 7, 7, 4, 1)	 & $952690031556127200$
\tabularnewline[-4pt] 
 (8, 7, 7, 5, 0)	 & $\llap{-}91914770340089280$ 	 & (8, 8, 4, 4, 3)	 & $3788700251223612606992$
\tabularnewline[-4pt] 
 (8, 8, 5, 3, 3)	 & $1485322356047272900832$ 	 & (8, 8, 5, 4, 2)	 & $429548694784566103072$
\tabularnewline[-4pt] 
 (8, 8, 5, 5, 1)	 & $998655920868343392$ 	 & (8, 8, 6, 3, 2)	 & $72150775131428807216$
\tabularnewline[-4pt] 
 (8, 8, 6, 4, 1)	 & $430155649709142688$ 	 & (8, 8, 6, 5, 0)	 & $\llap{-}40752668020556480$
\tabularnewline[-4pt] 
 (8, 8, 7, 2, 2)	 & $1469239641997479152$ 	 & (8, 8, 7, 3, 1)	 & $29808201189352224$
\tabularnewline[-4pt] 
 (8, 8, 7, 4, 0)	 & $\llap{-}7656245352109120$ 	 & (8, 8, 8, 2, 1)	 & $137150695673232$
\tabularnewline[-4pt] 
 (8, 8, 8, 3, 0)	 & $\llap{-}219322647849280$ 	 & (9, 5, 5, 4, 4)	 & $50915057095808531822560$
\tabularnewline[-4pt] 
 (9, 5, 5, 5, 3)	 & $20127314721021911676928$ 	 & (9, 6, 4, 4, 4)	 & $22390120482647854754272$
\tabularnewline[-4pt] 
 (9, 6, 5, 4, 3)	 & $8820756832744533115968$ 	 & (9, 6, 5, 5, 2)	 & $1003282043598184664064$
\tabularnewline[-4pt] 
 (9, 6, 6, 3, 3)	 & $1505336844006318854880$ 	 & (9, 6, 6, 4, 2)	 & $434985047971986769440$
\tabularnewline[-4pt] 
 (9, 6, 6, 5, 1)	 & $1011424106940266496$ 	 & (9, 6, 6, 6, 0)	 & $\llap{-}40752668020556480$
\tabularnewline[-4pt] 
 (9, 7, 4, 4, 3)	 & $1721332405243795563776$ 	 & (9, 7, 5, 3, 3)	 & $670374726927803695104$
\tabularnewline[-4pt] 
 (9, 7, 5, 4, 2)	 & $193221962896538278144$ 	 & (9, 7, 5, 5, 1)	 & $450524230700139520$
\tabularnewline[-4pt] 
 (9, 7, 6, 3, 2)	 & $31875712216477298048$ 	 & (9, 7, 6, 4, 1)	 & $192137049043395264$
\tabularnewline[-4pt] 
 (9, 7, 6, 5, 0)	 & $\llap{-}17802901998922240$ 	 & (9, 7, 7, 2, 2)	 & $623145599186074560$
\tabularnewline[-4pt] 
 (9, 7, 7, 3, 1)	 & $12940864211175424$ 	 & (9, 7, 7, 4, 0)	 & $\llap{-}3238317072282240$
\tabularnewline[-4pt] 
 (9, 8, 4, 3, 3)	 & $54398198798074388800$ 	 & (9, 8, 4, 4, 2)	 & $15534320958543602432$
\tabularnewline[-4pt] 
 (9, 8, 5, 3, 2)	 & $5907389929476004224$ 	 & (9, 8, 5, 4, 1)	 & $36360838742410016$
\tabularnewline[-4pt] 
 (9, 8, 5, 5, 0)	 & $\llap{-}3238317072282240$ 	 & (9, 8, 6, 2, 2)	 & $259636655223086848$
\tabularnewline[-4pt] 
 (9, 8, 6, 3, 1)	 & $5516943176462464$ 	 & (9, 8, 6, 4, 0)	 & $\llap{-}1345692401785920$
\tabularnewline[-4pt] 
 (9, 8, 7, 2, 1)	 & $60842853741632$ 	 & (9, 8, 7, 3, 0)	 & $\llap{-}85781801925696$
\tabularnewline[-4pt] 
 (9, 8, 8, 1, 1)	 & $\llap{-}425570551040$ 	 & (9, 8, 8, 2, 0)	 & $\llap{-}531223501536$
\tabularnewline[-4pt] 
 (9, 9, 3, 3, 3)	 & $595724740895599616$ 	 & (9, 9, 4, 3, 2)	 & $166286435635301824$
\tabularnewline[-4pt] 
 (9, 9, 4, 4, 1)	 & $1063377379790976$ 	 & (9, 9, 5, 2, 2)	 & $16773356868339520$
\tabularnewline[-4pt] 
 (9, 9, 5, 3, 1)	 & $381144946134016$ 	 & (9, 9, 5, 4, 0)	 & $\llap{-}85781801925696$
\tabularnewline[-4pt] 
 (9, 9, 6, 2, 1)	 & $10747142102848$ 	 & (9, 9, 6, 3, 0)	 & $\llap{-}12215408263200$
\tabularnewline[-4pt] 
 (9, 9, 7, 1, 1)	 & $\llap{-}126220459008$ 	 & (9, 9, 7, 2, 0)	 & $\llap{-}174588053440$
\tabularnewline[-4pt] 
 (9, 9, 8, 1, 0)	 & $\llap{-}76342880$ 	 & (10, 5, 4, 4, 4)	 & $928035242813679386880$
\tabularnewline[-4pt] 
 (10, 5, 5, 4, 3)	 & $356555435715410957376$ 	 & (10, 5, 5, 5, 2)	 & $38978014973683179264$
\tabularnewline[-4pt] 
 (10, 6, 4, 4, 3)	 & $149677043087457971504$ 	 & (10, 6, 5, 3, 3)	 & $56981868551941098752$
\tabularnewline[-4pt] 
 (10, 6, 5, 4, 2)	 & $16227227828645916976$ 	 & (10, 6, 5, 5, 1)	 & $38035145900721056$
\tabularnewline[-4pt] 
 (10, 6, 6, 3, 2)	 & $2517326301790872576$ 	 & (10, 6, 6, 4, 1)	 & $15666901809220000$
\tabularnewline[-4pt] 
 (10, 6, 6, 5, 0)	 & $\llap{-}1345692401785920$ 	 & (10, 7, 4, 3, 3)	 & $9881551554602229408$
\tabularnewline[-4pt] 
 (10, 7, 4, 4, 2)	 & $2792965386085320880$ 	 & (10, 7, 5, 3, 2)	 & $1040501263302552608$
\tabularnewline[-4pt] 
 (10, 7, 5, 4, 1)	 & $6536958768444736$ 	 & (10, 7, 5, 5, 0)	 & $\llap{-}548768158982912$
\tabularnewline[-4pt] 
 (10, 7, 6, 2, 2)	 & $42751365570060704$ 	 & (10, 7, 6, 3, 1)	 & $950472340606656$
\tabularnewline[-4pt] 
 (10, 7, 6, 4, 0)	 & $\llap{-}219322647849280$ 	 & (10, 7, 7, 2, 1)	 & $10748457996288$
\tabularnewline[-4pt] 
 (10, 7, 7, 3, 0)	 & $\llap{-}12215408263200$ 	 & (10, 8, 3, 3, 3)	 & $237188563748902272$
\tabularnewline[-4pt] 
 (10, 8, 4, 3, 2)	 & $65682988009064448$ 	 & (10, 8, 4, 4, 1)	 & $423655401709600$
\tabularnewline[-4pt] 
 (10, 8, 5, 2, 2)	 & $6466615047655104$ 	 & (10, 8, 5, 3, 1)	 & $150108480756288$
\tabularnewline[-4pt] 
 (10, 8, 5, 4, 0)	 & $\llap{-}32779917518560$ 	 & (10, 8, 6, 2, 1)	 & $4272120540288$
\tabularnewline[-4pt] 
 (10, 8, 6, 3, 0)	 & $\llap{-}4429601736480$ 	 & (10, 8, 7, 1, 1)	 & $\llap{-}35309984768$
\tabularnewline[-4pt] 
 (10, 8, 7, 2, 0)	 & $\llap{-}55127514240$ 	 & (10, 8, 8, 1, 0)	 & $\llap{-}16170272$
\tabularnewline[-4pt] 
 (10, 9, 3, 3, 2)	 & $418651558115072$ 	 & (10, 9, 4, 2, 2)	 & $110090172437152$
\tabularnewline[-4pt] 
 (10, 9, 4, 3, 1)	 & $2778335397408$ 	 & (10, 9, 4, 4, 0)	 & $\llap{-}531223501536$
\tabularnewline[-4pt] 
 (10, 9, 5, 2, 1)	 & $213298398784$ 	 & (10, 9, 5, 3, 0)	 & $\llap{-}174588053440$
\tabularnewline[-4pt] 
 (10, 9, 6, 1, 1)	 & $\llap{-}2202805408$ 	 & (10, 9, 6, 2, 0)	 & $\llap{-}4775506080$
\tabularnewline[-4pt] 
 (10, 9, 7, 1, 0)	 & $\llap{-}3031872$ 	 & (10, 10, 3, 2, 2)	 & $87650018048$
\tabularnewline[-4pt] 
 (10, 10, 3, 3, 1)	 & $2496782816$ 	 & (10, 10, 4, 2, 1)	 & $561090816$
\tabularnewline[-4pt] 
 (10, 10, 4, 3, 0)	 & $\llap{-}327015680$ 	 & (10, 10, 5, 1, 1)	 & $\llap{-}1114976$
\tabularnewline[-4pt] 
 (10, 10, 5, 2, 0)	 & $\llap{-}16170272$ 	 & (10, 10, 6, 1, 0)	 & $\llap{-}5600$
\tabularnewline[-4pt] 
 (11, 4, 4, 4, 4)	 & $5224733955268106112$ 	 & (11, 5, 4, 4, 3)	 & $1905129808949968992$
\tabularnewline[-4pt] 
 (11, 5, 5, 3, 3)	 & $689915910456635392$ 	 & (11, 5, 5, 4, 2)	 & $190939831236687552$
\tabularnewline[-4pt] 
 (11, 5, 5, 5, 1)	 & $442883019280896$ 	 & (11, 6, 4, 3, 3)	 & $265869597857942752$
\tabularnewline[-4pt] 
 (11, 6, 4, 4, 2)	 & $73138223025414832$ 	 & (11, 6, 5, 3, 2)	 & $25872896752378400$
\tabularnewline[-4pt] 
 (11, 6, 5, 4, 1)	 & $168513518883456$ 	 & (11, 6, 5, 5, 0)	 & $\llap{-}12215408263200$
\tabularnewline[-4pt] 
 (11, 6, 6, 2, 2)	 & $894838196834976$ 	 & (11, 6, 6, 3, 1)	 & $21688029832256$
\tabularnewline[-4pt] 
 (11, 6, 6, 4, 0)	 & $\llap{-}4429601736480$ 	 & (11, 7, 3, 3, 3)	 & $13183353406838784$
\tabularnewline[-4pt] 
 (11, 7, 4, 3, 2)	 & $3545177906512576$ 	 & (11, 7, 4, 4, 1)	 & $23383915823040$
\tabularnewline[-4pt] 
 (11, 7, 5, 2, 2)	 & $319546789488192$ 	 & (11, 7, 5, 3, 1)	 & $7905805097984$
\tabularnewline[-4pt] 
 (11, 7, 5, 4, 0)	 & $\llap{-}1559208760288$ 	 & (11, 7, 6, 2, 1)	 & $213298398784$
\tabularnewline[-4pt] 
 (11, 7, 6, 3, 0)	 & $\llap{-}174588053440$ 	 & (11, 7, 7, 1, 1)	 & $\llap{-}471020544$
\tabularnewline[-4pt] 
 (11, 7, 7, 2, 0)	 & $\llap{-}1292723968$ 	 & (11, 8, 3, 3, 2)	 & $47693058783296$
\tabularnewline[-4pt] 
 (11, 8, 4, 2, 2)	 & $12091316695232$ 	 & (11, 8, 4, 3, 1)	 & $318330381792$
\tabularnewline[-4pt] 
 (11, 8, 4, 4, 0)	 & $\llap{-}55127514240$ 	 & (11, 8, 5, 2, 1)	 & $23350187616$
\tabularnewline[-4pt] 
 (11, 8, 5, 3, 0)	 & $\llap{-}16642969280$ 	 & (11, 8, 6, 1, 1)	 & $\llap{-}87015936$
\tabularnewline[-4pt] 
 (11, 8, 6, 2, 0)	 & $\llap{-}327015680$ 	 & (11, 8, 7, 1, 0)	 & $\llap{-}61920$
\tabularnewline[-4pt] 
 (11, 9, 3, 2, 2)	 & $22327107072$ 	 & (11, 9, 3, 3, 1)	 & $646886400$
\tabularnewline[-4pt] 
 (11, 9, 4, 2, 1)	 & $138982240$ 	 & (11, 9, 4, 3, 0)	 & $\llap{-}76342880$
\tabularnewline[-4pt] 
 (11, 9, 5, 1, 1)	 & $75776$ 	 & (11, 9, 5, 2, 0)	 & $\llap{-}3031872$
\tabularnewline[-4pt] 
 (11, 9, 6, 1, 0)	 & $\llap{-}224$ 	 & (11, 10, 2, 2, 2)	 & $29808$
\tabularnewline[-4pt] 
 (11, 10, 3, 2, 1)	 & $928$ 	 & (11, 10, 3, 3, 0)	 & $\llap{-}224$
\tabularnewline[-4pt] 
 (12, 4, 4, 4, 3)	 & $1803381971700144$ 	 & (12, 5, 4, 3, 3)	 & $586171325733792$
\tabularnewline[-4pt] 
 (12, 5, 4, 4, 2)	 & $151342528026688$ 	 & (12, 5, 5, 3, 2)	 & $47693058783296$
\tabularnewline[-4pt] 
 (12, 5, 5, 4, 1)	 & $318330381792$ 	 & (12, 5, 5, 5, 0)	 & $\llap{-}16642969280$
\tabularnewline[-4pt] 
 (12, 6, 3, 3, 3)	 & $60862991224384$ 	 & (12, 6, 4, 3, 2)	 & $15219924472416$
\tabularnewline[-4pt] 
 (12, 6, 4, 4, 1)	 & $101366312448$ 	 & (12, 6, 5, 2, 2)	 & $1103727042528$
\tabularnewline[-4pt] 
 (12, 6, 5, 3, 1)	 & $30154035584$ 	 & (12, 6, 5, 4, 0)	 & $\llap{-}4775506080$
\tabularnewline[-4pt] 
 (12, 6, 6, 2, 1)	 & $554248640$ 	 & (12, 6, 6, 3, 0)	 & $\llap{-}327015680$
\tabularnewline[-4pt] 
 (12, 7, 3, 3, 2)	 & $385951211712$ 	 & (12, 7, 4, 2, 2)	 & $87650018048$
\tabularnewline[-4pt] 
 (12, 7, 4, 3, 1)	 & $2496782816$ 	 & (12, 7, 4, 4, 0)	 & $\llap{-}327015680$
\tabularnewline[-4pt] 
 (12, 7, 5, 2, 1)	 & $138982240$ 	 & (12, 7, 5, 3, 0)	 & $\llap{-}76342880$
\tabularnewline[-4pt] 
 (12, 7, 6, 1, 1)	 & $42560$ 	 & (12, 7, 6, 2, 0)	 & $\llap{-}484896$
\tabularnewline[-4pt] 
 (12, 8, 3, 2, 2)	 & $230549312$ 	 & (12, 8, 3, 3, 1)	 & $6953664$
\tabularnewline[-4pt] 
 (12, 8, 4, 2, 1)	 & $1194656$ 	 & (12, 8, 4, 3, 0)	 & $\llap{-}484896$
\tabularnewline[-4pt] 
 (12, 8, 5, 1, 1)	 & $928$ 	 & (12, 8, 5, 2, 0)	 & $\llap{-}5600$
\tabularnewline[-4pt] 
 (12, 9, 2, 2, 2)	 & $\llap{-}5600$ 	 & (12, 9, 3, 2, 1)	 & $\llap{-}224$
\tabularnewline[-4pt] 
 (13, 4, 4, 3, 3)	 & $2496782816$ 	 & (13, 4, 4, 4, 2)	 & $561090816$
\tabularnewline[-4pt] 
 (13, 5, 3, 3, 3)	 & $646886400$ 	 & (13, 5, 4, 3, 2)	 & $138982240$
\tabularnewline[-4pt] 
 (13, 5, 4, 4, 1)	 & $\llap{-}1114976$ 	 & (13, 5, 5, 2, 2)	 & $6414464$
\tabularnewline[-4pt] 
 (13, 5, 5, 3, 1)	 & $75776$ 	 & (13, 5, 5, 4, 0)	 & $\llap{-}61920$
\tabularnewline[-4pt] 
 (13, 6, 3, 3, 2)	 & $6953664$ 	 & (13, 6, 4, 2, 2)	 & $1194656$
\tabularnewline[-4pt] 
 (13, 6, 4, 3, 1)	 & $42560$ 	 & (13, 6, 4, 4, 0)	 & $\llap{-}5600$
\tabularnewline[-4pt] 
 (13, 6, 5, 2, 1)	 & $928$ 	 & (13, 6, 5, 3, 0)	 & $\llap{-}224$
\tabularnewline[-4pt] 
(13, 7, 3, 2, 2)	 & $928$	 &  	 &  
\tablepostambleInstantonNumbers \vskip-18pt 

\tablepreambleInstantonNumbersGone{28}
 (6, 6, 6, 5, 5)	 & $281396105446038829764570336$ 	 & (6, 6, 6, 6, 4)	 & $131098620977119640532677640$
\tabularnewline[-4pt] 
 (7, 6, 5, 5, 5)	 & $138237049377631956406444992$ 	 & (7, 6, 6, 5, 4)	 & $64187540696210944731760560$
\tabularnewline[-4pt] 
 (7, 6, 6, 6, 3)	 & $12132928161031878393585456$ 	 & (7, 7, 5, 5, 4)	 & $31280567724121492258032672$
\tabularnewline[-4pt] 
 (7, 7, 6, 4, 4)	 & $14433618175802344857159648$ 	 & (7, 7, 6, 5, 3)	 & $5877368216672968441640544$
\tabularnewline[-4pt] 
 (7, 7, 6, 6, 2)	 & $321577464094704294681456$ 	 & (7, 7, 7, 4, 3)	 & $1295336936098611323052960$
\tabularnewline[-4pt] 
 (7, 7, 7, 5, 2)	 & $153782147365152593438592$ 	 & (7, 7, 7, 6, 1)	 & $158207445568417102656$
\tabularnewline[-4pt] 
 (7, 7, 7, 7, 0)	 & $\llap{-}3435204329200397376$ 	 & (8, 5, 5, 5, 5)	 & $33565543023690642272679360$
\tabularnewline[-4pt] 
 (8, 6, 5, 5, 4)	 & $15461693507386297016774064$ 	 & (8, 6, 6, 4, 4)	 & $7101369488031789966243504$
\tabularnewline[-4pt] 
 (8, 6, 6, 5, 3)	 & $2882884240831835131741344$ 	 & (8, 6, 6, 6, 2)	 & $156242327334663646224888$
\tabularnewline[-4pt] 
 (8, 7, 5, 4, 4)	 & $3408941234035601841250272$ 	 & (8, 7, 5, 5, 3)	 & $1380187929485720284723200$
\tabularnewline[-4pt] 
 (8, 7, 6, 4, 3)	 & $627928503566826840282528$ 	 & (8, 7, 6, 5, 2)	 & $74184255759899839436832$
\tabularnewline[-4pt] 
 (8, 7, 6, 6, 1)	 & $76233400110735695520$ 	 & (8, 7, 7, 3, 3)	 & $53848312436074659256320$
\tabularnewline[-4pt] 
 (8, 7, 7, 4, 2)	 & $15741021131132583500640$ 	 & (8, 7, 7, 5, 1)	 & $35900356719080933088$
\tabularnewline[-4pt] 
 (8, 7, 7, 6, 0)	 & $\llap{-}1616941505273075616$ 	 & (8, 8, 4, 4, 4)	 & $353812311322644431070252$
\tabularnewline[-4pt] 
 (8, 8, 5, 4, 3)	 & $141959405252146760964096$ 	 & (8, 8, 5, 5, 2)	 & $16604759404714367599872$
\tabularnewline[-4pt] 
 (8, 8, 6, 3, 3)	 & $25428538397714935697520$ 	 & (8, 8, 6, 4, 2)	 & $7419452787535790690304$
\tabularnewline[-4pt] 
 (8, 8, 6, 5, 1)	 & $17001413171806593504$ 	 & (8, 8, 6, 6, 0)	 & $\llap{-}752251254607854720$
\tabularnewline[-4pt] 
 (8, 8, 7, 3, 2)	 & $599695012064780172192$ 	 & (8, 8, 7, 4, 1)	 & $3469894328671355136$
\tabularnewline[-4pt] 
 (8, 8, 7, 5, 0)	 & $\llap{-}345708325307878560$ 	 & (8, 8, 8, 2, 2)	 & $5923901626220768640$
\tabularnewline[-4pt] 
 (8, 8, 8, 3, 1)	 & $115480640308379520$ 	 & (8, 8, 8, 4, 0)	 & $\llap{-}30974226462442944$
\tabularnewline[-4pt] 
 (9, 5, 5, 5, 4)	 & $1771748988189311827248096$ 	 & (9, 6, 5, 4, 4)	 & $801414379844189789048640$
\tabularnewline[-4pt] 
 (9, 6, 5, 5, 3)	 & $322017472455505977946944$ 	 & (9, 6, 6, 4, 3)	 & $144746838738025811348976$
\tabularnewline[-4pt] 
 (9, 6, 6, 5, 2)	 & $16899949831877607148848$ 	 & (9, 6, 6, 6, 1)	 & $17294122346233404288$
\tabularnewline[-4pt] 
 (9, 7, 4, 4, 4)	 & $168965147904886287828768$ 	 & (9, 7, 5, 4, 3)	 & $67481808424198752531456$
\tabularnewline[-4pt] 
 (9, 7, 5, 5, 2)	 & $7838457970390311440448$ 	 & (9, 7, 6, 3, 3)	 & $11940345167441877911232$
\tabularnewline[-4pt] 
 (9, 7, 6, 4, 2)	 & $3475713613337883829632$ 	 & (9, 7, 6, 5, 1)	 & $8000113898763198720$
\tabularnewline[-4pt] 
 (9, 7, 6, 6, 0)	 & $\llap{-}345708325307878560$ 	 & (9, 7, 7, 3, 2)	 & $274043841200077763232$
\tabularnewline[-4pt] 
 (9, 7, 7, 4, 1)	 & $1603689039494806080$ 	 & (9, 7, 7, 5, 0)	 & $\llap{-}156843054827785632$
\tabularnewline[-4pt] 
 (9, 8, 4, 4, 3)	 & $6366888695682697842144$ 	 & (9, 8, 5, 3, 3)	 & $2506621507185094907424$
\tabularnewline[-4pt] 
 (9, 8, 5, 4, 2)	 & $726440453016521093136$ 	 & (9, 8, 5, 5, 1)	 & $1685025369541284480$
\tabularnewline[-4pt] 
 (9, 8, 6, 3, 2)	 & $123424186729923646080$ 	 & (9, 8, 6, 4, 1)	 & $730324510146863712$
\tabularnewline[-4pt] 
 (9, 8, 6, 5, 0)	 & $\llap{-}70199768003592720$ 	 & (9, 8, 7, 2, 2)	 & $2579059445631070176$
\tabularnewline[-4pt] 
 (9, 8, 7, 3, 1)	 & $51494523118178400$ 	 & (9, 8, 7, 4, 0)	 & $\llap{-}13461969999093600$
\tabularnewline[-4pt] 
 (9, 8, 8, 2, 1)	 & $230627260020528$ 	 & (9, 8, 8, 3, 0)	 & $\llap{-}405156007308576$
\tabularnewline[-4pt] 
 (9, 9, 4, 3, 3)	 & $94786453063647970848$ 	 & (9, 9, 4, 4, 2)	 & $27149794915803480576$
\tabularnewline[-4pt] 
 (9, 9, 5, 3, 2)	 & $10388937656332753536$ 	 & (9, 9, 5, 4, 1)	 & $63502596678612960$
\tabularnewline[-4pt] 
 (9, 9, 5, 5, 0)	 & $\llap{-}5758034709276000$ 	 & (9, 9, 6, 2, 2)	 & $465966091929471216$
\tabularnewline[-4pt] 
 (9, 9, 6, 3, 1)	 & $9751729607816928$ 	 & (9, 9, 6, 4, 0)	 & $\llap{-}2421429008571216$
\tabularnewline[-4pt] 
 (9, 9, 7, 2, 1)	 & $105019516952544$ 	 & (9, 9, 7, 3, 0)	 & $\llap{-}160791639748800$
\tabularnewline[-4pt] 
 (9, 9, 8, 1, 1)	 & $\llap{-}929847901440$ 	 & (9, 9, 8, 2, 0)	 & $\llap{-}1093125957120$
\tabularnewline[-4pt] 
 (9, 9, 9, 1, 0)	 & $\llap{-}203310240$ 	 & (10, 5, 5, 4, 4)	 & $39480418106981466533520$
\tabularnewline[-4pt] 
 (10, 5, 5, 5, 3)	 & $15578338962013218600864$ 	 & (10, 6, 4, 4, 4)	 & $17316166253014236259776$
\tabularnewline[-4pt] 
 (10, 6, 5, 4, 3)	 & $6808548519814415880000$ 	 & (10, 6, 5, 5, 2)	 & $772032475082783116752$
\tabularnewline[-4pt] 
 (10, 6, 6, 3, 3)	 & $1156070611877661822288$ 	 & (10, 6, 6, 4, 2)	 & $333685276507832498868$
\tabularnewline[-4pt] 
 (10, 6, 6, 5, 1)	 & $776764446454911264$ 	 & (10, 6, 6, 6, 0)	 & $\llap{-}30974226462442944$
\tabularnewline[-4pt] 
 (10, 7, 4, 4, 3)	 & $1320063534010559077776$ 	 & (10, 7, 5, 3, 3)	 & $512903625458504140800$
\tabularnewline[-4pt] 
 (10, 7, 5, 4, 2)	 & $147653937102100467360$ 	 & (10, 7, 5, 5, 1)	 & $344594558995733856$
\tabularnewline[-4pt] 
 (10, 7, 6, 3, 2)	 & $24205677459521479872$ 	 & (10, 7, 6, 4, 1)	 & $146449762016579712$
\tabularnewline[-4pt] 
 (10, 7, 6, 5, 0)	 & $\llap{-}13461969999093600$ 	 & (10, 7, 7, 2, 2)	 & $466497292363869360$
\tabularnewline[-4pt] 
 (10, 7, 7, 3, 1)	 & $9764259633947136$ 	 & (10, 7, 7, 4, 0)	 & $\llap{-}2421429008571216$
\tabularnewline[-4pt] 
 (10, 8, 4, 3, 3)	 & $41111092665529983696$ 	 & (10, 8, 4, 4, 2)	 & $11721366876475799760$
\tabularnewline[-4pt] 
 (10, 8, 5, 3, 2)	 & $4443062729668252800$ 	 & (10, 8, 5, 4, 1)	 & $27442949686074432$
\tabularnewline[-4pt] 
 (10, 8, 5, 5, 0)	 & $\llap{-}2421429008571216$ 	 & (10, 8, 6, 2, 2)	 & $193232919723748536$
\tabularnewline[-4pt] 
 (10, 8, 6, 3, 1)	 & $4137353643774720$ 	 & (10, 8, 6, 4, 0)	 & $\llap{-}1000067627325696$
\tabularnewline[-4pt] 
 (10, 8, 7, 2, 1)	 & $46040552189760$ 	 & (10, 8, 7, 3, 0)	 & $\llap{-}62415555336480$
\tabularnewline[-4pt] 
 (10, 8, 8, 1, 1)	 & $\llap{-}285485916336$ 	 & (10, 8, 8, 2, 0)	 & $\llap{-}368134868160$
\tabularnewline[-4pt] 
 (10, 9, 3, 3, 3)	 & $439198792163267520$ 	 & (10, 9, 4, 3, 2)	 & $122281608111146832$
\tabularnewline[-4pt] 
 (10, 9, 4, 4, 1)	 & $784225690196688$ 	 & (10, 9, 5, 2, 2)	 & $12238194838809408$
\tabularnewline[-4pt] 
 (10, 9, 5, 3, 1)	 & $280061917894368$ 	 & (10, 9, 5, 4, 0)	 & $\llap{-}62415555336480$
\tabularnewline[-4pt] 
 (10, 9, 6, 2, 1)	 & $7932337626384$ 	 & (10, 9, 6, 3, 0)	 & $\llap{-}8738280013680$
\tabularnewline[-4pt] 
 (10, 9, 7, 1, 1)	 & $\llap{-}83145266112$ 	 & (10, 9, 7, 2, 0)	 & $\llap{-}119442727776$
\tabularnewline[-4pt] 
 (10, 9, 8, 1, 0)	 & $\llap{-}46049040$ 	 & (10, 10, 3, 3, 2)	 & $294058742512224$
\tabularnewline[-4pt] 
 (10, 10, 4, 2, 2)	 & $76906534439280$ 	 & (10, 10, 4, 3, 1)	 & $1954674541824$
\tabularnewline[-4pt] 
 (10, 10, 4, 4, 0)	 & $\llap{-}368134868160$ 	 & (10, 10, 5, 2, 1)	 & $149497953456$
\tabularnewline[-4pt] 
 (10, 10, 5, 3, 0)	 & $\llap{-}119442727776$ 	 & (10, 10, 6, 1, 1)	 & $\llap{-}1334560224$
\tabularnewline[-4pt] 
 (10, 10, 6, 2, 0)	 & $\llap{-}3110590260$ 	 & (10, 10, 7, 1, 0)	 & $\llap{-}1679040$
\tabularnewline[-4pt] 
 (11, 5, 4, 4, 4)	 & $309132463141878510000$ 	 & (11, 5, 5, 4, 3)	 & $117552717975524482368$
\tabularnewline[-4pt] 
 (11, 5, 5, 5, 2)	 & $12636043517074729152$ 	 & (11, 6, 4, 4, 3)	 & $48619672420408782672$
\tabularnewline[-4pt] 
 (11, 6, 5, 3, 3)	 & $18293419124197908384$ 	 & (11, 6, 5, 4, 2)	 & $5175740190127316160$
\tabularnewline[-4pt] 
 (11, 6, 5, 5, 1)	 & $12132995150259168$ 	 & (11, 6, 6, 3, 2)	 & $777895843534310448$
\tabularnewline[-4pt] 
 (11, 6, 6, 4, 1)	 & $4906985498846880$ 	 & (11, 6, 6, 5, 0)	 & $\llap{-}405156007308576$
\tabularnewline[-4pt] 
 (11, 7, 4, 3, 3)	 & $3057298493530985568$ 	 & (11, 7, 4, 4, 2)	 & $857299960689895392$
\tabularnewline[-4pt] 
 (11, 7, 5, 3, 2)	 & $314475153857119104$ 	 & (11, 7, 5, 4, 1)	 & $2001252354617280$
\tabularnewline[-4pt] 
 (11, 7, 5, 5, 0)	 & $\llap{-}160791639748800$ 	 & (11, 7, 6, 2, 2)	 & $12276819589743504$
\tabularnewline[-4pt] 
 (11, 7, 6, 3, 1)	 & $281059260470880$ 	 & (11, 7, 6, 4, 0)	 & $\llap{-}62415555336480$
\tabularnewline[-4pt] 
 (11, 7, 7, 2, 1)	 & $3117411980160$ 	 & (11, 7, 7, 3, 0)	 & $\llap{-}3138370134624$
\tabularnewline[-4pt] 
 (11, 8, 3, 3, 3)	 & $67323025901888832$ 	 & (11, 8, 4, 3, 2)	 & $18421518301369920$
\tabularnewline[-4pt] 
 (11, 8, 4, 4, 1)	 & $120114507109632$ 	 & (11, 8, 5, 2, 2)	 & $1749317561578944$
\tabularnewline[-4pt] 
 (11, 8, 5, 3, 1)	 & $41786034116160$ 	 & (11, 8, 5, 4, 0)	 & $\llap{-}8738280013680$
\tabularnewline[-4pt] 
 (11, 8, 6, 2, 1)	 & $1179552933984$ 	 & (11, 8, 6, 3, 0)	 & $\llap{-}1093125957120$
\tabularnewline[-4pt] 
 (11, 8, 7, 1, 1)	 & $\llap{-}5776067616$ 	 & (11, 8, 7, 2, 0)	 & $\llap{-}11043084816$
\tabularnewline[-4pt] 
 (11, 8, 8, 1, 0)	 & $\llap{-}1679040$ 	 & (11, 9, 3, 3, 2)	 & $99761061359136$
\tabularnewline[-4pt] 
 (11, 9, 4, 2, 2)	 & $25630803734064$ 	 & (11, 9, 4, 3, 1)	 & $665062141248$
\tabularnewline[-4pt] 
 (11, 9, 4, 4, 0)	 & $\llap{-}119442727776$ 	 & (11, 9, 5, 2, 1)	 & $49806889344$
\tabularnewline[-4pt] 
 (11, 9, 5, 3, 0)	 & $\llap{-}37176746592$ 	 & (11, 9, 6, 1, 1)	 & $\llap{-}273996960$
\tabularnewline[-4pt] 
 (11, 9, 6, 2, 0)	 & $\llap{-}824199120$ 	 & (11, 9, 7, 1, 0)	 & $\llap{-}251520$
\tabularnewline[-4pt] 
 (11, 10, 3, 2, 2)	 & $14105356368$ 	 & (11, 10, 3, 3, 1)	 & $411633120$
\tabularnewline[-4pt] 
 (11, 10, 4, 2, 1)	 & $86694528$ 	 & (11, 10, 4, 3, 0)	 & $\llap{-}46049040$
\tabularnewline[-4pt] 
 (11, 10, 5, 1, 1)	 & $73824$ 	 & (11, 10, 5, 2, 0)	 & $\llap{-}1679040$
\tabularnewline[-4pt] 
 (11, 10, 6, 1, 0)	 & $\llap{-}48$ 	 & (11, 11, 2, 2, 2)	 & $528$
\tabularnewline[-4pt] 
 (12, 4, 4, 4, 4)	 & $597237294763420872$ 	 & (12, 5, 4, 4, 3)	 & $211913083229294304$
\tabularnewline[-4pt] 
 (12, 5, 5, 3, 3)	 & $74503089764268384$ 	 & (12, 5, 5, 4, 2)	 & $20254395759934128$
\tabularnewline[-4pt] 
 (12, 5, 5, 5, 1)	 & $46131775979616$ 	 & (12, 6, 4, 3, 3)	 & $27569906261747088$
\tabularnewline[-4pt] 
 (12, 6, 4, 4, 2)	 & $7431936473155680$ 	 & (12, 6, 5, 3, 2)	 & $2530955656217280$
\tabularnewline[-4pt] 
 (12, 6, 5, 4, 1)	 & $16731712682064$ 	 & (12, 6, 5, 5, 0)	 & $\llap{-}1093125957120$
\tabularnewline[-4pt] 
 (12, 6, 6, 2, 2)	 & $76906534439280$ 	 & (12, 6, 6, 3, 1)	 & $1954674541824$
\tabularnewline[-4pt] 
 (12, 6, 6, 4, 0)	 & $\llap{-}368134868160$ 	 & (12, 7, 3, 3, 3)	 & $1190848151512512$
\tabularnewline[-4pt] 
 (12, 7, 4, 3, 2)	 & $310831260169488$ 	 & (12, 7, 4, 4, 1)	 & $2072265197088$
\tabularnewline[-4pt] 
 (12, 7, 5, 2, 2)	 & $25630803734064$ 	 & (12, 7, 5, 3, 1)	 & $665062141248$
\tabularnewline[-4pt] 
 (12, 7, 5, 4, 0)	 & $\llap{-}119442727776$ 	 & (12, 7, 6, 2, 1)	 & $15766948032$
\tabularnewline[-4pt] 
 (12, 7, 6, 3, 0)	 & $\llap{-}11043084816$ 	 & (12, 7, 7, 1, 1)	 & $\llap{-}6126048$
\tabularnewline[-4pt] 
 (12, 7, 7, 2, 0)	 & $\llap{-}46049040$ 	 & (12, 8, 3, 3, 2)	 & $3154648420512$
\tabularnewline[-4pt] 
 (12, 8, 4, 2, 2)	 & $755118268080$ 	 & (12, 8, 4, 3, 1)	 & $20875131744$
\tabularnewline[-4pt] 
 (12, 8, 4, 4, 0)	 & $\llap{-}3110590260$ 	 & (12, 8, 5, 2, 1)	 & $1342319904$
\tabularnewline[-4pt] 
 (12, 8, 5, 3, 0)	 & $\llap{-}824199120$ 	 & (12, 8, 6, 1, 1)	 & $\llap{-}317232$
\tabularnewline[-4pt] 
 (12, 8, 6, 2, 0)	 & $\llap{-}9396672$ 	 & (12, 8, 7, 1, 0)	 & $\llap{-}48$
\tabularnewline[-4pt] 
 (12, 9, 3, 2, 2)	 & $652777584$ 	 & (12, 9, 3, 3, 1)	 & $19494816$
\tabularnewline[-4pt] 
 (12, 9, 4, 2, 1)	 & $3692400$ 	 & (12, 9, 4, 3, 0)	 & $\llap{-}1679040$
\tabularnewline[-4pt] 
 (12, 9, 5, 1, 1)	 & $4320$ 	 & (12, 9, 5, 2, 0)	 & $\llap{-}29136$
\tabularnewline[-4pt] 
 (12, 10, 2, 2, 2)	 & $\llap{-}2292$ 	 & (12, 10, 3, 2, 1)	 & $\llap{-}48$
\tabularnewline[-4pt] 
 (13, 4, 4, 4, 3)	 & $34155140507184$ 	 & (13, 5, 4, 3, 3)	 & $10348372749216$
\tabularnewline[-4pt] 
 (13, 5, 4, 4, 2)	 & $2545705442112$ 	 & (13, 5, 5, 3, 2)	 & $733831612704$
\tabularnewline[-4pt] 
 (13, 5, 5, 4, 1)	 & $4741984896$ 	 & (13, 5, 5, 5, 0)	 & $\llap{-}203310240$
\tabularnewline[-4pt] 
 (13, 6, 3, 3, 3)	 & $894337855968$ 	 & (13, 6, 4, 3, 2)	 & $208350582720$
\tabularnewline[-4pt] 
 (13, 6, 4, 4, 1)	 & $1342319904$ 	 & (13, 6, 5, 2, 2)	 & $12168742800$
\tabularnewline[-4pt] 
 (13, 6, 5, 3, 1)	 & $351706176$ 	 & (13, 6, 5, 4, 0)	 & $\llap{-}46049040$
\tabularnewline[-4pt] 
 (13, 6, 6, 2, 1)	 & $3692400$ 	 & (13, 6, 6, 3, 0)	 & $\llap{-}1679040$
\tabularnewline[-4pt] 
 (13, 7, 3, 3, 2)	 & $3347625888$ 	 & (13, 7, 4, 2, 2)	 & $652777584$
\tabularnewline[-4pt] 
 (13, 7, 4, 3, 1)	 & $19494816$ 	 & (13, 7, 4, 4, 0)	 & $\llap{-}1679040$
\tabularnewline[-4pt] 
 (13, 7, 5, 2, 1)	 & $679968$ 	 & (13, 7, 5, 3, 0)	 & $\llap{-}251520$
\tabularnewline[-4pt] 
 (13, 7, 6, 2, 0)	 & $\llap{-}48$ 	 & (13, 8, 3, 2, 2)	 & $104352$
\tabularnewline[-4pt] 
 (13, 8, 3, 3, 1)	 & $4320$ 	 & (13, 8, 4, 2, 1)	 & $528$
\tabularnewline[-4pt] 
 (13, 8, 4, 3, 0)	 & $\llap{-}48$ 	 & (14, 4, 4, 3, 3)	 & $\llap{-}46049040$
\tabularnewline[-4pt] 
 (14, 4, 4, 4, 2)	 & $\llap{-}9396672$ 	 & (14, 5, 3, 3, 3)	 & $\llap{-}9395616$
\tabularnewline[-4pt] 
 (14, 5, 4, 3, 2)	 & $\llap{-}1679040$ 	 & (14, 5, 4, 4, 1)	 & $\llap{-}29136$
\tabularnewline[-4pt] 
 (14, 5, 5, 2, 2)	 & $\llap{-}29136$ 	 & (14, 5, 5, 3, 1)	 & $\llap{-}2208$
\tabularnewline[-4pt] 
 (14, 6, 3, 3, 2)	 & $\llap{-}29136$ 	 & (14, 6, 4, 2, 2)	 & $\llap{-}2292$
\tabularnewline[-4pt] 
(14, 6, 4, 3, 1)	 & $\llap{-}48$	 &  	 &  
\tablepostambleInstantonNumbers \vskip-18pt 

\tablepreambleInstantonNumbersGone{29}
 (6, 6, 6, 6, 5)	 & $4210260616381832305777068000$ 	 & (7, 6, 6, 5, 5)	 & $2107392483854579518521492384$
\tabularnewline[-4pt] 
 (7, 6, 6, 6, 4)	 & $993268264214251658975065920$ 	 & (7, 7, 5, 5, 5)	 & $1050999213203945530932154368$
\tabularnewline[-4pt] 
 (7, 7, 6, 5, 4)	 & $493984816866806316213837888$ 	 & (7, 7, 6, 6, 3)	 & $95352673153203400660517280$
\tabularnewline[-4pt] 
 (7, 7, 7, 4, 4)	 & $114424618273312892366520384$ 	 & (7, 7, 7, 5, 3)	 & $47005257591061770949294080$
\tabularnewline[-4pt] 
 (7, 7, 7, 6, 2)	 & $2642910651915844284490560$ 	 & (7, 7, 7, 7, 1)	 & $1327111985411944965120$
\tabularnewline[-4pt] 
 (8, 6, 5, 5, 5)	 & $534167980184363652023678880$ 	 & (8, 6, 6, 5, 4)	 & $250174029024152132612941872$
\tabularnewline[-4pt] 
 (8, 6, 6, 6, 3)	 & $48000282449173462858848096$ 	 & (8, 7, 5, 5, 4)	 & $123320171802527109067861152$
\tabularnewline[-4pt] 
 (8, 7, 6, 4, 4)	 & $57440210392521975531302352$ 	 & (8, 7, 6, 5, 3)	 & $23538032787827317440430848$
\tabularnewline[-4pt] 
 (8, 7, 6, 6, 2)	 & $1313382488163858590583360$ 	 & (8, 7, 7, 4, 3)	 & $5310824139950279186413056$
\tabularnewline[-4pt] 
 (8, 7, 7, 5, 2)	 & $636966041838597182072352$ 	 & (8, 7, 7, 6, 1)	 & $655738370875420162368$
\tabularnewline[-4pt] 
 (8, 7, 7, 7, 0)	 & $\llap{-}15002076051597343200$ 	 & (8, 8, 5, 4, 4)	 & $13937339478667000119236400$
\tabularnewline[-4pt] 
 (8, 8, 5, 5, 3)	 & $5683140308380169857515456$ 	 & (8, 8, 6, 4, 3)	 & $2614878610961089413438336$
\tabularnewline[-4pt] 
 (8, 8, 6, 5, 2)	 & $312334294647361791229536$ 	 & (8, 8, 6, 6, 1)	 & $321538657422443256528$
\tabularnewline[-4pt] 
 (8, 8, 7, 3, 3)	 & $232464084388228505851296$ 	 & (8, 8, 7, 4, 2)	 & $68216655150306510669456$
\tabularnewline[-4pt] 
 (8, 8, 7, 5, 1)	 & $153930935927106810240$ 	 & (8, 8, 7, 6, 0)	 & $\llap{-}7217177104026255792$
\tabularnewline[-4pt] 
 (8, 8, 8, 3, 2)	 & $2772036316581728759040$ 	 & (8, 8, 8, 4, 1)	 & $15670595020056748992$
\tabularnewline[-4pt] 
 (8, 8, 8, 5, 0)	 & $\llap{-}1616941505273075616$ 	 & (9, 5, 5, 5, 5)	 & $66936916351993885652216832$
\tabularnewline[-4pt] 
 (9, 6, 5, 5, 4)	 & $30986533923874230605211360$ 	 & (9, 6, 6, 4, 4)	 & $14305111033066829808964272$
\tabularnewline[-4pt] 
 (9, 6, 6, 5, 3)	 & $5827961507523161595876576$ 	 & (9, 6, 6, 6, 2)	 & $319351886711731639092048$
\tabularnewline[-4pt] 
 (9, 7, 5, 4, 4)	 & $6913304732636292162876864$ 	 & (9, 7, 5, 5, 3)	 & $2809517795490346324727808$
\tabularnewline[-4pt] 
 (9, 7, 6, 4, 3)	 & $1285699665017929239033504$ 	 & (9, 7, 6, 5, 2)	 & $152785612691569068751872$
\tabularnewline[-4pt] 
 (9, 7, 6, 6, 1)	 & $157174083455108555904$ 	 & (9, 7, 7, 3, 3)	 & $112332804021170119948800$
\tabularnewline[-4pt] 
 (9, 7, 7, 4, 2)	 & $32906148327728452809888$ 	 & (9, 7, 7, 5, 1)	 & $74648911275455548416$
\tabularnewline[-4pt] 
 (9, 7, 7, 6, 0)	 & $\llap{-}3435204329200397376$ 	 & (9, 8, 4, 4, 4)	 & $733085308674475362988800$
\tabularnewline[-4pt] 
 (9, 8, 5, 4, 3)	 & $295443957082726234710912$ 	 & (9, 8, 5, 5, 2)	 & $34790520976213893550464$
\tabularnewline[-4pt] 
 (9, 8, 6, 3, 3)	 & $53548008570062637417696$ 	 & (9, 8, 6, 4, 2)	 & $15659653975155880831536$
\tabularnewline[-4pt] 
 (9, 8, 6, 5, 1)	 & $35705455486282123392$ 	 & (9, 8, 6, 6, 0)	 & $\llap{-}1616941505273075616$
\tabularnewline[-4pt] 
 (9, 8, 7, 3, 2)	 & $1296069103156616952768$ 	 & (9, 8, 7, 4, 1)	 & $7411795886398986624$
\tabularnewline[-4pt] 
 (9, 8, 7, 5, 0)	 & $\llap{-}752251254614400960$ 	 & (9, 8, 8, 2, 2)	 & $13404209975769169392$
\tabularnewline[-4pt] 
 (9, 8, 8, 3, 1)	 & $254920513927767648$ 	 & (9, 8, 8, 4, 0)	 & $\llap{-}70199768003592720$
\tabularnewline[-4pt] 
 (9, 9, 4, 4, 3)	 & $13740363716226750753600$ 	 & (9, 9, 5, 3, 3)	 & $5442063622506471346176$
\tabularnewline[-4pt] 
 (9, 9, 5, 4, 2)	 & $1581805356183712228896$ 	 & (9, 9, 5, 5, 1)	 & $3655384754255841792$
\tabularnewline[-4pt] 
 (9, 9, 6, 3, 2)	 & $273179253014427444000$ 	 & (9, 9, 6, 4, 1)	 & $1598269886739062400$
\tabularnewline[-4pt] 
 (9, 9, 6, 5, 0)	 & $\llap{-}156843054827785632$ 	 & (9, 9, 7, 2, 2)	 & $5921797441909136064$
\tabularnewline[-4pt] 
 (9, 9, 7, 3, 1)	 & $115413671534048256$ 	 & (9, 9, 7, 4, 0)	 & $\llap{-}30974226462689184$
\tabularnewline[-4pt] 
 (9, 9, 8, 2, 1)	 & $486039827347776$ 	 & (9, 9, 8, 3, 0)	 & $\llap{-}1000067627051904$
\tabularnewline[-4pt] 
 (9, 9, 9, 1, 1)	 & $\llap{-}2891432289792$ 	 & (9, 9, 9, 2, 0)	 & $\llap{-}3138370134624$
\tabularnewline[-4pt] 
 (10, 5, 5, 5, 4)	 & $1771748988189311827248096$ 	 & (10, 6, 5, 4, 4)	 & $801414379844189789048640$
\tabularnewline[-4pt] 
 (10, 6, 5, 5, 3)	 & $322017472455505977946944$ 	 & (10, 6, 6, 4, 3)	 & $144746838738025811348976$
\tabularnewline[-4pt] 
 (10, 6, 6, 5, 2)	 & $16899949831877607148848$ 	 & (10, 6, 6, 6, 1)	 & $17294122346233404288$
\tabularnewline[-4pt] 
 (10, 7, 4, 4, 4)	 & $168965147904886287828768$ 	 & (10, 7, 5, 4, 3)	 & $67481808424198752531456$
\tabularnewline[-4pt] 
 (10, 7, 5, 5, 2)	 & $7838457970390311440448$ 	 & (10, 7, 6, 3, 3)	 & $11940345167441877911232$
\tabularnewline[-4pt] 
 (10, 7, 6, 4, 2)	 & $3475713613337883829632$ 	 & (10, 7, 6, 5, 1)	 & $8000113898763198720$
\tabularnewline[-4pt] 
 (10, 7, 6, 6, 0)	 & $\llap{-}345708325307878560$ 	 & (10, 7, 7, 3, 2)	 & $274043841200077763232$
\tabularnewline[-4pt] 
 (10, 7, 7, 4, 1)	 & $1603689039494806080$ 	 & (10, 7, 7, 5, 0)	 & $\llap{-}156843054827785632$
\tabularnewline[-4pt] 
 (10, 8, 4, 4, 3)	 & $6366888695682697842144$ 	 & (10, 8, 5, 3, 3)	 & $2506621507185094907424$
\tabularnewline[-4pt] 
 (10, 8, 5, 4, 2)	 & $726440453016521093136$ 	 & (10, 8, 5, 5, 1)	 & $1685025369541284480$
\tabularnewline[-4pt] 
 (10, 8, 6, 3, 2)	 & $123424186729923646080$ 	 & (10, 8, 6, 4, 1)	 & $730324510146863712$
\tabularnewline[-4pt] 
 (10, 8, 6, 5, 0)	 & $\llap{-}70199768003592720$ 	 & (10, 8, 7, 2, 2)	 & $2579059445631070176$
\tabularnewline[-4pt] 
 (10, 8, 7, 3, 1)	 & $51494523118178400$ 	 & (10, 8, 7, 4, 0)	 & $\llap{-}13461969999093600$
\tabularnewline[-4pt] 
 (10, 8, 8, 2, 1)	 & $230627260020528$ 	 & (10, 8, 8, 3, 0)	 & $\llap{-}405156007308576$
\tabularnewline[-4pt] 
 (10, 9, 4, 3, 3)	 & $94786453063647970848$ 	 & (10, 9, 4, 4, 2)	 & $27149794915803480576$
\tabularnewline[-4pt] 
 (10, 9, 5, 3, 2)	 & $10388937656332753536$ 	 & (10, 9, 5, 4, 1)	 & $63502596678612960$
\tabularnewline[-4pt] 
 (10, 9, 5, 5, 0)	 & $\llap{-}5758034709276000$ 	 & (10, 9, 6, 2, 2)	 & $465966091929471216$
\tabularnewline[-4pt] 
 (10, 9, 6, 3, 1)	 & $9751729607816928$ 	 & (10, 9, 6, 4, 0)	 & $\llap{-}2421429008571216$
\tabularnewline[-4pt] 
 (10, 9, 7, 2, 1)	 & $105019516952544$ 	 & (10, 9, 7, 3, 0)	 & $\llap{-}160791639748800$
\tabularnewline[-4pt] 
 (10, 9, 8, 1, 1)	 & $\llap{-}929847901440$ 	 & (10, 9, 8, 2, 0)	 & $\llap{-}1093125957120$
\tabularnewline[-4pt] 
 (10, 9, 9, 1, 0)	 & $\llap{-}203310240$ 	 & (10, 10, 3, 3, 3)	 & $439198792163267520$
\tabularnewline[-4pt] 
 (10, 10, 4, 3, 2)	 & $122281608111146832$ 	 & (10, 10, 4, 4, 1)	 & $784225690196688$
\tabularnewline[-4pt] 
 (10, 10, 5, 2, 2)	 & $12238194838809408$ 	 & (10, 10, 5, 3, 1)	 & $280061917894368$
\tabularnewline[-4pt] 
 (10, 10, 5, 4, 0)	 & $\llap{-}62415555336480$ 	 & (10, 10, 6, 2, 1)	 & $7932337626384$
\tabularnewline[-4pt] 
 (10, 10, 6, 3, 0)	 & $\llap{-}8738280013680$ 	 & (10, 10, 7, 1, 1)	 & $\llap{-}83145266112$
\tabularnewline[-4pt] 
 (10, 10, 7, 2, 0)	 & $\llap{-}119442727776$ 	 & (10, 10, 8, 1, 0)	 & $\llap{-}46049040$
\tabularnewline[-4pt] 
 (11, 5, 5, 4, 4)	 & $18266693407737182177184$ 	 & (11, 5, 5, 5, 3)	 & $7166819934960635255808$
\tabularnewline[-4pt] 
 (11, 6, 4, 4, 4)	 & $7947181852777331886432$ 	 & (11, 6, 5, 4, 3)	 & $3105932030883925870176$
\tabularnewline[-4pt] 
 (11, 6, 5, 5, 2)	 & $348836588057363657472$ 	 & (11, 6, 6, 3, 3)	 & $519156967290582783648$
\tabularnewline[-4pt] 
 (11, 6, 6, 4, 2)	 & $149323799470083244320$ 	 & (11, 6, 6, 5, 1)	 & $348635537576206944$
\tabularnewline[-4pt] 
 (11, 6, 6, 6, 0)	 & $\llap{-}13461969999093600$ 	 & (11, 7, 4, 4, 3)	 & $590236080316778614944$
\tabularnewline[-4pt] 
 (11, 7, 5, 3, 3)	 & $227671001933190288384$ 	 & (11, 7, 5, 4, 2)	 & $65291193554279696544$
\tabularnewline[-4pt] 
 (11, 7, 5, 5, 1)	 & $152724917794335744$ 	 & (11, 7, 6, 3, 2)	 & $10495212981639920256$
\tabularnewline[-4pt] 
 (11, 7, 6, 4, 1)	 & $64196301394125984$ 	 & (11, 7, 6, 5, 0)	 & $\llap{-}5758034709276000$
\tabularnewline[-4pt] 
 (11, 7, 7, 2, 2)	 & $193409949629545344$ 	 & (11, 7, 7, 3, 1)	 & $4142934152742912$
\tabularnewline[-4pt] 
 (11, 7, 7, 4, 0)	 & $\llap{-}1000067627051904$ 	 & (11, 8, 4, 3, 3)	 & $17566723688872737408$
\tabularnewline[-4pt] 
 (11, 8, 4, 4, 2)	 & $4983325664622060480$ 	 & (11, 8, 5, 3, 2)	 & $1869909171436016640$
\tabularnewline[-4pt] 
 (11, 8, 5, 4, 1)	 & $11668379661795360$ 	 & (11, 8, 5, 5, 0)	 & $\llap{-}1000067627051904$
\tabularnewline[-4pt] 
 (11, 8, 6, 2, 2)	 & $78667271480859264$ 	 & (11, 8, 6, 3, 1)	 & $1722815728418688$
\tabularnewline[-4pt] 
 (11, 8, 6, 4, 0)	 & $\llap{-}405156007308576$ 	 & (11, 8, 7, 2, 1)	 & $19460158766688$
\tabularnewline[-4pt] 
 (11, 8, 7, 3, 0)	 & $\llap{-}23657221999872$ 	 & (11, 8, 8, 1, 1)	 & $\llap{-}83145266112$
\tabularnewline[-4pt] 
 (11, 8, 8, 2, 0)	 & $\llap{-}119442727776$ 	 & (11, 9, 3, 3, 3)	 & $173735334568551936$
\tabularnewline[-4pt] 
 (11, 9, 4, 3, 2)	 & $47974716780673152$ 	 & (11, 9, 4, 4, 1)	 & $310320577356576$
\tabularnewline[-4pt] 
 (11, 9, 5, 2, 2)	 & $4682747289739200$ 	 & (11, 9, 5, 3, 1)	 & $109488862752768$
\tabularnewline[-4pt] 
 (11, 9, 5, 4, 0)	 & $\llap{-}23657221999872$ 	 & (11, 9, 6, 2, 1)	 & $3117411980160$
\tabularnewline[-4pt] 
 (11, 9, 6, 3, 0)	 & $\llap{-}3138370134624$ 	 & (11, 9, 7, 1, 1)	 & $\llap{-}22739284992$
\tabularnewline[-4pt] 
 (11, 9, 7, 2, 0)	 & $\llap{-}37176746592$ 	 & (11, 9, 8, 1, 0)	 & $\llap{-}9395616$
\tabularnewline[-4pt] 
 (11, 10, 3, 3, 2)	 & $99761061359136$ 	 & (11, 10, 4, 2, 2)	 & $25630803734064$
\tabularnewline[-4pt] 
 (11, 10, 4, 3, 1)	 & $665062141248$ 	 & (11, 10, 4, 4, 0)	 & $\llap{-}119442727776$
\tabularnewline[-4pt] 
 (11, 10, 5, 2, 1)	 & $49806889344$ 	 & (11, 10, 5, 3, 0)	 & $\llap{-}37176746592$
\tabularnewline[-4pt] 
 (11, 10, 6, 1, 1)	 & $\llap{-}273996960$ 	 & (11, 10, 6, 2, 0)	 & $\llap{-}824199120$
\tabularnewline[-4pt] 
 (11, 10, 7, 1, 0)	 & $\llap{-}251520$ 	 & (11, 11, 3, 2, 2)	 & $3347625888$
\tabularnewline[-4pt] 
 (11, 11, 3, 3, 1)	 & $99761664$ 	 & (11, 11, 4, 2, 1)	 & $19494816$
\tabularnewline[-4pt] 
 (11, 11, 4, 3, 0)	 & $\llap{-}9395616$ 	 & (11, 11, 5, 1, 1)	 & $30720$
\tabularnewline[-4pt] 
 (11, 11, 5, 2, 0)	 & $\llap{-}251520$ 	 & (12, 5, 4, 4, 4)	 & $56439747241501122192$
\tabularnewline[-4pt] 
 (12, 5, 5, 4, 3)	 & $21104661093843211008$ 	 & (12, 5, 5, 5, 2)	 & $2206547301748229184$
\tabularnewline[-4pt] 
 (12, 6, 4, 4, 3)	 & $8521174156401735360$ 	 & (12, 6, 5, 3, 3)	 & $3144478089605250720$
\tabularnewline[-4pt] 
 (12, 6, 5, 4, 2)	 & $879997438952410320$ 	 & (12, 6, 5, 5, 1)	 & $2055320972401920$
\tabularnewline[-4pt] 
 (12, 6, 6, 3, 2)	 & $125469245902996512$ 	 & (12, 6, 6, 4, 1)	 & $805991883499728$
\tabularnewline[-4pt] 
 (12, 6, 6, 5, 0)	 & $\llap{-}62415555336480$ 	 & (12, 7, 4, 3, 3)	 & $494845153306899264$
\tabularnewline[-4pt] 
 (12, 7, 4, 4, 2)	 & $136879891485939456$ 	 & (12, 7, 5, 3, 2)	 & $48909204818311680$
\tabularnewline[-4pt] 
 (12, 7, 5, 4, 1)	 & $316717033197408$ 	 & (12, 7, 5, 5, 0)	 & $\llap{-}23657221999872$
\tabularnewline[-4pt] 
 (12, 7, 6, 2, 2)	 & $1749317561578944$ 	 & (12, 7, 6, 3, 1)	 & $41786034116160$
\tabularnewline[-4pt] 
 (12, 7, 6, 4, 0)	 & $\llap{-}8738280013680$ 	 & (12, 7, 7, 2, 1)	 & $428753909184$
\tabularnewline[-4pt] 
 (12, 7, 7, 3, 0)	 & $\llap{-}368134832160$ 	 & (12, 8, 3, 3, 3)	 & $9449678610817056$
\tabularnewline[-4pt] 
 (12, 8, 4, 3, 2)	 & $2530955656217280$ 	 & (12, 8, 4, 4, 1)	 & $16731712682064$
\tabularnewline[-4pt] 
 (12, 8, 5, 2, 2)	 & $225463832566752$ 	 & (12, 8, 5, 3, 1)	 & $5619611350656$
\tabularnewline[-4pt] 
 (12, 8, 5, 4, 0)	 & $\llap{-}1093125957120$ 	 & (12, 8, 6, 2, 1)	 & $149497953456$
\tabularnewline[-4pt] 
 (12, 8, 6, 3, 0)	 & $\llap{-}119442727776$ 	 & (12, 8, 7, 1, 1)	 & $\llap{-}273996960$
\tabularnewline[-4pt] 
 (12, 8, 7, 2, 0)	 & $\llap{-}824199120$ 	 & (12, 8, 8, 1, 0)	 & $\llap{-}29136$
\tabularnewline[-4pt] 
 (12, 9, 3, 3, 2)	 & $10348372749216$ 	 & (12, 9, 4, 2, 2)	 & $2545705442112$
\tabularnewline[-4pt] 
 (12, 9, 4, 3, 1)	 & $68863079616$ 	 & (12, 9, 4, 4, 0)	 & $\llap{-}11043084816$
\tabularnewline[-4pt] 
 (12, 9, 5, 2, 1)	 & $4741984896$ 	 & (12, 9, 5, 3, 0)	 & $\llap{-}3110582880$
\tabularnewline[-4pt] 
 (12, 9, 6, 1, 1)	 & $\llap{-}6126048$ 	 & (12, 9, 6, 2, 0)	 & $\llap{-}46049040$
\tabularnewline[-4pt] 
 (12, 9, 7, 1, 0)	 & $\llap{-}2208$ 	 & (12, 10, 3, 2, 2)	 & $652777584$
\tabularnewline[-4pt] 
 (12, 10, 3, 3, 1)	 & $19494816$ 	 & (12, 10, 4, 2, 1)	 & $3692400$
\tabularnewline[-4pt] 
 (12, 10, 4, 3, 0)	 & $\llap{-}1679040$ 	 & (12, 10, 5, 1, 1)	 & $4320$
\tabularnewline[-4pt] 
 (12, 10, 5, 2, 0)	 & $\llap{-}29136$ 	 & (12, 11, 2, 2, 2)	 & $\llap{-}48$
\tabularnewline[-4pt] 
 (13, 4, 4, 4, 4)	 & $29809312235610960$ 	 & (13, 5, 4, 4, 3)	 & $10159668608774304$
\tabularnewline[-4pt] 
 (13, 5, 5, 3, 3)	 & $3417190702574592$ 	 & (13, 5, 5, 4, 2)	 & $903625742797728$
\tabularnewline[-4pt] 
 (13, 5, 5, 5, 1)	 & $1974181959168$ 	 & (13, 6, 4, 3, 3)	 & $1190848151512512$
\tabularnewline[-4pt] 
 (13, 6, 4, 4, 2)	 & $310831260169488$ 	 & (13, 6, 5, 3, 2)	 & $99761061359136$
\tabularnewline[-4pt] 
 (13, 6, 5, 4, 1)	 & $665062141248$ 	 & (13, 6, 5, 5, 0)	 & $\llap{-}37176746592$
\tabularnewline[-4pt] 
 (13, 6, 6, 2, 2)	 & $2461712752416$ 	 & (13, 6, 6, 3, 1)	 & $66382892544$
\tabularnewline[-4pt] 
 (13, 6, 6, 4, 0)	 & $\llap{-}11043084816$ 	 & (13, 7, 3, 3, 3)	 & $41704406393856$
\tabularnewline[-4pt] 
 (13, 7, 4, 3, 2)	 & $10348372749216$ 	 & (13, 7, 4, 4, 1)	 & $68863079616$
\tabularnewline[-4pt] 
 (13, 7, 5, 2, 2)	 & $733831612704$ 	 & (13, 7, 5, 3, 1)	 & $20194851840$
\tabularnewline[-4pt] 
 (13, 7, 5, 4, 0)	 & $\llap{-}3110582880$ 	 & (13, 7, 6, 2, 1)	 & $351706176$
\tabularnewline[-4pt] 
 (13, 7, 6, 3, 0)	 & $\llap{-}203310240$ 	 & (13, 7, 7, 1, 1)	 & $30720$
\tabularnewline[-4pt] 
 (13, 7, 7, 2, 0)	 & $\llap{-}251520$ 	 & (13, 8, 3, 3, 2)	 & $65707393920$
\tabularnewline[-4pt] 
 (13, 8, 4, 2, 2)	 & $14105356368$ 	 & (13, 8, 4, 3, 1)	 & $411633120$
\tabularnewline[-4pt] 
 (13, 8, 4, 4, 0)	 & $\llap{-}46049040$ 	 & (13, 8, 5, 2, 1)	 & $19494816$
\tabularnewline[-4pt] 
 (13, 8, 5, 3, 0)	 & $\llap{-}9395616$ 	 & (13, 8, 6, 1, 1)	 & $4320$
\tabularnewline[-4pt] 
 (13, 8, 6, 2, 0)	 & $\llap{-}29136$ 	 & (13, 9, 3, 2, 2)	 & $679968$
\tabularnewline[-4pt] 
 (13, 9, 3, 3, 1)	 & $30720$ 	 & (13, 9, 4, 2, 1)	 & $4320$
\tabularnewline[-4pt] 
 (13, 9, 4, 3, 0)	 & $\llap{-}2208$ 	 & (14, 4, 4, 4, 3)	 & $20875131744$
\tabularnewline[-4pt] 
 (14, 5, 4, 3, 3)	 & $5848333440$ 	 & (14, 5, 4, 4, 2)	 & $1342319904$
\tabularnewline[-4pt] 
 (14, 5, 5, 3, 2)	 & $351706176$ 	 & (14, 5, 5, 4, 1)	 & $\llap{-}6126048$
\tabularnewline[-4pt] 
 (14, 5, 5, 5, 0)	 & $\llap{-}251520$ 	 & (14, 6, 3, 3, 3)	 & $411633120$
\tabularnewline[-4pt] 
 (14, 6, 4, 3, 2)	 & $86694528$ 	 & (14, 6, 4, 4, 1)	 & $\llap{-}317232$
\tabularnewline[-4pt] 
 (14, 6, 5, 2, 2)	 & $3692400$ 	 & (14, 6, 5, 3, 1)	 & $73824$
\tabularnewline[-4pt] 
 (14, 6, 5, 4, 0)	 & $\llap{-}29136$ 	 & (14, 6, 6, 2, 1)	 & $528$
\tabularnewline[-4pt] 
 (14, 6, 6, 3, 0)	 & $\llap{-}48$ 	 & (14, 7, 3, 3, 2)	 & $679968$
\tabularnewline[-4pt] 
 (14, 7, 4, 2, 2)	 & $104352$ 	 & (14, 7, 4, 3, 1)	 & $4320$
\tabularnewline[-4pt] 
 (14, 7, 4, 4, 0)	 & $\llap{-}48$ 	 & (14, 8, 3, 2, 2)	 & $\llap{-}48$
\tablepostambleInstantonNumbers \vskip-18pt 
\newpage
\subsection{The list of webs}\label{sect:weblist}
\addtocounter{table}{-24}
\begin{table}[H]
\begin{center}
\capt{5.5in}{tab:webtab}{The positive webs, together with $\IW_{+}$, and the instanton numbers $n_W$ and $d_W$,  for source vectors up to degree 19. $\deg(W)$ denotes the degree of the source vector 
$\mathring{\bm{I}}$ of the corresponding web, and $h(W)$ is defined as $h(W)=h\big(\mathring{\bm{I}}\big)$.}
\end{center}
\end{table}
\vspace*{-25pt}
\begin{center}
\renewcommand*{\arraystretch}{1.06}
\begin{longtable}{|>{\hskip3pt}c<{\hskip-3pt}|c|c|>{\hspace*{-2pt}}l<{\hspace*{-2pt}}|>{\hspace*{-2pt}}l<{\hspace*{-2pt}}|}
% firsthead
\hline
\vrule height13pt depth5pt width0pt Web & deg$(W)$ & $h(W)$ & \hfil $n_W$ & \hfil $d_W$ \\
\hline\hline
&&&& \\[-10pt]
\endfirsthead

% general head
\multicolumn{5}{l}{\tablename\ \thetable{} -- continued} \\[3pt]
\hline 
\multicolumn{1}{|c|}{\vrule height 12pt depth5pt width0pt Web}  & \multicolumn{1}{c|}{deg$(W)$}  & \multicolumn{1}{c|}{$h(W)$} & \multicolumn{1}{c|}{$n_W$} & \multicolumn{1}{c|}{$d_W$} \\ 
\hline\hline 
&&&& \\[-10pt]
\endhead

% general foot
\hline\hline 
\multicolumn{5}{|r|}{{\vrule height 12pt depth5pt width0pt Continued on next page}} \\ 
\hline
\endfoot

% last foot
\hline \hline
\multicolumn{5}{|c|}{{\vrule height 12pt depth5pt width0pt}}\\
\hline
\endlastfoot

 {}\hskip-6.5pt$\IW_{+}$[(1,0,0,0,0)] & 1 & \+1& 24 & \+0 \\
 W[(1,1,1,0,0)] & 3 & \+0& 112 & \+0 \\
 W[(1,1,1,1,0)] & 4 & -2 & 1104 & \+0 \\
 W[(1,1,1,1,1)] & 5 & -5 & 19200 & \+0 \\
 W[(2,1,1,1,1)] & 6 & -6 & 45408 & \+0 \\
 W[(2,2,1,1,0)] & 6 & -3 & 2800 & \+0 \\
 W[(2,2,1,1,1)] & 7 & -8 & 212880 & \+0 \\
 W[(2,2,2,0,0)] & 6 & \+0& 80 & \+4 \\
\hline
\end{longtable}
\end{center}

\newpage
%%%%%%%%%%%%%%%%%%%%%%%%%%%%%%%%%%%%%%%%%
% Bibliography%
%%%%%%%%%%%%%%%%%%%%%%%%%%%%%%%%%%%%%%%%%
\newpage
\phantomsection
\addcontentsline{toc}{section}{References}
\bibliographystyle{JHEP}
\bibliography{Mirror_HV}
%%%%%%%%%%%%%%%%%%%%%%%%%%%%%%%%%%%%%%%%%
%%%%%%%%%%%%%%%%%%%%%%%%%%%%%%%%%%%%%%%%%

%%%%%%%%%%%%%%%%%%%%%%%%%%%%%%%%%%%%
%%%%%%%%%%%%%%%%%%%%%%%%%%%%%%%%%%%%
\end{document}